\documentclass[12pt]{report}

\usepackage{algorithm}
\usepackage{algpseudocode}
\usepackage{amsfonts}
\usepackage{amsmath}
\usepackage{amssymb}
\usepackage{amsthm}
\usepackage{authblk}
\usepackage[font=footnotesize,labelfont=bf]{caption}
\usepackage{enumitem}
\usepackage{epsfig}
\usepackage{gensymb}
\usepackage[margin=3.0cm]{geometry}
\usepackage{graphicx}
\usepackage[colorlinks=true, allcolors=blue]{hyperref}
\usepackage{latexsym}
\usepackage{lineno}
\usepackage{listings}
\usepackage{mathrsfs} 
\usepackage{placeins}
\usepackage{subcaption}
\usepackage[most]{tcolorbox}
\usepackage{textcomp}
\usepackage[colorinlistoftodos]{todonotes}
\usepackage{rotating}
\usepackage{xcolor}

\definecolor{codegreen}{rgb}{0,0.6,0}
\definecolor{codegray}{rgb}{0.5,0.5,0.5}
\definecolor{codepurple}{rgb}{0.58,0,0.82}
\definecolor{backcolour}{rgb}{0.95,0.95,0.92}
\definecolor{backterminal}{rgb}{0.0,0.0,0.0}
\definecolor{codeblue}{rgb}{0.13, 0.67, 0.8}
\definecolor{applegreen}{rgb}{0.24, 0.82, 0.44}
\definecolor{davysgrey}{rgb}{0.21, 0.21, 0.21}

\newcounter{codeCnt}[chapter]
\lstdefinestyle{PythonStyle}{
    backgroundcolor=\color{backcolour},
    commentstyle=\color{codegreen},
    frame=single,
    keywordstyle=\color{magenta},
    numberstyle=\tiny\color{codegray},
    stringstyle=\color{codepurple},
    basicstyle=\ttfamily\footnotesize,
    breakatwhitespace=false,
    breaklines=true,
    captionpos=b,
    numbers=left,                    
    numbersep=5pt
}
\lstnewenvironment{Code}[1][]{
  \setcounter{lstlisting}{\value{codeCnt}}
  \lstset{caption={#1}, label={#1}, style=PythonStyle, #1}
}{\addtocounter{codeCnt}{1}}

\newcounter{terminalCnt}[chapter]
\lstdefinestyle{TerminalStyle}{
    basicstyle=\linespread{1.1}\color{white}\ttfamily\small,
    captionpos=b,
    frame = none,
    literate={tilde} {{{\color{codeblue} $\sim$}}}1,
    classoffset=0,
    morekeywords={user@mypc},
    keywordstyle={\bf \color{applegreen}},
    breaklines=true
}
\lstnewenvironment{Terminal}[1][]{
  \setcounter{lstlisting}{\value{terminalCnt}}
  \lstset{language=bash, caption={#1}, label={#1}, escapechar=ä,numbers=none, framerule=0.0pt, style=TerminalStyle, #1}
}{\addtocounter{terminalCnt}{1}}

\title{\textbf{FLUIDITY} \\ Indoor-outdoor exchanges \\ Urban environment simulations \\ \small{Version 1.0.a}}

\author[a]{\textbf{Carolanne Vouriot}}
\author[b,c]{\textbf{Laetitia Mottet}}

\affil[a]{Dept. of Civil Engineering, Imperial College London, UK}
\affil[b]{Dept. of Earth Science $\&$ Engineering, Imperial College London, UK}
\affil[c]{Dept. of Architecture, Cambridge University, UK}

\date{February 14, 2019}

\begin{document}
    \maketitle
    
    \tableofcontents{}

    \chapter{Introduction}

\section{Purpose and aim of the document}
The aim of this document is to show how \textbf{Fluidity} can be used to set up typical indoor simulations under a range of conditions. Further information can also be found in the \textbf{Fluidity} manual~\cite{AMCG2015} and online \url{http://fluidityproject.github.io/}.

\noindent The purpose of this document is to allow a new \textbf{Fluidity} user to quickly become independent and start running simulations early on. Despite its length, it should be read carefully and followed step by step. Concrete examples will be used and developed along this document, all sections being equally important. In addition, the tricks and information given might be directly experience based and as such might not appear in the \textbf{Fluidity} manual~\cite{AMCG2015}.

\section{Getting and installing \textbf{Fluidity}} 
The following is more or less copied from the \textbf{Fluidity} manual~\cite{AMCG2015} and describes how to get and install \textbf{Fluidity}. The user is advised to refer to the manual~\cite{AMCG2015} for a detailed description. The following described how to install \textbf{Fluidity} on an Ubuntu machine (\textbf{Fluidity} does not work under Windows). When this manual was written, all the following were working properly up to Ubuntu 16.04. Some troubles happen on Ubuntu 18.04: the package \textit{fluidity} was not yet available, while \textit{fluidity-dev} was.

\subsection{Getting binary source of \textbf{Fluidity}: for user only} \label{Sec:FluidityBinary}
In that section, \textbf{Fluidity} will be installed on the computer but the installation will be transparent for the user, i.e. the code cannot be changed. Add the package archive to the system, update it and install \textbf{Fluidity} along with the required supporting software by typing:
\begin{Terminal}[caption={Getting and installing Fluidity from binary sources.}]
ä\colorbox{davysgrey}{
\parbox{435pt}{
\color{applegreen} \textbf{user@mypc}\color{white}\textbf{:}\color{codeblue}$\sim$
\color{white}\$ sudo apt-add-repository -y ppa:fluidity-core/ppa
\newline
\color{applegreen} \textbf{user@mypc}\color{white}\textbf{:}\color{codeblue}$\sim$
\color{white}\$ sudo apt-get update
\newline
\color{applegreen} \textbf{user@mypc}\color{white}\textbf{:}\color{codeblue}$\sim$
\color{white}\$ sudo apt-get -y install fluidity
}}
\end{Terminal}

\noindent \textbf{Fluidity} is now installed on the computer. Ready to be used. Typing \texttt{fluidity} in a terminal to ensure that it works.

\subsection{Getting source code of \textbf{Fluidity}: for developer}
To develop or locally build \textbf{Fluidity}, the \textit{fluidity-dev} package need to be install, which depends on all the other software required for building \textbf{Fluidity} (see Appendix C of the \textbf{Fluidity} manual~\cite{AMCG2015} for more details about all the other required software and libraries). \textbf{To minimise the number of potential errors and libraries to install manually, it is recommended to firstly install the `standard' \textbf{Fluidity} as explain in the section~\ref{Sec:FluidityBinary}}: this will automatically install lot of libraries that \textbf{Fluidity} needs.

\noindent Add the package archive to the system, update it and install the developer version of \textbf{Fluidity} by typing:
\begin{Terminal}[caption={Getting and installing Fluidity from sources.}]
ä\colorbox{davysgrey}{
\parbox{435pt}{
\color{applegreen} \textbf{user@mypc}\color{white}\textbf{:}\color{codeblue}$\sim$
\color{white}\$ sudo apt-add-repository -y ppa:fluidity-core/ppa
\newline
\color{applegreen} \textbf{user@mypc}\color{white}\textbf{:}\color{codeblue}$\sim$
\color{white}\$ sudo apt-get update
\newline
\color{applegreen} \textbf{user@mypc}\color{white}\textbf{:}\color{codeblue}$\sim$
\color{white}\$ sudo apt-get -y install fluidity-dev
}}
\end{Terminal}

\noindent If GitHub is not already install on your system, install it (first line of Command~\ref{Lst:GitHub}). Clone a copy of the latest correct and usable version of \textbf{Fluidity}, then change the right of the \texttt{fluidity} folder created by typing:
\begin{Terminal}[caption={GitHub.}, label={Lst:GitHub}]
ä\colorbox{davysgrey}{
\parbox{435pt}{
\color{applegreen} \textbf{user@mypc}\color{white}\textbf{:}\color{codeblue}$\sim$
\color{white}\$ sudo apt-get install github
\newline
\color{applegreen} \textbf{user@mypc}\color{white}\textbf{:}\color{codeblue}$\sim$
\color{white}\$ git clone https://github.com/FluidityProject/fluidity.git
\newline
\color{applegreen} \textbf{user@mypc}\color{white}\textbf{:}\color{codeblue}$\sim$
\color{white}\$ sudo chmod -R a+rwx fluidity/
}}
\end{Terminal}

\noindent The build process for \textbf{Fluidity} then comprises a configuration stage and a compile stage. Go in the directory containing your local source code (the \texttt{fluidity} folder downloaded by the GitHub command above), denoted \texttt{<<FluiditySourcePath>>} here, and run:
\begin{Terminal}[caption={Installing Fluidity by hand from source code.}, label={Lst:BuildFluidity}]
ä\colorbox{davysgrey}{
\parbox{435pt}{
\color{applegreen} \textbf{user@mypc}\color{white}\textbf{:}\color{codeblue}$\sim$
\color{white}\$ cd <<FluiditySourcePath>> 
\newline
\color{applegreen} \textbf{user@mypc}\color{white}\textbf{:}\color{codeblue}$\sim$
\color{white}\$ sudo ./configure --enable-sam
\newline
\color{applegreen} \textbf{user@mypc}\color{white}\textbf{:}\color{codeblue}$\sim$
\color{white}\$ sudo make
\newline
\color{applegreen} \textbf{user@mypc}\color{white}\textbf{:}\color{codeblue}$\sim$
\color{white}\$ sudo make install
}}
\end{Terminal}

\noindent The above was tested on a `blank' computer and two libraries where missing: \textbf{PETSc} and \textbf{ParMetis}. If you don't yet have them on your system, following are how to install them. Once they are install re-do the whole commands in Command~\ref{Lst:BuildFluidity}. If other libraries are missing, please refer to the Annex C of the \textbf{Fluidity} manual~\cite{AMCG2015}.

\subsubsection{PETSc installation}
To install \textbf{PETSc}, type:
\begin{Terminal}[caption={Installing PETSc.}]
ä\colorbox{davysgrey}{
\parbox{435pt}{
\color{applegreen} \textbf{user@mypc}\color{white}\textbf{:}\color{codeblue}$\sim$
\color{white}\$ sudo apt-get install petsc-dev
}}
\end{Terminal}

\subsubsection{ParMetis installation}
Note that \textbf{Fluidity} will NOT work correctly with versions of \textbf{ParMetis} higher than 3.2. \textbf{ParMetis} can be downloaded from \url{http://glaros.dtc.umn.edu/gkhome/fsroot/sw/parmetis/OLD}. Once downloaded, \textbf{ParMetis} is built in the source directory typing:
\begin{Terminal}[caption={Installing ParMetis.}]
ä\colorbox{davysgrey}{
\parbox{435pt}{
\color{applegreen} \textbf{user@mypc}\color{white}\textbf{:}\color{codeblue}$\sim$
\color{white}\$ make
}}
\end{Terminal}

\noindent It is then recommended to copy-paste the libraries generated not only in the \texttt{fluidity} folder as suggested in the \textbf{Fluidity} manual~\cite{AMCG2015} but also in the \texttt{/usr/} folder, typing:
\begin{Terminal}[caption={Copying ParMetis libraries.}]
ä\colorbox{davysgrey}{
\parbox{435pt}{
\color{applegreen} \textbf{user@mypc}\color{white}\textbf{:}\color{codeblue}$\sim$
\color{white}\$ sudo cp lib*.a /usr/lib
\newline
\color{applegreen} \textbf{user@mypc}\color{white}\textbf{:}\color{codeblue}$\sim$
\color{white}\$ sudo cp parmetis.h /usr/include
\newline
\color{applegreen} \textbf{user@mypc}\color{white}\textbf{:}\color{codeblue}$\sim$
\color{white}\$ sudo cp lib*.a <<FluiditySourcePath>>/lib
\newline
\color{applegreen} \textbf{user@mypc}\color{white}\textbf{:}\color{codeblue}$\sim$
\color{white}\$ sudo cp parmetis.h <<FluiditySourcePath>>/include
}}
\end{Terminal}

\section{Quick start}
Running a simulation with \textbf{Fluidity} consists of several systematic step described below:
\begin{itemize}
    \item \textbf{Step 1:} Create the geometry (\textit{Box.geo}) using \textbf{GMSH}. To visualise the geometry, run \texttt{gmsh Box.geo \&}
    \item \textbf{Step 2:} Create the 3D mesh (\textit{Box.msh}) running \texttt{gmsh -3 Box.geo}. To visualise the mesh, run \texttt{gmsh Box.msh \&}
    \item \textbf{Step 3:} Check the mesh consistency using \texttt{gmsh -check Box.msh} and \texttt{checkmesh Box}. If none of the two commands return errors: the mesh is done.
    \item \textbf{Step 4:} Set up the \textbf{Fluidity} options in \textit{Box.flml} using the graphical interface \textbf{Diamond} running \texttt{diamond Box.flml \&}
    \item \textbf{Step 5:} Run \textbf{Fluidity} using \texttt{<<FluiditySourcePath>>/bin/fluidity -l -v3 Box.flml \&}
    \item \textbf{Step 6:} Visualise the output (\textit{Box.vtu}) using \textbf{ParaView}.
    \item \textbf{Step 7:} Post-process the output using python scripts.
\end{itemize}

    \chapter{Geometry and mesh}\label{Sec:GeometryMesh}
\section{Introduction}
The software used to generate the geometry (\textit{*.geo}) and the mesh (\textit{*.msh}) is \textbf{GMSH} ~\cite{GMSH2009}. \textbf{GMSH} is a mesh generator freely available at \url{http://geuz.org/gmsh/}. Both the geometry and the mesh can be created by \textbf{GMSH} using the graphical interface and a number of tutorials can be found online. However, in this document, the choice was made to describe how to write a geometry file by hand and generate the mesh directly from this file. This approach gives more flexibility for the geometry and mesh generation. The file \textit{Box.geo} is provided with this document and is used as a example in the following sections.

\section{Quick start}
\begin{itemize}
    \item \textbf{Step 1:} Create the geometry (\textit{Box.geo}). To visualise the geometry, run \texttt{gmsh Box.geo \&}
    \item \textbf{Step 2:} Create the 3D mesh (\textit{Box.msh}) running \texttt{gmsh -3 Box.geo}. To visualise the mesh, run \texttt{gmsh Box.msh \&}
    \item \textbf{Step 3:} Check the mesh consistency using \texttt{gmsh -check Box.msh} and \texttt{checkmesh Box}. If none of the two commands return errors: the mesh is done.
\end{itemize}

\noindent The surface IDs needed in \textbf{Fluidity} to prescribe the boundary conditions are assigned in the \textit{*.geo} file.

\section{Geometry}
\begin{figure}
    \centering
    \begin{subfigure}{0.45\textwidth}
        \includegraphics[width=\textwidth]{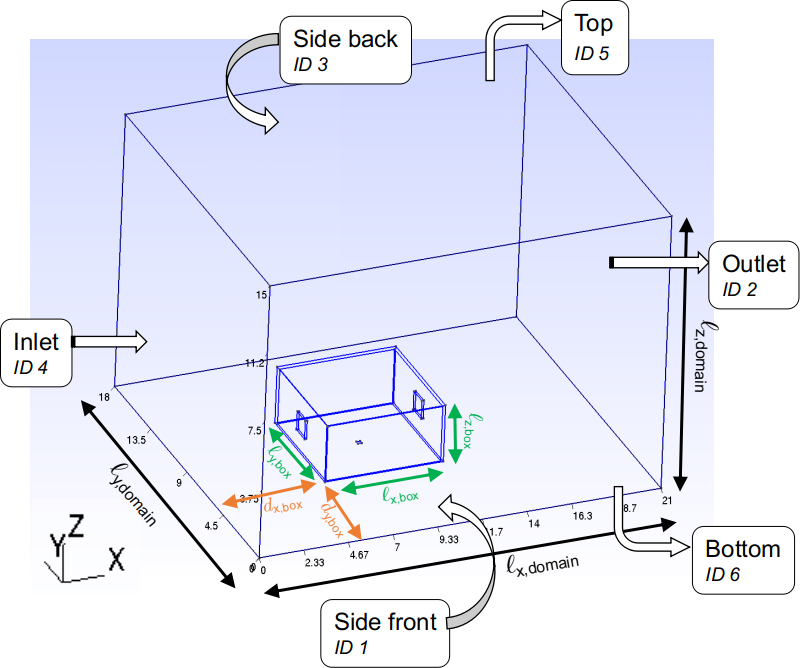}
        \caption{ }
        \label{Fig:GeometryDomain}
    \end{subfigure}
    \begin{subfigure}{0.46\textwidth}
        \includegraphics[width=\textwidth]{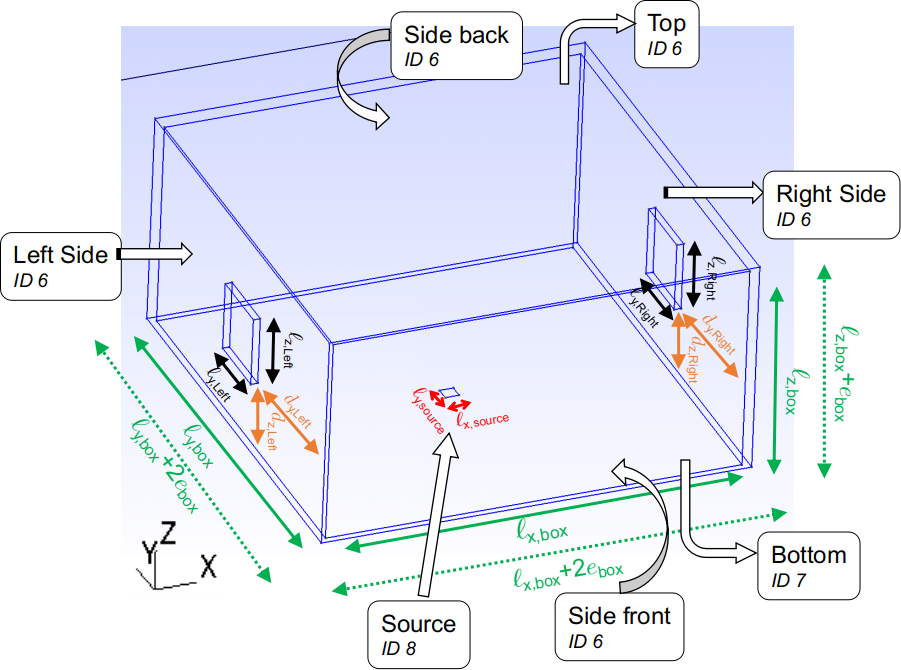}
        \caption{ }
        \label{Fig:GeometryBox}
    \end{subfigure}
    \caption{Geometry used in \textit{Box.geo} example.}
    \label{Fig:Geometry}
\end{figure}

The extension of the geometry file is \textit{*.geo}. A \textit{*.geo} file is a text file that can be written and/or edited by hand using your favourite text editor. The file \textit{Box.geo} is provided with this document and a number of comments (lines started with $//$) were added into the file to help the user. In the following, the geometry file \textit{Box.geo} is explained step by step and the geometry itself can be seen in Figure~\ref{Fig:Geometry}. The user should pay particular attention to  section~\ref{Sec:PhysicalIDs} as it is one of the most important when creating a geometry.

\subsection{Checking the consistency of the geometry}
At any time, the user can visualise the geometry using the graphical interface of \textbf{GMSH} (Command~\ref{Lst:VisualiseGeo}) and check its consistency by running Command~\ref{Lst:CheckGeo} in a terminal. 

\begin{Terminal}[caption={Visualising the geometry in \textbf{GMSH}.}, label=Lst:VisualiseGeo]
ä\colorbox{davysgrey}{
\parbox{435pt}{
\color{applegreen} \textbf{user@mypc}\color{white}\textbf{:}\color{codeblue}$\sim$
\color{white}\$ gmsh Box.geo
}}
\end{Terminal}

\begin{Terminal}[caption={Checking the geometry in \textbf{GMSH}}, label=Lst:CheckGeo]
ä\colorbox{davysgrey}{
\parbox{435pt}{
\color{applegreen} \textbf{user@mypc}\color{white}\textbf{:}\color{codeblue}$\sim$
\color{white}\$ gmsh -check Box.geo
}}
\end{Terminal}

\subsection{Parameters of the geometry}\label{Sec:GeomParam}
At the beginning of the file (see Code~\ref{Lst:GeoParameters}), several parameters are defined to provide a certain flexibility in the creation of the geometry.
\begin{itemize}
    \item If desired, the inner box can be rotated by an angle \texttt{theta} (in degree).
    \item \texttt{m\_domain\_top}, \texttt{m\_domain\_bottom}, \texttt{m\_box}, \texttt{m\_opening} and \texttt{m\_source} define the size of the elements in the final mesh (units are in meters). It should be noted that different size can be assigned in different part of the domain. Decreasing these values will increase the number of elements in the final mesh.
    \item Finally, all the geometric parameters are defined, including the size and the position of the domain, the inner box and the openings. All units are in meters and the user can refer to Figure~\ref{Fig:Geometry} to interpret each parameter.
\end{itemize}

\begin{Code}[language=C, caption={Parameters defining the geometry - from \textit{*.geo} file}, label={Lst:GeoParameters}]
//--------------------------------------//
// Parameters defining the geometry     //
//--------------------------------------//
// NB: All the following parameters are used to define the geometry.

//--> Rotation of the geometry
theta = 0.0; // in degree

//--> Elements size (edge length) of the mesh
m_domain_top     = 3.0;
m_domain_bottom  = 0.5;
m_box            = 0.5;
m_opening        = 0.2;
m_source         = 0.5;

//--> Position and size of the inner box. Unit in meters.
// The size is defined as the size of the inner volume of the box.
dx_box = 6.0; // x Position 
dy_box = 6.0; // y Position

lx_box = 6.0; // Length in the x-direction
ly_box = 6.0; // Length in the y-direction
lz_box = 3.0; // Length in the z-direction

e_box  = 0.1; // Thickness of the walls 10cm
\end{Code}

\subsection{Defining the geometry}
\subsubsection{General idea}
Defining a geometry can be broken down into the following 6 steps:
\begin{itemize}
    \item \textbf{Point:} A point is defined as \texttt{Point(idP)=\{x,y,z,m\};}, where \texttt{x}, \texttt{y} and \texttt{z} define the coordinates of the point having the ID \texttt{idP}. Each \texttt{idP} needs to be unique. \texttt{m} is the element size expected by the user near this point.
    \item \textbf{Line:} A line is defined as \texttt{Line(idL)=\{idP1,idP2\};}, where \texttt{idL} is the ID of the line starting at point \texttt{idP1} and ending at point \texttt{idP2}. Each \texttt{idL} needs to be unique.
    \item \textbf{Line Loop:} A line loop is defined as \texttt{Line Loop(idLL)=\{idL1,idL2,idL3\};}, where \texttt{idLL} is the ID of the line loop composed by the lines having the IDs \texttt{idL1}, \texttt{idL2} and \texttt{idL3}. Each \texttt{idLL} needs to be unique. Line loops are oriented and define how each lines are connected to each other to create a surface (at least 3 lines need to be provided). As line loops are oriented, a minus sign `-' needs to be added in front of lines ID if needed (see line 101 in Code~\ref{Lst:GeoDomain} for example).
    \item \textbf{Plane Surface:} A plane surface is defined as \texttt{Plane Surface(idS)=\{idLL\};}, where \texttt{idS} is the ID of the plane surface defined by the line loop having the ID \texttt{idLL}. Each \texttt{idS} needs to be unique. Note that surfaces need to be plane. The plane surfaces define the 2D surfaces that need to be meshed.
    \item \textbf{Surface Loop:} A surface loop is defined as \texttt{Surface Loop(idSL)=\{idS1,idS2, \ldots\};}, where \texttt{idSL} is the ID of the surface loop created with the surfaces \texttt{idS1}, \texttt{idS2}, \ldots. Each \texttt{idSL} needs to be unique. The surface loop groups all the surfaces defining a closed geometry (i.e a volume).
    \item \textbf{Volume:} A volume is defined as \texttt{Volume(idV)=\{idSL\};}, where \texttt{idV} is the ID of the volume defined by the surface loop having the ID \texttt{idSL}. Each \texttt{idV} needs to be unique. The volumes define the 3D space that needs to be meshed.
\end{itemize}

\subsubsection{Useful trick}
Even for simple geometries, it is easy to get confused with the IDs of points, lines, surfaces... These IDs can be seen on the graphical interface of \textbf{GMSH} for convenience.
In a terminal, run \texttt{gmsh Box.geo \&} to open the geometry. In \textbf{GMSH}, under \texttt{Tools/Options/Geometry/Visibility/}, the \texttt{Point labels}, \texttt{Line labels} and \texttt{Surface labels} can be easily displayed.

\subsubsection{Example}
\begin{figure}
    \centering
    \begin{subfigure}[t]{0.40\textwidth}
        \includegraphics[width=\textwidth]{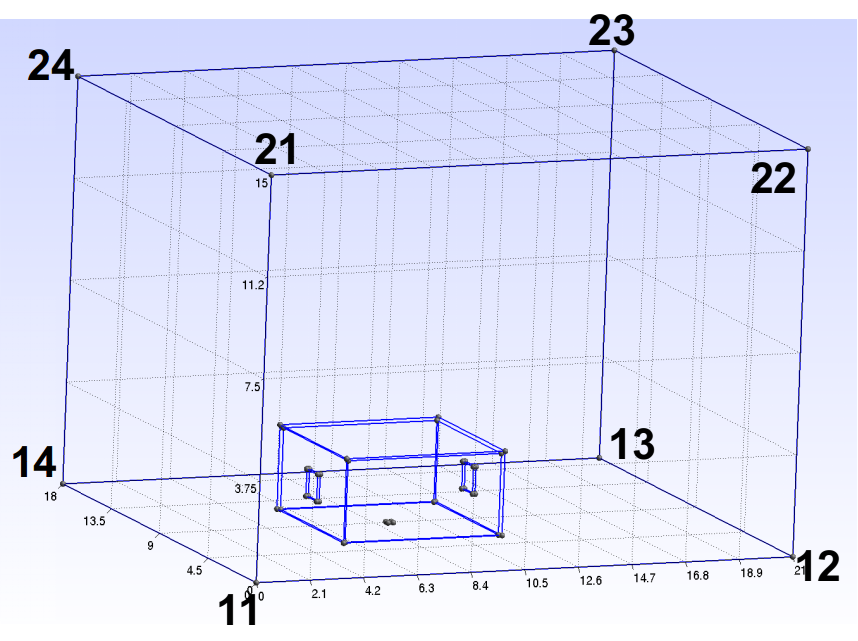}
        \caption{ }
        \label{Fig:GeometryIDDomain}
    \end{subfigure}
    \begin{subfigure}[t]{0.5\textwidth}
        \includegraphics[width=\textwidth]{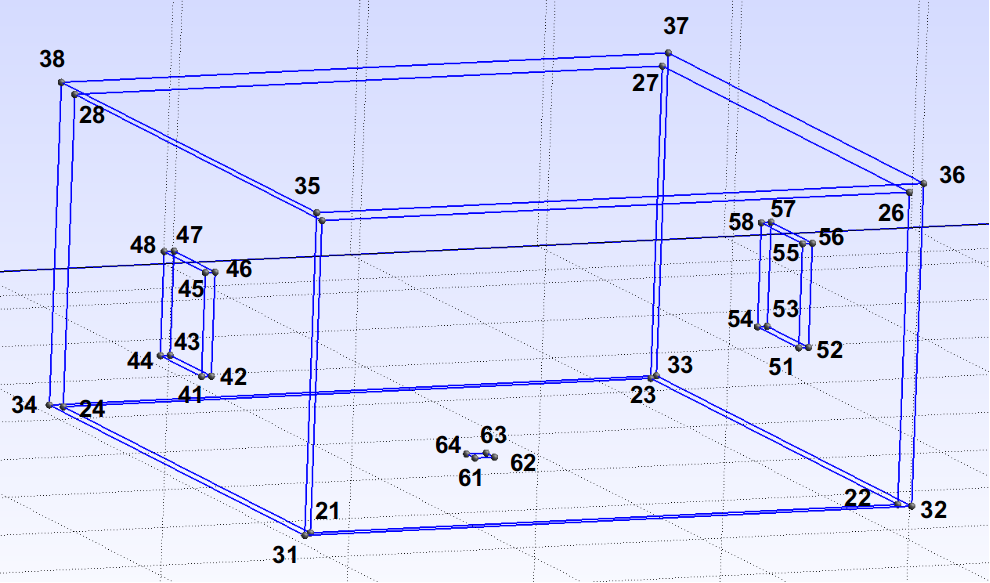}
        \caption{ }
        \label{Fig:GeometryIDBox}
    \end{subfigure}
    \caption{IDs of points used in example \textit{Box.geo}.}
    \label{Fig:GeometryID}
\end{figure}

In the example provided (\textit{Box.geo}), the geometry of the domain is defined as in Code~\ref{Lst:GeoDomain}. The final \texttt{Surface Loop} and \texttt{Volume} are defined as shown in Code~\ref{Lst:GeoBottomWalls}. The user can check, using any text editor, that the inner box, the opening and the source are defined like the domain. To help the user, the IDs of the points are given in Figure~\ref{Fig:GeometryID}.

\begin{Code}[language=C, firstnumber=58, caption={Defining the geometry of the domain - from \textit{*.geo} file.}, label={Lst:GeoDomain}]
//--------------------------------------------//
//   ----   OUTER BOX - Domain     ----       //
//--------------------------------------------//
//--------------//
//    POINTS    //
//--------------//
// - Bottom points
Point(11) =  {0.0,       0.0,       0.0, m_domain_bottom};
Point(12) =  {lx_domain, 0.0,       0.0, m_domain_bottom}; 
Point(13) =  {lx_domain, ly_domain, 0.0, m_domain_bottom}; 
Point(14) =  {0.0,       ly_domain, 0.0, m_domain_bottom}; 

// - Top points
Point(15) =  {0.0,       0.0,       lz_domain,  m_domain_top}; 
Point(16) =  {lx_domain, 0.0,       lz_domain,  m_domain_top}; 
Point(17) =  {lx_domain, ly_domain, lz_domain,  m_domain_top}; 
Point(18) =  {0.0,       ly_domain, lz_domain,  m_domain_top}; 


//--------------//
//    LINES     //
//--------------//
// - Horizontal lines
Line(101) = {11,12};
Line(102) = {12,13};
Line(103) = {13,14};
Line(104) = {14,11};

Line(105) = {15,16};
Line(106) = {16,17};
Line(107) = {17,18};
Line(108) = {18,15};

// - Vertical lines
Line(109) = {11,15};
Line(110) = {12,16};
Line(111) = {13,17};
Line(112) = {14,18};

//------------------//
//    LINE LOOP     //
//------------------//
// NB: Line loop are oriented that is why you can see a sign "-" in  front of some lines ID.
Line Loop(1001) = {109,  105, -110, -101}; // Side front
Line Loop(1002) = {110,  106, -111, -102}; // Outlet
Line Loop(1003) = {111,  107, -112, -103}; // Side back
Line Loop(1004) = {112,  108, -109, -104}; // Inlet

Line Loop(1005) = {105,  106,  107,  108}; // Top
Line Loop(1006) = {101,  102,  103,  104}; // Bottom

//-----------------//
//    SURFACES     //
//-----------------//
// NB: Only the inlet, outlet, top and sides surfaces are defined. The bottom surface will be defined later because we need to subtract the surface of the inner box.
Plane Surface(10001) = {1001}; // Side front
Plane Surface(10002) = {1002}; // Outlet
Plane Surface(10003) = {1003}; // Side back
Plane Surface(10004) = {1004}; // Inlet
Plane Surface(10005) = {1005}; // Top
\end{Code}

\noindent As shown in Code~\ref{Lst:Rotation}, if desired the inner box can be rotated, while keeping the external domain fixed.
\begin{Code}[language=C, firstnumber=203, caption={Rotation of points based on the centre of the box - from \textit{*.geo} file.}, label={Lst:Rotation}]
Rotate{{0,0,1}, {x_center, y_center, z_center}, theta * Pi/180.}{Point{31, 32, 33, 34, 35, 36, 37, 38};}
\end{Code}

\subsection{Surfaces with holes}
\subsubsection{General idea}
A \texttt{Plane Surface} with holes is defined as:
\begin{itemize}
    \item \texttt{Plane Surface(idS)=\{idLL1,idLL2,idLL3 \ldots\};}, where the surface with the ID \texttt{idS} is a plane surface defined by the line loop \texttt{idLL1} minus the surfaces defined by the line loops \texttt{idLL2}, \texttt{idLL3}, \ldots
\end{itemize}

\noindent This is particularly useful in two cases:
\begin{itemize}
    \item if a surface has actually a real hole (like the openings on the walls)
    \item if different boundary conditions have to be assigned to different regions (in this example, we would like to assign a different boundary condition for the domain's bottom and the box's bottom, see section~\ref{Sec:PhysicalIDs} for further details).
\end{itemize}

\subsubsection{Example}
Examining \textit{Box.geo} file, one can see that some surfaces are not directly defined as \texttt{Plane Surface} if they contains holes: it is the case of the walls with openings and the bottom of the ground for example. As shown in Code~\ref{Lst:GeoBottomWalls} (line 410), the bottom of the domain is the surface defined by the line loop with ID 1006, minus the surface defined by the line loop with ID 3006 (corresponding to the bottom line loop defined by the external walls).

\begin{Code}[language=C, firstnumber=407, caption={Defining the surfaces with holes, the surface loop and the volume - from \textit{*.geo} file.}, label={Lst:GeoBottomWalls}]
//---------------------------------------------------------//
//----> Bottom and wall surfaces with openings    ---      //
//---------------------------------------------------------//
Plane Surface(10006) = {1006, 3006}; // Bottom of the domain
Plane Surface(20006) = {2006, 6001}; // Bottom of the box without the source

Plane Surface(20002) = {2002,5004}; // Box, Inner wall: Opening right 
Plane Surface(30002) = {3002,5002}; // Box, Outer wall: Opening right 

Plane Surface(20004) = {2004,4002}; // Box, Inner wall: Opening left 
Plane Surface(30004) = {3004,4004}; // Box, Outer wall: Opening left

//----------------------//
//    DEFINE VOLUME     //
//----------------------//
Surface Loop(100001) = {10001, 10002, 10003, 10004, 10005, 10006, 20001, 20002, 20003, 20004, 20005, 20006, 30001, 30002, 30003, 30004, 30005, 40001, 40003, 40005, 40006, 50001, 50003, 50005, 50006, 60001};
Volume(1000001) = {100001};
\end{Code}

\subsection{Defining the physical IDs}\label{Sec:PhysicalIDs}
Surface IDs are used in \textbf{Fluidity} to mark different parts of the boundary of the computational domain so that different boundary conditions can be associated with them. In three dimensions, surface IDs are defined by assigning \texttt{Physical Surface} IDs in \textbf{GMSH}. The IDs are defined at the end of \textit{Box.geo} as shown in Code~\ref{Lst:GeoIDs}. They are also summarised in the Figure~\ref{Fig:Geometry}.

\begin{Code}[language=C, firstnumber=425, caption={Defining the IDs of surfaces and volumes - from \textit{*.geo} file.}, label={Lst:GeoIDs}]
//----------------------//
//    Physical ID       //
//----------------------//
// NB: These IDs are the one used in Fluidity
Physical Surface(1) = {10001};    // Side front
Physical Surface(2) = {10002};    // Outlet
Physical Surface(3) = {10003};    // Side back
Physical Surface(4) = {10004};    // Inlet
Physical Surface(5) = {10005};    // Top
Physical Surface(6) = {10006, 20001, 20002, 20003, 20004, 20005, 30001, 30002, 30003, 30004, 30005, 40001, 40003, 40005, 40006, 50001, 50003, 50005, 50006}; // Bottom and Walls excepted the ground of the box and the source
Physical Surface(7) = {20006}; // Ground of the box
Physical Surface(8) = {60001}; // Source
Physical Volume(1000002) = {1000001}; // Volume to be meshed
\end{Code}

\subsection{Advice}
It is recommended to add the line in Code~\ref{Lst:GeoCoherence} at the end of any \textit{*.geo} file. This function removes all duplicate elementary geometrical entities (e.g., points having identical coordinates).

\begin{Code}[language=C, firstnumber=439, caption={Coherence - from \textit{*.geo} file}, label={Lst:GeoCoherence}]
Coherence;
\end{Code}

\section{Mesh}

\subsection{Generating the mesh}\label{Sec:GenerateMesh}
\subsubsection{Method}
The extension of the mesh file is \textit{*.msh}. To generate the mesh \textit{Box.msh} associated with the geometry \textit{Box.geo}, run the Command~\ref{Lst:GenerateMesh} in a terminal. The option \texttt{-3} in Command~\ref{Lst:GenerateMesh} means that the geometry is in 3 dimensions (2 should be used instead if the geometry is 2D).
\begin{Terminal}[caption={Generating the geometry using \textbf{GMSH}.}, label={Lst:GenerateMesh}]
ä\colorbox{davysgrey}{
\parbox{435pt}{
\color{applegreen} \textbf{user@mypc}\color{white}\textbf{:}\color{codeblue}$\sim$
\color{white}\$ gmsh -3 Box.geo
}}
\end{Terminal}

\begin{figure}
    \centering
    \begin{subfigure}{0.5\textwidth}
        \includegraphics[width=\textwidth]{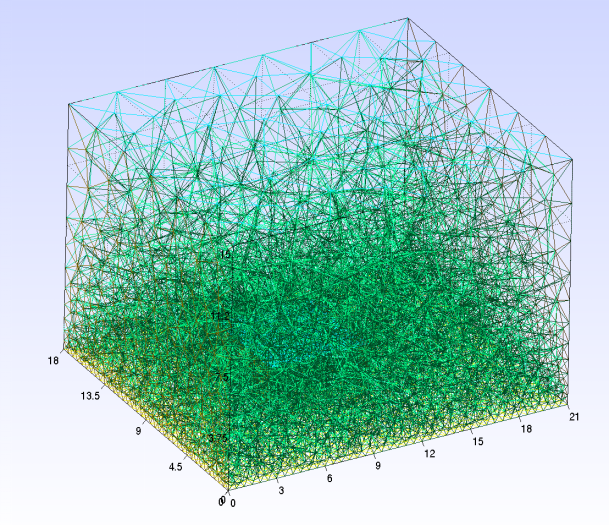}
        \caption{3D mesh}
        \label{Fig:MeshVolume}
    \end{subfigure}
    \newline
    \begin{subfigure}{0.24\textwidth}
        \includegraphics[width=\textwidth]{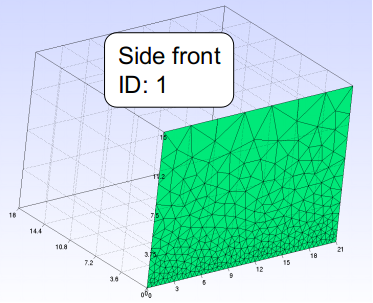}
        \caption{ }
        \label{Fig:Mesh1}
    \end{subfigure}
    \begin{subfigure}{0.24\textwidth}
        \includegraphics[width=\textwidth]{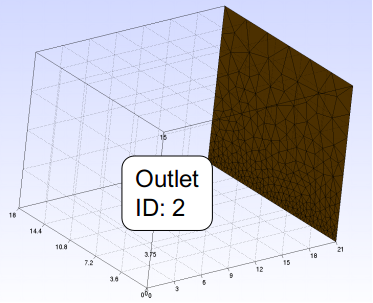}
        \caption{ }
        \label{Fig:Mesh2}
    \end{subfigure}
    \begin{subfigure}{0.24\textwidth}
        \includegraphics[width=\textwidth]{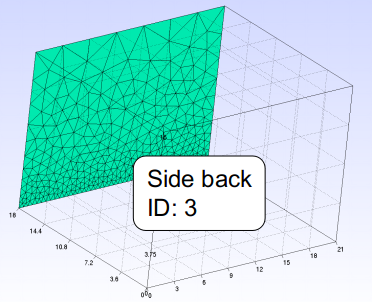}
        \caption{ }
        \label{Fig:Mesh3}
    \end{subfigure}
    \begin{subfigure}{0.24\textwidth}
        \includegraphics[width=\textwidth]{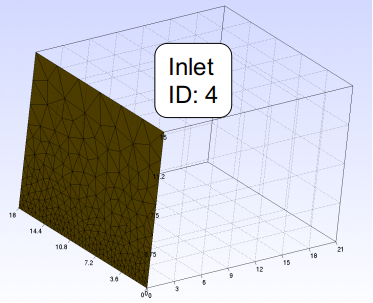}
        \caption{ }
        \label{Fig:Mesh4}
    \end{subfigure}    
    \begin{subfigure}{0.24\textwidth}
        \includegraphics[width=\textwidth]{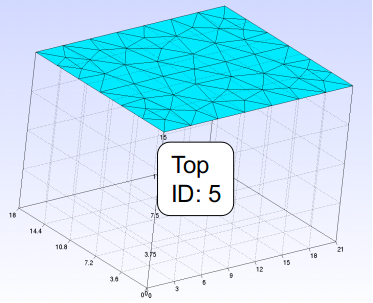}
        \caption{ }
        \label{Fig:Mesh5}
    \end{subfigure}    
    \begin{subfigure}{0.24\textwidth}
        \includegraphics[width=\textwidth]{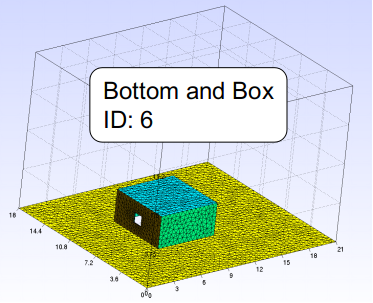}
        \caption{ }
        \label{Fig:Mesh6}
    \end{subfigure}    
    \begin{subfigure}{0.24\textwidth}
        \includegraphics[width=\textwidth]{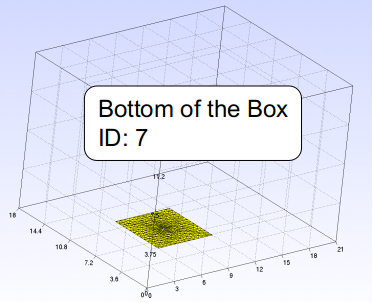}
        \caption{ }
        \label{Fig:Mesh7}
    \end{subfigure}
    \begin{subfigure}{0.24\textwidth}
        \includegraphics[width=\textwidth]{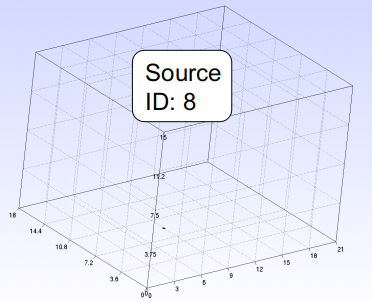}
        \caption{ }
        \label{Fig:Mesh8}
    \end{subfigure}    
    \caption{Mesh generated from the geometry \textit{Box.geo}.}
    \label{Fig:Mesh}
\end{figure}

\noindent A file named \textit{Box.msh} will be created. To visualise the mesh in \textbf{GMSH}, run the Command~\ref{Lst:VisualiseMesh} in a terminal. In \textbf{GMSH}, under \texttt{Tools/Options/Mesh/Visibility}, the \texttt{Surface faces} or the \texttt{Volume edges} can be displayed as shown in Figure~\ref{Fig:Mesh}. By default, \textbf{GMSH} displays the  \texttt{Surface edges} only. For convenience, the user can also choose to only display some part of the mesh. In \textbf{GMSH}, every surface and volume with respectively a \texttt{Physical Surface} or \texttt{Physical Volume} ID is listed under \texttt{Tools/}\texttt{Visibility/} \texttt{List browser}. The user can select one or several surfaces at the same time to display in \textbf{GMSH} as done in Figure~\ref{Fig:Mesh}.

\begin{Terminal}[caption={Visualising the geometry in \textbf{GMSH}.}, label={Lst:VisualiseMesh}]
ä\colorbox{davysgrey}{
\parbox{435pt}{
\color{applegreen} \textbf{user@mypc}\color{white}\textbf{:}\color{codeblue}$\sim$
\color{white}\$ gmsh Box.msh
}}
\end{Terminal}

\subsubsection{Common errors}
The main common errors that occur when trying to generate the mesh are:
\begin{itemize}
    \item \textit{No tetrahedra in region idV}: this happens mainly if the \texttt{Surface Loop} is not closed.
    \item \textit{Found two facets intersect each other}: two surfaces in \texttt{Surface Loop} are overlapping.
    \item \textit{Found two duplicated facets}: a surface in \texttt{Surface Loop} is defined twice.
\end{itemize}

\subsection{Checking the consistency of the mesh}
\subsubsection{Method}
At this stage, the mesh is created. Before running \textbf{Fluidity}, it is recommended to check if the generated mesh is consistent and suitable to run simulations. The check can be done in the following 3 steps:
\begin{itemize}
    \item \textbf{Step 1:} The first check is visual: open the mesh into \textbf{GMSH} using the Command~\ref{Lst:VisualiseMesh} and check the visualise aspect of the mesh using the tricks to display the mesh explained in section~\ref{Sec:GenerateMesh}. Are the finer elements where expected? Do they seem fine enough?
    \item \textbf{Step 2:} The second step is to check if the mesh is consistent on a ``\textbf{GMSH}" point of view. To this end, run the Command~\ref{Lst:CheckGMSH} in a terminal.
    \begin{Terminal}[caption={Check the consistency of the mesh using \textbf{GMSH} tool.}, label={Lst:CheckGMSH}]
ä\colorbox{davysgrey}{
\parbox{400pt}{
\color{applegreen} \textbf{user@mypc}\color{white}\textbf{:}\color{codeblue}$\sim$
\color{white}\$ gmsh -check Box.msh
}}
    \end{Terminal}

    \item \textbf{Step 3:} The third step consists of using a \textbf{Fluidity} tool to test if the mesh is consistent on a ``\textbf{Fluidity}" point of view. This can be done with the Command~\ref{Lst:CheckFluidity}.
    \begin{Terminal}[caption={Check the consistency of the mesh using \textbf{Fluidity} tool.}, label={Lst:CheckFluidity}]
ä\colorbox{davysgrey}{
\parbox{400pt}{
\color{applegreen} \textbf{user@mypc}\color{white}\textbf{:}\color{codeblue}$\sim$
\color{white}\$ checkmesh Box
}}
    \end{Terminal}
\end{itemize}

\noindent If all 3 steps suggested above are successful, then you are ready to run \textbf{Fluidity} simulations.

\subsubsection{Common errors}
Below are listed common errors that Command~\ref{Lst:CheckGMSH} and Command~\ref{Lst:CheckFluidity} can raise, as well as ideas on how to fix them:
\begin{itemize}
    \item From \textbf{GMSH} tool (Command~\ref{Lst:CheckGMSH})
    \begin{itemize}
        \item \textit{duplicate vertex - duplicate elements}: A surface is defined twice and different \texttt{Physical Surface} ID were assigned to them.
    \end{itemize}
    \item From \textbf{Fluidity} tool (Command~\ref{Lst:CheckFluidity}):
        \begin{itemize}
        \item \textit{Degenerate surface element found}: if you have this error it is probably because you already had errors when generating the mesh. This is mainly due to the fact that the \texttt{Surface Loop} is not closed.
        \item \textit{Degenerate volume element found}: this error occurs if you have forgotten to define a \texttt{Volume} or to assign a \texttt{Physical Volume} ID to the volume.
        \item \textit{Surface element does not exist in the mesh}: A surface defined by a \texttt{Physical} \texttt{Surface} ID is not present in the \texttt{Surface Loop}.
        \item \textit{WARNING: an incomplete surface mesh has been provided. This will not work in parallel. All parts of the domain boundary need to be marked with a (physical) surface id}: The error is quite clear: each surface defining the \texttt{Surface Loop} has to have a \texttt{Physical Surface} ID and at least one surface does not.
    \end{itemize}
\end{itemize}

\subsubsection{Useful mesh statistics}
Note that the tools using in Command~\ref{Lst:CheckGMSH} and Command~\ref{Lst:CheckFluidity} give also interesting information and statistics concerning the mesh:
\begin{itemize}
    \item From \textbf{GMSH} tool (Command~\ref{Lst:CheckGMSH}): Number of nodes and number of elements...
    \item From \textbf{Fluidity} tool (Command~\ref{Lst:CheckFluidity}): Number of nodes, number of elements, minimum edge length and maximum edge length...
\end{itemize}

\section{Playing around}
It is recommended that the user plays around with the parameters defined at the beginning of the file, to change the position and/or size of the openings for example. The user can also rotate the box if desired. Interested exercises would consist of modifying the \textit{Box.geo} file to have the openings as ``doors" instead of ``windows"; or having the box in the middle of the domain instead of clipped to the ground. To create these geometries, the \textit{Box.geo} will need a number of changes (the bottom surface will be defined differently for example), but the \textit{Box.geo} provided can easily be used as a starting point.

\noindent Other options to experiment with are the parameters defining the elements size. Hence, the user can see the impact of reducing or increasing these values on the mesh aspect.

    \chapter{Equations and numerical methods}
The aim of this chapter is not to give extended overview of the equations and their implementations but only a brief idea of them. The reader can refer to \cite{AMCG2015,Pain2001,Aristodemou2009,Pavlidis2010,Aristodemou2018} for more details.

\section{Equations}
\subsection{Navier-Stokes and Large Eddy Simulation (LES)}
The Large Eddy Simulation (LES) formulation implemented in \textbf{Fluidity} describes turbulent flows based on the filtered (three dimensional) incompressible Navier-Stokes equations (continuity of mass and momentum equation, see equation~\ref{Eq:Continuity} and equation~\ref{Eq:Momentum}):

\begin{equation}
\nabla . \overline{u} = 0 
\label{Eq:Continuity}
\end{equation}

\begin{equation}
\frac{\partial\overline{u}}{\partial t} +  \overline{u} . \nabla  \overline{u}  = - \frac{1}{\rho} \nabla\overline{p} + \nabla . \Big[ \left( \nu + \nu_{\tau} \right) \nabla \overline{u} \Big] 
\label{Eq:Momentum}
\end{equation}

\noindent where $\overline{u}$ is the resolved velocity (m/s), $\overline{p}$ is the resolved pressure (Pa), $\rho$ is the fluid density (kg/m\textsuperscript{3}), $\nu$ is the kinematic viscosity (m\textsuperscript{2}/s) and $\nu_{\tau}$ is the anisotropic eddy viscosity (m\textsuperscript{2}/s).

\noindent A novel component in the implementation of the standard LES equations within \textbf{Fluidity} is the anisotropic eddy viscosity tensor $\nu_{\tau}$ defined by equation~\ref{Eq:SmagorinsgkyViscosity}.

\begin{equation}
\nu_{\tau} = C_{S}^{2}l^{2}|\overline{S}|
\label{Eq:SmagorinsgkyViscosity}
\end{equation}

\noindent $C_{S}$ is the Smagorinsky coefficient (usually taken equal to $0.1$), $l$ is the Smagorinsky length-scale which depends on the local element size and $|\overline{S}|$  is the strain rate expressed as in equation~\ref{Eq:StrainRate}.

\begin{equation}
|\overline{S}| = (2 \overline{S}_{ij} \overline{S}_{ij})^{1/2}
\label{Eq:StrainRate}
\end{equation}

\noindent where $ \overline{S}_{ij}$ is the local strain rate defined by equation~\ref{Eq:LocalStrainRate}.

\begin{equation}
\overline{S}_{ij}=\frac{1}{2}\left(\frac{\partial\overline{u}_{i}}{\partial x_{j}}+\frac{\partial\overline{u}_{i}}{\partial x_{i}}\right)
\label{Eq:LocalStrainRate}
\end{equation}

\subsection{Advection-Diffusion equation}
The transport of a scalar field $c$ (i.e, a passive tracer) having the unit $unit_{c}$ is expressed using a classic advection-diffusion equation with a source term (equation~\ref{Eq:TracerEq}).

\begin{equation}
    \frac{\partial c}{\partial t}+\nabla.(\mathbf{u}c)=\nabla.\left(\overline{\overline{\kappa}}\nabla c\right)+F
    \label{Eq:TracerEq}
\end{equation}

\noindent where $\mathbf{u}$ is the velocity vector (m/s), $\overline{\overline{\kappa}}$ is the diffusivity tensor (m\textsuperscript{2}/s) and $F$ represents the source terms ($unit_{c}$/s).

\noindent In the case where the scalar $c$ is the temperature in Kelvin, then the source term $F$ is expressed by equation~\ref{Eq:SourceTemperature}

\begin{equation}
    F = \frac{Q}{\rho c_{p}}
    \label{Eq:SourceTemperature}
\end{equation}

\noindent where $Q$ is a power density expressed in W/m\textsuperscript{3}.

\noindent In the case where the scalar $c$ is the species concentration in $kg/m^{3}$, then the source term $F$ is expressed by equation~\ref{Eq:SourceConcentrationMass} or equation~\ref{Eq:SourceConcentrationVol}:

\begin{equation}
    F = \frac{\dot{m}}{V}
    \label{Eq:SourceConcentrationMass}
\end{equation}

\noindent where $\dot{m}$ is a mass flow rate expressed in $kg/s$ and $V$ is the volume of the source in $m^{3}$;

\begin{equation}
    F = \frac{Q \rho}{V}
    \label{Eq:SourceConcentrationVol}
\end{equation}

\noindent where $Q$ is a volumetric flow rate expressed in $m^{3}/s$ and $V$ is the volume of the source in $m^{3}$.

\subsection{Boussinesq approximation}
Under certain conditions, one can assume that density does not vary greatly about a mean reference density, that is, the density at a position $\textbf{x}$ can be written as: 

\begin{equation}
\rho(\textbf{x},t) = \rho_0 + \rho'(\textbf{x},t)
\end{equation}

\noindent  where $\rho'\ll\rho_0$. Such an approximation is named the Boussinesq approximation. This assumption ignores density differences except when they are multiplied by $g$, the acceleration due to gravity.

\section{Numerical methods}
\subsection{Discretisation}
For indoor-outdoor exchange simulations or urban environment simulations, the following discretisation are recommended:
\begin{itemize}
    \item \textbf{Navier-Stokes equations:} The Navier-Stokes equations, under the Boussinesq approximation, are solved using a continuous Galerkin finite element discretisation, while a Crank-Nicolson time discretisation approach is adopted.
    \item \textbf{Advection-diffusion equation:} The advection-diffusion is solved using a control volume - finite element space discretisation, while a Crank-Nicolson time discretisation approach is adopted.
\end{itemize}

\subsection{CFL number} \label{Sec:CFLNumber}
To avoid crashes in simulations, the time step can be adaptive using a Courant-Friedrichs-Lewy (CFL) condition. This option allows the time step $\Delta t$ to vary throughout the run, depending on the CFL number. The maximum CFL number $C_{\max}$ is set by the user as shown in Figure~\ref{Fig:AdaptCFL}. The CFL condition is a necessary condition for convergence while solving certain partial differential equations and has the following form:

\begin{equation} \label{Eq:CFLNumber}
    C = \frac{u \Delta t}{\Delta x} \leq C_{\max}
\end{equation}

\noindent where C is the dimensionless CFL number, $C_{\max}$ is the maximum CFL number given by the user, $u$ is the magnitude of the velocity (m/s), $\Delta t$ is the time step (s) and $\Delta x$ is the length interval, i.e taken as the edge element in the mesh (m).

\noindent It is commonly say that the maximum value of $C_{\max}$ should be $1$. However, in \textbf{Fluidity}, this value can be increased further (until 5-10) without convergence issues. However, it is recommended to start the simulations with a value of $1$ and then increase this value incrementally checking if the accuracy of the results are not affected. A too high CFL number might prematurely kill the turbulence.

\subsection{Solver: \textbf{PETSc}}
The solver used in \textbf{Fluidity} is the open-source solver toolbox \textbf{PETSc}. The user can refer to \url{https://www.mcs.anl.gov/petsc/} for more information.

    \chapter{Boundary and initial conditions}\label{Sec:ChapterBC}
\section{Introduction}
This section shows examples of boundary conditions that can be used to set up a wide range of indoor simulations. They all rely on the geometry described previously which is a box with two openings in the middle of a wider computational domain. In this chapter, the fluid is air, with properties defined in Table~\ref{Tab:AirPorperties}. Gravity ($g=$ 9.81 m/s\textsuperscript{2}) is taken into account. The Navier-Stokes equations, under the Boussinesq approximation, are solved for the fluid, while the advection-diffusion equation is solved for the heat transfers. The mesh and the time step are fixed and the simulations are run in serial.

\noindent \textbf{Important note:} The simulations with constant time steps work with the geometry file provided, i.e. if the element sizes have not been changed (see Section~\ref{Sec:GeomParam}). If the mesh has been refined (i.e. the element size decreased), the simulations may crash. In that case, the time step has to be decreased under the option \texttt{timestepping/timestep} or the user can use an adaptive time step (option \texttt{timestepping/adaptive\_timestep}) and use a CFL condition (see Section~\ref{Sec:CFLNumber}).

\begin{table}
    \centering
    \begin{tabular}{l|r}
        Thermal Diffusivity & $  \kappa = 2.12\times10^{-5}$ m\textsuperscript{2}/s\\
        Thermal Conductivity & $ \lambda = 2.597\times10^{-2}$ W/m/K\\
        Kinematic Viscosity & $ \nu = 1.5\times10^{-5}$ m\textsuperscript{2}/s \\
        Reference Density & $\rho_{0} = 1.225$ kg/m\textsuperscript{3}\\
        Thermal Expansion coefficient & $\alpha = 3.43\times 10^{-3} $K\textsuperscript{-1}\\ & at $T=293$ K\\
        Specific Heat Capacity & $c_{p} = 1000$ J/kg/K
    \end{tabular}
    \caption{\label{Tab:AirPorperties}Properties of the air used in the simulations.}
\end{table}

\noindent In the next sections, different common and useful boundary conditions will be tested. The following will however remain fixed:
\begin{itemize}
    \item \textbf{Outlet (Dirichlet boundary condition):} A zero stress conditions is imposed at the outlet which sets $p=0$. It must noted that at least one pressure boundary condition is required as a reference for \textbf{Fluidity} to run.  See Section~\ref{Sec:RefPressure}.
    \item \textbf{Sides and top of the domain (Dirichlet boundary condition):} A perfect slip boundary condition is imposed on the sides and top of the domain which sets the normal component of velocity equal to zero.
    \item \textbf{Ground of the domain and walls (Dirichlet boundary condition):} A zero velocity boundary condition is prescribed on the floor of the domain, the floor of the box and the walls of the box which sets the three components of velocity equal to zero. This boundary condition is discussed in details in Section~\ref{Sec:BCWalls}.
\end{itemize}

\noindent Table~\ref{Tab:SimuCases} summarises the different simulations presented in the next sections.

\begin{table}
    \centering
    \begin{tabular}{ |p{1.1cm}||p{1.1cm}|p{2.8cm}|p{3.0cm}|p{1.2cm}|p{1.8cm}|p{1.5cm}|  }
         \hline
         \textbf{Case \newline Nbr} & \textbf{Time \newline Step} & \bf{$T_{init}$} & \textbf{Floor BC} & \bf{$u_{init}$ \newline (m/s)}   & \textbf{Inlet \newline velocity \newline (m/s)} & \textbf{Section}\\
         \hline
         1a & 1s & 293 K & Dirichlet BC \newline $T_{f}=298 $ K  & 0 & 0 & ~\ref{Sec:TempDirichletUni}\\\hline
         1b & 1s & 293 K  & Dirichlet BC \newline $T_{f}=f(x)$ & 0 & 0 &  ~\ref{Sec:TempDirichletPython}\\\hline
         1c & 1s & 293 K  & Dirichlet BC \newline $T_{f}=f(t)$ & 0 & 0 & ~\ref{Sec:TempDirichletPython}\\\hline
         1d & 1s & 293 K  & Dirichlet BC \newline $T_{f}=f(x,t)$ & 0 & 0 & ~\ref{Sec:TempDirichletPython}\\\hline
         2a & 1s & 293 K & Neumann BC \newline $\phi_{f}=10$ W/m\textsuperscript{2} & 0 & 0 & ~\ref{Sec:TempNeumannUniFloor}\\\hline
         2b & 1s & 293 K & Neumann BC \newline $\phi_{s}=10^3$ W/m\textsuperscript{2} & 0 & 0 & ~\ref{Sec:TempNeumannUniSource}\\\hline
         2c & 1s & 293 K  & Neumann BC \newline $\phi_{f}=f(x,t)$ & 0 & 0 & ~\ref{Sec:TempNeumannPython}\\\hline
         3 & 1s & 293 K & Robin BC \newline $T_{\infty}=25 $ \degree C  \newline $h=5$ W/m\textsuperscript{2}/K & 0 & 0 & ~\ref{Sec:TempRobin}\\\hline
         4 & 1s & Outside: 293 K \newline Inside: 298 K & / & 0 & 0 & ~\ref{Sec:TempInit}\\\hline
         5a & 1s & Outside: 293 K \newline Inside: 298 K & / & 1 & 1 & ~\ref{Sec:VelDiricUni}\\\hline
         5b & 1s & Outside: 293 K \newline Inside: 298 K & / & Log- \newline profile & Log-\newline profile & ~\ref{Sec:VelDiricPython}\\\hline
         5c & 1s & Outside: 293 K \newline Inside: 298 K & / & 1 & Turbulent \newline inlet & ~\ref{Sec:VelTurbUni}\\\hline
         5d & 1s & Outside: 293 K \newline Inside: 298 K & / & Log \newline profile & Turbulent\newline  inlet & ~\ref{Sec:VelTurbPython}\\\hline  
    \end{tabular}
    \caption{\label{Tab:SimuCases}Summary of the simulations presented in the following sections of Chapter~\ref{Sec:ChapterBC}. Subscript $f$ stands for \textit{floor} (ground of the box + the source) and subscript $s$ stands for \textit{source} only. All these simulations are for a fixed mesh.}
\end{table}

\section{Quick start}
\subsubsection{Running a simulation}
Running a simulation with \textbf{Fluidity} consists of the following steps: 
\begin{itemize}
    \item \textbf{Step 1:} Set up the \textbf{Fluidity} options in \textit{3d\_Case.flml} using the graphical interface \textbf{Diamond} running \texttt{diamond 3d\_Case.flml \&}
    \item \textbf{Step 2:} Run \textbf{Fluidity} using \texttt{<<FluiditySourcePath>>/bin/fluidity -l -v3 3d\_Case.flml \&}
    \item \textbf{Step 3:} Visualise the \textbf{Fluidity} log file during the simulation using the command \texttt{tail -f fluidity.log-0} in a terminal.
    \item \textbf{Step 4:} Open the \textbf{Fluidity} error file using the command \texttt{gedit fluidity.err-0 \&} in a terminal.
\end{itemize}

\noindent Command~\ref{Lst:RunFluidity} is the basic command line to run a simulation with \textbf{Fluidity}. One can notice two options:
\begin{itemize}
    \item \texttt{-l}: This option writes the terminal log and errors in the files \texttt{fluidity.log-0} and \texttt{fluidity.err-0}, respectively, instead of writing them directly in the terminal.
    \item \texttt{-v3}: This is the degree of verbosity that the user wants. The user can choose between \texttt{-v1}, \texttt{-v2} and \texttt{-v3}, where \texttt{-v1} will be less verbose than \texttt{-v3}.
\end{itemize}

\begin{Terminal}[caption={Command to run a simulation with \textbf{Fluidity}.}, label={Lst:RunFluidity}]
ä\colorbox{davysgrey}{
\parbox{435pt}{
\color{applegreen} \textbf{user@mypc}\color{white}\textbf{:}\color{codeblue}$\sim$
\color{white}\$ <<FluiditySourcePath>>/bin/fluidity -l -v3 3d\_Case.flml \&
}}
\end{Terminal}

\subsubsection{Killing a simulation}
At any time, to kill a simulation, the following can be done in a terminal:
\begin{itemize}
    \item Run the command \texttt{top}. All the processes currently running on the machine are listed, including the \textbf{Fluidity} simulations. Note the \texttt{ProcID} that you want to kill, then press the key \texttt{q} to quit.
    \item As the user can have several simulations running in different folders, the command \texttt{pwdx ProcID} can be use to determine where the \texttt{ProcID} is currently running. This command will provide the path where the simulation was launched.
    \item Once the user is sure that the simulation to kill has the ID \texttt{ProcID}, then the command \texttt{kill -9 ProcID} can be run in a terminal to kill the simulation.
\end{itemize}

\section{Thermal boundary conditions}
\subsection{Dirichlet boundary condition: Constant temperature}
\subsubsection{Constant floor temperature}\label{Sec:TempDirichletUni}
\begin{figure}
    \centering
    \begin{subfigure}{0.49\textwidth}
        \includegraphics[width=\textwidth]{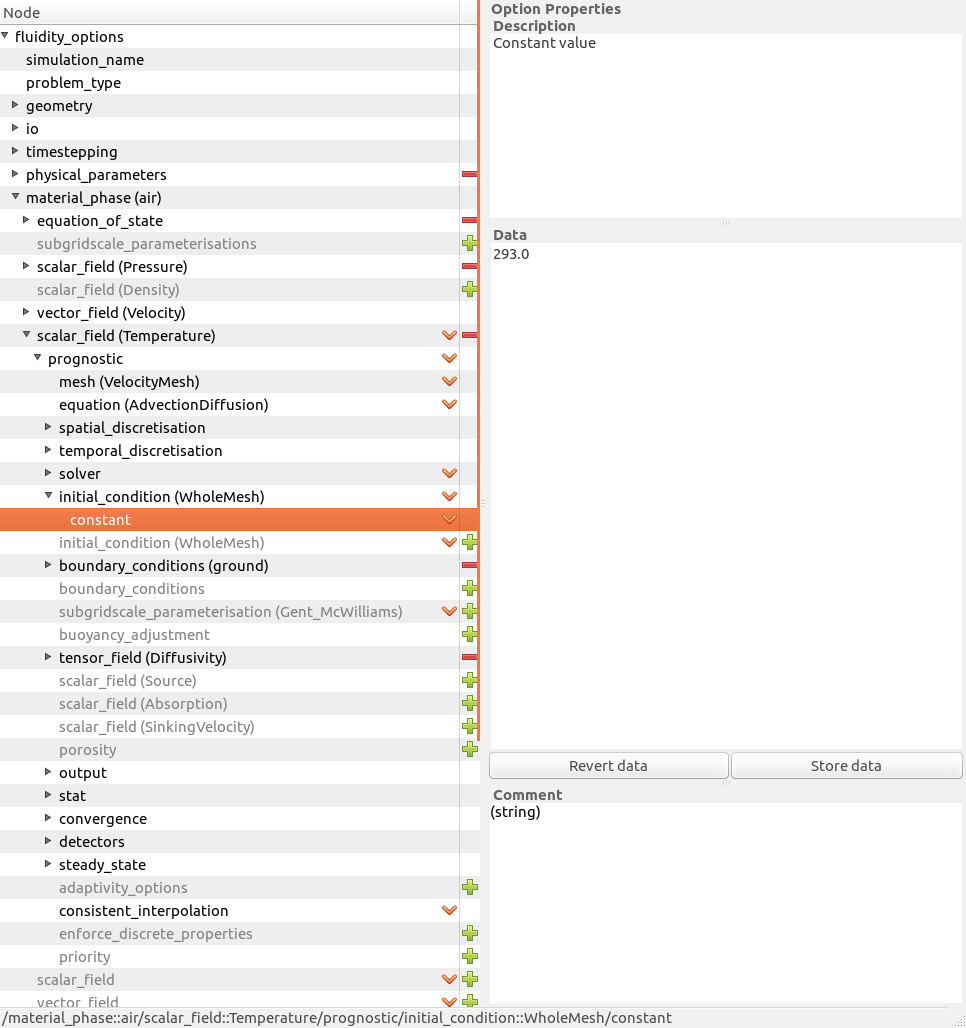}
        \caption{ }
        \label{Fig:Case1a_TempInit}
    \end{subfigure}
    \begin{subfigure}{0.49\textwidth}
        \includegraphics[width=\textwidth]{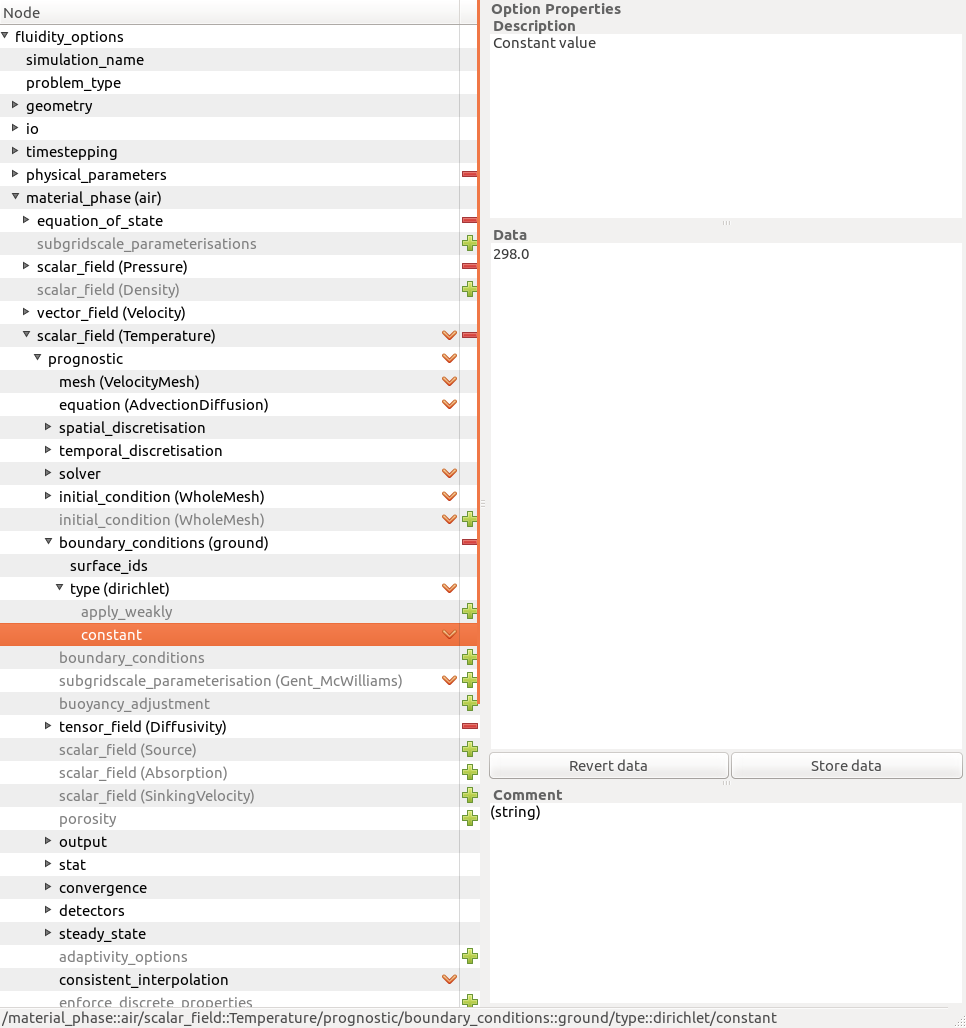}
        \caption{ }
        \label{Fig:Case1a_TempBC}
    \end{subfigure}
    \begin{subfigure}{0.49\textwidth}
        \includegraphics[width=\textwidth]{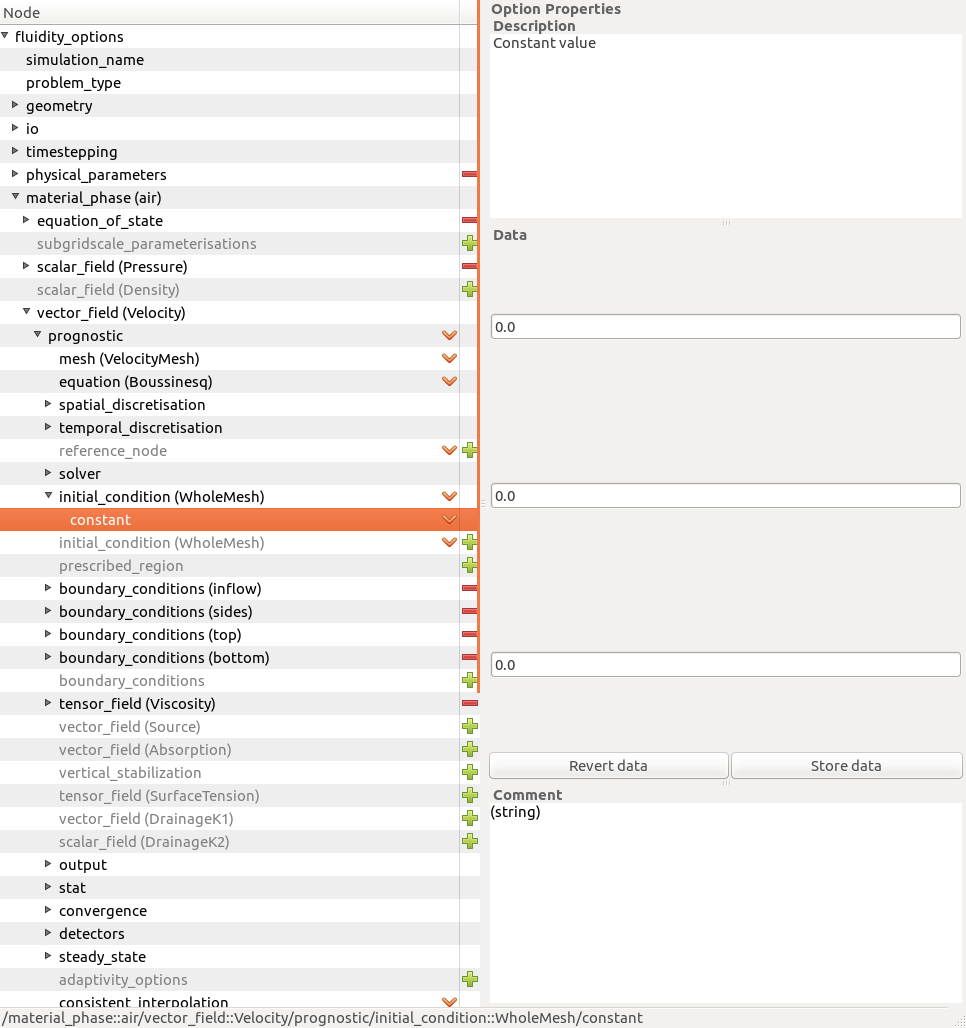}
        \caption{ }
        \label{Fig:Case1a_VelInit}
    \end{subfigure}
    \begin{subfigure}{0.49\textwidth}
        \includegraphics[width=\textwidth]{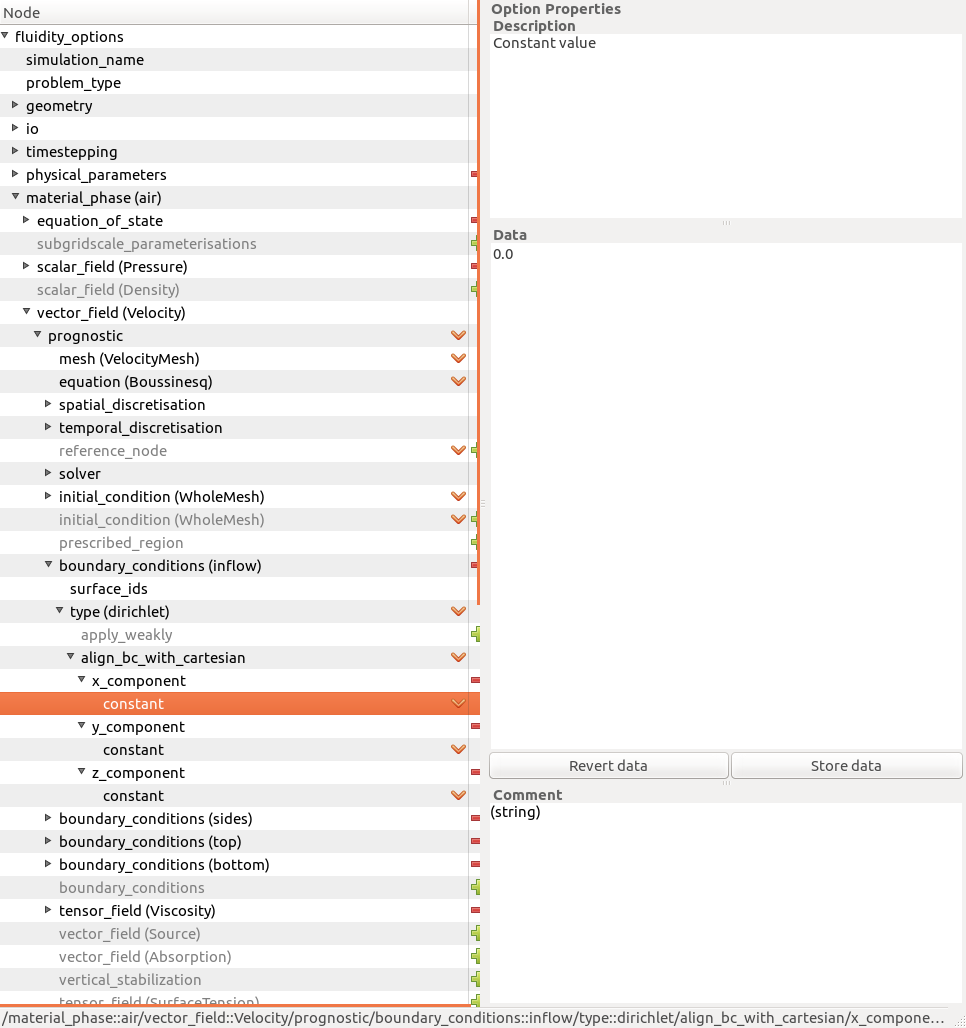}
        \caption{ }
        \label{Fig:Case1a_VelBC}
    \end{subfigure}
    \caption{Options used in \textit{3dBox\_Case1a.flml}. (a) Initial temperature, (b) Thermal boundary condition for the floor, (c) Initial velocity and (d) Inlet boundary condition for the velocity.}
    \label{Fig:Case1a}
\end{figure}

In \textit{3dBox\_Case1a.flml}, the floor of the box is set to a constant temperature of 298K (Figure~\ref{Fig:Case1a_TempBC}) using a Dirichlet boundary condition. The initial and the ambient temperatures are set to 293 K (Figure~\ref{Fig:Case1a_TempInit}). The initial and inlet velocity are set to $0$ m/s (Figure~\ref{Fig:Case1a_VelInit} and Figure~\ref{Fig:Case1a_VelBC}). This simulation can be run using the command: 
\begin{Terminal}[]
ä\colorbox{davysgrey}{
\parbox{435pt}{
\color{applegreen} \textbf{user@mypc}\color{white}\textbf{:}\color{codeblue}$\sim$
\color{white}\$ <<FluiditySourcePath>>/bin/fluidity -l -v3 3dBox\_Case1a.flml \&
}}
\end{Terminal}

\begin{figure}
    \centering
    \includegraphics[scale=0.25]{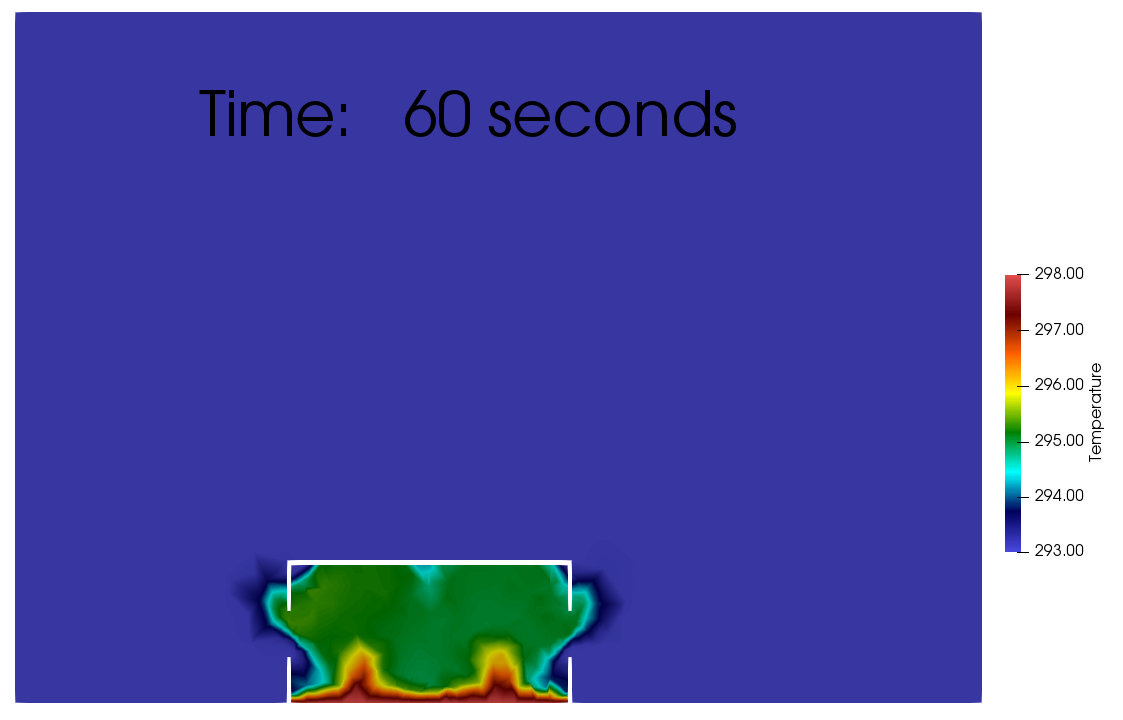}
    \caption{Temperature field within the box in example \textit{3dBox\_Case1a.flml} with ground temperature set at 298 K.}
    \label{Fig:Case1a_Results}
\end{figure}

\noindent A snapshot of the result obtained at 60 s is shown in Figure~\ref{Fig:Case1a_Results}. Go to Chapter~\ref{Sec:PostProcessing} to learn how to visualise the results using \textbf{ParaView}.

\subsubsection{Floor temperature as a function of space and/or time: python script}\label{Sec:TempDirichletPython}
\begin{figure}
    \centering
    \begin{subfigure}{0.3\textwidth}
        \includegraphics[width=\textwidth]{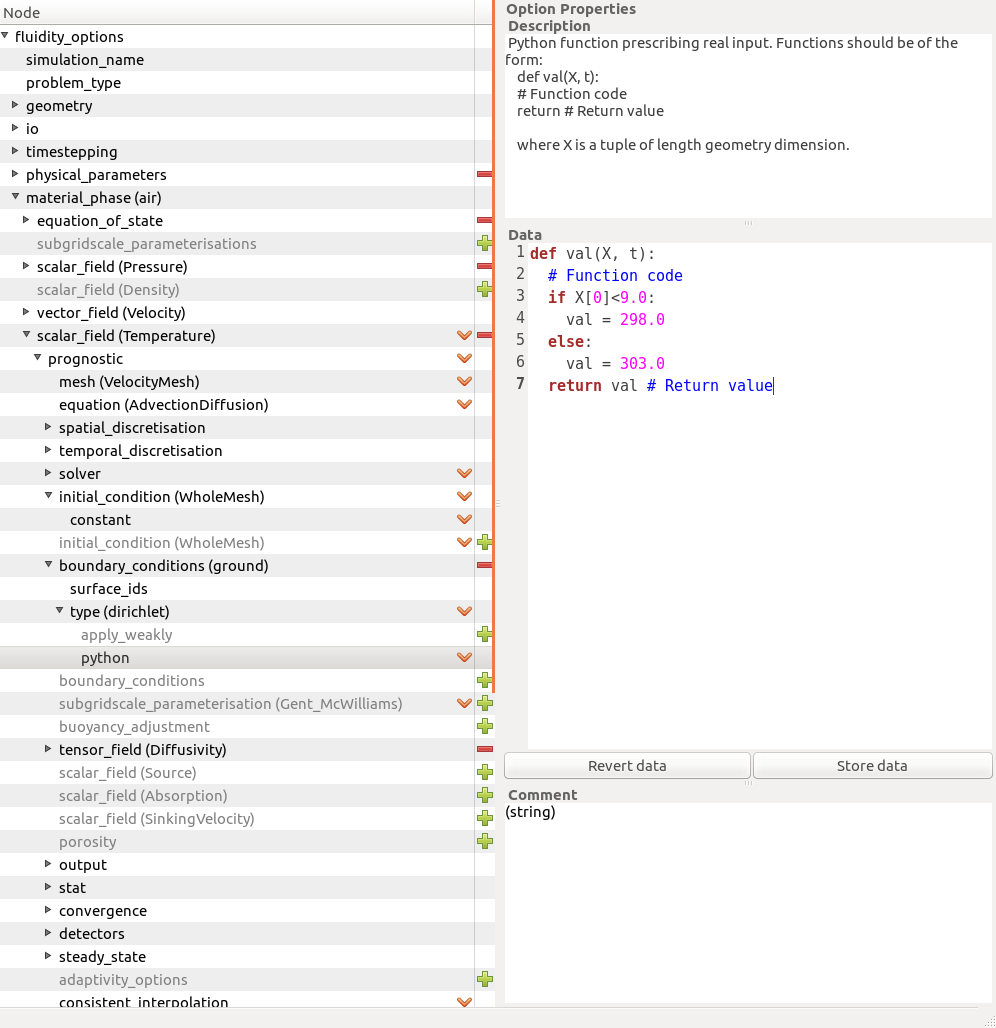}
        \caption{\textit{3dBox\_Case1b.flml}}
        \label{Fig:Case1b_TempBC}
    \end{subfigure}
    \begin{subfigure}{0.3\textwidth}
        \includegraphics[width=\textwidth]{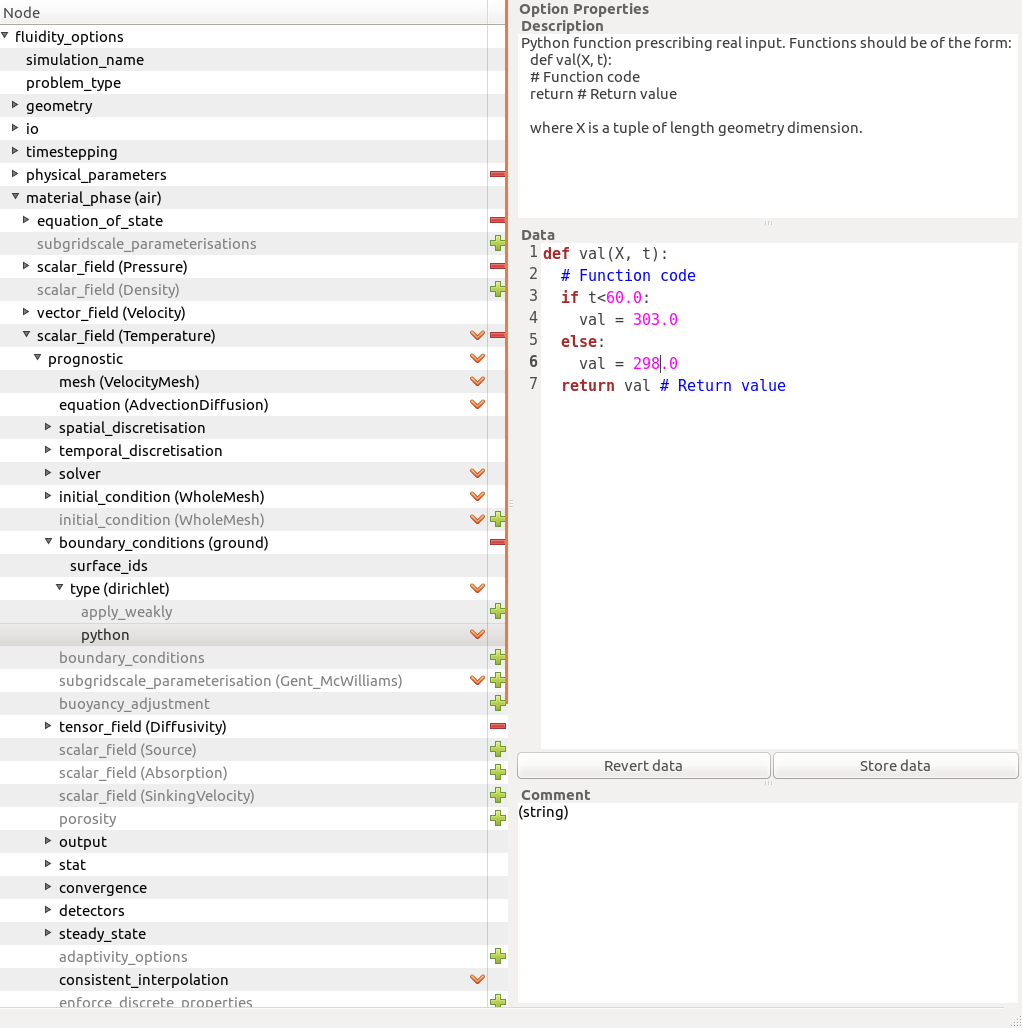}
        \caption{\textit{3dBox\_Case1c.flml}}
        \label{Fig:Case1c_TempBC}
    \end{subfigure}
    \begin{subfigure}{0.3\textwidth}
        \includegraphics[width=\textwidth]{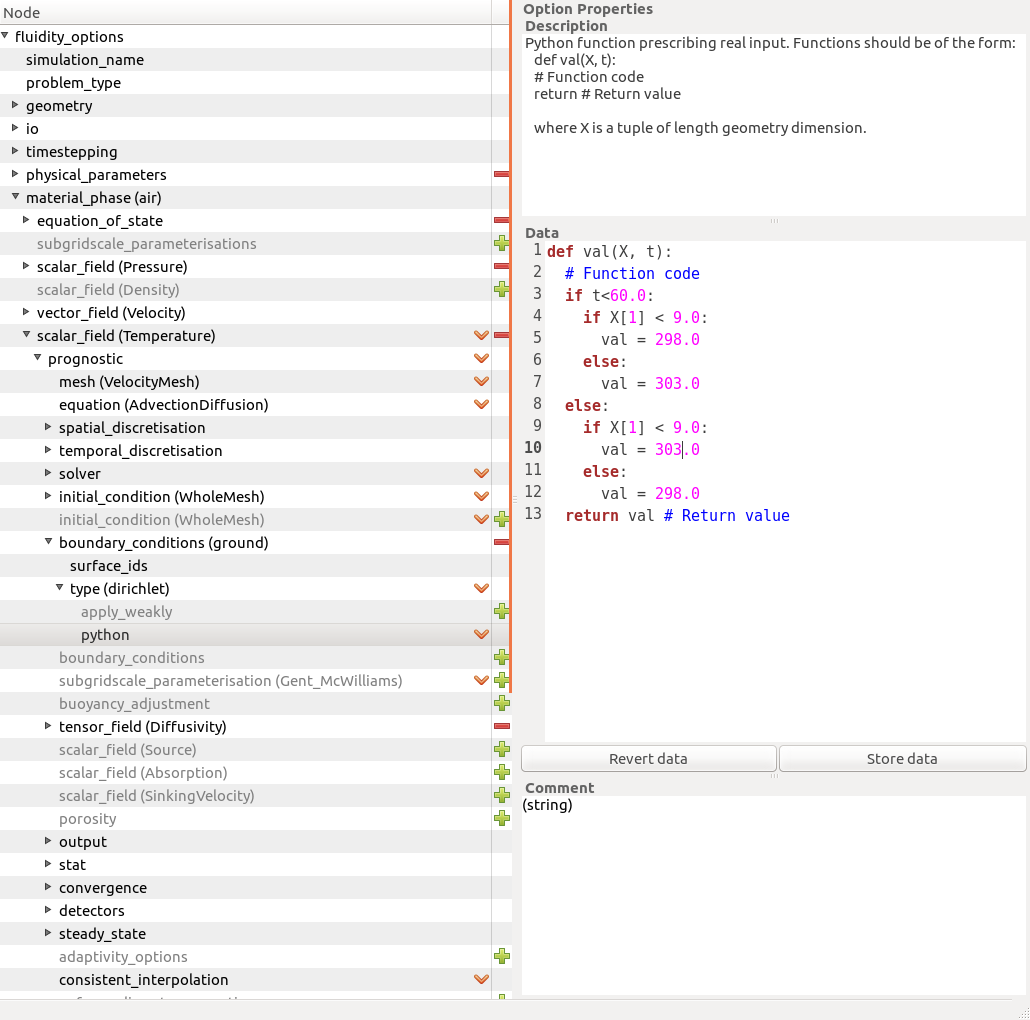}
        \caption{\textit{3dBox\_Case1d.flml}}
        \label{Fig:Case1d_TempBC}
    \end{subfigure}
    \caption{Space and/or time dependent Dirichlet boundary condition for the temperature. (a) Space dependent boundary condition, (b) time dependent boundary condition and (c) space and time dependent boundary condition.}
    \label{Fig:Case1_TempBC}
\end{figure}

In some cases, the user might want to prescribe a Dirichlet boundary that is space and/or time dependent. This can be done in \textbf{Diamond} using a python script. The following examples show how to prescribe these kinds of boundary conditions.
\begin{itemize}
    \item \textbf{Space dependent boundary condition:} In \textit{3dBox\_Case1b.flml}, as shown in Figure~\ref{Fig:Case1b_TempBC}, the python script Code~\ref{Lst:TempBCTx} is used.
    \begin{Code}[language=python, caption={Space dependent Dirichlet boundary condition for temperature.}, label={Lst:TempBCTx}]
def val(X, t):
  # Function code
  if X[0]<9.0:
    val = 298.0
  else:
    val = 303.0
  return val # Return value
    \end{Code}
    \item \textbf{Time dependent boundary condition:} In \textit{3dBox\_Case1c.flml}, as shown in Figure~\ref{Fig:Case1c_TempBC}, the python script Code~\ref{Lst:TempBCTt} is used.
    \begin{Code}[language=python, caption={Time dependent Dirichlet boundary condition for temperature.}, label={Lst:TempBCTt}]
def val(X, t):
  # Function code
  if t<60.0:
    val = 303.0
  else:
    val = 298.0
  return val # Return value
    \end{Code}
    \item \textbf{Space and time dependent boundary condition:} In \textit{3dBox\_Case1d.flml}, as shown in Figure~\ref{Fig:Case1d_TempBC}, the python script Code~\ref{Lst:TempBCTxt} is used.
    \begin{Code}[language=python, caption={Space and time dependent Dirichlet boundary condition for temperature.}, label={Lst:TempBCTxt}]
def val(X, t):
  # Function code
  if t<60.0:
    if X[1] < 9.0:
      val = 298.0
    else:
      val = 303.0
  else:
    if X[1] < 9.0:
      val = 303.0
    else:
      val = 298.0
  return val # Return value
\end{Code}
\end{itemize}

\noindent These examples can be run using the commands:
\begin{Terminal}[]
ä\colorbox{davysgrey}{
\parbox{435pt}{
\color{applegreen} \textbf{user@mypc}\color{white}\textbf{:}\color{codeblue}$\sim$
\color{white}\$ <<FluiditySourcePath>>/bin/fluidity -l -v3 3dBox\_Case1b.flml \&
\newline
\color{applegreen} \textbf{user@mypc}\color{white}\textbf{:}\color{codeblue}$\sim$
\color{white}\$ <<FluiditySourcePath>>/bin/fluidity -l -v3 3dBox\_Case1c.flml \&
\newline
\color{applegreen} \textbf{user@mypc}\color{white}\textbf{:}\color{codeblue}$\sim$
\color{white}\$ <<FluiditySourcePath>>/bin/fluidity -l -v3 3dBox\_Case1d.flml \&
}}
\end{Terminal}

\begin{figure}
    \centering
    \begin{subfigure}{0.32\textwidth}
        \includegraphics[width=\textwidth]{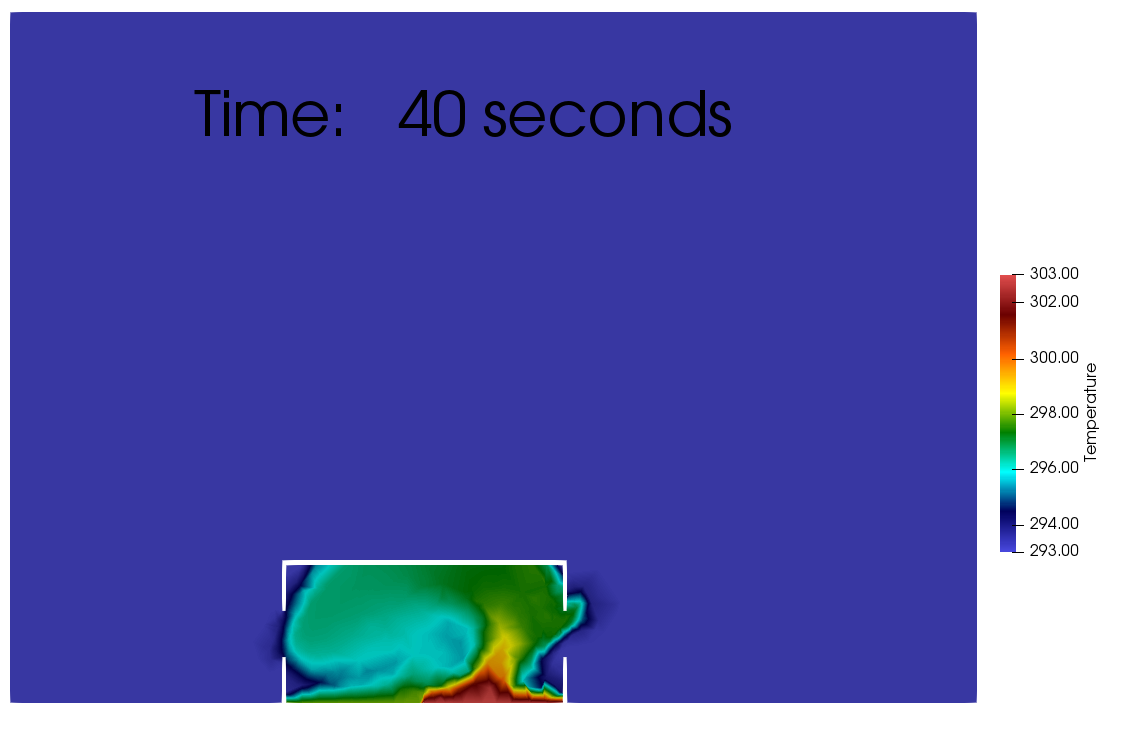}
        \caption{\textit{3dBox\_Case1b.flml}}
        \label{Fig:Case1b_Results}
    \end{subfigure}
    \begin{subfigure}{0.32\textwidth}
        \includegraphics[width=\textwidth]{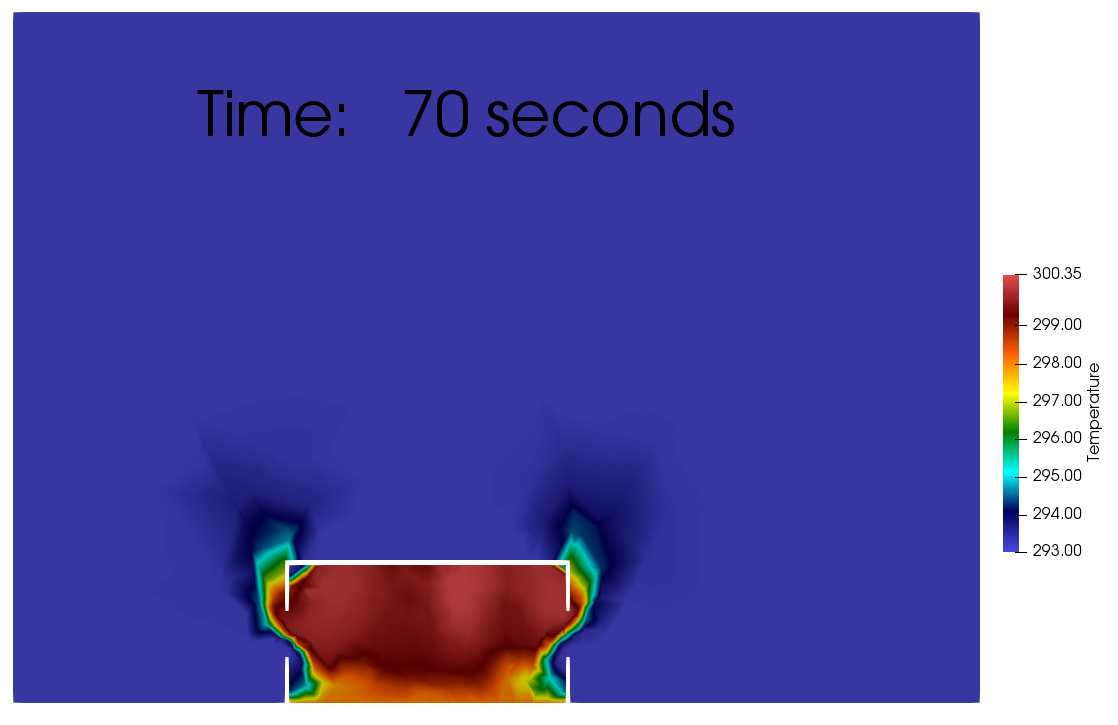}
        \caption{\textit{3dBox\_Case1c.flml}}
        \label{Fig:Case1c_Results}
    \end{subfigure}
    \begin{subfigure}{0.32\textwidth}
        \includegraphics[width=\textwidth]{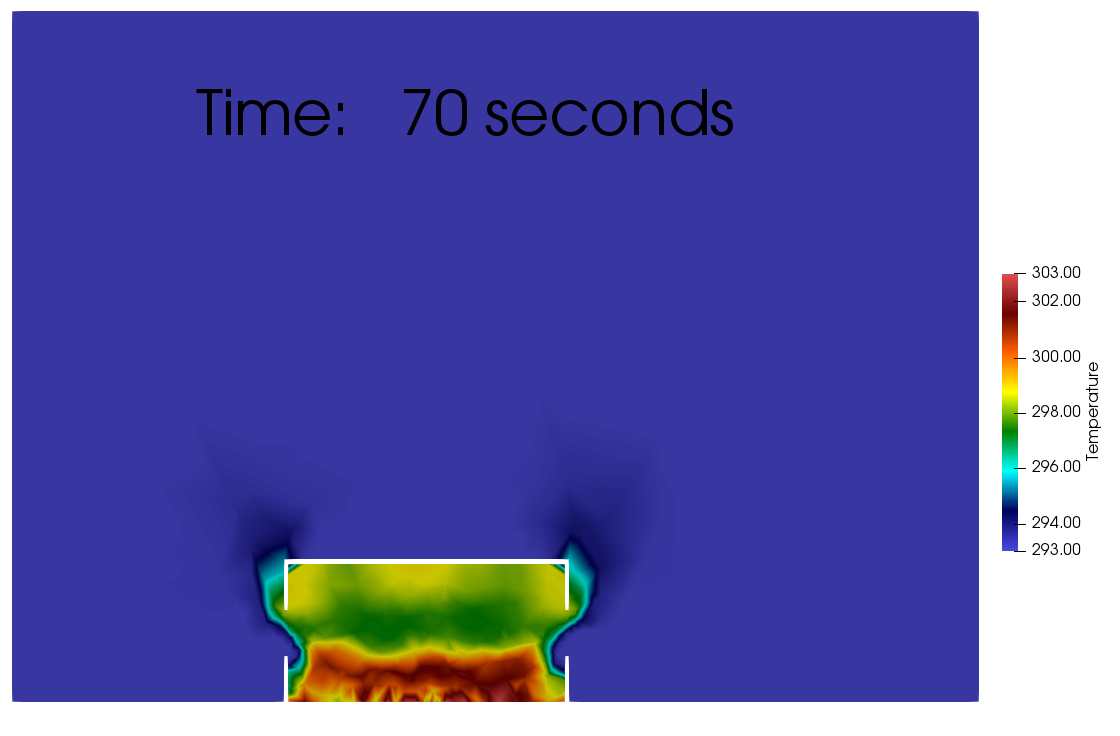}
        \caption{\textit{3dBox\_Case1d.flml}}
        \label{Fig:Case1d_Results}
    \end{subfigure}
    \caption{Temperature field within the box when using (a) a space dependent, (b) a time dependent or (c) a space and time dependent Dirichlet boundary condition for the ground of the box.}
    \label{Fig:Case1_Results}
\end{figure}

\noindent Snapshots of the results obtained from these three simulations are shown in Figure~\ref{Fig:Case1_Results}. Go to Chapter~\ref{Sec:PostProcessing} to learn how to visualise the results using \textbf{ParaView}.

\subsection{Neumann boundary condition: Heat flux}
\subsubsection{Heat flux from the ground}\label{Sec:TempNeumannUniFloor}
\begin{figure}
    \centering
    \begin{subfigure}{0.49\textwidth}
        \includegraphics[width=\textwidth]{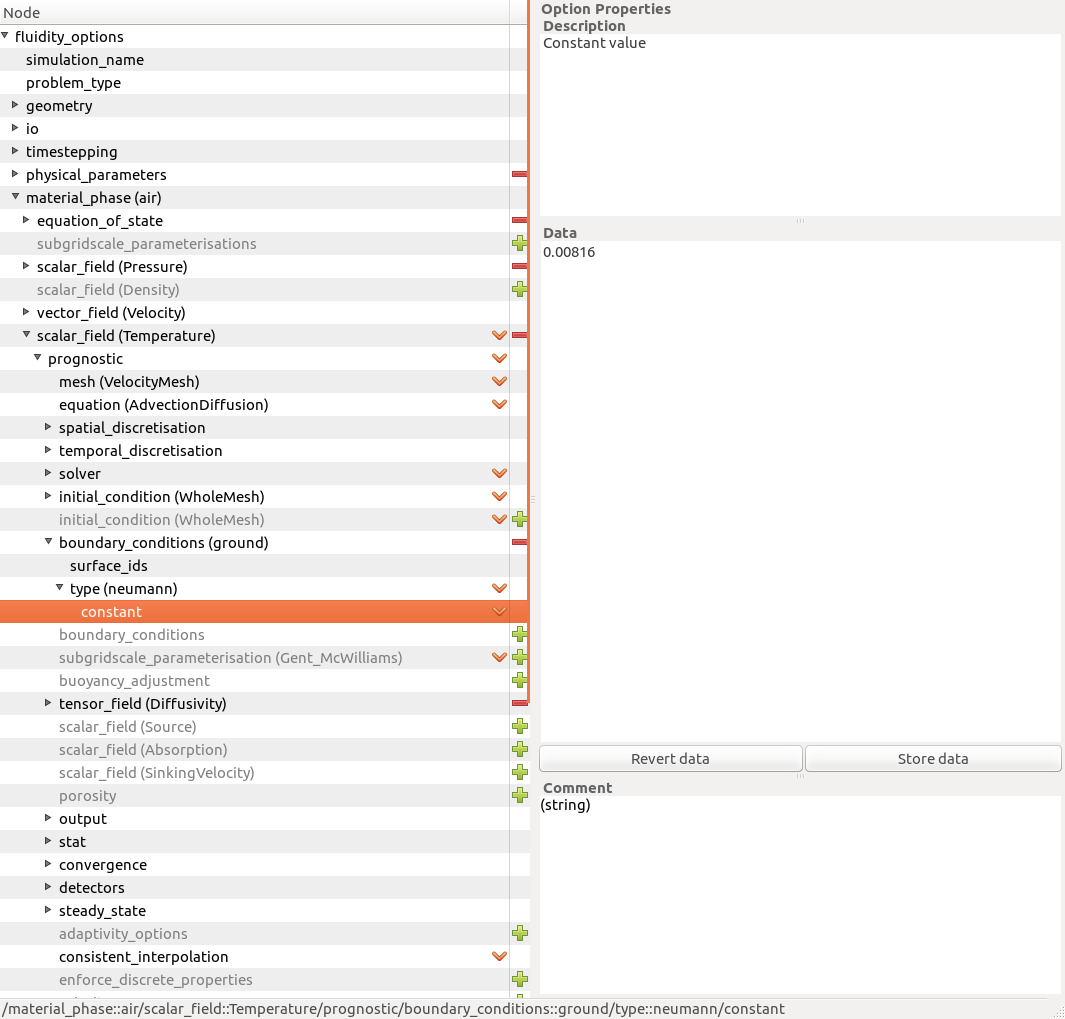}
        \caption{\textit{3dBox\_Case2a.flml}}
        \label{Fig:Case2a_TempBCFlux}
    \end{subfigure}
    \begin{subfigure}{0.49\textwidth}
        \includegraphics[width=\textwidth]{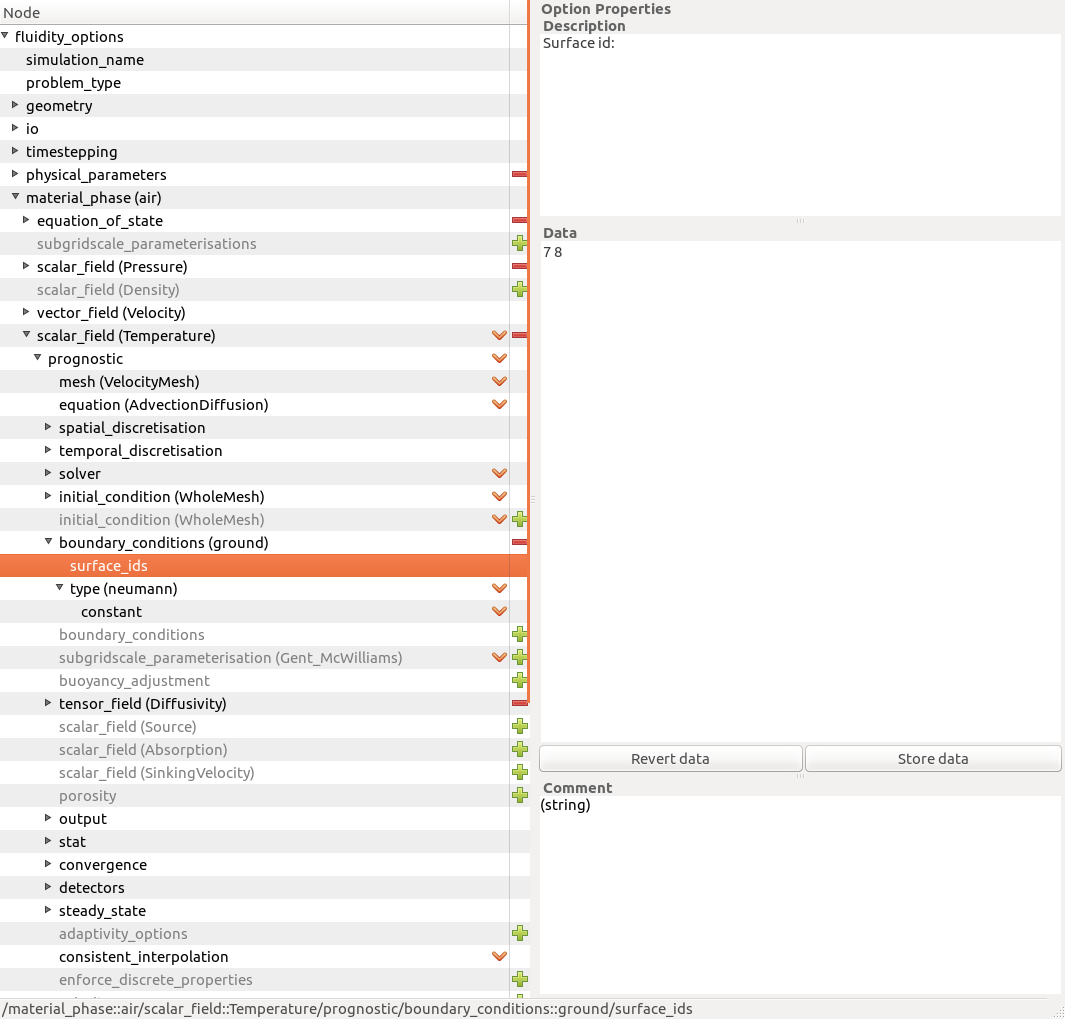}
        \caption{\textit{3dBox\_Case2a.flml}}
        \label{Fig:Case2a_TempBCID}
    \end{subfigure}
    \begin{subfigure}{0.49\textwidth}
        \includegraphics[width=\textwidth]{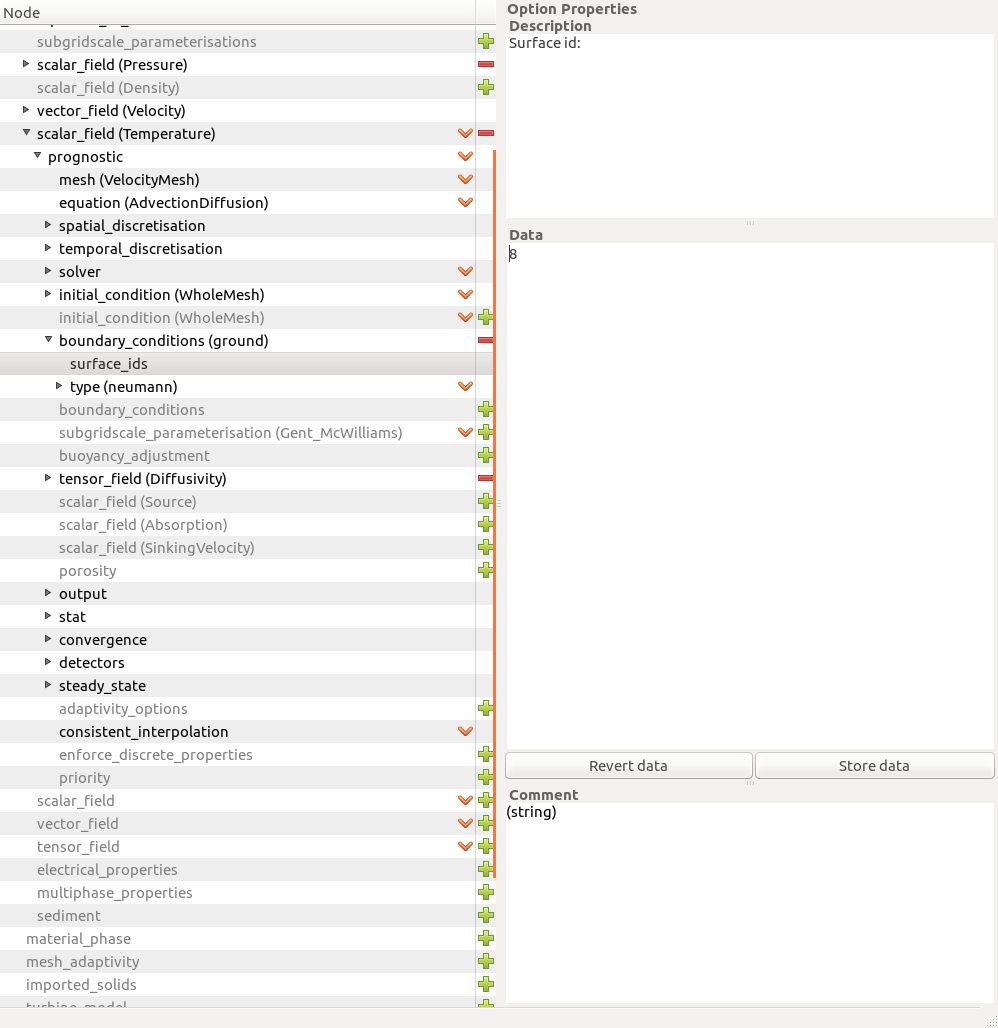}
        \caption{\textit{3dBox\_Case2b.flml}}
        \label{Fig:Case2b_TempBCID}
    \end{subfigure}
        \begin{subfigure}{0.49\textwidth}
        \includegraphics[width=\textwidth]{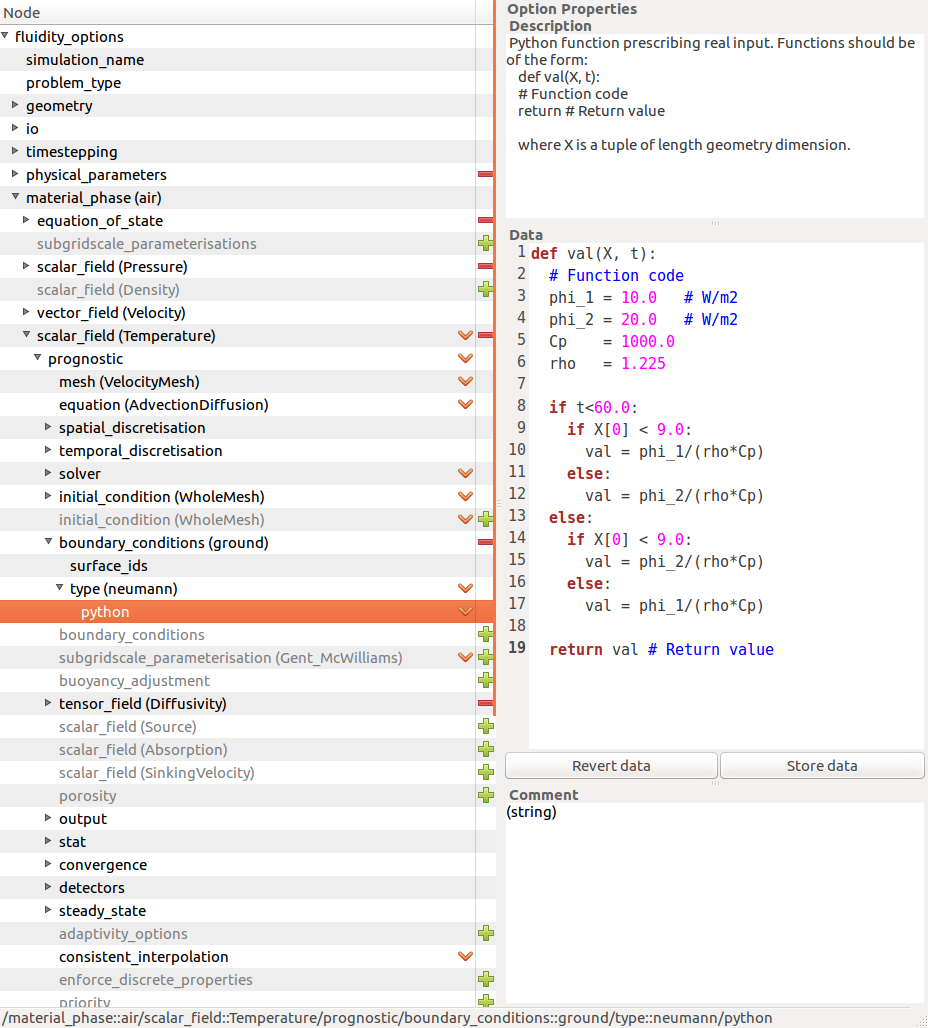}
        \caption{\textit{3dBox\_Case2c.flml}}
        \label{Fig:Case2c_TempBCID}
    \end{subfigure}
    \caption{Neumann boundary condition (a) Value of $\phi_{fluidity}$, (b) List of IDs to prescribe a heat flux on the overall floor box, (c) The heat flux is applied at the source only and (d) The heat flux is space and time dependent and is prescribed using a python script.}
    \label{Fig:Case2_TempBC}
\end{figure}

Case \textit{3dBox\_Case2a.flml} is similar to the previous one, except a heat flux is now applied at the bottom, effectively changing the Dirichlet boundary condition to a Neumann one as shown in Figure~\ref{Fig:Case2a_TempBCFlux}. The ambient and the initial temperatures are set to 293 K and $u$ is set equal to $0$ m/s. In \textbf{Fluidity}, the value of the heat flux, specified in \textbf{Diamond}, is given by equation~\ref{Eq:FluxFluidity}:

\begin{equation}\label{Eq:FluxFluidity}
    \phi_{fluidity} = \frac{\phi_{floor}}{\rho_{0} c_{p}}
\end{equation}

\noindent where $\phi_{fluidity}$ (Km/s) is the value that needs to be set into \textbf{Diamond}, $\phi_{floor}$ (W/m\textsuperscript{2}) is the actual heat flux that the user wants to prescribe, $\rho_{0}$ (kg/m\textsuperscript{3}) is the reference density and $c_p$ (J/kg/K) is the heat capacity of the fluid.

\noindent In example \textit{3dBox\_Case2a.flml}, the heat flux $\phi_{floor}$ is equal to 10 W/m\textsuperscript{2}, hence the value $\frac{\phi_{floor}}{\rho_{0} c_{p}}=\frac{10}{1,225\times1000}=0.00816$ is set up in \textbf{Diamond} as shown in Figure~\ref{Fig:Case2a_TempBCFlux}. The 10 W/m\textsuperscript{2} heat flux is applied to the box's floor having a surface equal to $6\times6=36$ m\textsuperscript{2}: the flux is then equal to $360$ W.

\noindent This example can be run using the command:
\begin{Terminal}[]
ä\colorbox{davysgrey}{
\parbox{435pt}{
\color{applegreen} \textbf{user@mypc}\color{white}\textbf{:}\color{codeblue}$\sim$
\color{white}\$ <<FluiditySourcePath>>/bin/fluidity -l -v3 3dBox\_Case2a.flml \&
}}
\end{Terminal}

\begin{figure}
    \centering
    \begin{subfigure}{0.32\textwidth}
        \includegraphics[width=\textwidth]{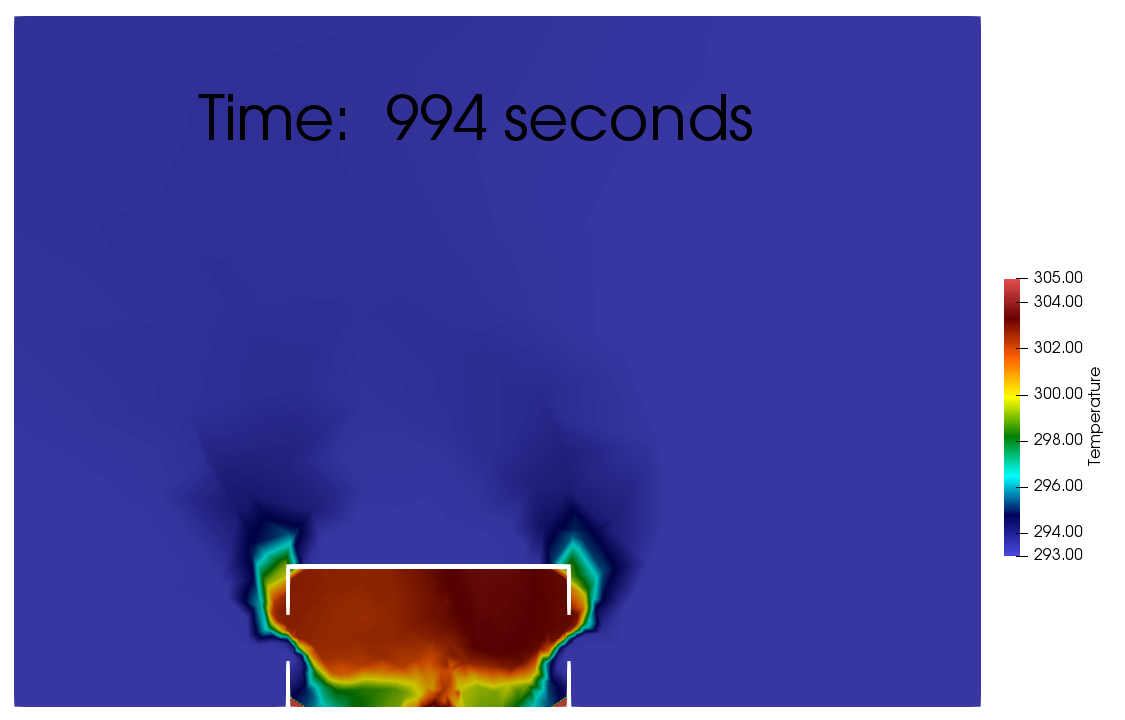}
        \caption{\textit{3dBox\_Case2a.flml}}
        \label{Fig:Case2a_Results}
    \end{subfigure}
    \begin{subfigure}{0.32\textwidth}
        \includegraphics[width=\textwidth]{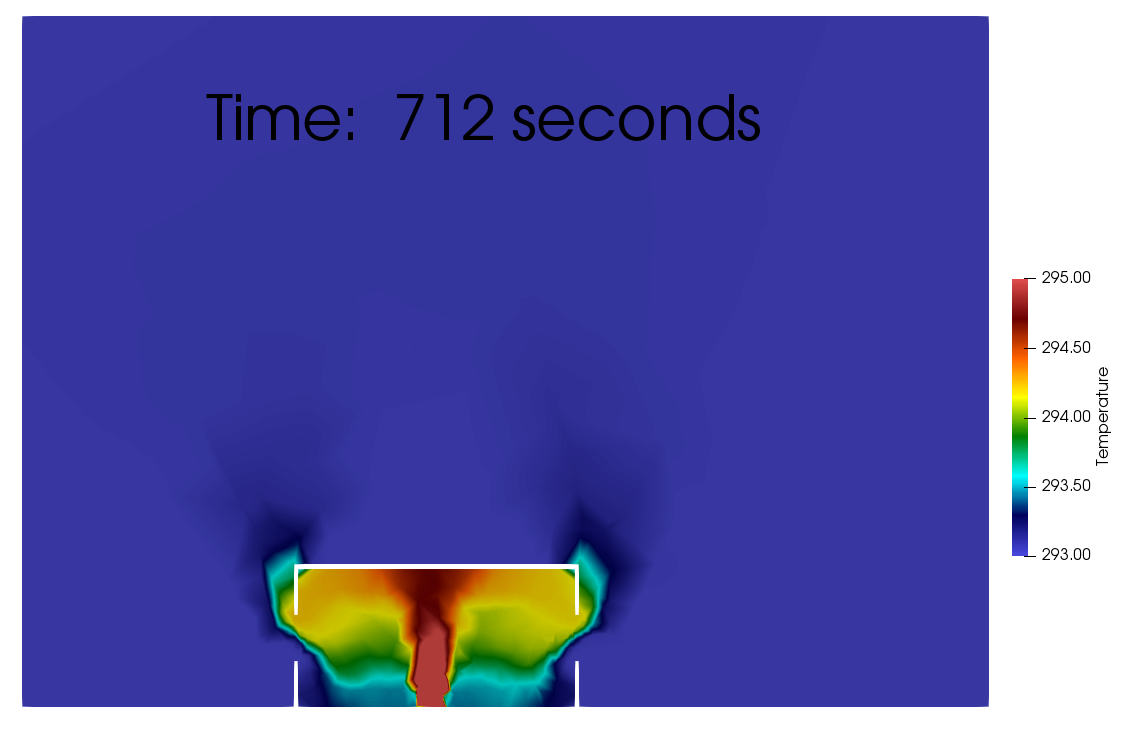}
        \caption{\textit{3dBox\_Case2b.flml}}
        \label{Fig:Case2b_Results}
    \end{subfigure}
    \begin{subfigure}{0.32\textwidth}
        \includegraphics[width=\textwidth]{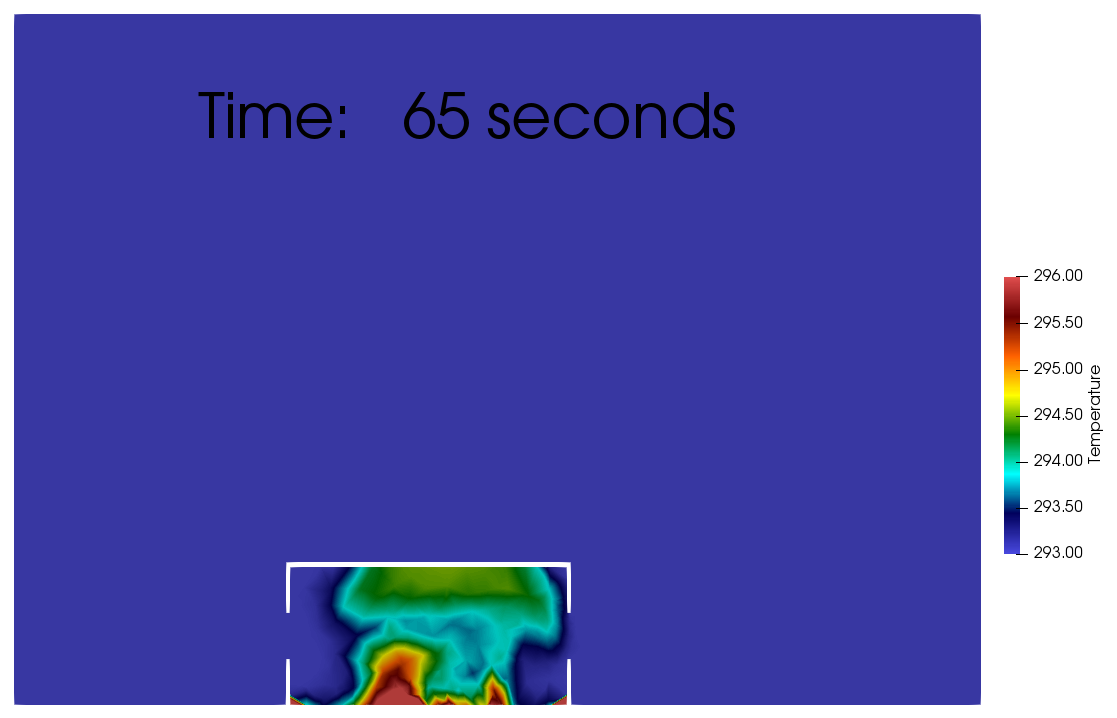}
        \caption{\textit{3dBox\_Case2c.flml}}
        \label{Fig:Case2c_Results}
    \end{subfigure}
    \caption{Temperature field within the box when using (a) heat flux on the ground, (b) a heat flux in the middle of the box only and (c) a space and time dependent heat flux on the ground.}
    \label{Fig:Case2_Results}
\end{figure}

\noindent A snapshot of the result obtained at 994 s is shown in Figure~\ref{Fig:Case2a_Results}. Go to Chapter~\ref{Sec:PostProcessing} to learn how to visualise the results using \textbf{ParaView}.

\subsubsection{Heat flux from the source only}\label{Sec:TempNeumannUniSource}
Using the physical IDs previously defined (see Section~\ref{Sec:PhysicalIDs}), the heat flux boundary condition can be changed to include only a section of the floor, defined as the source. Based on Section~\ref{Sec:PhysicalIDs}, the ID of the floor of the box (without the source) is 7 and the ID of the source only is 8. In the previous example \textit{3dBox\_Case2a.flml}, the surface IDs, where the Neumann boundary condition were applied, are 7 (the floor) and 8 (the source) (see Figure~\ref{Fig:Case2a_TempBCID}): in that case, the heat flux is applied everywhere on the floor. In example \textit{3dBox\_Case2b.flml}, the surface ID where the Neumann boundary condition is applied is 8 only (the source) (see Figure~\ref{Fig:Case2b_TempBCID}). Therefore the heat flux is applied at the source only. 

\noindent In case \textit{3dBox\_Case2b.flml}, the initial and the ambient temperatures are set to 293 K and $u$ is set equal to $0$ m/s. The heat flux applied at the source $\phi_{source}$ is equal to 1000 W/m\textsuperscript{2}, hence $\phi_{fluidity}$ set up in \textbf{Diamond} is equal to 0.8163 according to equation~\ref{Eq:FluxFluidity}. The surface of the heat source is $0.2 \times 0.2$ m\textsuperscript{2}: the flux applied at the source is then equal to 40 W.

\noindent This example can be run using the command:
\begin{Terminal}[]
ä\colorbox{davysgrey}{
\parbox{435pt}{
\color{applegreen} \textbf{user@mypc}\color{white}\textbf{:}\color{codeblue}$\sim$
\color{white}\$ <<FluiditySourcePath>>/bin/fluidity -l -v3 3dBox\_Case2b.flml \&
}}
\end{Terminal}

\noindent A snapshot of the result obtained at 712 s is shown in Figure~\ref{Fig:Case2b_Results}. Go to Chapter~\ref{Sec:PostProcessing} to learn how to visualise the results using \textbf{ParaView}.

\subsubsection{Heat flux as a function of space or time: python script}\label{Sec:TempNeumannPython}
In some cases, the user might want to prescribe a Neumann boundary that is space and/or time dependent. This can be done in \textbf{Diamond} using a python script as shown in Figure~\ref{Fig:Case2c_TempBCID}.

\noindent In example \textit{3dBox\_Case2c.flml}, the python script Code~\ref{Lst:TempBCNeumann} is used. Note that in this example, the choice was made to calculate the heat flux $\phi_{fluidity}$ directly in the python script.

\begin{Code}[language=python, caption={Space and time dependent Neumann boundary condition for temperature field in example \textit{3dBox\_Case2c.flml}.}, label={Lst:TempBCNeumann}]
def val(X, t):
  # Function code
  phi_1 = 10.0   # W/m2
  phi_2 = 20.0   # W/m2
  Cp    = 1000.0
  rho   = 1.225
  
  if t<60.0:
    if X[0] < 9.0:
      val = phi_1/(rho*Cp)
    else:
      val = phi_2/(rho*Cp)
  else:
    if X[0] < 9.0:
      val = phi_2/(rho*Cp)
    else:
      val = phi_1/(rho*Cp)
  
  return val # Return value
\end{Code}

\noindent This example can be run using the command: 
\begin{Terminal}[]
ä\colorbox{davysgrey}{
\parbox{435pt}{
\color{applegreen} \textbf{user@mypc}\color{white}\textbf{:}\color{codeblue}$\sim$
\color{white}\$ <<FluiditySourcePath>>/bin/fluidity -l -v3 3dBox\_Case2c.flml \&
}}
\end{Terminal}

\noindent A snapshot of the result obtained at 65 s is shown in Figure~\ref{Fig:Case2c_Results}. Go to Chapter~\ref{Sec:PostProcessing} to learn how to visualise the results using \textbf{ParaView}.

\subsection{Robin boundary condition}\label{Sec:TempRobin}
\begin{figure}
    \centering
    \begin{subfigure}{0.45\textwidth}
        \includegraphics[width=\textwidth]{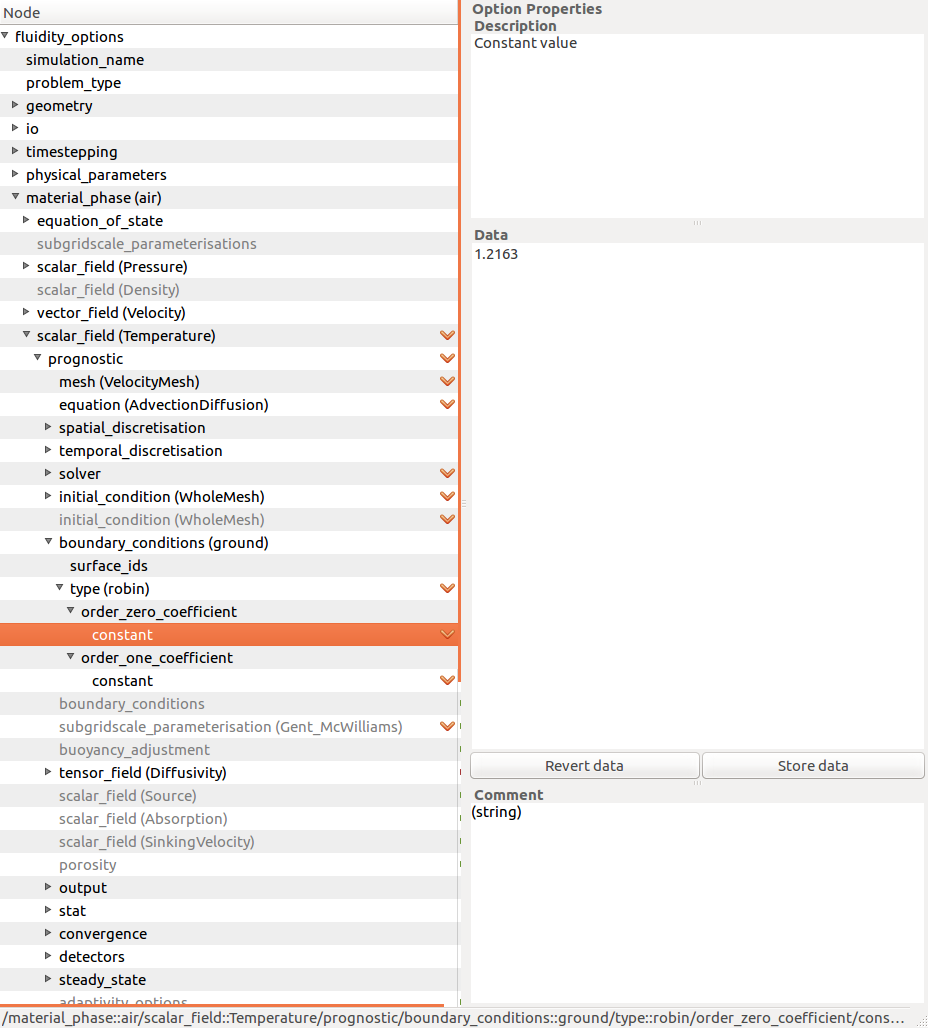}
        \caption{$C_0$ coefficient}
        \label{Fig:Case3_RobinC0}
    \end{subfigure}
    \begin{subfigure}{0.45\textwidth}
        \includegraphics[width=\textwidth]{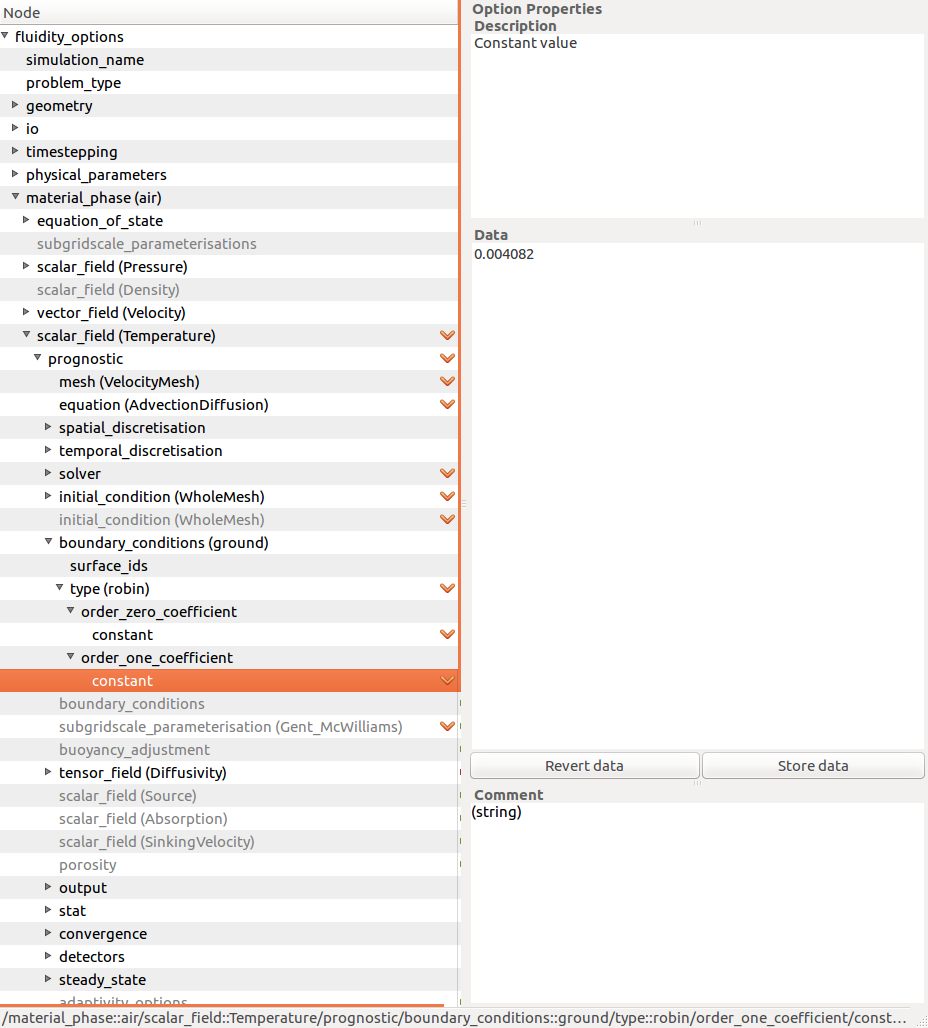}
        \caption{$C_1$ coefficient}
        \label{Fig:Case3_RobinC1}
    \end{subfigure}
    \caption{Robin boundary condition used in example \textit{3dBox\_Case3.flml}.}
    \label{Fig:Case3_Robin}
\end{figure}

In example \textit{3dBox\_Case3.flml}, a Robin boundary condition (equation~\ref{Eq:RobinBC}) is applied between the box's floor and the air as shown in Figure~\ref{Fig:Case3_Robin}. Generally this boundary condition is used to model the heat exchange by convection between a fluid and a solid surface.

\begin{equation}\label{Eq:RobinBC}
    -(\overline{\overline{\kappa}} \nabla T)\cdot\textbf{n} = \frac{h}{\rho_{0} c_{p}} (T - T_{\infty})
\end{equation}

\noindent where $\overline{\overline{\kappa}}$ is the thermal diffusivity (m\textsuperscript{2}/s) of the fluid, $h$ is the convective heat transfer coefficient (W/m\textsuperscript{2}/K) and $T_{\infty}$ is the ground temperature (K).

\noindent In \textbf{Fluidity}, the Robin boundary condition is specified using the equation~\ref{Eq:Robin}.

\begin{equation}\label{Eq:Robin}
    C_{1} T + \textbf{n}\cdot(\overline{\overline{\kappa}} \nabla T) = C_{0}
\end{equation}

\noindent Hence, the coefficients $C_0$ and $C_1$ are given by equation~\ref{Eq:RobinC0} and equation~\ref{Eq:RobinC1}, respectively.

\begin{equation}\label{Eq:RobinC0}
    C_{0} = \frac{h}{\rho_{0} c_{p}}T_{\infty}
\end{equation}

\begin{equation}\label{Eq:RobinC1}
    C_{1} = \frac{h}{\rho_{0} c_{p}}
\end{equation}

\noindent where $C_{0}$ is in Km/s and $C_{1}$ is in m/s. Note that if $C_{1}$ is equal to 0, then the Robin boundary condition leads to a Neumann boundary condition.

\noindent In example \textit{3dBox\_Case3.flml}, the heat transfer coefficient between the ground and the air is assumed to be equal to 5 W/m\textsuperscript{2}/K and the ground temperature $T_{\infty}$ is taken equal to 298 K. Hence, the value of $C_{0}$ and $C_{1}$ are equal to 1.2163 Km/s and 0.004082 m/s, respectively as shown in Figure~\ref{Fig:Case3_Robin}. The initial temperature is set to 293 K and $u$ is set equal to $0$ m/s.

\noindent This example can be run using the command:
\begin{Terminal}[]
ä\colorbox{davysgrey}{
\parbox{435pt}{
\color{applegreen} \textbf{user@mypc}\color{white}\textbf{:}\color{codeblue}$\sim$
\color{white}\$ <<FluiditySourcePath>>/bin/fluidity -l -v3 3dBox\_Case3.flml \&
}}
\end{Terminal}

\begin{figure}
    \centering
    \includegraphics[scale=0.14]{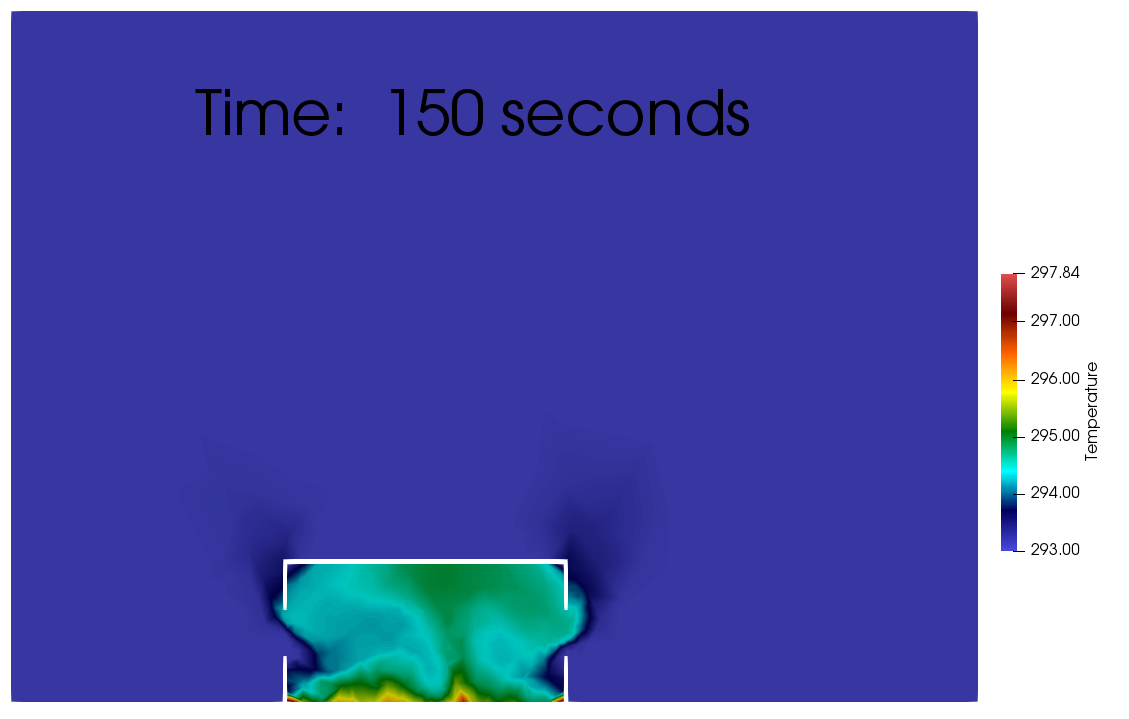}
    \caption{Temperature field within the box in example \textit{3dBox\_Case3.flml} using a Robin condition.}
    \label{Fig:Case3_Results}
\end{figure}

\noindent A snapshot of the result obtained at 150 s is shown in Figure~\ref{Fig:Case3_Results}. Go to Chapter~\ref{Sec:PostProcessing} to learn how to visualise the results using \textbf{ParaView}.

\section{Initial conditions for temperature}\label{Sec:TempInit}
\begin{figure}
    \centering
    \includegraphics[scale=0.2]{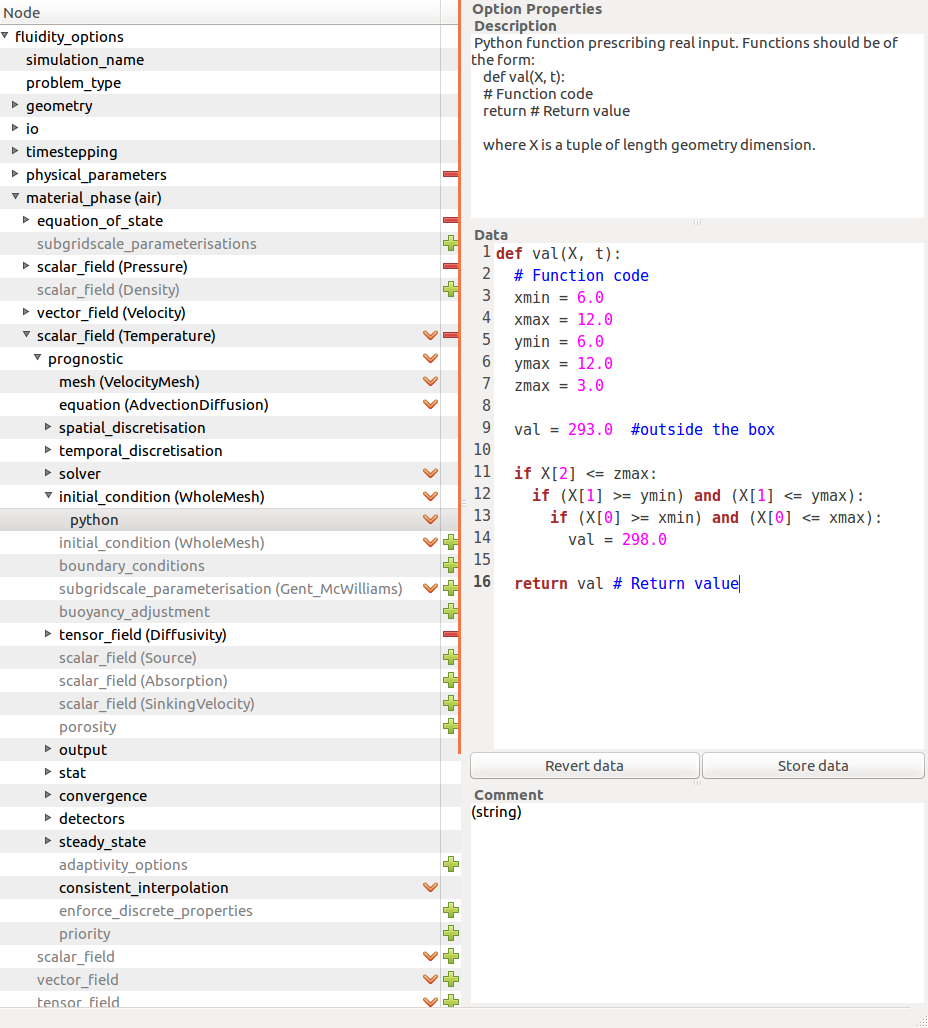}
    \caption{Example \textit{3dBox\_Case4.flml}: python script to prescribe different initial temperatures.}
    \label{Fig:Case4_Init}
\end{figure}

The initial temperature can be set using a python script to prescribe different initial values in different regions. In example \textit{3dBox\_Case4.flml}, the interior of the box is set to 298 K  while the outside remains at ambient temperature, i.e. 293 K. The python script in Code~\ref{Lst:InitTemp} is used as shown in Figure~\ref{Fig:Case4_Init}. No particular thermal boundary condition is prescribed for the floor of the box and $u$ is set to $0$ m/s.

\begin{Code}[language=python, caption={Python script to prescribe different initial temperatures inside and outside the box.}, label={Lst:InitTemp}]
def val(X, t):
  # Function code
  xmin = 6.0
  xmax = 12.0
  ymin = 6.0
  ymax = 12.0
  zmax = 3.0
  
  val = 293.0  # outside of the box
  
  if X[2] <= zmax:
    if (X[1] >= ymin) and (X[1] <= ymax):
      if (X[0] >= xmin) and (X[0] <= xmax):
        val = 298.0  # inside of the box
  
  return val # Return value
\end{Code}

\noindent This example can be run using the command:
\begin{Terminal}[]
ä\colorbox{davysgrey}{
\parbox{435pt}{
\color{applegreen} \textbf{user@mypc}\color{white}\textbf{:}\color{codeblue}$\sim$
\color{white}\$ <<FluiditySourcePath>>/bin/fluidity -l -v3 3dBox\_Case4.flml \&
}}
\end{Terminal}

\begin{figure}
    \centering
    \includegraphics[scale=0.15]{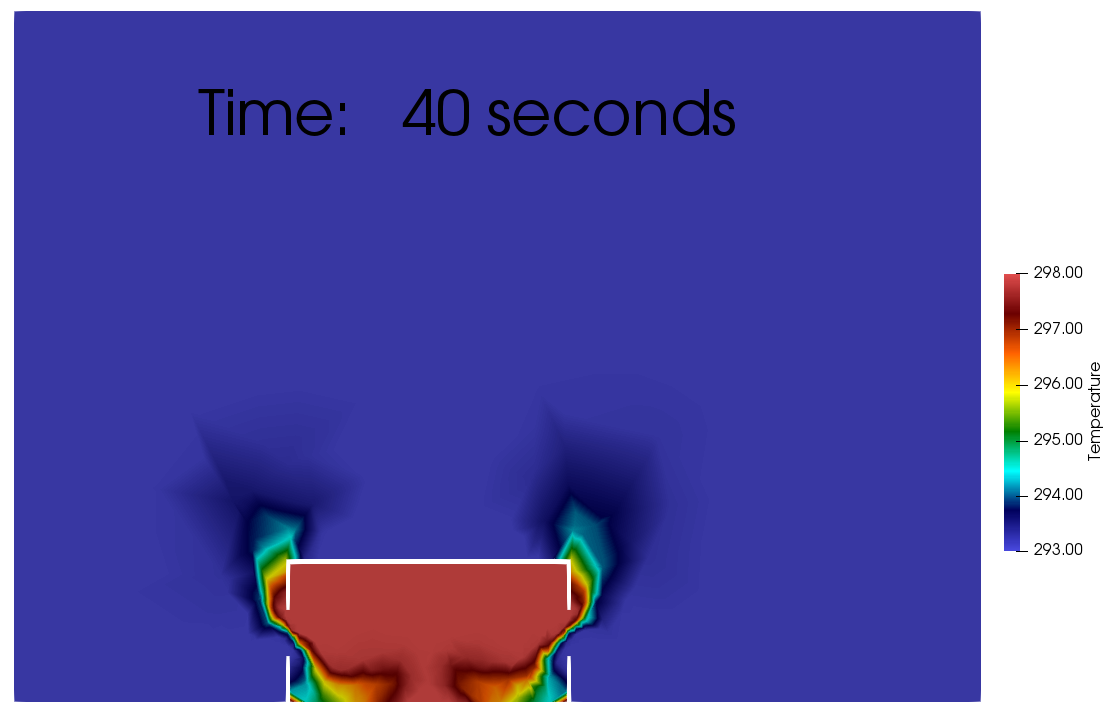}
    \caption{Temperature field in example \textit{3dBox\_Case4.flml} with an initial temperature in the box.}
    \label{Fig:Case4_Results}
\end{figure}

\noindent A snapshot of the result obtained at 40 s is shown in Figure~\ref{Fig:Case4_Results}. Go to Chapter~\ref{Sec:PostProcessing} to learn how to visualise the results using \textbf{ParaView}.

\section{Velocity boundary conditions}
\subsection{Dirichlet boundary condition: Constant and uniform inlet wind}\label{Sec:VelDiricUni}
\subsubsection{Uniform inlet velocity}
\begin{figure}
    \centering
    \begin{subfigure}{0.44\textwidth}
        \includegraphics[width=\textwidth]{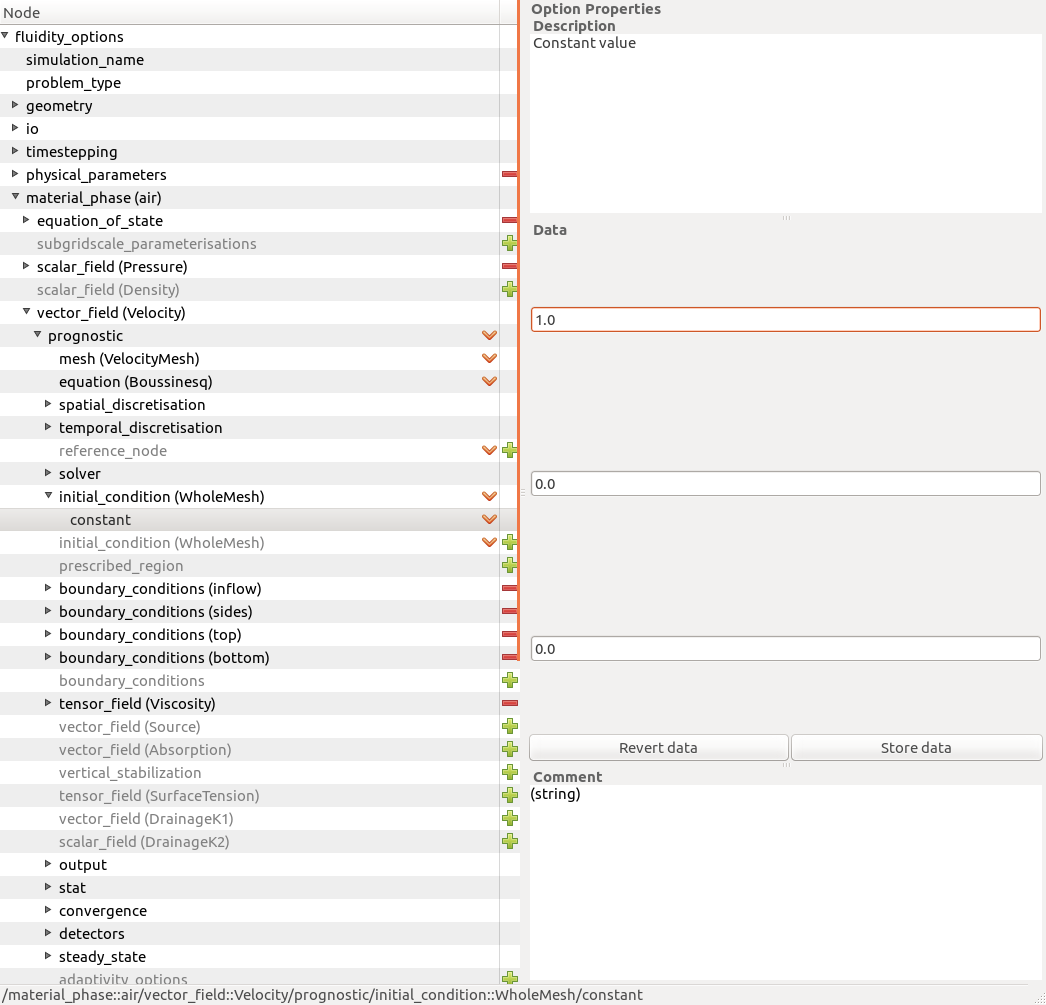}
        \caption{Initial velocity}
        \label{Fig:Case5a_VelInit}
    \end{subfigure}
    \begin{subfigure}{0.44\textwidth}
        \includegraphics[width=\textwidth]{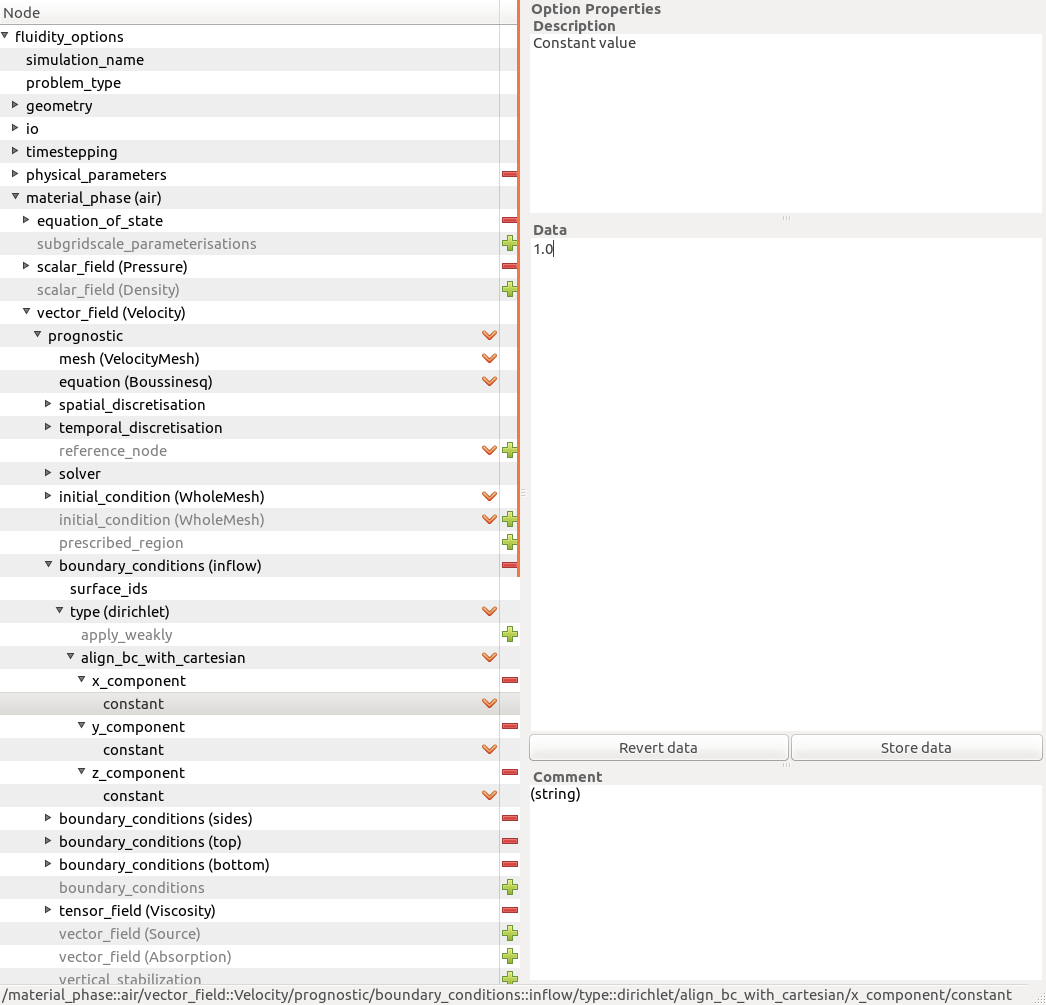}
        \caption{Inlet velocity}
        \label{Fig:Case5a_VelBC}
    \end{subfigure}
    \caption{(a) Initial and (b) inlet velocity prescribed in example \textit{3dBox\_Case5a.flml}.}
    \label{Fig:Case5a_VelConstant}
\end{figure}

In addition to setting boundary or initial values of temperature, the boundary condition for velocity can also be specified. In example \textit{3dBox\_Case5a.flml}, the initial and inlet velocities $(u,v,w)$ are set to $(1,0,0)$ m/s (Figure~\ref{Fig:Case5a_VelConstant}) and the interior of the box is set to 298 K while the outside remains at ambient temperature 293 K (Code~\ref{Lst:InitTemp}). The time step is kept equal to 1 s in this example. However, the results of the first 10 time steps are not totally accurate. To avoid this, a value of 0.01 s is recommended.

\noindent This example can be run using the command:
\begin{Terminal}[]
ä\colorbox{davysgrey}{
\parbox{435pt}{
\color{applegreen} \textbf{user@mypc}\color{white}\textbf{:}\color{codeblue}$\sim$
\color{white}\$ <<FluiditySourcePath>>/bin/fluidity -l -v3 3dBox\_Case5a.flml \&
}}
\end{Terminal}

\begin{figure}
    \centering
    \begin{subfigure}{0.35\textwidth}
        \includegraphics[width=\textwidth]{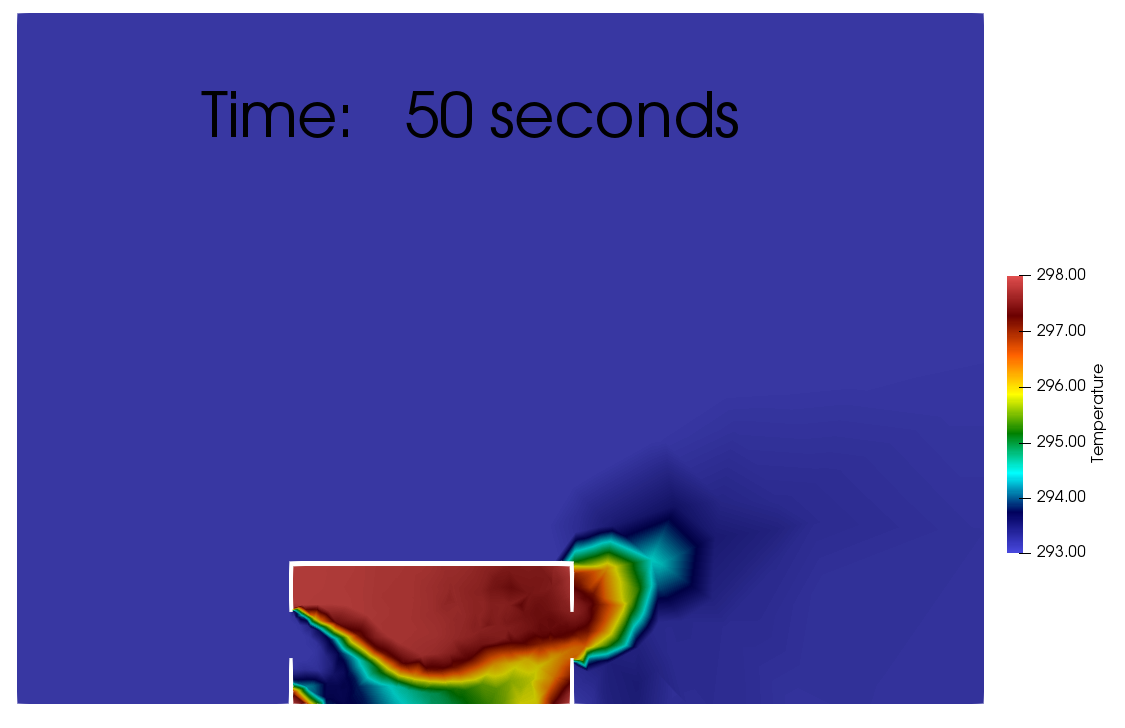}
        \caption{\textit{3dBox\_Case5a.flml}}
        \label{Fig:Case5a_Results}
    \end{subfigure}
    \begin{subfigure}{0.35\textwidth}
        \includegraphics[width=\textwidth]{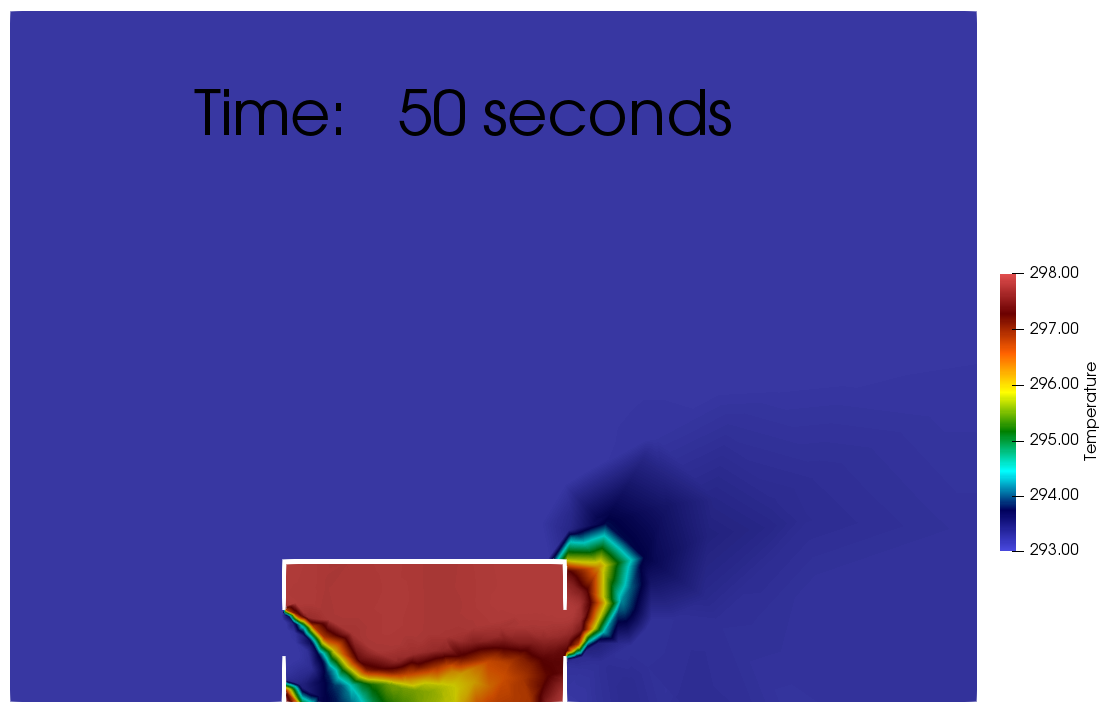}
        \caption{\textit{3dBox\_Case5b.flml}}
        \label{Fig:Case5b_Results}
    \end{subfigure}
    \begin{subfigure}{0.35\textwidth}
        \includegraphics[width=\textwidth]{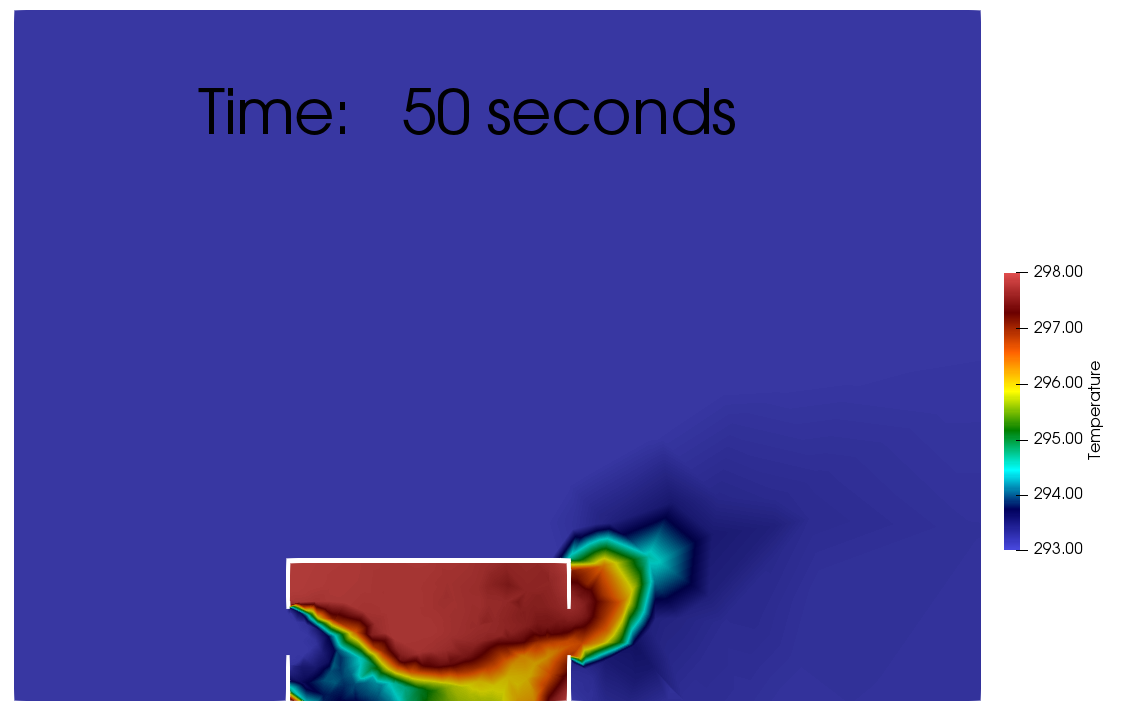}
        \caption{\textit{3dBox\_Case5c.flml}}
        \label{Fig:Case5c_Results}
    \end{subfigure}
    \begin{subfigure}{0.35\textwidth}
        \includegraphics[width=\textwidth]{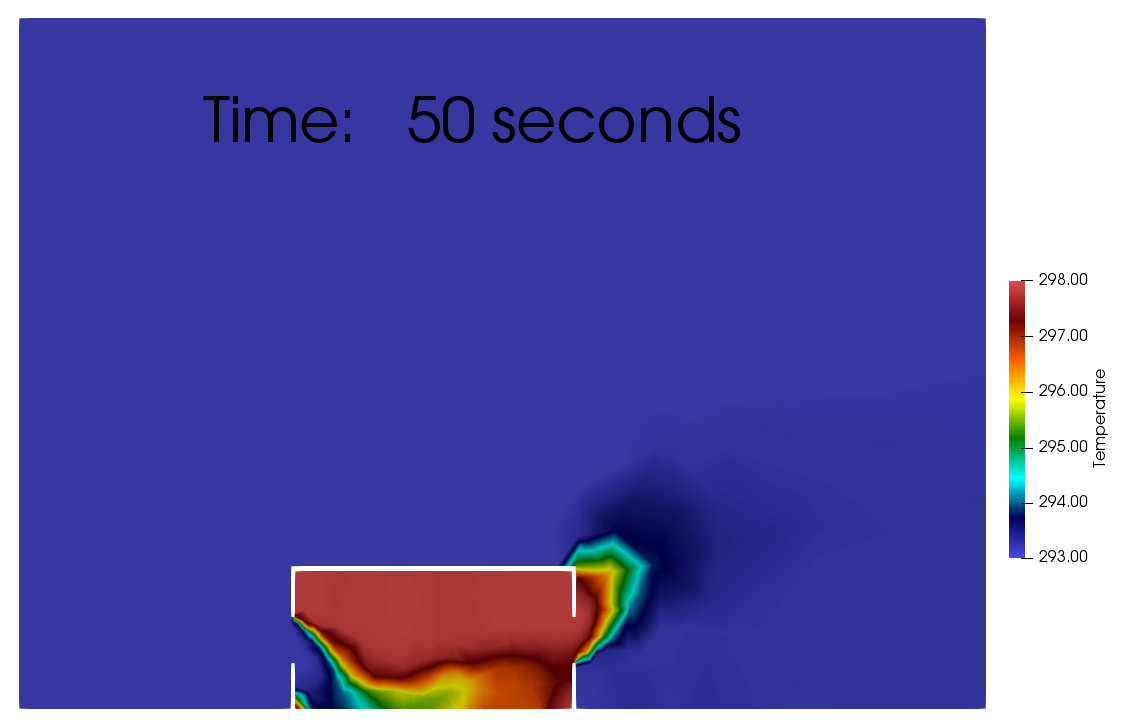}
        \caption{\textit{3dBox\_Case5d.flml}}
        \label{Fig:Case5d_Results}
    \end{subfigure}
    \caption{Temperature field at 50 s when using (a) a uniform and constant inlet velocity, (b) a log-profile inlet velocity, (c) a turbulent inlet velocity with uniform profiles and (d) a turbulent inlet velocity with a log-profile mean velocity.}
    \label{Fig:Case5_Results}
\end{figure}

\noindent A snapshot of the result obtained at 50 s is shown in Figure~\ref{Fig:Case5a_Results}. Go to Chapter~\ref{Sec:PostProcessing} to learn how to visualise the results using \textbf{ParaView}.

\subsubsection{Prescribing a velocity profile: python script}\label{Sec:VelDiricPython}
\begin{figure}
    \centering
    \begin{subfigure}{0.49\textwidth}
        \includegraphics[width=\textwidth]{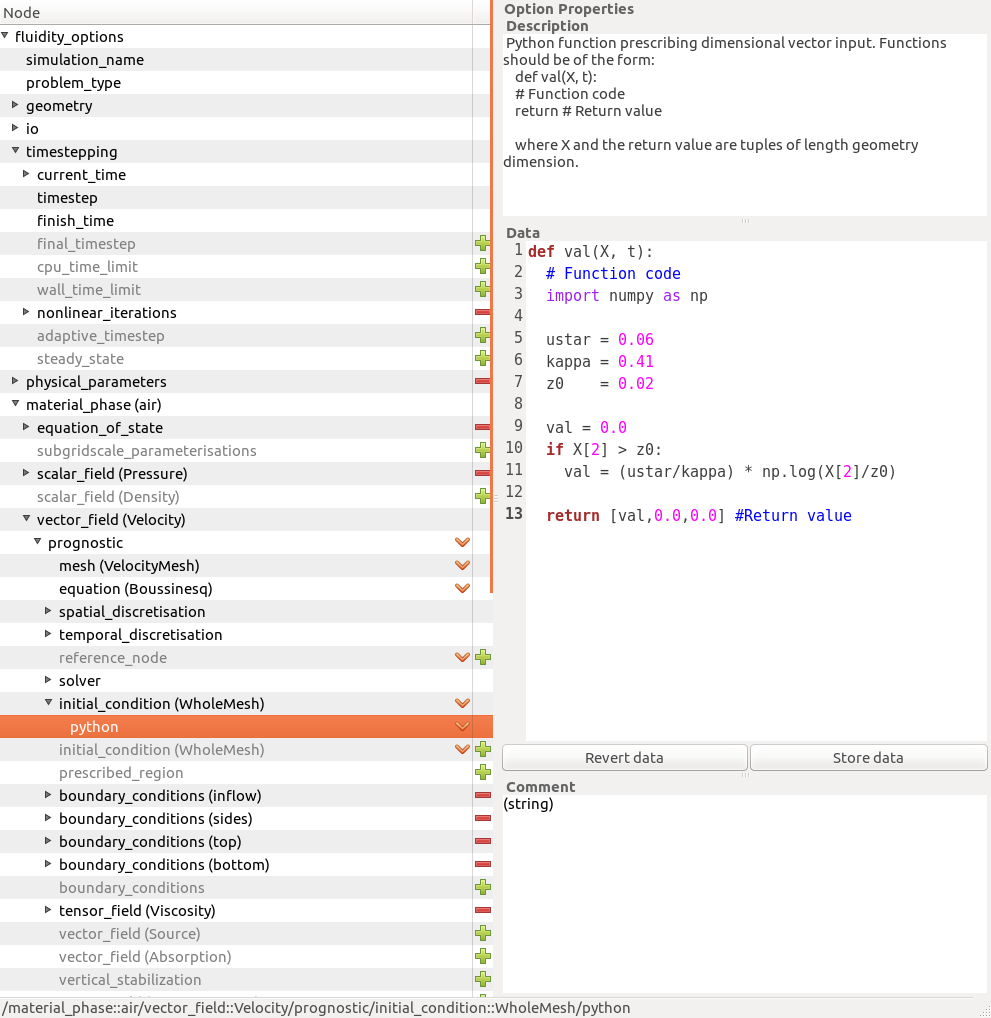}
        \caption{Initial velocity}
        \label{Fig:Case5b_VelInit}
    \end{subfigure}
    \begin{subfigure}{0.49\textwidth}
        \includegraphics[width=\textwidth]{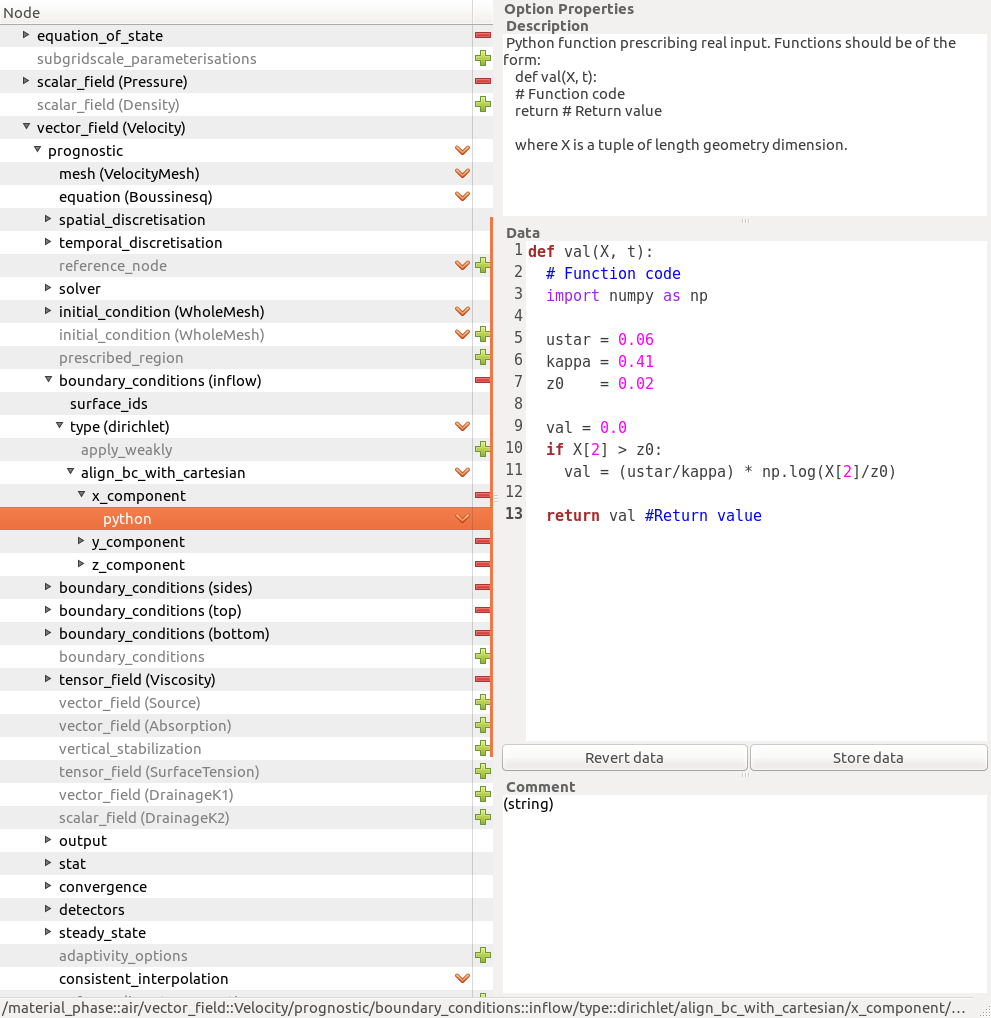}
        \caption{Inlet velocity}
        \label{Fig:Case5b_VelBC}
    \end{subfigure}
    \caption{(a) Initial and (b) inlet log-profile velocity in example \textit{3dBox\_Case5b.flml} prescribed using python scripts.}
    \label{Fig:Case5_VelBC}
\end{figure}

In example \textit{3dBox\_Case5b.flml}, the $u$-component of the initial and inlet velocity are set using a log-profile and the python scripts in Code~\ref{Lst:VelInitPython} and Code~\ref{Lst:VelBCPython} are used (Figure~\ref{Fig:Case5_VelBC}).The interior of the box is set to 298 K while the outside remains at ambient temperature 293 K (Code~\ref{Lst:InitTemp}).

\begin{Code}[language=python, caption={Python script to prescribe an initial log-profile for the velocity.}, label={Lst:VelInitPython}]
def val(X, t):
  # Function code
  import numpy as np
  
  ustar = 0.06
  kappa = 0.41
  z0    = 0.02
  
  val = 0.0
  if X[2] > z0:
    val = (ustar/kappa) * np.log(X[2]/z0)

  return [val,0.0,0.0] #Return value
\end{Code}

\begin{Code}[language=python, caption={Python script to prescribe an inlet log-profile for the velocity.}, label={Lst:VelBCPython}]
def val(X, t):
  # Function code
  import numpy as np
  
  ustar = 0.06
  kappa = 0.41
  z0    = 0.02

  val = 0.0
  if X[2] > z0:
    val = (ustar/kappa) * np.log(X[2]/z0)

  return val #Return value
\end{Code}

\noindent This example can be run using the command:
\begin{Terminal}[]
ä\colorbox{davysgrey}{
\parbox{435pt}{
\color{applegreen} \textbf{user@mypc}\color{white}\textbf{:}\color{codeblue}$\sim$
\color{white}\$ <<FluiditySourcePath>>/bin/fluidity -l -v3 3dBox\_Case5b.flml \&
}}
\end{Terminal}

\noindent A snapshot of the result obtained at 50 s is shown in Figure~\ref{Fig:Case5b_Results}. Go to Chapter~\ref{Sec:PostProcessing} to learn how to visualise the results using \textbf{ParaView}.

\subsection{Synthetic eddy method: Turbulent inlet velocity}
\subsubsection{Constant values}\label{Sec:VelTurbUni}
Finally, the last interesting velocity boundary condition is called the \texttt{Synthetic eddy method} and mimics a turbulent inlet velocity. For more details, the user can refer to~\cite{Pavlidis2010}. This boundary condition is used to reproduced the behaviour of the atmospheric boundary layer and the 4 following variables  need to be defined by the user for each velocity component:
\begin{itemize}
    \item \textbf{Number of eddies:} this number has to be large enough to ensure the Gaussian behaviour of the fluctuating component. Usually this value is taken to be $4000$.
    \item \textbf{Turbulence lengthscale:} the turbulence lengthscale is in meters and is defined by equation~\ref{Eq:Lengthscale}. 
    \item \textbf{Mean profile:} the mean velocity profiles of each velocity component in m/s.
    \item \textbf{Reynolds stresses profile:} the Reynolds stresses profile of the components $\overline{u'u'}$, $\overline{v'v'}$ and $\overline{w'w'}$ (in m\textsuperscript{2}/s\textsuperscript{2}) are prescribed assuming that the remaining stresses are negligible as in equation~\ref{Eq:ReynoldsStresses}.
\end{itemize}

\begin{equation}\label{Eq:Lengthscale}
    \textbf{L} = \left( \begin{array}{ccc} L_{u} & 0 & 0 \\ 0 & L_{v} & 0 \\ 0 & 0& L_{w} \end{array}\right)
\end{equation}

\begin{equation}\label{Eq:ReynoldsStresses}
    \textbf{Re} = \left( \begin{array}{ccc} \overline{u'u'} & 0 & 0 \\ 0 & \overline{v'v'} & 0 \\ 0 & 0& \overline{w'w'} \end{array}\right)
\end{equation}

\begin{figure}
    \centering
    \begin{subfigure}{0.49\textwidth}
        \includegraphics[width=\textwidth]{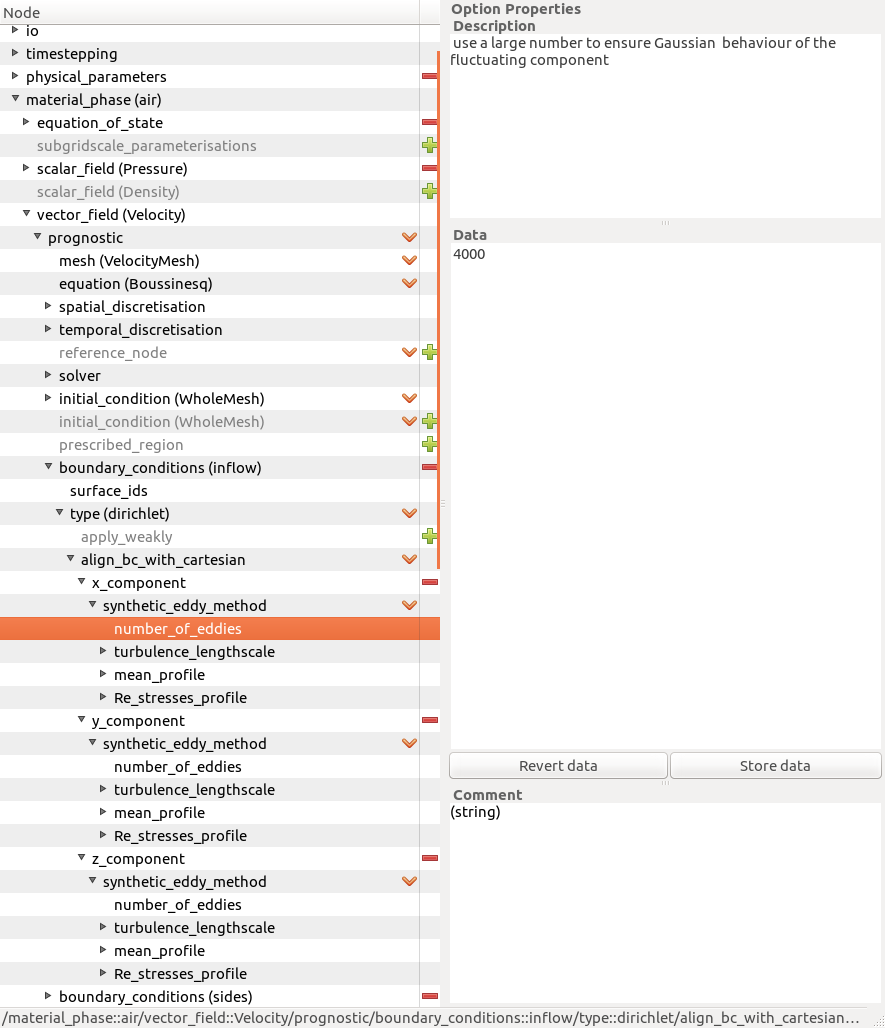}
        \caption{Number of eddies}
        \label{Fig:Case5c_Nu}
    \end{subfigure}
    \begin{subfigure}{0.49\textwidth}
        \includegraphics[width=\textwidth]{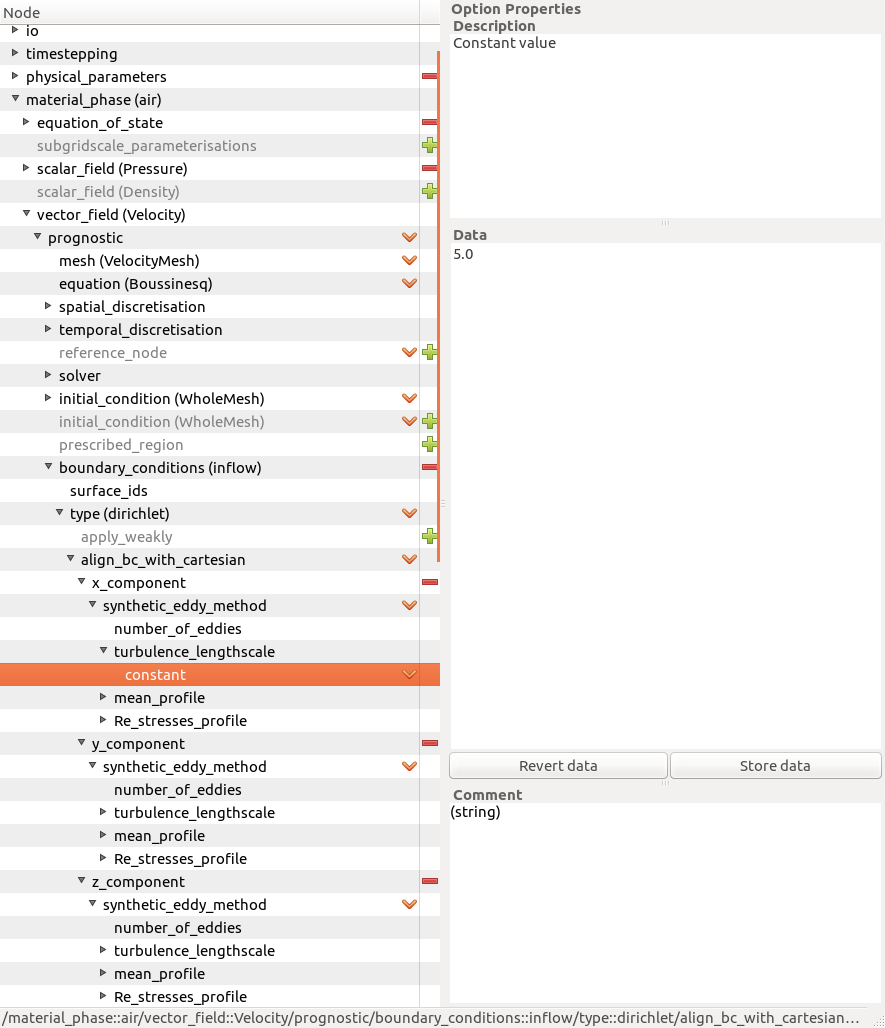}
        \caption{Lengthscale}
        \label{Fig:Case5c_Lu}
    \end{subfigure}
    \begin{subfigure}{0.49\textwidth}
        \includegraphics[width=\textwidth]{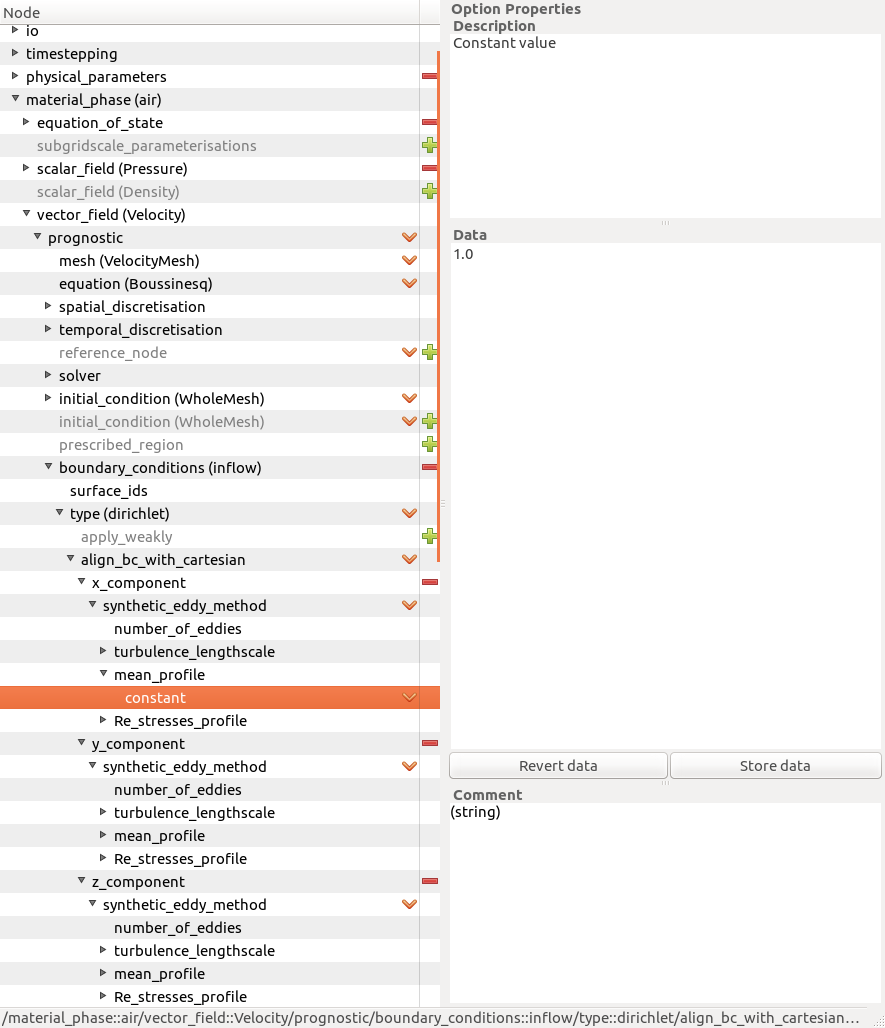}
        \caption{Mean profile}
        \label{Fig:Case5c_u}
    \end{subfigure}
    \begin{subfigure}{0.49\textwidth}
        \includegraphics[width=\textwidth]{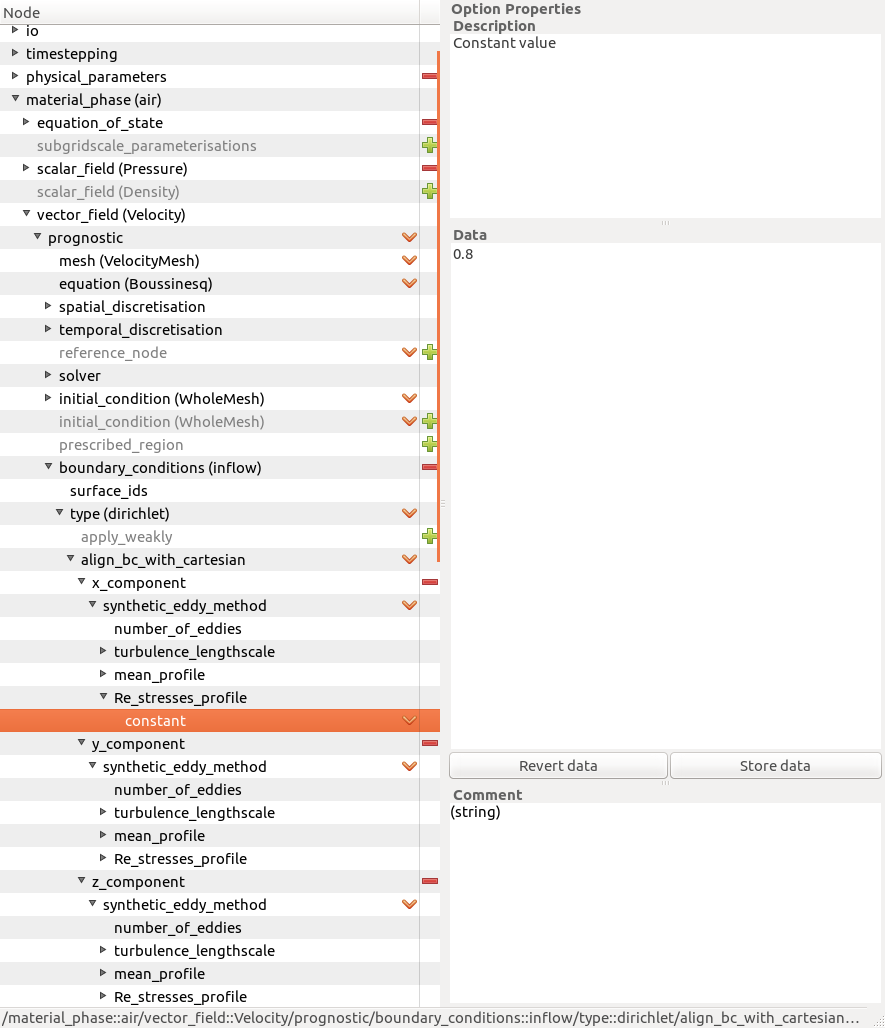}
        \caption{Reynolds stresses}
        \label{Fig:Case5c_uu}
    \end{subfigure}
    \caption{Options used for the turbulent inlet velocity in example \textit{3dBox\_Case5c.flml}.}
    \label{Fig:Case5c}
\end{figure}

\noindent In example \textit{3dBox\_Case5c.flml}, the interior of the box is set to 298 K while the outside remains at ambient temperature 293 K (Code~\ref{Lst:InitTemp}). The initial velocity and the inlet velocity $(u,v,w)$ are set to $(1,0,0) $ m/s. The number of eddies is taken as $4000$ and the turbulence lengthscale is equal to $5$ m. The $\overline{u'u'}$-component of the Reynolds stresses is taken to be to $0.8$, while $\overline{v'v'}$ and $\overline{w'w'}$ are equal to $0.3$. For brevity, the options set up for the $u$-component only are shown in Figure~\ref{Fig:Case5c}.

\noindent This example can be run using the command:
\begin{Terminal}[]
ä\colorbox{davysgrey}{
\parbox{435pt}{
\color{applegreen} \textbf{user@mypc}\color{white}\textbf{:}\color{codeblue}$\sim$
\color{white}\$ <<FluiditySourcePath>>/bin/fluidity -l -v3 3dBox\_Case5c.flml \&
}}
\end{Terminal}

\noindent A snapshot of the result obtained at 50 s is shown in Figure~\ref{Fig:Case5c_Results}. Go to Chapter~\ref{Sec:PostProcessing} to learn how to visualise the results using \textbf{ParaView}.

\subsubsection{Using python scripts}\label{Sec:VelTurbPython}
\begin{figure}
    \centering
    \begin{subfigure}{0.3\textwidth}
        \includegraphics[width=\textwidth]{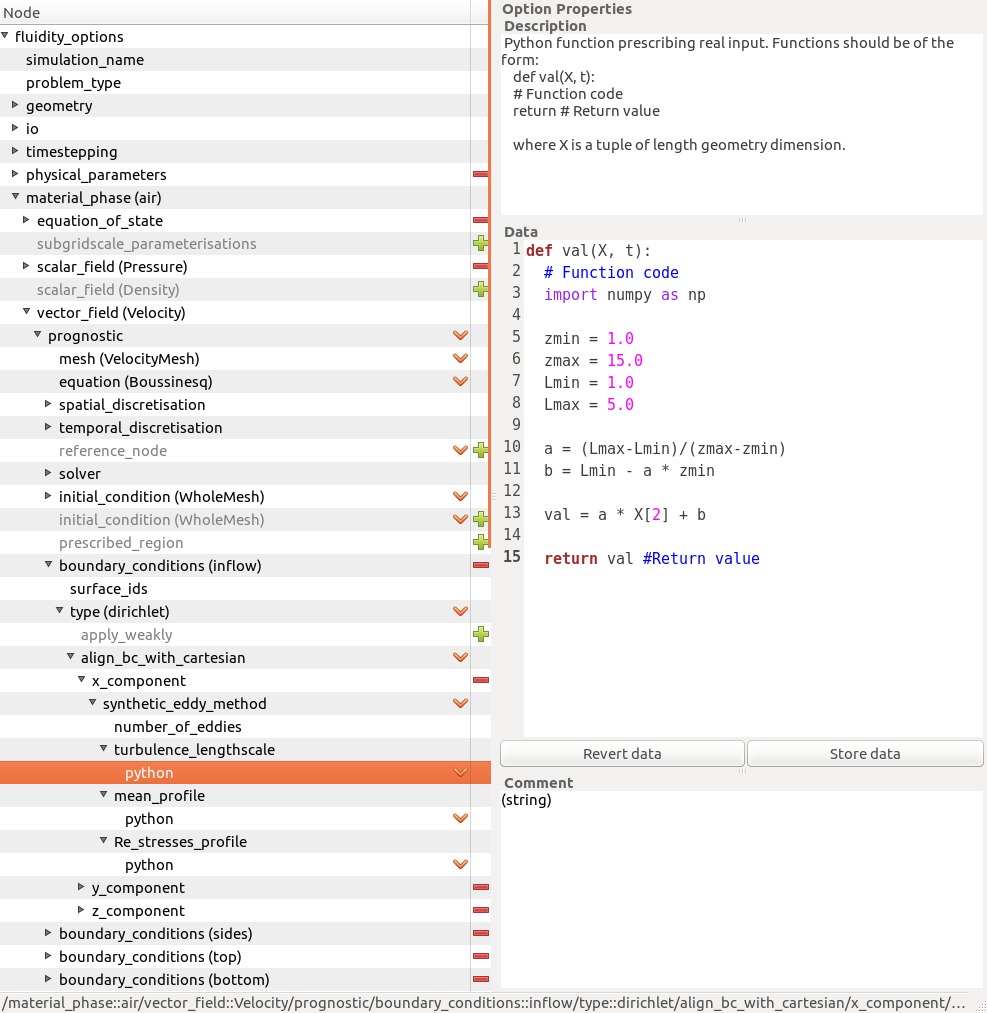}
        \caption{Lengthscale}
        \label{Fig:Case5d_Lengthscale}
    \end{subfigure}
    \begin{subfigure}{0.3\textwidth}
        \includegraphics[width=\textwidth]{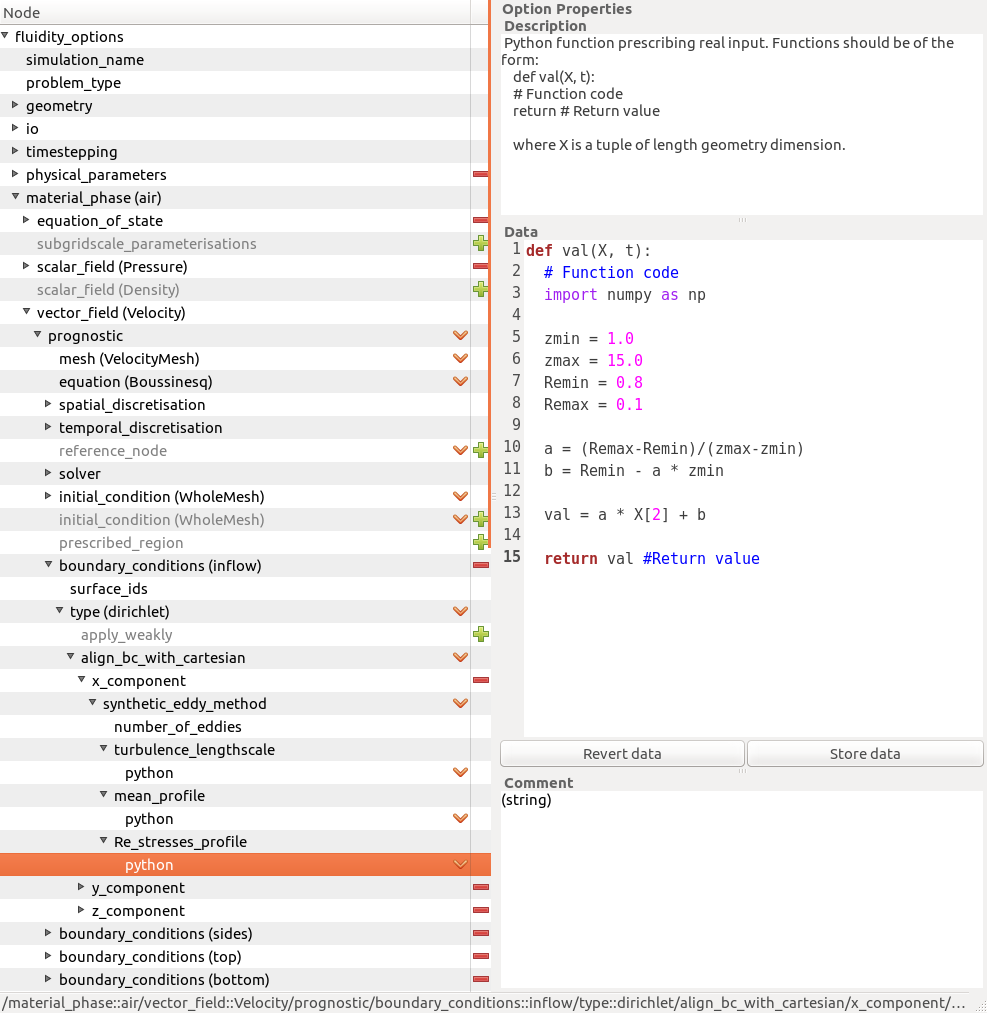}
        \caption{Reynolds stresses $\overline{u'u'}$}
        \label{Fig:Case5d_Reynoldsu}
    \end{subfigure}
    \begin{subfigure}{0.3\textwidth}
        \includegraphics[width=\textwidth]{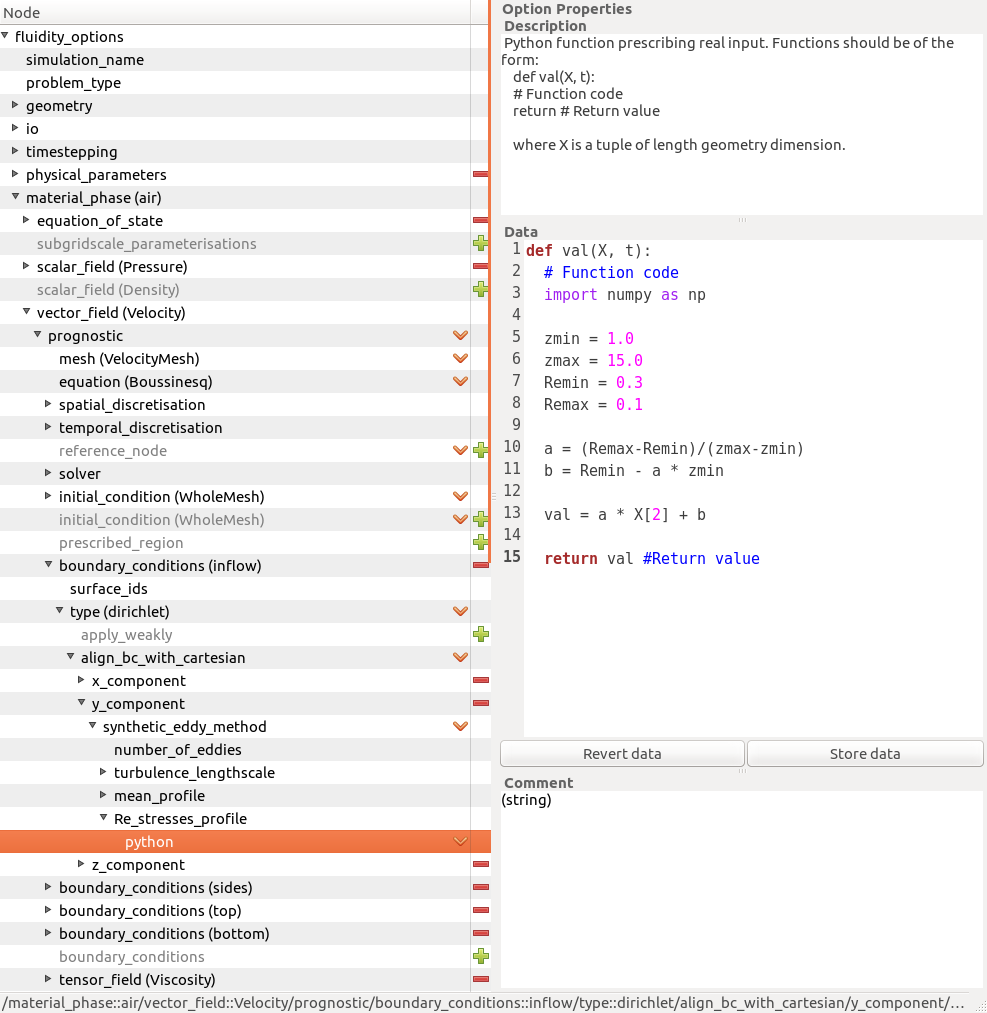}
        \caption{Reynolds stresses $\overline{v'v'}$}
        \label{Fig:Case5d_Reynoldsv}
    \end{subfigure}
    \caption{Python scripts used to prescribe a turbulent inlet velocity in \textit{3dBox\_Case5d.flml}.}
    \label{Fig:Case5d}
\end{figure}

The initial velocity, the turbulence lengthscale, the mean velocity and the Reynolds stresses profiles can also be prescribed using python scripts if the user wants to use real profiles found in literature. In example \textit{3dBox\_Case5d.flml}, the interior of the box is set to 298 K  while the outside remains at ambient temperature 293 K (Code~\ref{Lst:InitTemp}). The initial velocity is prescribed using Code~\ref{Lst:VelInitPython}. The turbulent inlet options are defined as follows (see also Figure~\ref{Fig:Case5d}):
\begin{itemize}
    \item \textbf{Turbulence lengthscale:} Assuming a linear relationship between the lengthscale and the height, the python script in Code~\ref{Lst:TurbLengthscale} is used for the three velocity components as shown in Figure~\ref{Fig:Case5d_Lengthscale}.
            \begin{Code}[language=python, caption={Python script to prescribe a turbulence lengthscale profile.}, label={Lst:TurbLengthscale}]
def val(X, t):
  # Function code
  import numpy as np
  
  zmin = 1.0
  zmax = 15.0
  Lmin = 1.0
  Lmax = 5.0
  
  a = (Lmax-Lmin)/(zmax-zmin)
  b = Lmin - a * zmin

  val = a * X[2] + b

  return val #Return value
        \end{Code}
    \item \textbf{Mean velocity: } Assuming a log-profile, the python script in Code~\ref{Lst:VelBCPython} is prescribed to the $u$-component of the velocity, while $0$ m/s is prescribed to the $v$ and $w$-components.
    \item \textbf{Reynolds Stresses:} The $\overline{u'u'}$-component of the Reynolds stresses are prescribed using the python script in Code~\ref{Lst:Reynodlsu} (Figure~\ref{Fig:Case5d_Reynoldsu}). The $\overline{v'v'}$ and $\overline{w'w'}$ components  of the Reynolds stresses is prescribed using the python script in Code~\ref{Lst:Reynodlsv} (Figure~\ref{Fig:Case5d_Reynoldsv}). A linear relationship between the Reynolds stresses and the height is assumed.
                \begin{Code}[language=python, caption={Python script to prescribe the Reynolds stresses $\overline{u'u'}$.}, label={Lst:Reynodlsu}]
def val(X, t):
  # Function code
  import numpy as np
  
  zmin = 1.0
  zmax = 15.0
  Remin = 0.8
  Remax = 0.1
  
  a = (Remax-Remin)/(zmax-zmin)
  b = Remin - a * zmin

  val = a * X[2] + b

  return val #Return value
        \end{Code}
                    \begin{Code}[language=python, caption={Python script to prescribe the Reynolds stresses $\overline{v'v'}$ and $\overline{w'w'}$.}, label={Lst:Reynodlsv}]
def val(X, t):
  # Function code
  import numpy as np
  
  zmin = 1.0
  zmax = 15.0
  Remin = 0.3
  Remax = 0.1
  
  a = (Remax-Remin)/(zmax-zmin)
  b = Remin - a * zmin

  val = a * X[2] + b

  return val #Return value
        \end{Code}
\end{itemize}

\noindent This example can be run using the command:
\begin{Terminal}[]
ä\colorbox{davysgrey}{
\parbox{435pt}{
\color{applegreen} \textbf{user@mypc}\color{white}\textbf{:}\color{codeblue}$\sim$
\color{white}\$ <<FluiditySourcePath>>/bin/fluidity -l -v3 3dBox\_Case5d.flml \&
}}
\end{Terminal}

\noindent A snapshot of the result obtained at 50 s is shown in Figure~\ref{Fig:Case5d_Results}. Go to Chapter~\ref{Sec:PostProcessing} to learn how to visualise the results using \textbf{ParaView}.

\subsection{Dirichlet boundary condition on solids}\label{Sec:BCWalls}
\textbf{Note: This section is particularly important and it is recommended to read it carefully.} This section is also summarised in Section~\ref{Sec:BCTricks} due to its importance. The option \texttt{align\_bc\_with\_surface} in the Dirichlet type boundary condition is not well-implemented in \textbf{Fluidity} and not always work properly. This section describes in details in which case it works or not. In any case, this option should be avoided if possible. The simulations presented in that section are based on \textit{3dBox\_Case4.flml} and are summarised in Table~\ref{Tab:DirichletBCWalls}.

\begin{table}
    \centering
    \begin{tabular}{ |p{1.1cm}||p{3.8cm}|p{1.4cm}|p{2.0cm}|p{2.5cm}|p{1.4cm}|  }
         \hline
         \textbf{Case \newline Nbr} & \bf{BC type} & \textbf{From} & \textbf{Aligned \newline with} & \textbf{BCs aligned \newline with surface} & \textbf{Work?} \\
         \hline
         4   & Dirichlet: No-slip   & Strong & Cartesian & N/A & YES \\\hline
         13a & Dirichlet: No-slip   & Strong & Surface   & 1   & YES \\\hline
         13b & Dirichlet: No-slip   & Strong & Surface   & 2   & NO  \\\hline
         13c & Dirichlet: No-slip   & Weak   & Cartesian & N/A & YES \\\hline
         13d & Dirichlet: No-slip   & Weak   & Surface   & 1   & YES \\\hline
         13e & Dirichlet: No-slip   & Weak   & Surface   & 2   & YES \\\hline
         13f & Dirichlet: Slip      & Strong & Surface   & 1   & YES \\\hline
         13g & Dirichlet: Slip      & Weak   & Cartesian & N/A & YES \\\hline
         13h & Dirichlet: Slip      & Weak   & Surface   & 1   & NO  \\\hline
         13i & No normal flow: Slip & Weak   & N/A       & N/A & YES \\\hline
    \end{tabular}
    \caption{\label{Tab:DirichletBCWalls}Summary of the simulations using different Dirichlet velocity boundary condition for solid surfaces. }
\end{table}

\subsubsection{Boundary condition \texttt{align\_bc\_with\_cartesian} or \texttt{align\_bc\_with\_surface}?}
It exists two options available to apply a Dirichlet boundary condition in \textbf{Fluidity}:
\begin{itemize}
    \item \texttt{align\_bc\_with\_cartesian}: the three components of the velocity are assigned.
    \item \texttt{align\_bc\_with\_surface}: the normal and the two tangential components of the velocity are assigned.
\end{itemize}
The second option can be really useful when the surfaces of the geometry are not aligned with Cartesian coordinates system for very complex geometry. However, this functionality does not always work properly in \textbf{Fluidity} as detailed in the following:
\begin{itemize}
    \item Example \textit{3dBox\_Case4.flml} uses the option \texttt{align\_bc\_with\_cartesian} to apply a no-slip boundary condition (the three components $x$, $y$ and $z$ of the velocity are equal to zero) on the solid surfaces.
    \item Example \textit{3dBox\_Case13a.flml} is a replicate of example \textit{3dBox\_Case4.flml}, excepted that the velocity boundary condition on the solid surfaces is now applied using the \texttt{align\_bc\_with\_surface} option, i.e. the normal and the two tangential components of the velocity are now equal to zero.
    \begin{itemize}[label=$\boldsymbol{\Rightarrow}$]
        \item Examples \textit{3dBox\_Case4.flml} and \textit{3dBox\_Case13a.flml} run correctly and give the same results - that was actually expected...
    \end{itemize}
    \item In \textit{3dBox\_Case13b.flml}, the no-slip boundary condition \texttt{align\_bc\_with\_surface} is now dissociated into two no-slip boundary conditions \texttt{align\_bc\_with\_surface}: one for the ground of the domain and the walls; and one for the ground of the box only.
    \begin{itemize}[label=$\boldsymbol{\Rightarrow}$]
        \item If the user runs example \textit{3dBox\_Case13b.flml}, the simulation will crash and the error in Command~\ref{Lst:ErrorAlignedSurface} will be raised.
    \end{itemize}
\end{itemize}

\begin{Terminal}[caption={Error occurring when two strongly applied Dirichlet boundary conditions using \texttt{align\_bc\_with\_surface} are defined.}, label={Lst:ErrorAlignedSurface}]
ä\colorbox{davysgrey}{
\parbox{435pt}{
\color{applegreen} \textbf{user@mypc}\color{white}\textbf{:}\color{codeblue}$\sim$
\color{white}\$ Inside create\_rotation\_matrix
\newline
*** ERROR ***
\newline
Error message: Two rotated boundary condition specifications for the same node.
}}
\end{Terminal}

\noindent Unfortunately, if several strong boundary conditions (see next section for discussion about the strong form) \texttt{align\_bc\_with\_surface} are really wanted, there is not tricks to avoid this error in \textbf{Fluidity}. Only one \texttt{align\_bc\_with\_surface} Dirichlet boundary condition strongly applied is allowed by \textbf{Fluidity}, which can be problematic when dealing with complex geometries.

\subsubsection{Boundary condition applied strongly or weakly?}
The only way to avoid the error previously described (when more than one \texttt{align\_bc\_} \texttt{with\_surface} no-slip Dirichlet boundary condition is used) is to apply the boundary conditions weakly.

\noindent It is to be noted that when boundary conditions are applying weakly, the discrete solution will not satisfy the boundary condition exactly. Instead the solution will converge to the correct boundary condition along with the solution in the interior as the mesh is refined. An alternative way of implementing boundary conditions is to strongly imposed boundary conditions. Although this guarantees that the Dirichlet boundary condition will be satisfied exactly, it does not at all mean that the discrete solution converges to the exact continuous solution more quickly than it would with weakly imposed boundary conditions. Strongly imposed boundary conditions may sometimes be necessary if the boundary condition needs to be imposed strictly for physical reasons. Unlike the strong form of the Dirichlet conditions, weak Dirichlet conditions do not force the solution on the boundary to be point-wise equal to the boundary condition.

\noindent If boundary conditions are applied weakly, then the following options need to be turned on in \textbf{Diamond}:
\begin{itemize}
    \item Under the Pressure field: \texttt{spatial\_discretisation/continuous\_galerkin/} \newline
    \texttt{integrate\_continuity\_by\_parts}
    \item Under the Velocity field: \texttt{spatial\_discretisation/continuous\_galerkin/} \newline
    \texttt{advection\_terms/integrate\_advection\_by\_parts}
\end{itemize}

\begin{figure}
    \centering
    \begin{subfigure}{0.30\textwidth}
        \includegraphics[width=\textwidth]{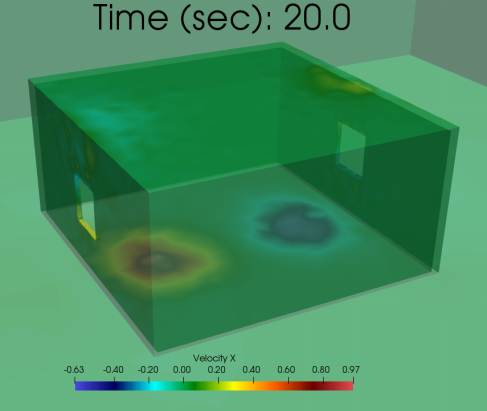}
        \caption{$x$-velocity}
        \label{Fig:Case13c_VelocityX}
    \end{subfigure}
    \begin{subfigure}{0.30\textwidth}
        \includegraphics[width=\textwidth]{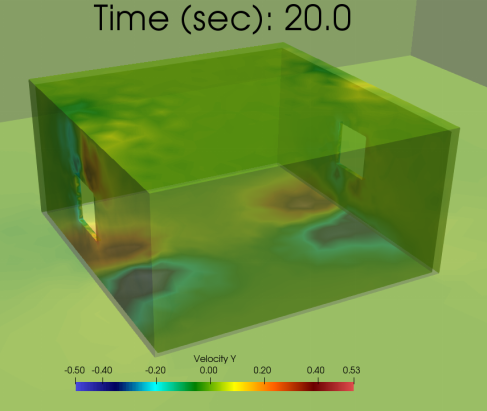}
        \caption{$y$-velocity}
        \label{Fig:Case13c_VelocityY}
    \end{subfigure}
    \begin{subfigure}{0.30\textwidth}
        \includegraphics[width=\textwidth]{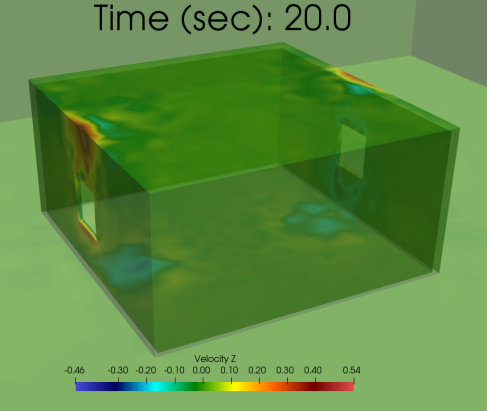}
        \caption{$z$-velocity}
        \label{Fig:Case13c_VelocityZ}
    \end{subfigure}
    \caption{(a) $x-$, (b) $y-$ and (c) $z-$ components of the velocity in the box. A weak no-slip boundary condition is applied on solid surfaces. Example \textit{3dBox\_Case13c.flml}.}
    \label{Fig:Case13cVelocity}
\end{figure}

\noindent Examples \textit{3dBox\_Case13c.flml}, \textit{3dBox\_Case13d.flml} and \textit{3dBox\_Case13e.flml} use a no-slip boundary condition applied weakly. The time-step was reduce to $0.01$ second to avoid divergence of the simulation and the options \texttt{integrate\_*\_by\_parts} are turned on.
\begin{itemize}
    \item Example \textit{3dBox\_Case13c.flml} is equivalent to \textit{3dBox\_Case4.flml}, excepted that the velocity boundary condition on solid surfaces is now applied weakly instead of strongly, still using \texttt{align\_bc\_with\_cartesian}.
    \begin{itemize}[label=$\boldsymbol{\Rightarrow}$]
        \item As shown in Figure~\ref{Fig:Case13cVelocity}, the velocity is not equal to zero on solid walls for the reason explained above, i.e. because the boundary condition is applied weakly.
    \end{itemize}
    \item Example \textit{3dBox\_Case13d.flml} is the same than \textit{3dBox\_Case13c.flml}, excepted that the velocity boundary condition on solid surfaces, still applied weakly, is now \texttt{align\_bc\_with\_surface}. 
    \begin{itemize}[label=$\boldsymbol{\Rightarrow}$]
        \item The results obtained from example \textit{3dBox\_Case13d.flml} are the same than the ones from \textit{3dBox\_Case13c.flml}.
    \end{itemize}    
    \item Finally, example \textit{3dBox\_Case13e.flml} is the same than \textit{3dBox\_Case13b.flml} (which was previously crashing because of two strong \texttt{align\_bc\_with\_surface} boundary type), excepted that the two velocity boundary conditions are now applied weakly.
    \begin{itemize}[label=$\boldsymbol{\Rightarrow}$]
        \item Contrary to \textit{3dBox\_Case13b.flml}, example \textit{3dBox\_Case13e.flml} runs and does not crashed. As expected, results are the same than in \textit{3dBox\_Case13c.flml} and \textit{3dBox\_Case13d.flml}.
    \end{itemize}    
\end{itemize}

\subsubsection{No-slip or slip boundary condition on solid?}
This section will discuss the use of slip or no-slip boundary condition on solid surfaces. One can argue that a no-slip is more appropriate, while other will prone the use of a slip boundary condition. A no-slip boundary condition is defined by the three components of the velocity being equal to zero, while a slip boundary condition corresponds to the normal component of the velocity only being equal to zero.

\begin{figure}
    \centering
    \begin{subfigure}{0.35\textwidth}
        \includegraphics[width=\textwidth]{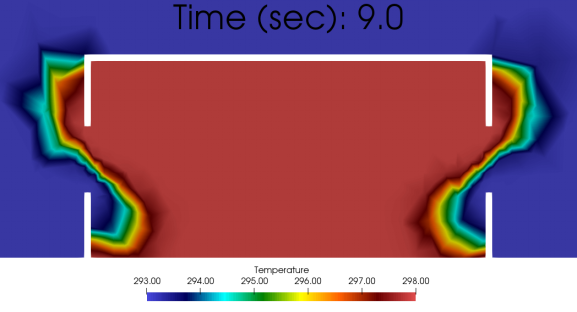}
        \caption{\textit{3dBox\_Case13f.flml}}
        \label{Fig:Case13f_Temperature}
    \end{subfigure}
    \begin{subfigure}{0.35\textwidth}
        \includegraphics[width=\textwidth]{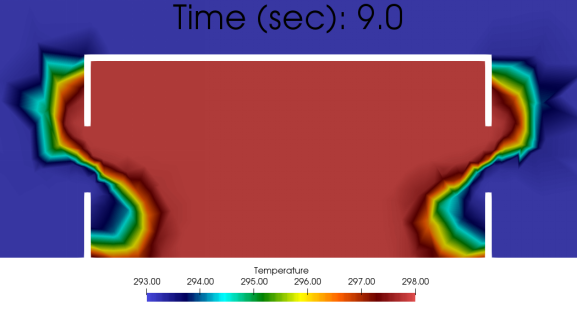}
        \caption{\textit{3dBox\_Case13g.flml}}
        \label{Fig:Case13g_Temperature}
    \end{subfigure}
    \begin{subfigure}{0.35\textwidth}
        \includegraphics[width=\textwidth]{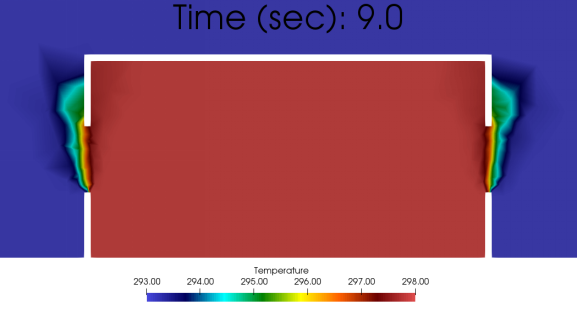}
        \caption{\textit{3dBox\_Case13h.flml}}
        \label{Fig:Case13h_Temperature}
    \end{subfigure}
    \begin{subfigure}{0.35\textwidth}
        \includegraphics[width=\textwidth]{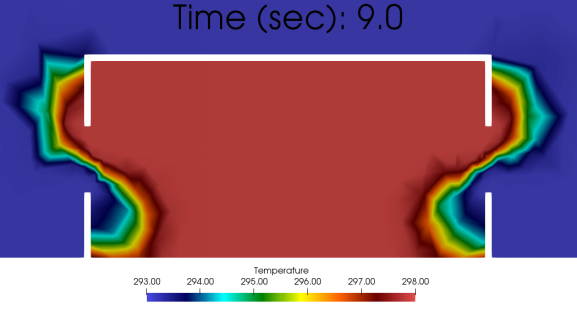}
        \caption{\textit{3dBox\_Case13i.flml}}
        \label{Fig:Case13i_Temperature}
    \end{subfigure}
    \caption{Temperature field in the box when a slip boundary condition is applied on solid walls. The boundary condition is applied (a) strongly \texttt{align\_bc\_with\_surface}; (b) weakly \texttt{align\_bc\_with\_cartesian}, (c) weakly \texttt{align\_bc\_with\_surface} and (d) weakly using \texttt{no\_normal\_flow}. Note that example (c) \textit{3dBox\_Case13h.flml} does not work properly.}
    \label{Fig:Case13_SlipBCTemp}
\end{figure}

\begin{figure}
    \centering
    \begin{subfigure}{0.35\textwidth}
        \includegraphics[width=\textwidth]{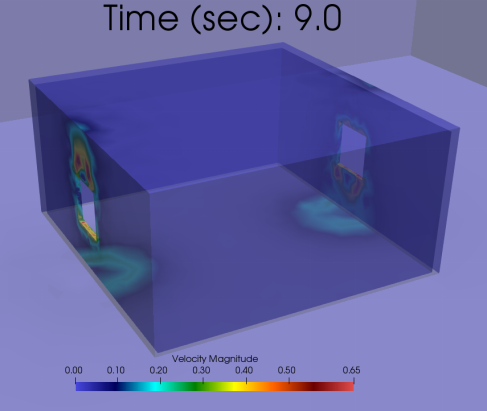}
        \caption{\textit{3dBox\_Case13f.flml}}
        \label{Fig:Case13f_VelocityMag}
    \end{subfigure}
    \begin{subfigure}{0.35\textwidth}
        \includegraphics[width=\textwidth]{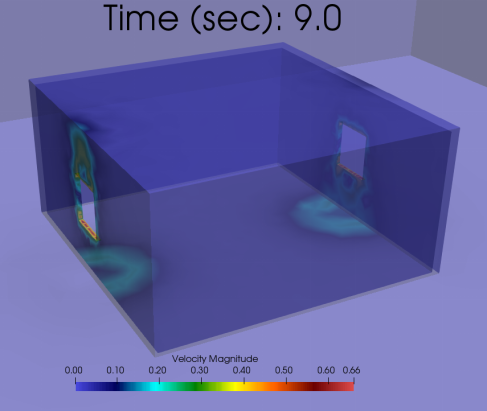}
        \caption{\textit{3dBox\_Case13g.flml}}
        \label{Fig:Case13g_VelocityMag}
    \end{subfigure}
    \begin{subfigure}{0.35\textwidth}
        \includegraphics[width=\textwidth]{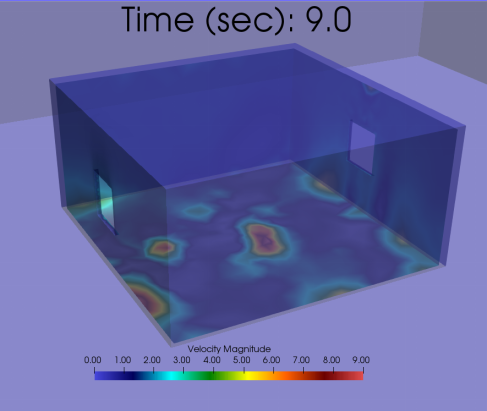}
        \caption{\textit{3dBox\_Case13h.flml}}
        \label{Fig:Case13h_VelocityMag}
    \end{subfigure}
    \begin{subfigure}{0.35\textwidth}
        \includegraphics[width=\textwidth]{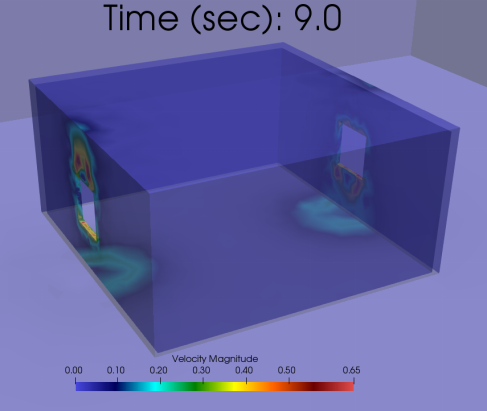}
        \caption{\textit{3dBox\_Case13i.flml}}
        \label{Fig:Case13i_VelocityMag}
    \end{subfigure}
    \caption{Velocity magnitude when a slip boundary condition is applied on solid surfaces of the domain. The boundary condition is applied (a) strongly \texttt{align\_bc\_with\_surface}; (b) weakly \texttt{align\_bc\_with\_cartesian}; (c) weakly \texttt{align\_bc\_with\_surface} and (d) weakly using \texttt{no\_normal\_flow}. Note that example (c) \textit{3dBox\_Case13h.flml} does not work properly.}
    \label{Fig:Case13_SlipBCVelMAg}
\end{figure}

\begin{figure}
    \centering
    \begin{subfigure}{0.35\textwidth}
        \includegraphics[width=\textwidth]{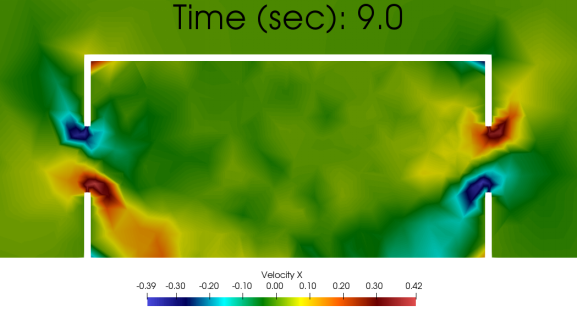}
        \caption{\textit{3dBox\_Case13f.flml}}
        \label{Fig:Case13f_VelocityX}
    \end{subfigure}
    \begin{subfigure}{0.35\textwidth}
        \includegraphics[width=\textwidth]{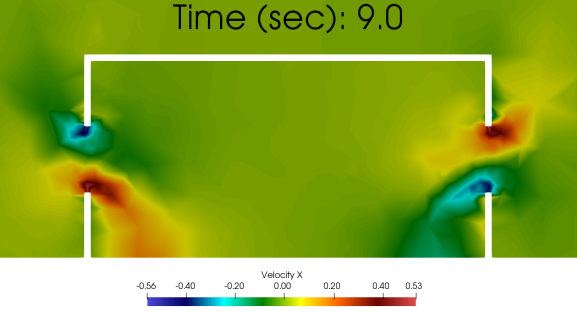}
        \caption{\textit{3dBox\_Case13g.flml}}
        \label{Fig:Case13g_VelocityX}
    \end{subfigure}
    \begin{subfigure}{0.35\textwidth}
        \includegraphics[width=\textwidth]{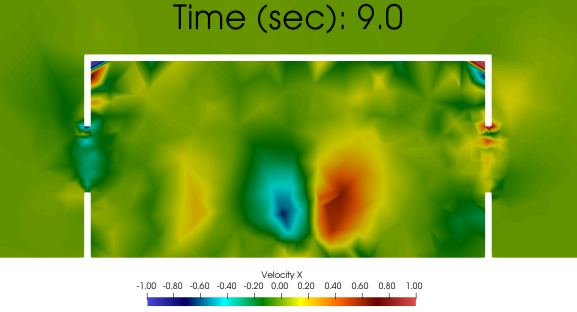}
        \caption{\textit{3dBox\_Case13h.flml}}
        \label{Fig:Case13h_VelocityX}
    \end{subfigure}
    \begin{subfigure}{0.35\textwidth}
        \includegraphics[width=\textwidth]{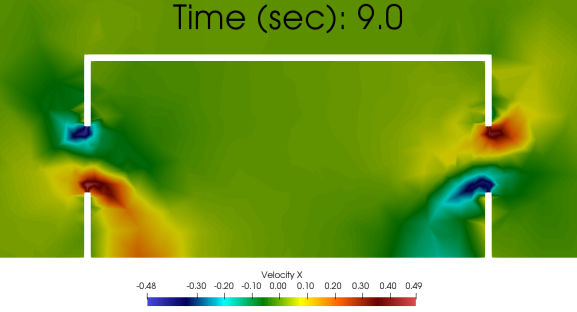}
        \caption{\textit{3dBox\_Case13i.flml}}
        \label{Fig:Case13i_VelocityX}
    \end{subfigure}
    \caption{$x$-component of the velocity when a slip boundary condition is applied on solid surfaces. The boundary condition is applied (a) strongly \texttt{align\_bc\_with\_surface}; (b) weakly \texttt{align\_bc\_with\_cartesian}; (c) weakly \texttt{align\_bc\_with\_surface} and (d) weakly using \texttt{no\_normal\_flow}. Note that example (c) \textit{3dBox\_Case13h.flml} does not work properly.}
    \label{Fig:Case13_SlipBCVelComp}
\end{figure}

\noindent The following examples use a slip boundary condition on solid surfaces and for comparison results are shown in Figure~\ref{Fig:Case13_SlipBCTemp}, Figure~\ref{Fig:Case13_SlipBCVelMAg} and Figure~\ref{Fig:Case13_SlipBCVelComp}.

\begin{itemize}
    \item Example \textit{3dBox\_Case13f.flml} prescribes a slip boundary condition on solid surfaces applied strongly using a Dirichlet type \texttt{align\_bc\_with\_surface}.
    \begin{itemize}[label=$\boldsymbol{\Rightarrow}$]
        \item This case works perfectly as long as only one \texttt{align\_bc\_with\_surface} is used. Results are shown in Figure~\ref{Fig:Case13f_Temperature}, Figure~\ref{Fig:Case13f_VelocityMag} and Figure~\ref{Fig:Case13f_VelocityX}.
    \end{itemize}
    \item Examples \textit{3dBox\_Case13g.flml} and \textit{3dBox\_Case13h.flml} prescribe a slip boundary condition applied weakly using the option \texttt{align\_bc\_with\_cartesian} and \texttt{align\_bc\_with\_surface}, respectively.
    \begin{itemize}[label=$\boldsymbol{\Rightarrow}$]
        \item While \textit{3dBox\_Case13g.flml} runs like a charm (Figure~\ref{Fig:Case13g_Temperature}, Figure~\ref{Fig:Case13g_VelocityMag} and Figure~\ref{Fig:Case13g_VelocityX}), example \textit{3dBox\_Case13h.flml} gives weird results (Figure~\ref{Fig:Case13h_Temperature}, Figure~\ref{Fig:Case13h_VelocityMag} and Figure~\ref{Fig:Case13h_VelocityX}) and finally crashes. Indeed, the option \texttt{align\_bc\_with\_surface}, when a slip boundary condition is weakly applied, does not work in \textbf{Fluidity} and the user should instead use the \texttt{no\_normal\_flow} option.
    \end{itemize}
    \item The option \texttt{no\_normal\_flow} is used in example \textit{3dBox\_Case13i.flml}: this option apply automatically a weak slip boundary condition on any surfaces specified.
    \begin{itemize}[label=$\boldsymbol{\Rightarrow}$]
        \item This case runs and results  are shown in Figure~\ref{Fig:Case13i_Temperature}, Figure~\ref{Fig:Case13i_VelocityMag} and Figure~\ref{Fig:Case13i_VelocityX}.
    \end{itemize}    
\end{itemize}

\subsubsection{In summary}
In summary:
\begin{itemize}
    \item The option \texttt{align\_bc\_with\_cartesian} should always be preferred if possible.
    \item Only one no-slip Dirichlet \texttt{align\_bc\_with\_surface} applied strongly is allowed, while several can be used when applied weakly.
    \item For a slip boundary condition weakly applied, \texttt{align\_bc\_with\_cartesian} type for simple geometry or \texttt{no\_normal\_flow} type for any geometry should be used. The Dirichlet \texttt{align\_bc\_with\_surface} type does not work.
\end{itemize}

\begin{figure}
    \centering
    \begin{subfigure}{0.35\textwidth}
        \includegraphics[width=\textwidth]{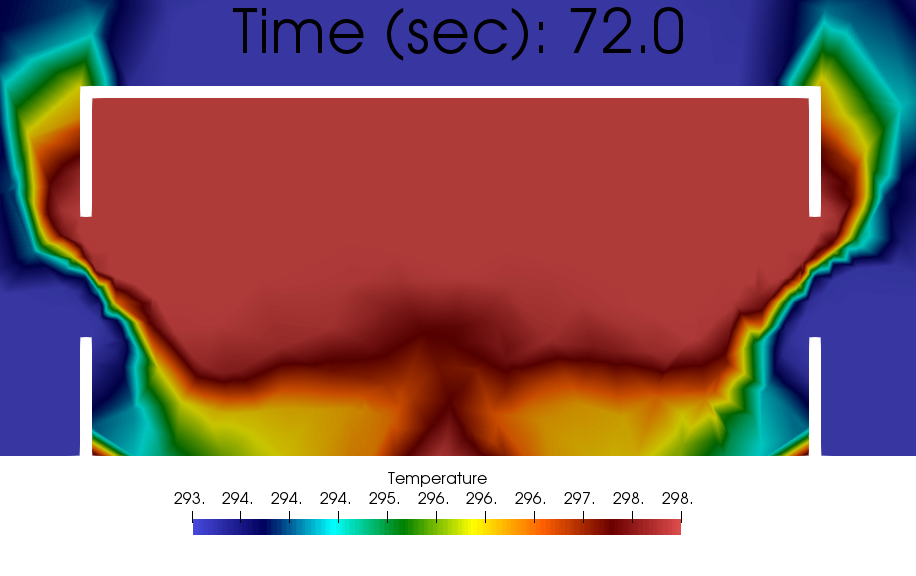}
        \caption{\textit{3dBox\_Case4.flml}}
        \label{Fig:Case4_Temp}
    \end{subfigure}
    \begin{subfigure}{0.35\textwidth}
        \includegraphics[width=\textwidth]{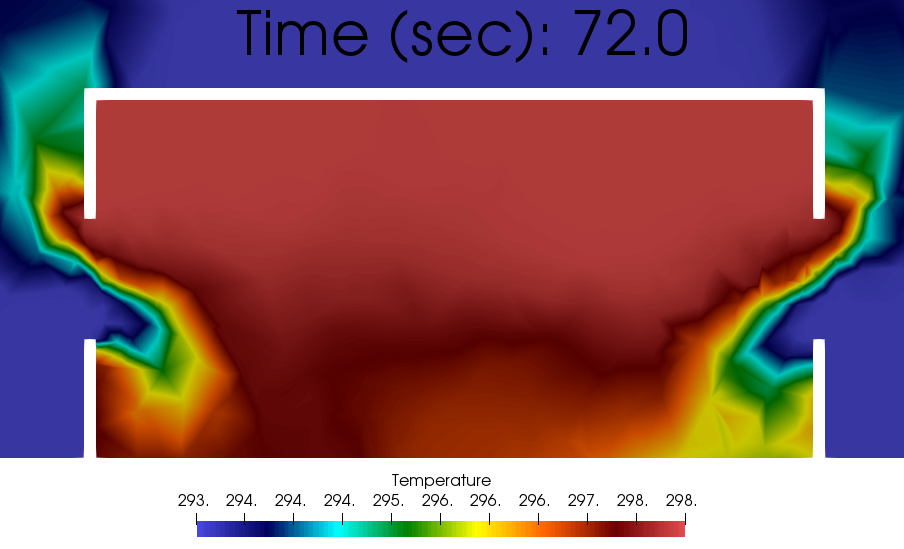}
        \caption{\textit{3dBox\_Case13c.flml}}
        \label{Fig:Case13c_Temp}
    \end{subfigure}
    \begin{subfigure}{0.35\textwidth}
        \includegraphics[width=\textwidth]{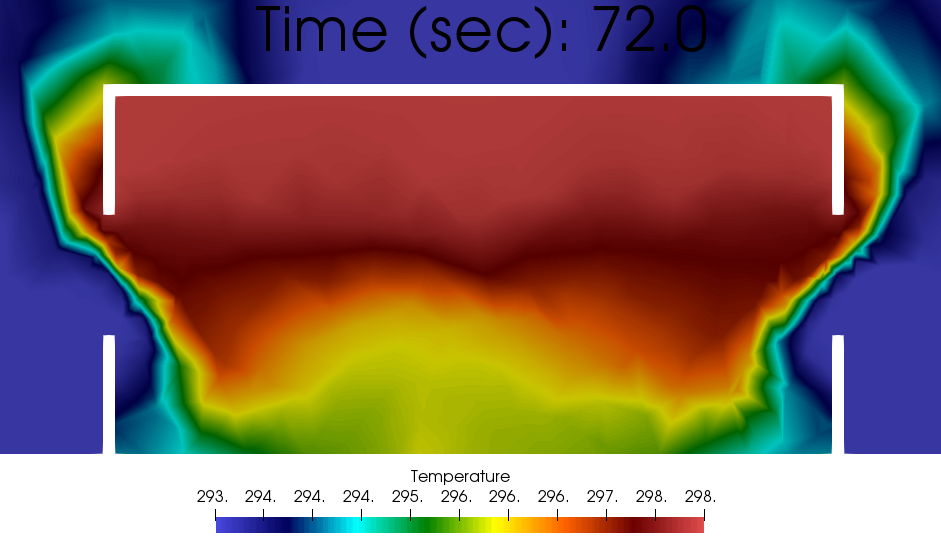}
        \caption{\textit{3dBox\_Case13f.flml}}
        \label{Fig:Case13f_Temp}
    \end{subfigure}
    \begin{subfigure}{0.35 \textwidth}
        \includegraphics[width=\textwidth]{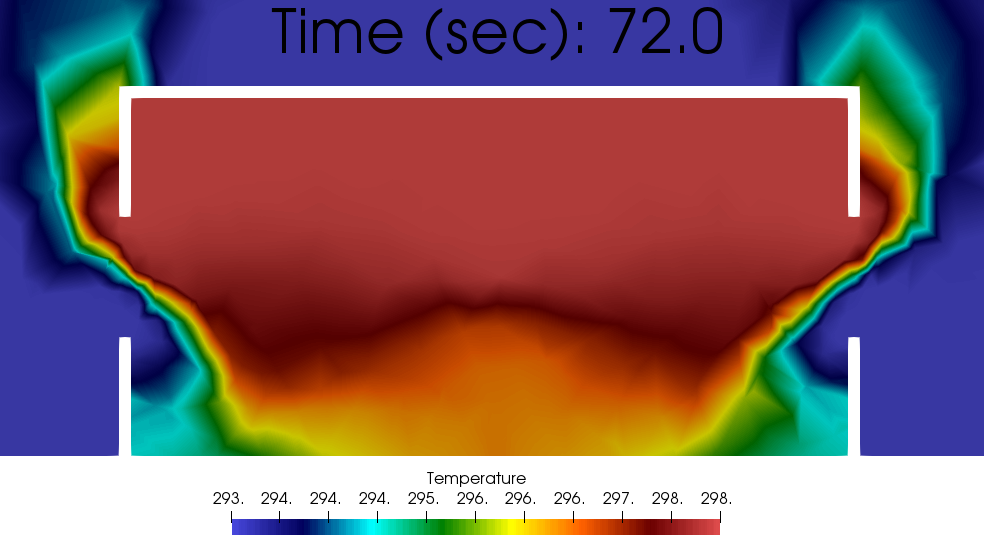}
        \caption{\textit{3dBox\_Case13g.flml}}
        \label{Fig:Case13g_Temp}
    \end{subfigure}
    \caption{Temperature field in the box with a (a) no-slip strongly applied; (b) no-slip weakly applied; (c) slip strongly applied and (d) slip weakly applied boundary condition on solid surfaces.}
    \label{Fig:Case13_Temperature}
\end{figure}

\noindent Finally, when defining the velocity boundary conditions at a wall, it is recommended to use a slip boundary condition (normal component only equal to 0) instead of a no-slip condition (all components are set to 0 at the wall) if the boundary layer is not going to be fully resolved with the chosen mesh. Using a no-slip condition can notably be problematic near a heat source and will generate wide variations in the expected temperature. The temperature fields obtained from examples \textit{3dBox\_Case4.flml}, \textit{3dBox\_Case13c.flml}, \textit{3dBox\_Case13f.flml} and \textit{3dBox\_Case13g.flml} are shown in Figure~\ref{Fig:Case13_Temperature}. One can noticed that the temperature stays hot in the lower corners of the box when a no-slip strongly applied boundary condition is used, while this hot spots disappear when a slip boundary condition is used or when the boundary condition is applied weakly (which is basically more or less the same...).

\section{Reference pressure}\label{Sec:RefPressure}
\begin{figure}
    \centering
    \begin{subfigure}{0.3\textwidth}
        \includegraphics[width=\textwidth]{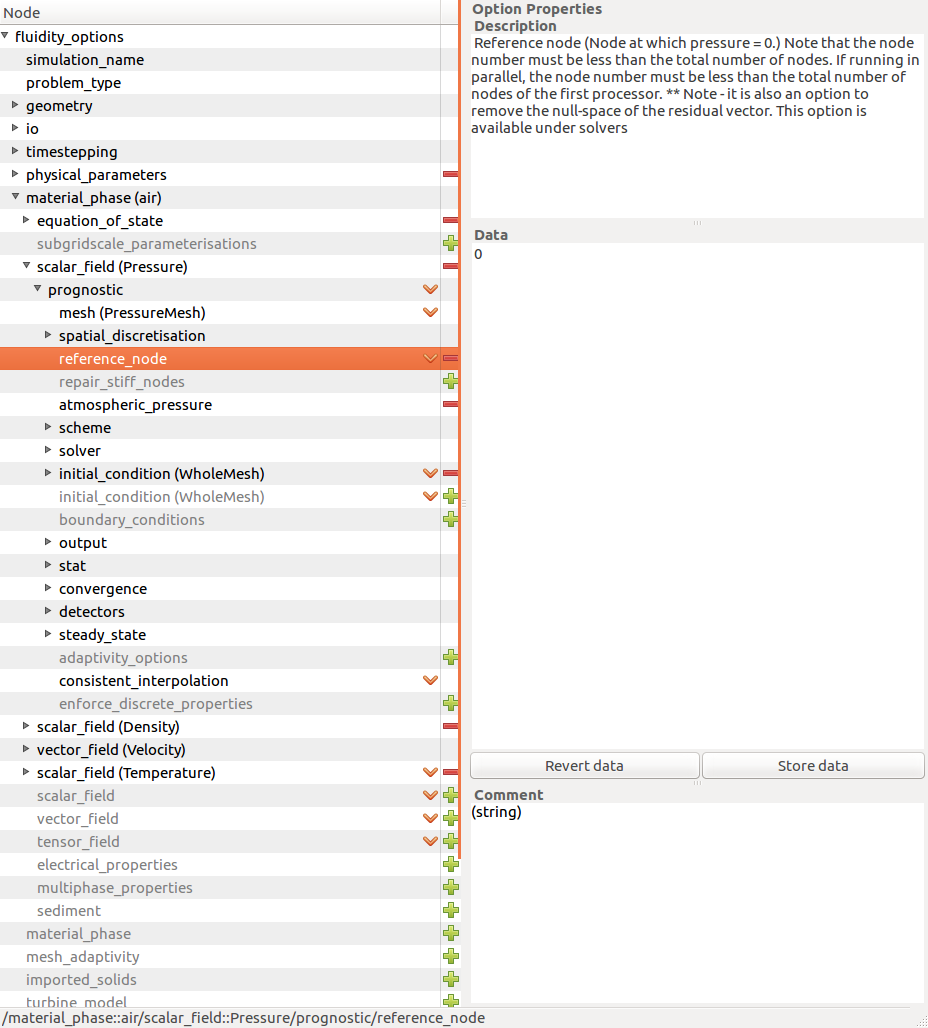}
        \caption{$P_{ref}$ at node ID}
        \label{Fig:PressureBC_NodeID}
    \end{subfigure}
    \begin{subfigure}{0.3\textwidth}
        \includegraphics[width=\textwidth]{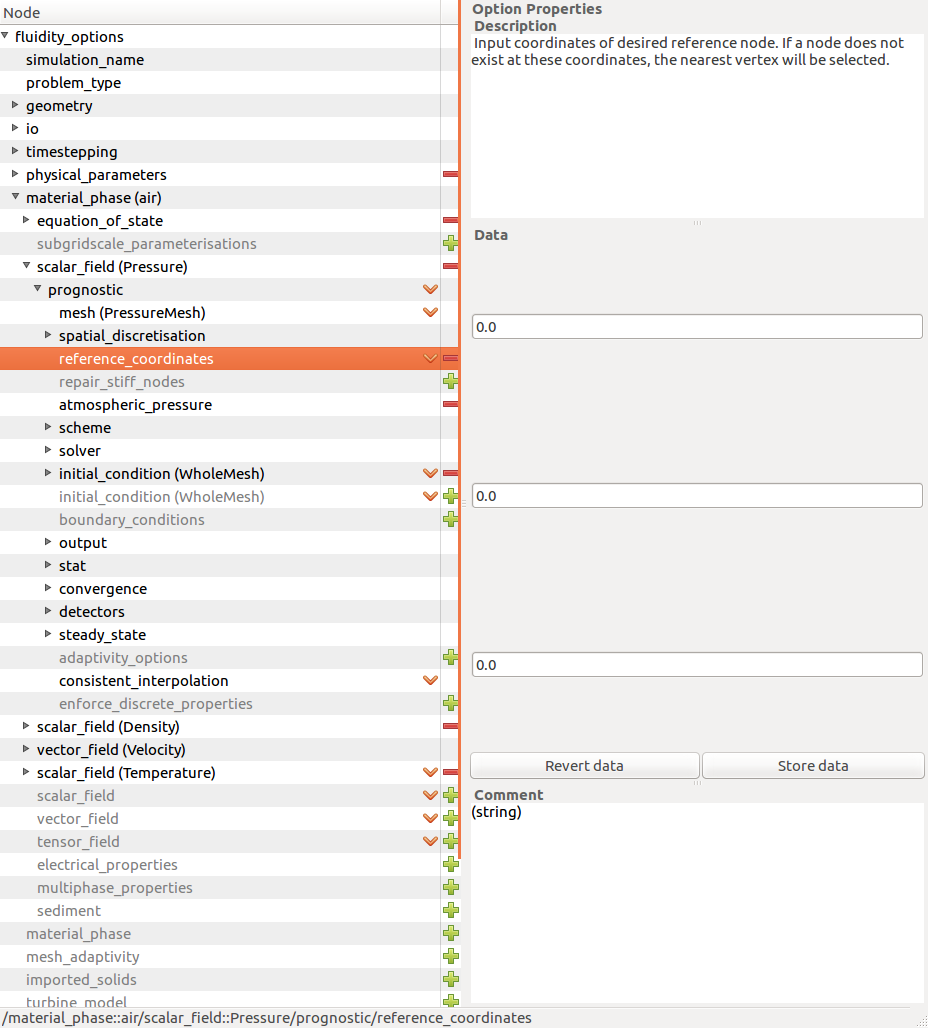}
        \caption{$P_{ref}$ at given coordinates}
        \label{Fig:PressureBC_NodeCoord}
    \end{subfigure}
    \begin{subfigure}{0.3\textwidth}
        \includegraphics[width=\textwidth]{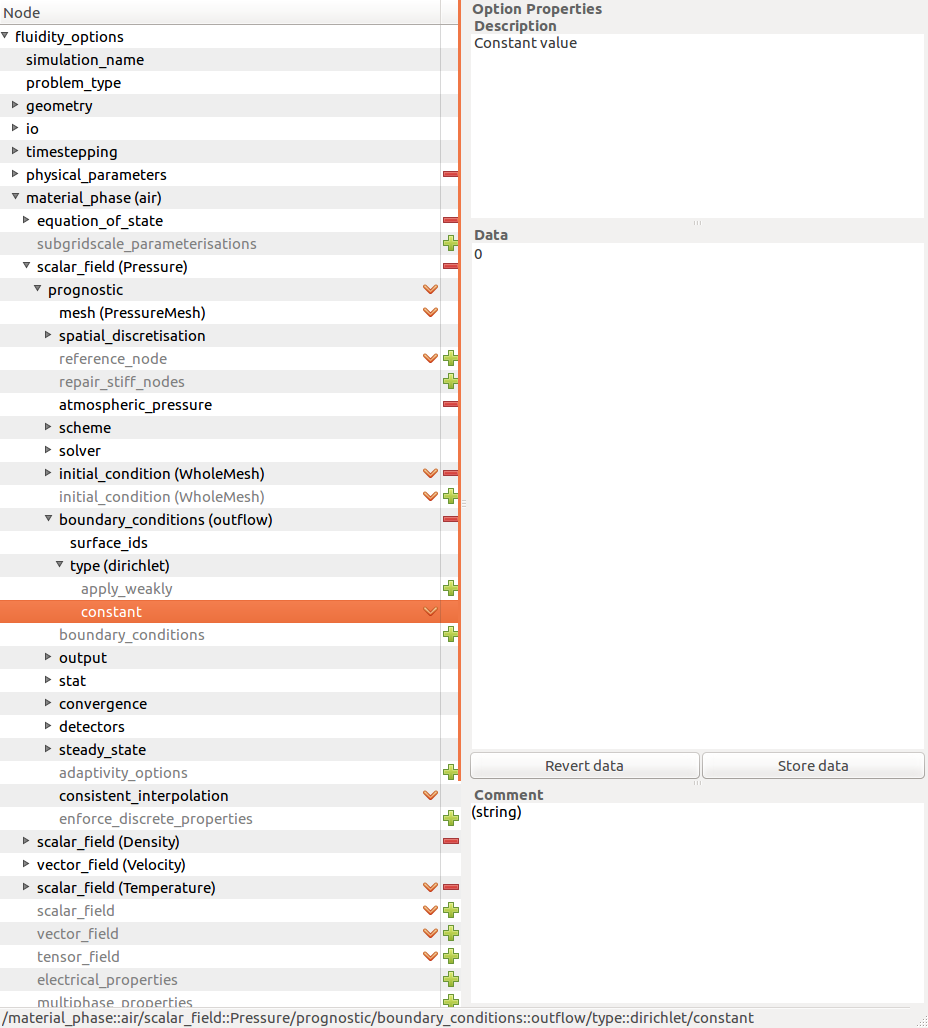}
        \caption{$P_{ref}$ on outlet surface}
        \label{Fig:PressureBC_Surface}
    \end{subfigure}
    \caption{The three different ways to assign a reference pressure in \textbf{Fluidity}.}
    \label{Fig:PressureBC}
\end{figure}

In every simulations, a reference pressure needs to be given. In \textbf{Fluidity}, there are three different ways to assign the reference pressure:
\begin{itemize}
    \item The reference pressure is given at a specific node
    \begin{itemize}
        \item defined by one node's ID \texttt{reference\_node} (Figure~\ref{Fig:PressureBC_NodeID})
        \item defined by one node's coordinates \texttt{reference\_}\texttt{coordinates} (Figure~\ref{Fig:PressureBC_NodeCoord})
    \end{itemize}
    \item The reference pressure is given as a \texttt{boundary\_conditions} using a Dirichlet type, usually imposed equal to 0 at the outlet surface (Figure~\ref{Fig:PressureBC_Surface}).
\end{itemize}

\noindent The last option (reference pressure given as a boundary condition) is recommended.

\section{Common errors}
If the simulation crashes, the user can have a look at the file \texttt{fluidity-err.0}, which will give information about the reasons for the crash. Usually, the main and recurrent errors are:
\begin{itemize}
    \item \textbf{Mesh errors}: The mesh is not consistent and this might be caused by the user not following the steps in Chapter~\ref{Sec:GeometryMesh}.
    \item \textbf{Python script errors}: The Python language is sensitive to indentation. As such the indentation needs to be checked in the python scripts.
    \item \textbf{Non-convergence of the solver}: This is frequently caused by the time step being too large. The user should reduce the time step and/or the CFL number.
\end{itemize}

See also section \ref{Sec:BCTricks} for other tricks.

    \chapter{Mesh adaptivity}\label{Sec:MeshAdaptivity}
\section{Explanation and tricks}
\subsection{Explanations}
One of the key aspects of \textbf{Fluidity} is its mesh adaptivity capability with unstructured meshes, making it a unique tool that enhances and provides detailed and accurate information at high resolutions within the computational domain. The aim of this section is not to describe the theory behind mesh adaptivity but to explain, from a user point of view, how to set up mesh adaptivity options. The user can refer to ~\cite{AMCG2015} and ~\cite{Pain2001} for more details regarding the theory. Different mesh adaptivity algorithms exist and the one used in this document is the \texttt{hr-adaptivity} one based on the change of the connectivity of the mesh and the relocation of the vertices of the mesh while retaining that same connectivity, as described in~\cite{Pain2001}.

\begin{figure}
    \centering
        \begin{subfigure}{0.49\textwidth}
        \includegraphics[width=\textwidth]{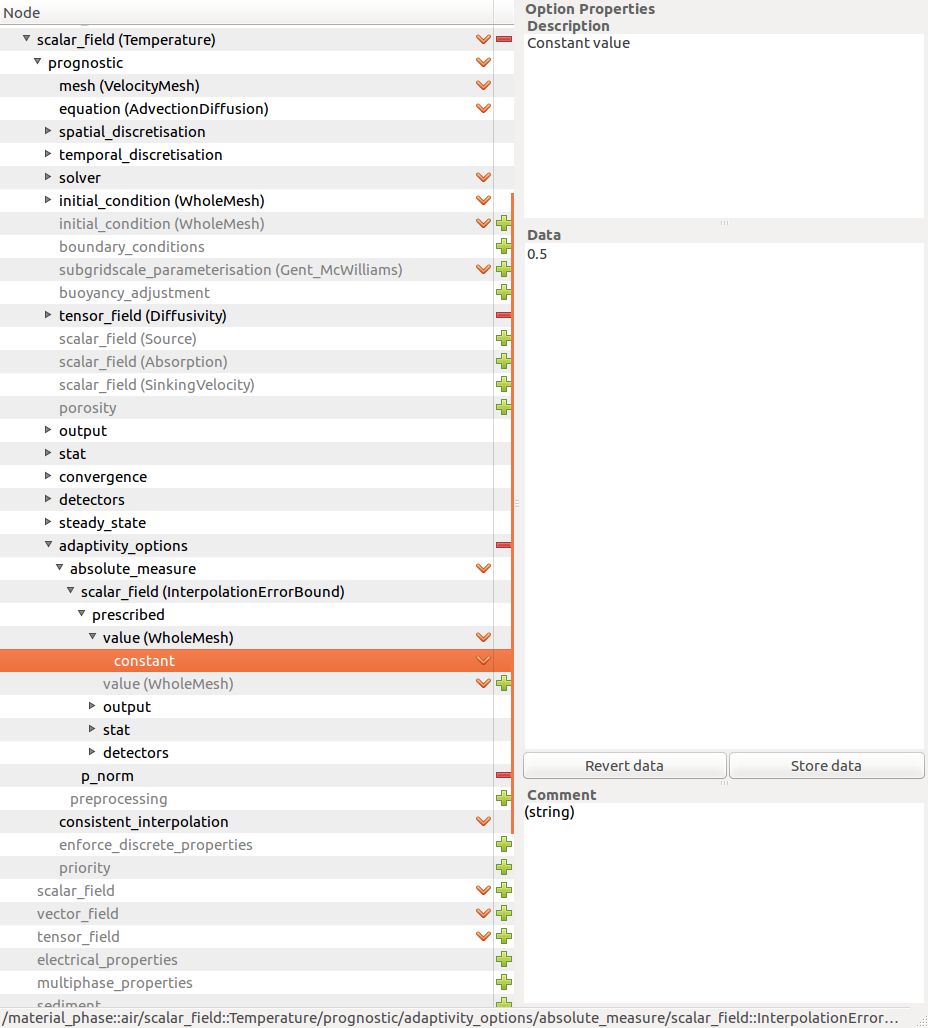}
        \caption{ }
        \label{Fig:MeshAdapt_Field}
    \end{subfigure}
    \begin{subfigure}{0.49\textwidth}
        \includegraphics[width=\textwidth]{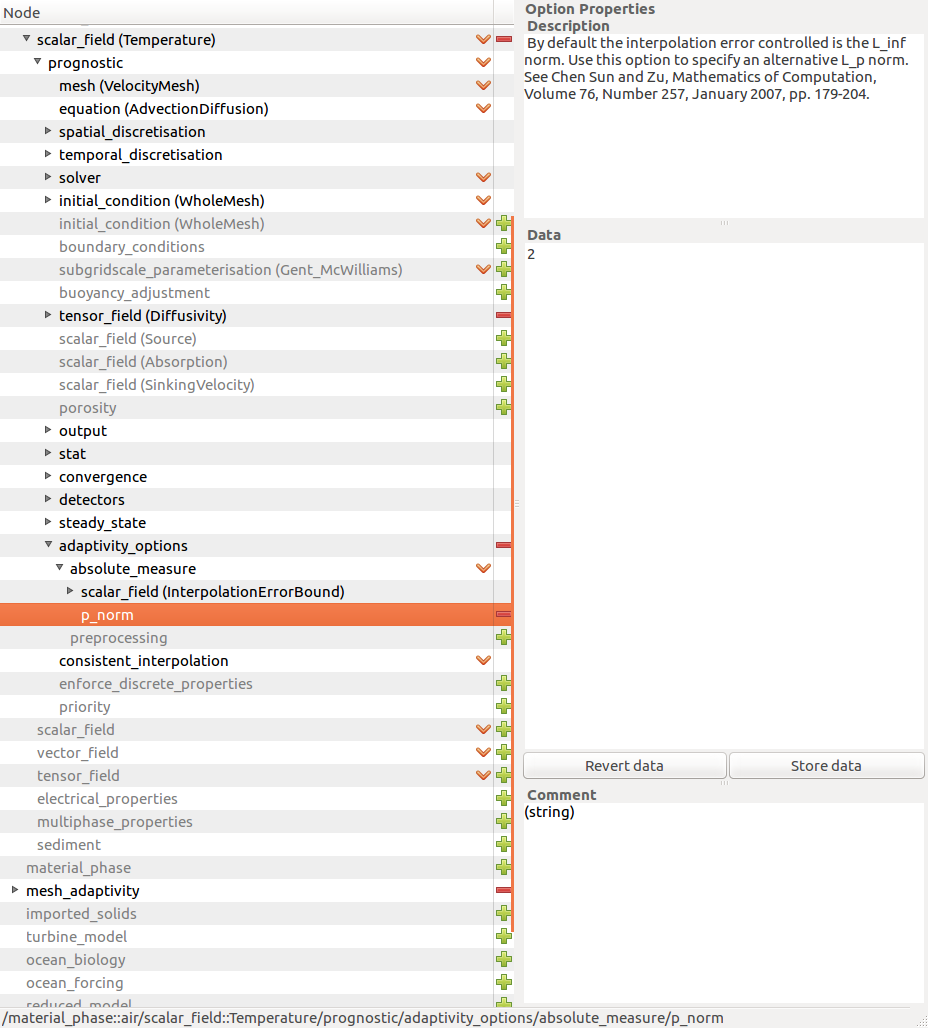}
        \caption{ }
        \label{Fig:MeshAdapt_pnorm}
    \end{subfigure}
    \begin{subfigure}{0.49\textwidth}
        \includegraphics[width=\textwidth]{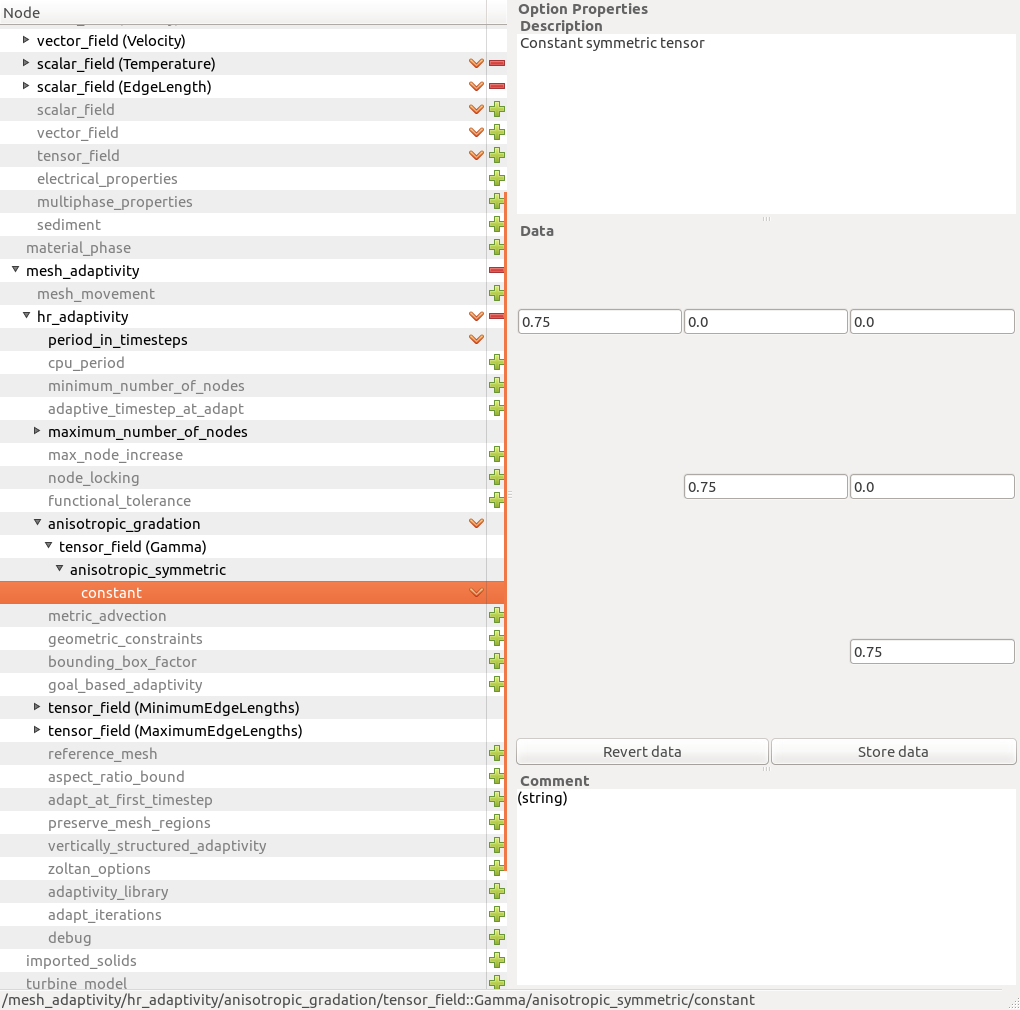}
        \caption{ }
        \label{Fig:MeshAdapt_Gradation}
    \end{subfigure}
    \begin{subfigure}{0.49\textwidth}
        \includegraphics[width=\textwidth]{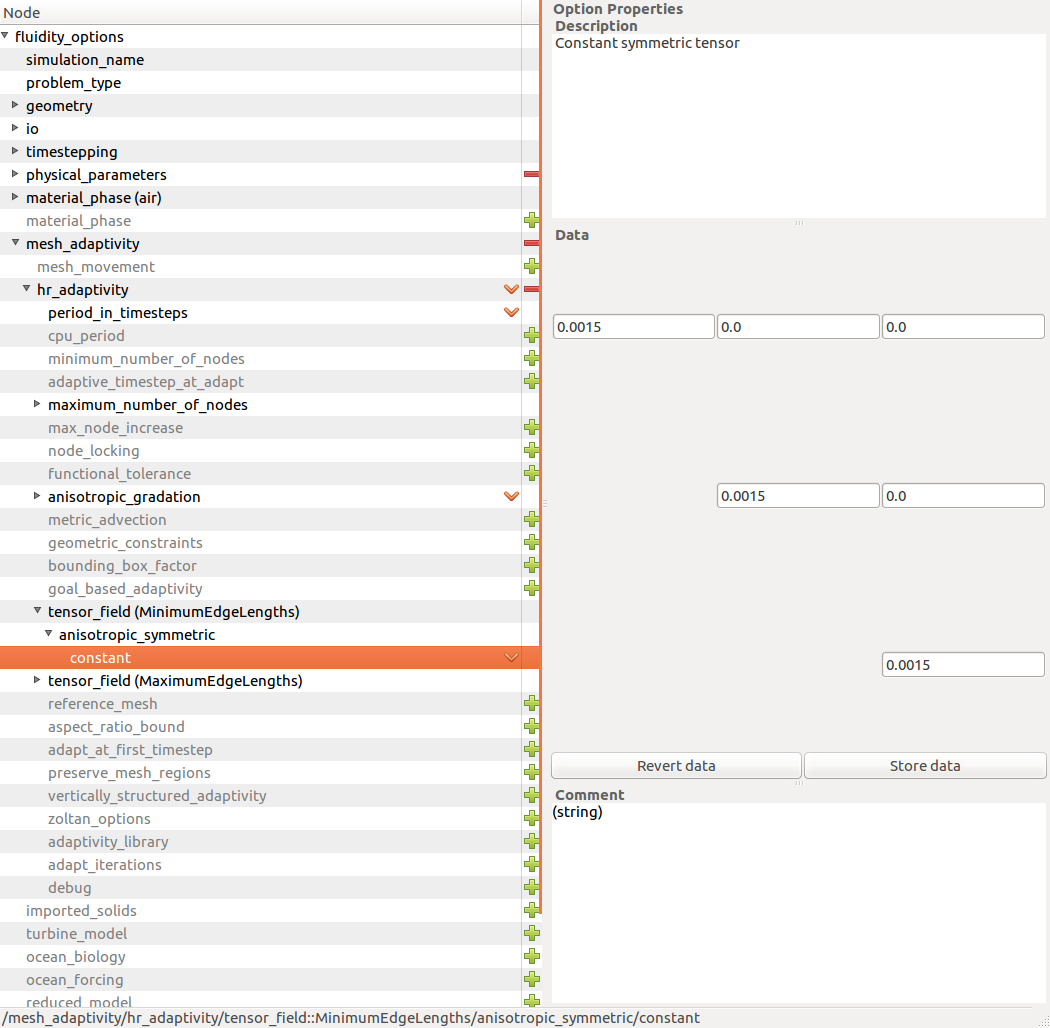} 
        \caption{ }
        \label{Fig:MeshAdapt_EdgeLength}
    \end{subfigure}
    \caption{Options that need to be turned on in \textbf{Diamond} for mesh adaptivity.}
    \label{Fig:MeshAdap}
\end{figure}

\noindent The mesh adaptivity process refines automatically the mesh in regions where significant physical processes are happening, which implies that the mesh adaptivity process is field-specific. The mandatory options that need to be turned on for mesh adaptivity are the following:
\begin{itemize}
    \item \textbf{In the \texttt{Field} of interest:} As the mesh adaptivity is field-specific, the option \texttt{adaptivity}\texttt{\_options} in the field of interest needs to be turned on as shown in Figure~\ref{Fig:MeshAdapt_Field} and the \texttt{error\_bound\_interpolation} value has to be set (see Section~\ref{Sec:TricksOptions}). Moreover, the option \texttt{p\_norm} can also be enabled and set to 2. Historically, the interpolation error was first controlled in the L$_{\infty}$ norm. The metric formulation which controls the L$_{\infty}$ norm is the simplest, and remains the default in \textbf{Fluidity} (option \texttt{p\_norm} is turned off by default). However the L$_{\infty}$ norm can have a tendency to focus the resolution entirely on the dynamics with the largest magnitude. Therefore, the L$_p$ norm, which also includes the influence of the dynamics with smaller magnitudes, can be used. Empirical experience indicates that choosing $p = 2$, and hence the L$_2$ norm, generally gives better results. For that reason we recommend it as default for all adaptivity configurations (Figure~\ref{Fig:MeshAdapt_pnorm}). However, this option can also tend to focus excessively the resolution on dynamics of very small magnitudes. The user should do a prior sensibility analysis to figure out which option is more appropriate for the case considered.  See sections 7.5.1 and 7.5.2 of the \textbf{Fluidity} manual~\cite{AMCG2015} for more details. 
    \item \textbf{In the \texttt{Mesh\_adaptivity} options:} 
    \begin{itemize}
        \item \textbf{Period:} defines how often the mesh should be adapted. This can be set in number of simulation seconds \texttt{period}, or in number of time steps \texttt{period\_in\_timesteps}. Note that mesh adaptivity has a certain computational cost and a trade-off has to be found between how often the mesh is adapted and the total simulation time. It is recommended that adaptation happens every 10-20 time steps.
        \item \textbf{Maximum number of nodes:} sets the maximum possible number of nodes \texttt{maximum\_number\_of\_nodes} in the domain (see Section~\ref{Sec:TricksOptions} to know how). In parallel, by default, this is the global maximum number of nodes. If the mesh adaptivity algorithm wants to place more nodes than this, the desired mesh is coarsened everywhere in space until it fits within this limit. In general, the error tolerances should be set so that this is never reached; it should only be a safety catch. If the simulation runs in parallel, make sure that the maximum number of nodes specified is at least $Nbr_{Proc} \times 50,000$ nodes.
        \item \textbf{Gradation:} In numerical simulations, a smooth transition from small elements to large elements is generally important for mesh quality. Therefore, a mesh gradation algorithm is applied to smooth out sudden variations in the mesh sizing function. Various mesh gradation algorithms have been introduced to solve this problem and the one recommended is the \texttt{anisotropic\_gradation}, with $0.75$ prescribed on the diagonal and $0$ otherwise, as shown in Figure~\ref{Fig:MeshAdapt_Gradation}.
        \item \textbf{Minimum edge length:} is the minimum edge length of an element allowed in the mesh (in meters). The input to this quantity is a tensor allowing one to impose different limits in different directions (see Figure~\ref{Fig:MeshAdapt_EdgeLength}). See Section~\ref{Sec:TricksOptions} to know how to find the appropriate value. This condition is not a hard constraint and the user may observe the constraint being (slightly) broken in places.
        \item \textbf{Maximum edge length:} is the maximum edge length of an element allowed in the mesh (in meters). The input to this quantity is a tensor allowing one to impose different limits in different directions. See Section~\ref{Sec:TricksOptions} to know how to find the appropriate value. This condition is not a hard constraint and the user may observe the constraint being (slightly) broken in places.
    \end{itemize}
\end{itemize}

\noindent The minimum and maximum edge lengths and the maximum number of nodes, in combination with the interpolation error bound, will define how the resolution of the mesh varies with adaptation.

\subsection{Tricks to set up mesh adaptivity}\label{Sec:TricksOptions}
Setting up the correct parameters needed for mesh adaptivity is not trivial and there are no universal rules. The user needs to experiment with the three main parameters which are: \texttt{Interpolation\_error\_bound}, \texttt{Minimum\_edge\_length} and \texttt{Maximum\_edge\_length}. However, here are some basic rules that can be followed.

\begin{itemize}
    \item \textbf{Maximum number of nodes:} As a start, the user is suggested to use a relatively small number of maximum nodes to test all the options: between $100,000$ and $200,000$ is recommended. This value can be increased later.
    \item \textbf{Minimum edge length:} The minimum edge length should be quite large to start with. The value $H/1000$, where $H$ is the height of the domain OR the value $L/100$, where $L$ is a characteristic length of your geometry (here, the height of the openings for example), are recommended. This value can be progressively decreased afterwards to reach the resolution needed.
    \item \textbf{Maximum edge length:} The maximum edge length should be quite large to start with and the value $H/10$, where $H$ is the height of the domain, is recommended.
    \item \textbf{Interpolation error bound:} The general advice would be to start with a high interpolation error, 10$\%$ of the range of the field considered being a good rule of thumb. As this value will certainly be too high, it is then recommended to reduce this value by small increments until reaching a refinement able to represent the desired dynamics. In any case, the interpolation error bound value should be the main parameter to vary to control the resolution of the adapted meshes. The value of \texttt{p\_norm} should be equal to $2$ as a first try. However, using $p=2$ might also tend to focus the resolution on dynamics of too small magnitudes. The user should do a prior sensibility analysis with and without the \texttt{p\_norm} function to figure out which option is more appropriate for the case considered.
\end{itemize}

\noindent Python scripts can be used to vary the specified lengths over the domain, allowing finer resolutions over areas of interest for instance. This will be discussed in Section~\ref{Sec:AdaptPython}.

\noindent \textbf{Important note:} The velocity field and/or the temperature (or any scalar field) can be adapted. Several field can be taken into account during adaptation. However, it is not recommended to adapt the pressure field: for cases involving indoor-outdoor exchanges and/or the urban environment, one might even say it is forbidden!

\section{Summary of the simulations}
\noindent Table~\ref{Tab:SimuCasesAdapt} summarises the different simulations presented in the next sections.

\begin{table}
    \centering
    \begin{tabular}{ |p{1.1cm}||p{1.35cm}|p{3.0cm}|p{3cm}|p{3cm}|p{1.5cm}|  }
         \hline
         \textbf{Case \newline Nbr} & \textbf{Initial \newline Test \newline Case \newline Nbr} & \textbf{Field \newline adapted} & \bf{Interpolation \newline error \newline bound} & \bf{Advected \newline mesh} & \textbf{Section}\\
         \hline
         6a & 5a & Temperature & 0.5 & No & ~\ref{Sec:AdaptTemp}\\\hline
         6b & 5a & Temperature & 0.3 & No & ~\ref{Sec:AdaptTemp}\\\hline
         6c & 5a & Temperature & 0.1 & No & ~\ref{Sec:AdaptTemp}\\\hline
         6d & 5a & Temperature & 0.05 & No & ~\ref{Sec:AdaptTemp}\\\hline
         7a & 5a & Velocity    & 0.5 & No & ~\ref{Sec:AdaptVel}\\\hline
         7b & 5a & Velocity    & 0.25 & No & ~\ref{Sec:AdaptVel}\\\hline
         7c & 5a & Velocity    & 0.15 & No & ~\ref{Sec:AdaptVel}\\\hline
         7d & 5a & Velocity    & 0.1 & No & ~\ref{Sec:AdaptVel}\\\hline
         8  & 5a & Temperature \newline Velocity & 0.15 \newline 0.15 & No & ~\ref{Sec:AdaptVelTemp}\\\hline
         9  & 2b & Temperature \newline Velocity & 075 \newline 0.045 & No & ~\ref{Sec:AdaptPython}\\\hline
         10 & 5a & Temperature & 0.1 & Yes & ~\ref{Sec:AdaptAdvec}\\\hline
    \end{tabular}
    \caption{\label{Tab:SimuCasesAdapt}Summary of the simulations presented in the following sections of the Chapter~\ref{Sec:MeshAdaptivity}. }
\end{table}

\section{Field specific adaptation}
\subsection{Set-up of examples}
\subsubsection{Boundary conditions}
In examples \textit{3dBox\_Case6a.flml} to \textit{3dBox\_Case7d.flml}, the initial velocity and the inlet velocity $(u,v,w)$ are set to $(1,0,0)$ m/s and the interior of the box is set to 298 K, while the outside remains at ambient temperature 293 K. These examples are similar to example \textit{3dBox\_Case5a.flml} which is set up without mesh adaptivity, i.e on a fixed mesh.

\subsubsection{CFL number}
\begin{figure}
    \begin{center}
        \includegraphics[scale=0.2]{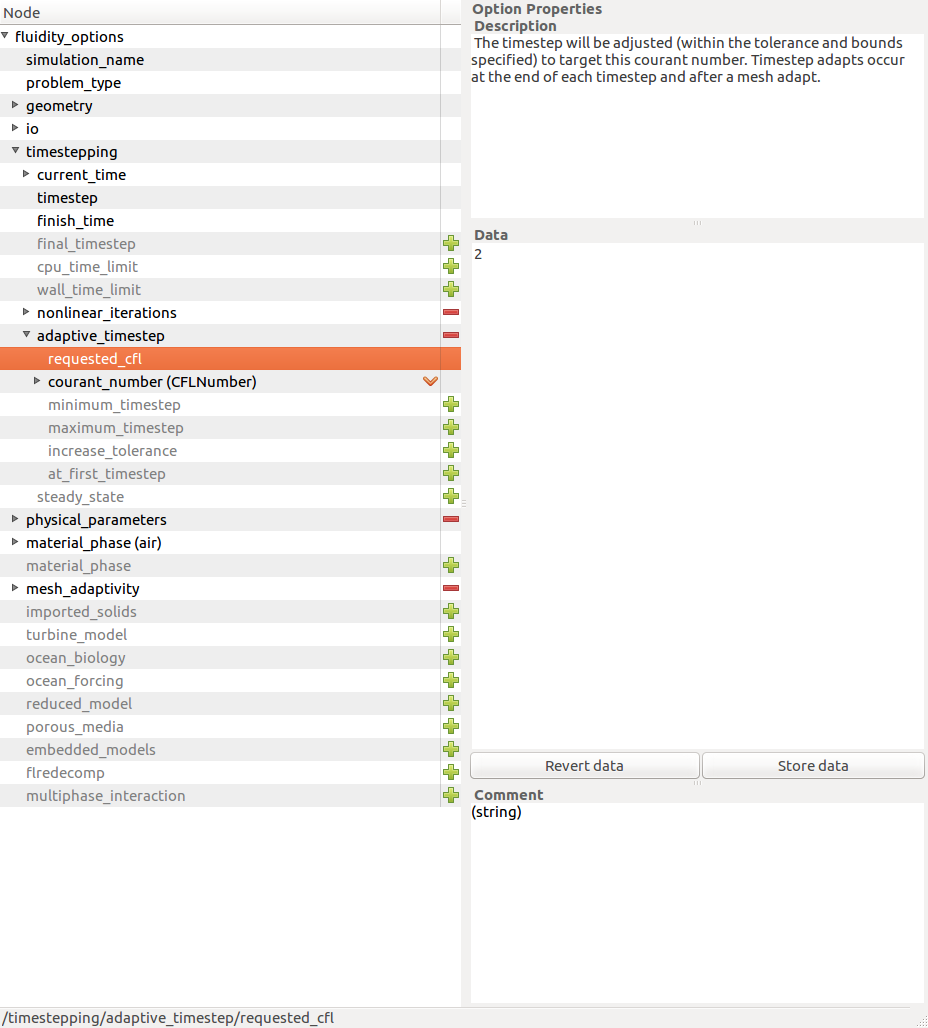}
        \caption{Adaptive time step.}
        \label{Fig:AdaptCFL}
    \end{center}
\end{figure}

To avoid any crash of the simulations, the time step is now also adaptive using a CFL condition as described in Section~\ref{Sec:CFLNumber} and the CFL number is taken equal to 2 in the following examples (Figure~\ref{Fig:AdaptCFL}). This value can be increased later on to speed up the simulations. However, it is recommended to the user to do a sensibility analysis of the results as a function of the CFL number to ensure that important information is not missing.

\subsubsection{General mesh adaptivity options}
In the following sections, mesh adaptivity will be performed based on the temperature field only (Section~\ref{Sec:AdaptTemp}), the velocity field only (Section~\ref{Sec:AdaptVel}) and both the velocity and temperature fields (Section~\ref{Sec:AdaptVelTemp}). When the user wants to adapt several fields, it is recommended firstly to analyse each field independently as done in the following sections.

\noindent The maximum number of nodes is set equal to $200,000$ and the mesh is adapted every 10 time steps. 

\noindent The characteristic length $L$ of the domain is taken to be the windows opening, i.e. 1 m: the minimum edge length is set up to $L/100$, i.e. 0.01 m. The height $H$ of the domain is 21 m: the maximum edge length is set up to $H/10$, i.e 2.1 m.

\subsection{Adaptation based on the temperature field}\label{Sec:AdaptTemp}
In this section, the mesh adaptivity process is prescribed based on the temperature field only.

\subsubsection{Mesh adaptivity options}
The range of the temperature is between 293 K and 298 K, i.e. a difference of 5 K. The \texttt{error\_bound\_interpolation} is in a first run set up at 10$\%$ of this temperature range: the value $5\times10\%=0.5$ is used in example \textit{3dBox\_Case6a.flml}. Then the \texttt{error\_bound\_interpolation} is progressively decreased to values equal to 0.3 (6$\%$ of the temperature range) in \textit{3dBox\_Case6b.flml}, 0.1 (2$\%$ of the temperature range) in \textit{3dBox\_Case6c.flml} and 0.05 (1$\%$ of the temperature range) in \textit{3dBox\_Case6d.flml}.

\noindent These examples can be run using the commands:
\begin{Terminal}[]
ä\colorbox{davysgrey}{
\parbox{435pt}{
\color{applegreen} \textbf{user@mypc}\color{white}\textbf{:}\color{codeblue}$\sim$
\color{white}\$ <<FluiditySourcePath>>/bin/fluidity -l -v3 3dBox\_Case6a.flml \&
\newline
\color{applegreen} \textbf{user@mypc}\color{white}\textbf{:}\color{codeblue}$\sim$
\color{white}\$ <<FluiditySourcePath>>/bin/fluidity -l -v3 3dBox\_Case6b.flml \&
\newline
\color{applegreen} \textbf{user@mypc}\color{white}\textbf{:}\color{codeblue}$\sim$
\color{white}\$ <<FluiditySourcePath>>/bin/fluidity -l -v3 3dBox\_Case6c.flml \&
\newline
\color{applegreen} \textbf{user@mypc}\color{white}\textbf{:}\color{codeblue}$\sim$
\color{white}\$ <<FluiditySourcePath>>/bin/fluidity -l -v3 3dBox\_Case6d.flml \&
}}
\end{Terminal}

\subsubsection{Results and discussion}
\begin{figure}[t]
    \centering
    \begin{subfigure}{0.4\textwidth}
        \includegraphics[width=\textwidth]{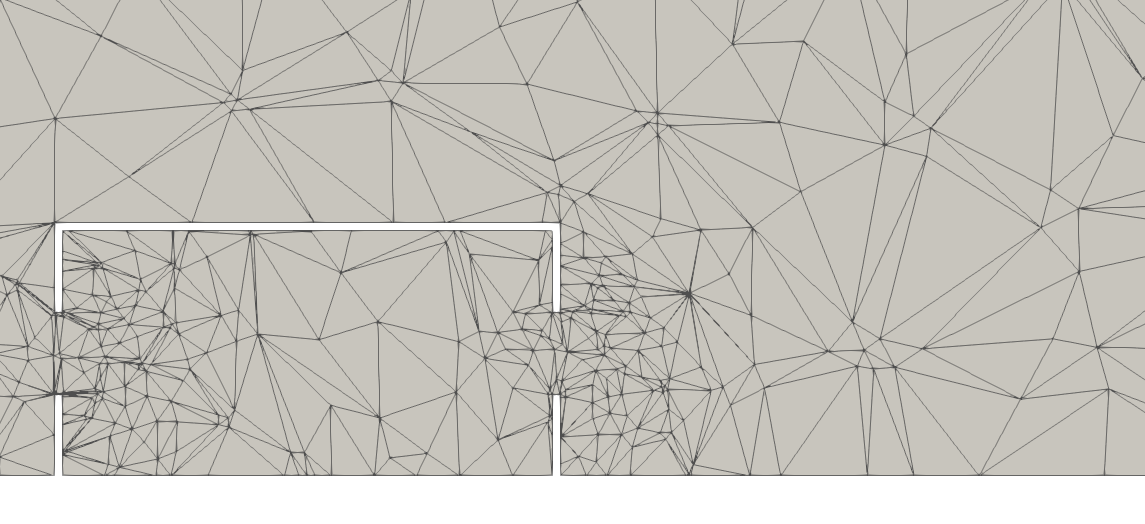}
        \caption{2 s}
        \label{Fig:Case6a_Mesh_2sec}
    \end{subfigure}
    \begin{subfigure}{0.41\textwidth}
        \includegraphics[width=\textwidth]{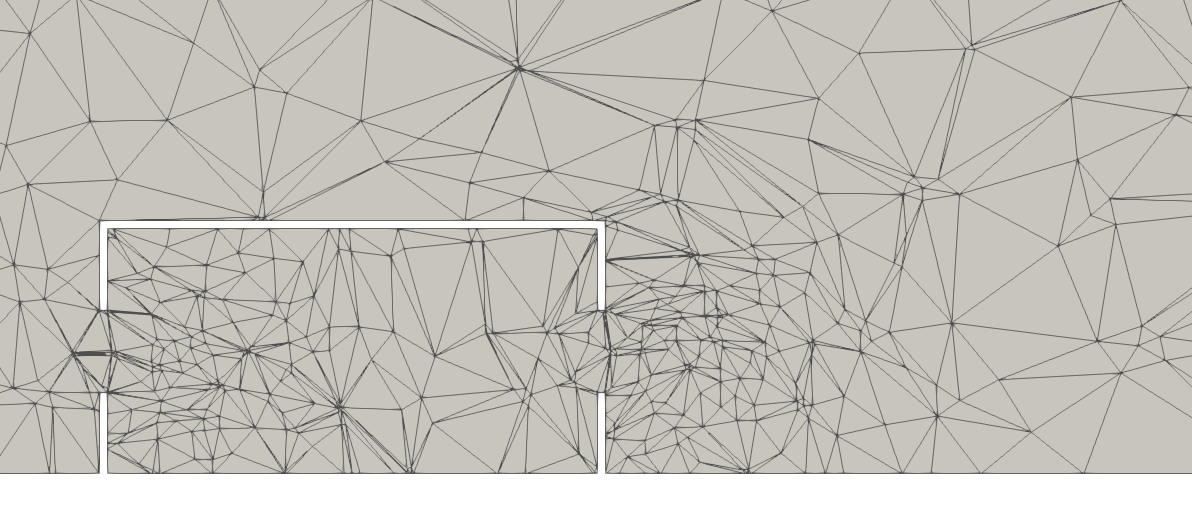}
        \caption{6 s}
        \label{Fig:Case6a_Mesh_6sec}
    \end{subfigure}
    \begin{subfigure}{0.4\textwidth}
        \includegraphics[width=\textwidth]{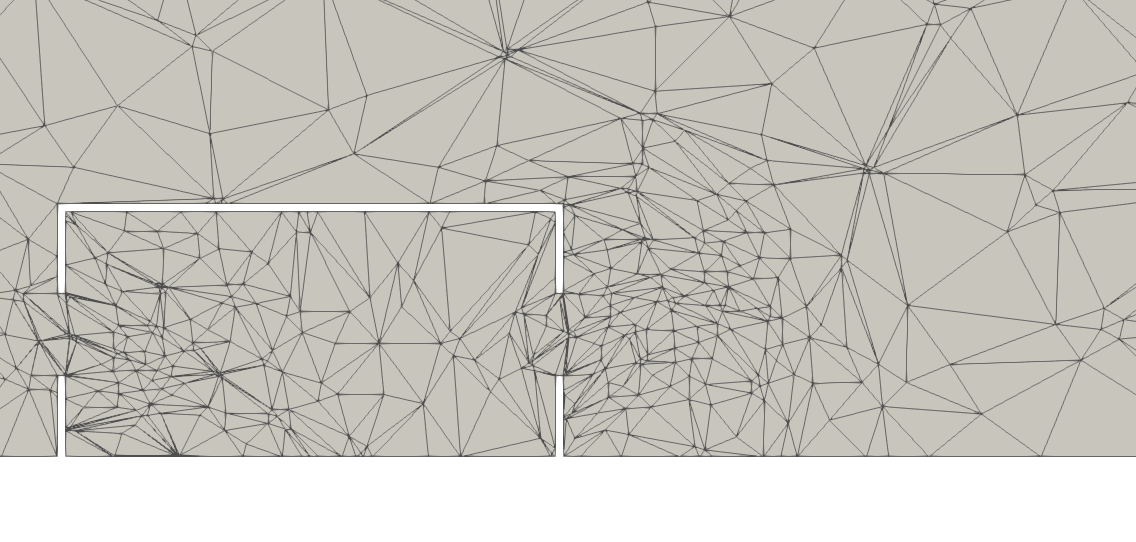}
        \caption{9 s}
        \label{Fig:Case6a_Mesh_9sec}
    \end{subfigure}
    \begin{subfigure}{0.4\textwidth}
        \includegraphics[width=\textwidth]{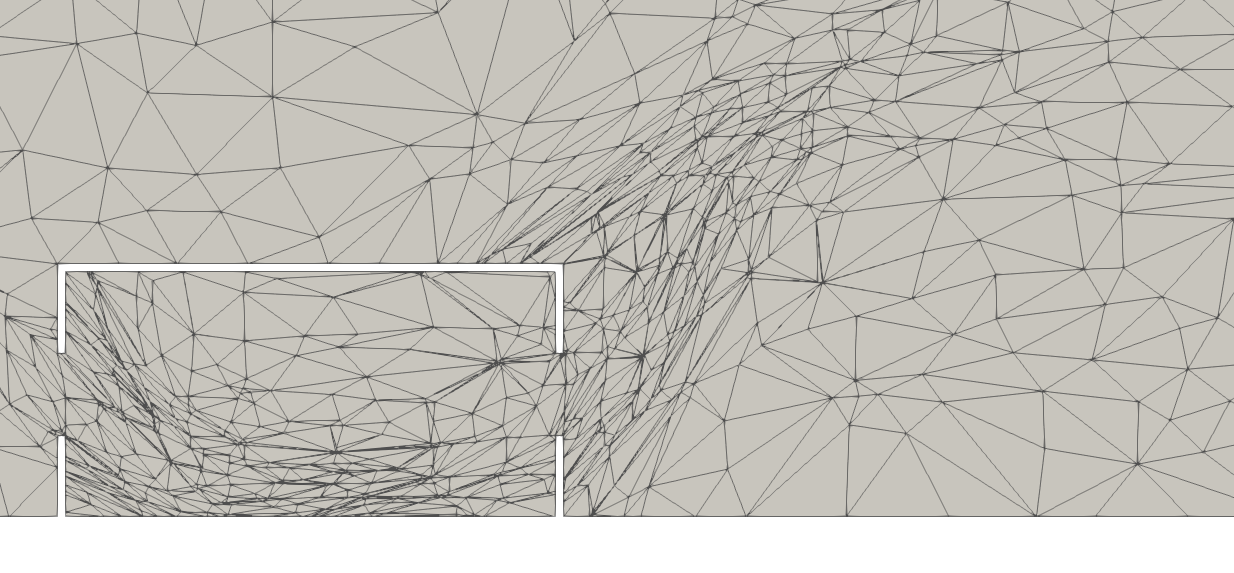}
        \caption{40 s}
        \label{Fig:Case6a_Mesh_40sec}
    \end{subfigure}
    \caption{Meshes at different instant for example \textit{3dBox\_Case6a.flml}. The temperature field only is adapted with an \texttt{error\_bound\_interpolation} equal to 0.5.}
    \label{Fig:Case6a_Mesh}
\end{figure}

\begin{figure}
    \centering    
    \begin{subfigure}{0.4\textwidth}
        \includegraphics[width=\textwidth]{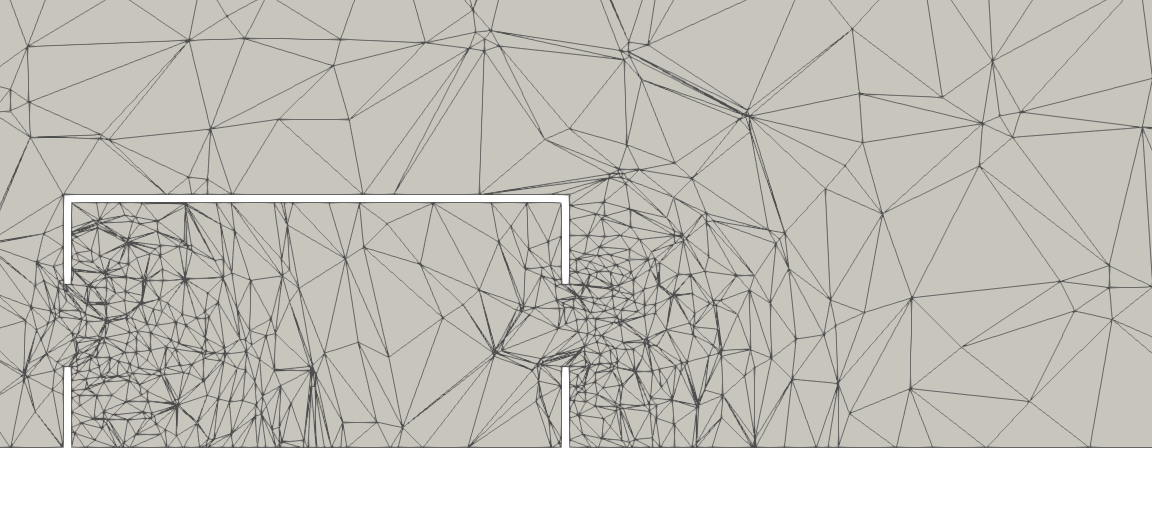}
        \caption{2 s}
        \label{Fig:Case6b_Mesh_2sec}
    \end{subfigure}
    \begin{subfigure}{0.41\textwidth}
        \includegraphics[width=\textwidth]{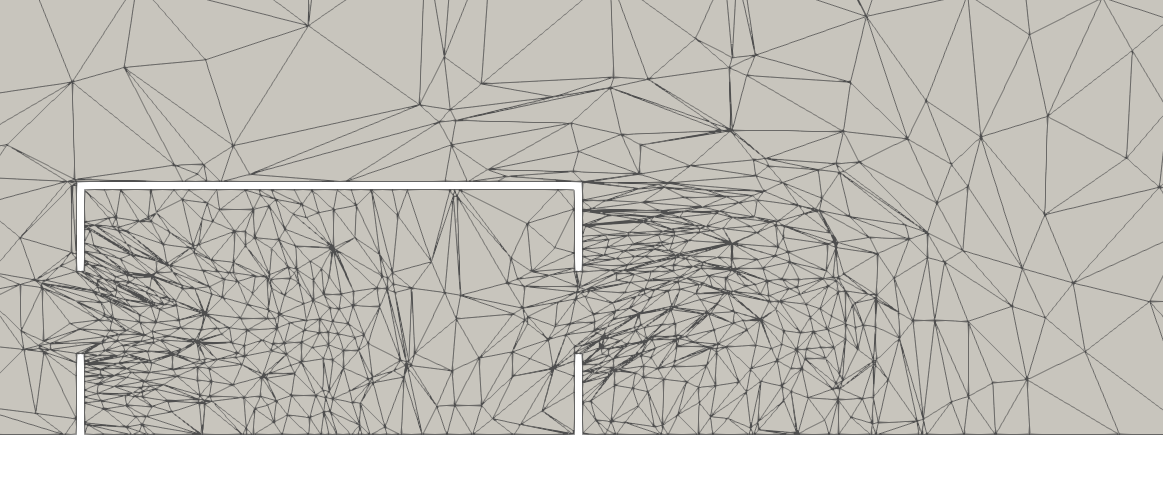}
        \caption{6 s}
        \label{Fig:Case6b_Mesh_6sec}
    \end{subfigure}
    \begin{subfigure}{0.41\textwidth}
        \includegraphics[width=\textwidth]{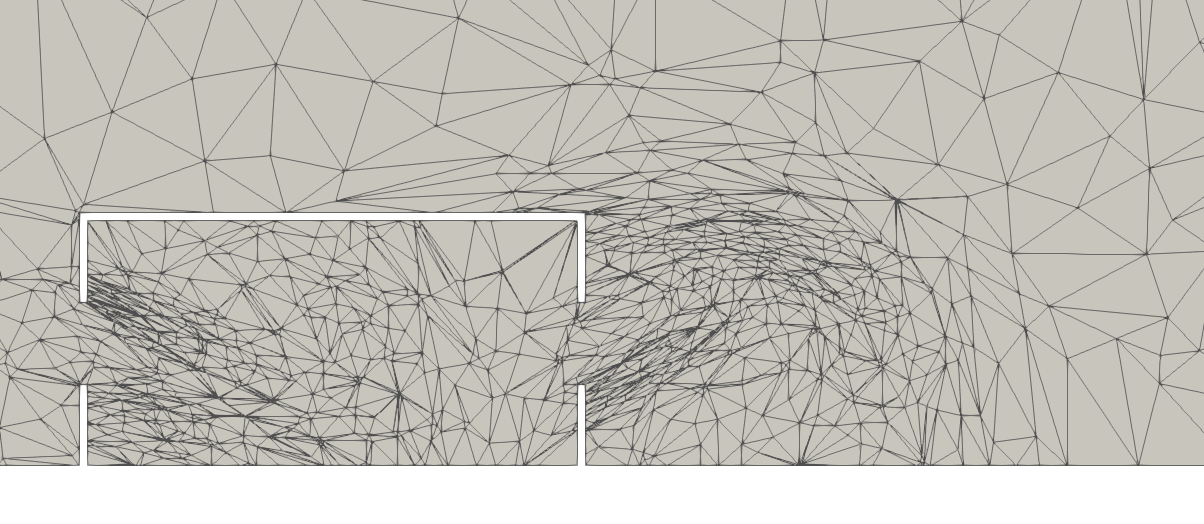}
        \caption{9 s}
        \label{Fig:Case6b_Mesh_9sec}
    \end{subfigure}
    \begin{subfigure}{0.4\textwidth}
        \includegraphics[width=\textwidth]{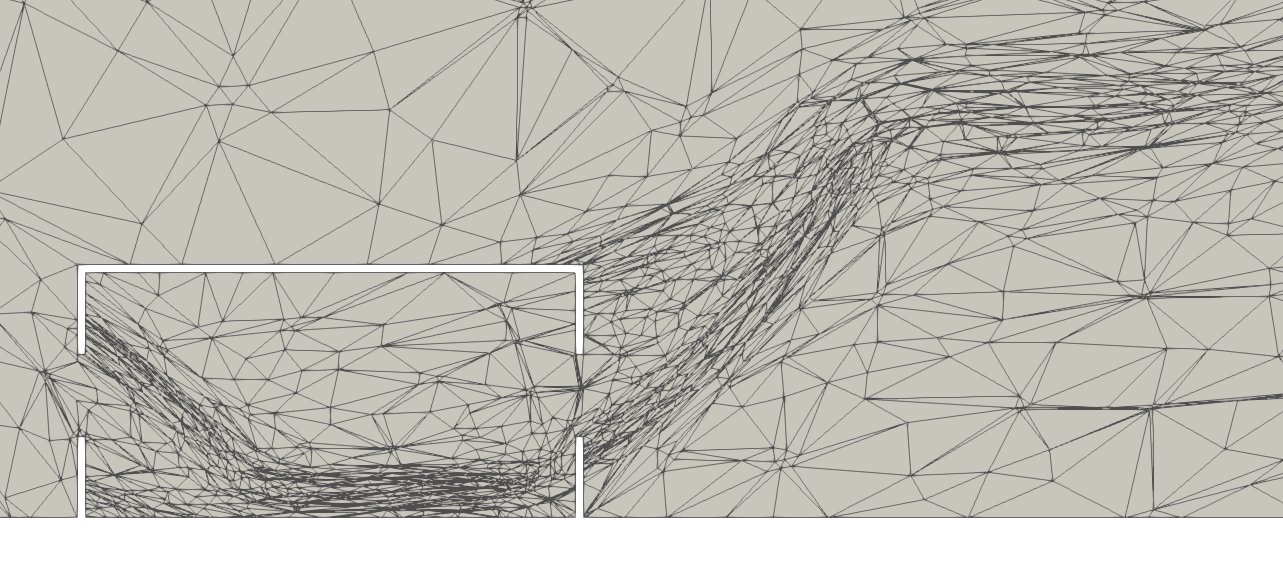}
        \caption{40 s}
        \label{Fig:Case6b_Mesh_40sec}
    \end{subfigure}
    \caption{Meshes at different instant for example \textit{3dBox\_Case6b.flml}. The temperature field only is adapted with an \texttt{error\_bound\_interpolation} equal to 0.3.}
    \label{Fig:Case6b_Mesh}
\end{figure}

\begin{figure}
    \centering
    \begin{subfigure}{0.4\textwidth}
        \includegraphics[width=\textwidth]{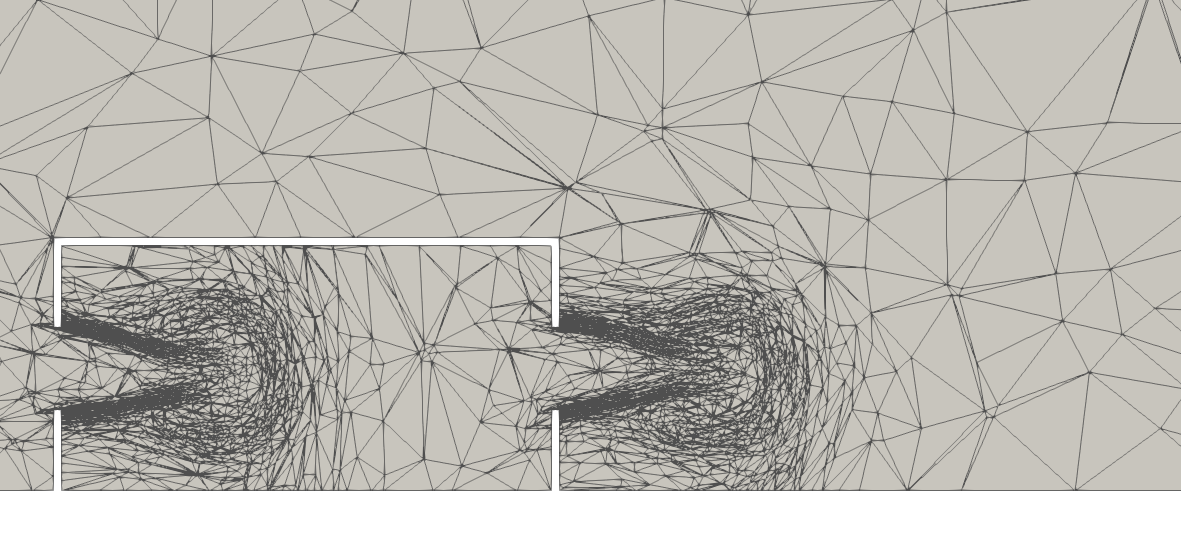}
        \caption{2 s}
        \label{Fig:Case6c_Mesh_2sec}
    \end{subfigure}
    \begin{subfigure}{0.41\textwidth}
        \includegraphics[width=\textwidth]{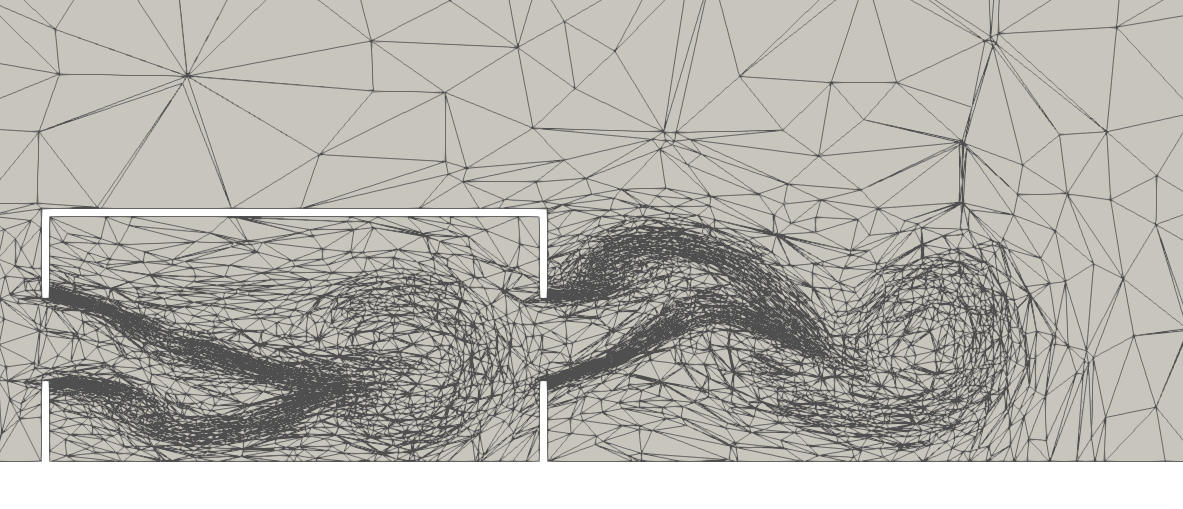}
        \caption{6 s}
        \label{Fig:Case6c_Mesh_6sec}
    \end{subfigure}
    \begin{subfigure}{0.4\textwidth}
        \includegraphics[width=\textwidth]{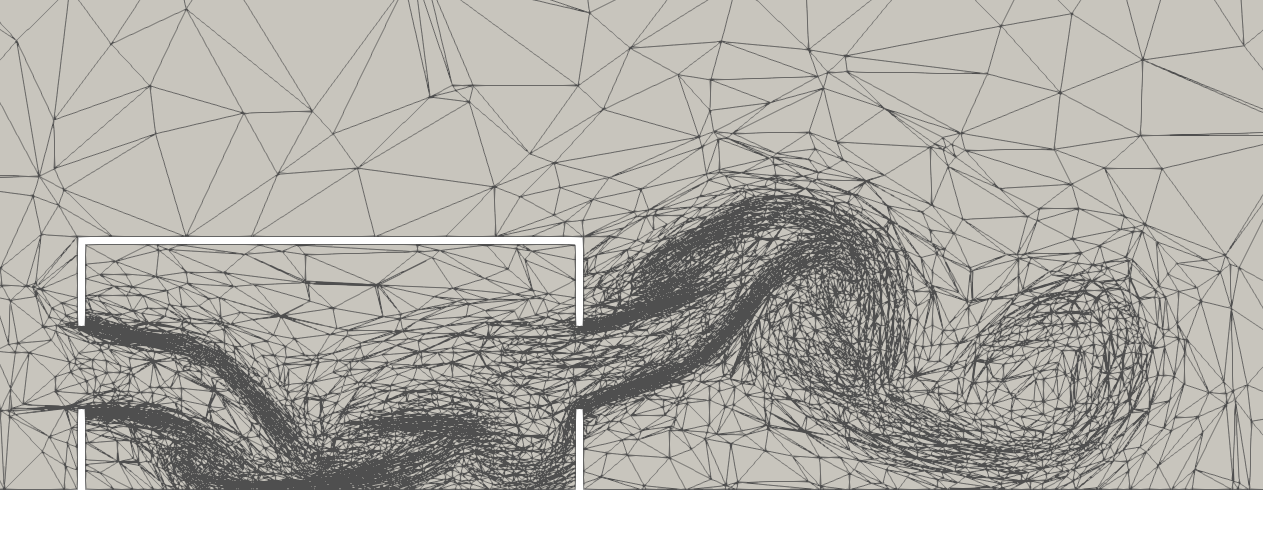}
        \caption{9 s}
        \label{Fig:Case6c_Mesh_9sec}
    \end{subfigure}
    \begin{subfigure}{0.4\textwidth}
        \includegraphics[width=\textwidth]{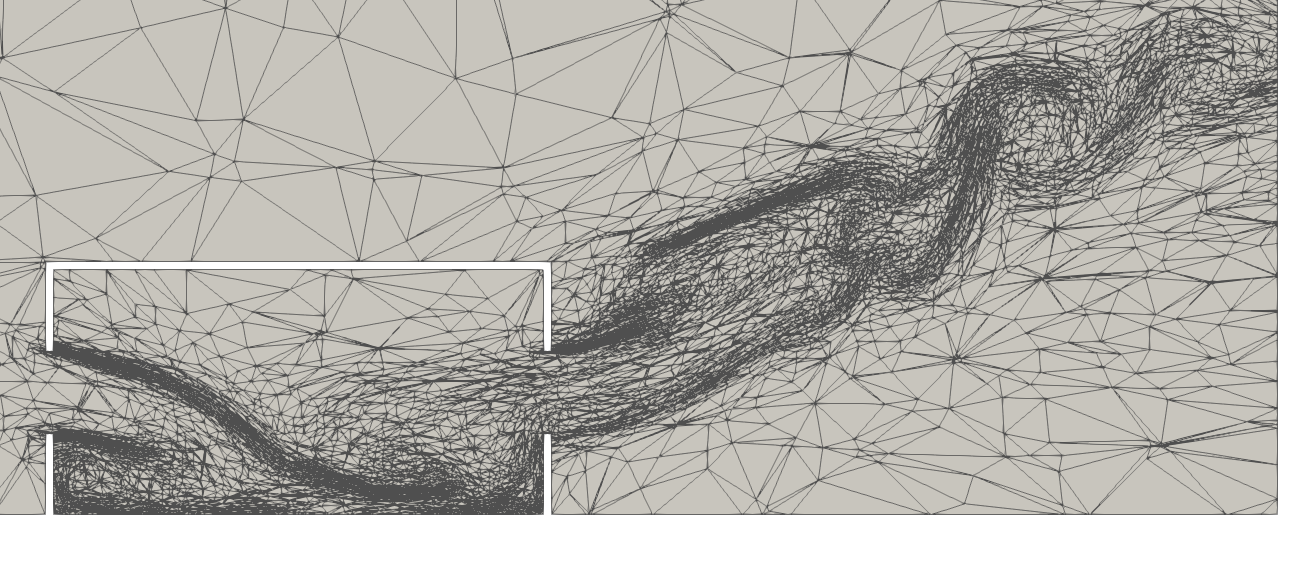}
        \caption{40 s}
        \label{Fig:Case6c_Mesh_40sec}
    \end{subfigure}
    \caption{Meshes at different instant for example \textit{3dBox\_Case6c.flml}. The temperature field only is adapted with an \texttt{error\_bound\_interpolation} equal to 0.1.}
    \label{Fig:Case6c_Mesh}
\end{figure}

\begin{figure}
    \centering
    \begin{subfigure}{0.4\textwidth}
        \includegraphics[width=\textwidth]{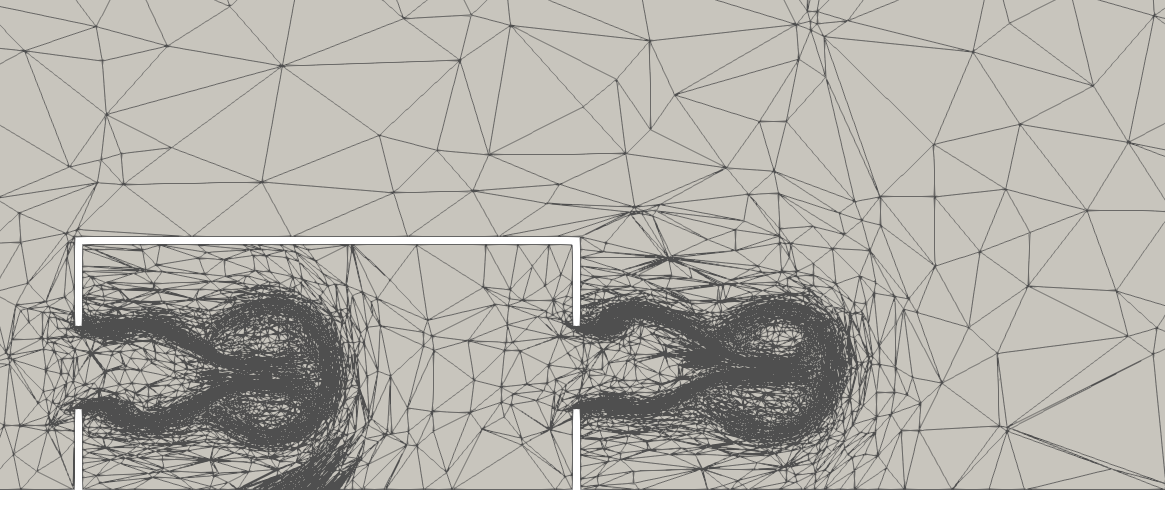}
        \caption{2 s}
        \label{Fig:Case6d_Mesh_2sec}
    \end{subfigure}
    \begin{subfigure}{0.42\textwidth}
        \includegraphics[width=\textwidth]{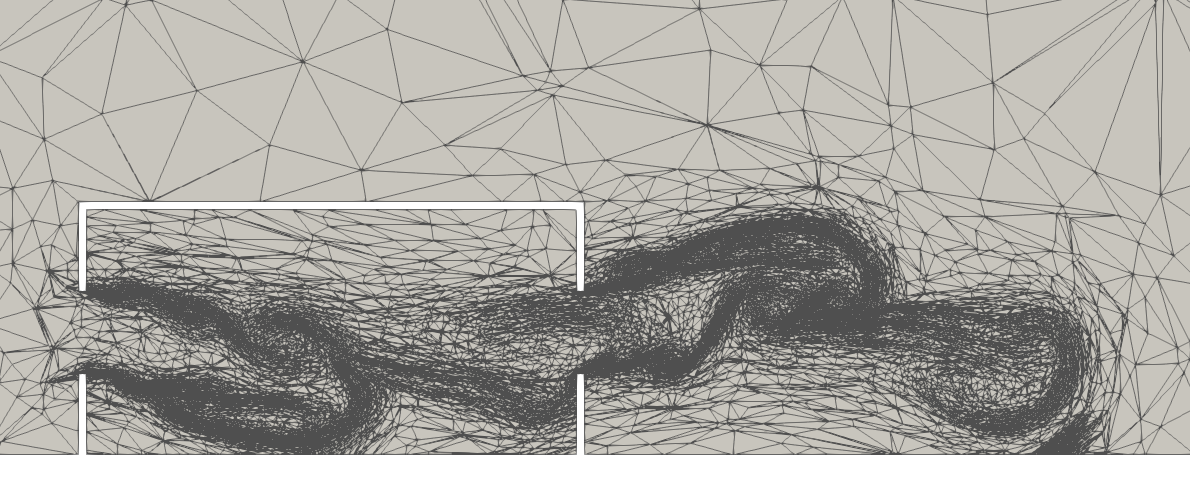}
        \caption{6 s}
        \label{Fig:Case6d_Mesh_6sec}
    \end{subfigure}
    \begin{subfigure}{0.42\textwidth}
        \includegraphics[width=\textwidth]{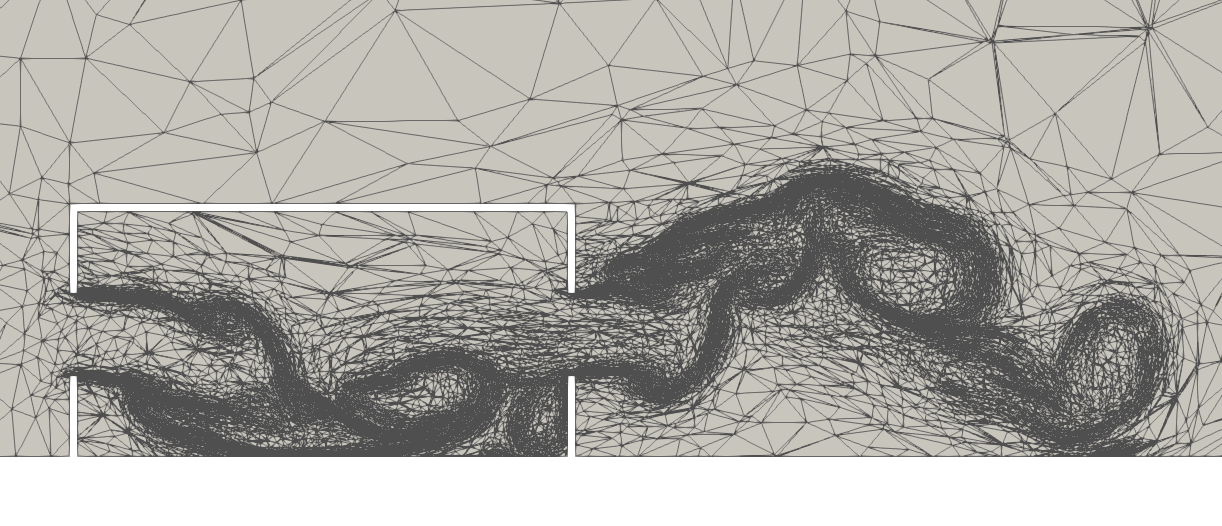}
        \caption{9 s}
        \label{Fig:Case6d_Mesh_9sec}
    \end{subfigure}
    \begin{subfigure}{0.4\textwidth}
        \includegraphics[width=\textwidth]{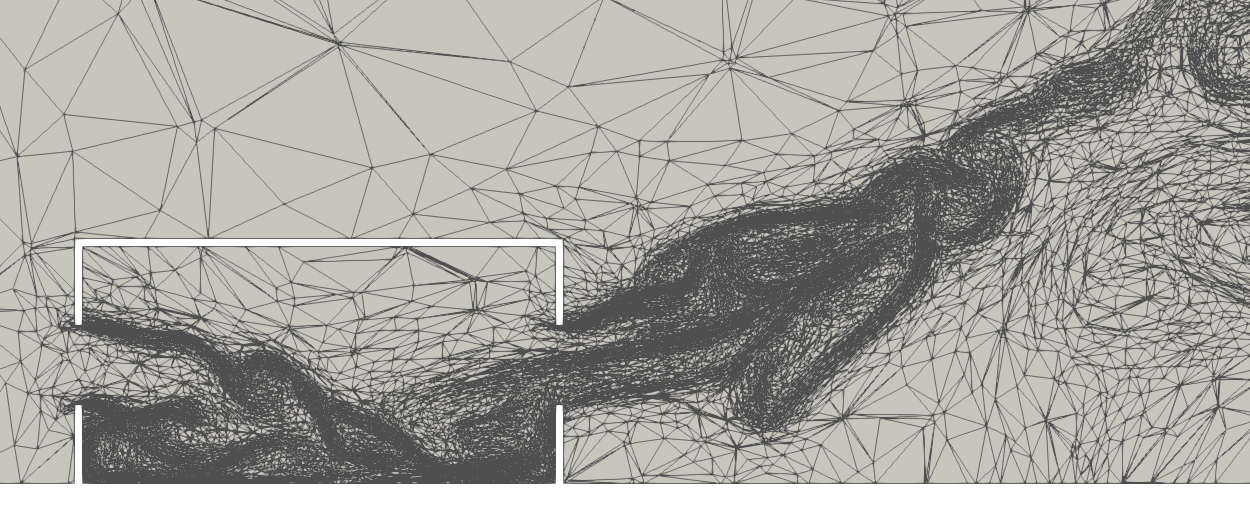}
        \caption{30 s}
        \label{Fig:Case6d_Mesh_40sec}
    \end{subfigure}
    \caption{Meshes at different instant for example \textit{3dBox\_Case6d.flml}. The temperature field only is adapted with an \texttt{error\_bound\_interpolation} equal to 0.05.}
    \label{Fig:Case6d_Mesh}
\end{figure}

\begin{figure}
    \centering
    \begin{subfigure}{0.4\textwidth}
        \includegraphics[width=\textwidth]{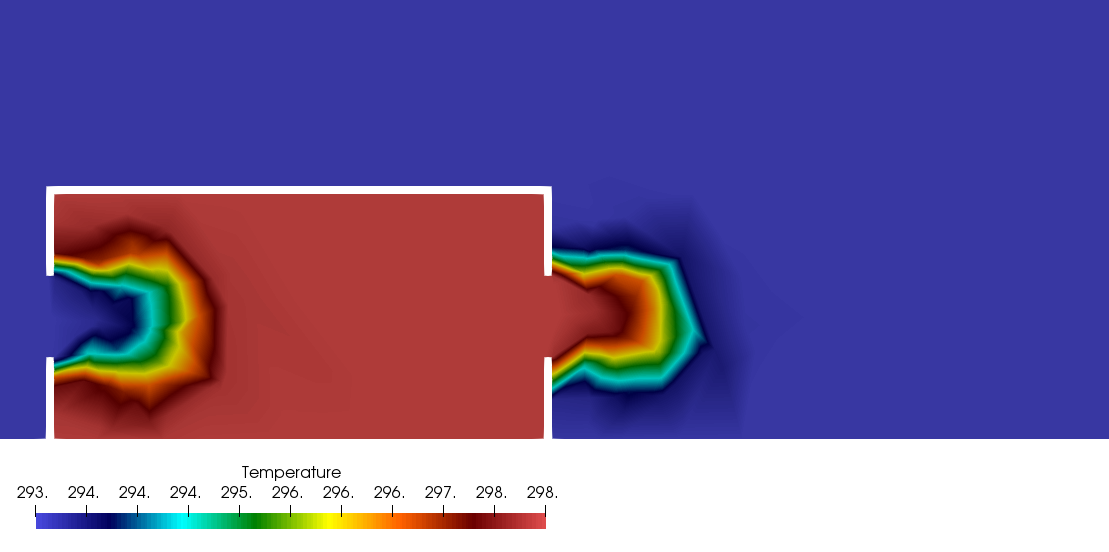}
        \caption{2 s}
        \label{Fig:Case6a_Temp_2sec}
    \end{subfigure}
    \begin{subfigure}{0.42\textwidth}
        \includegraphics[width=\textwidth]{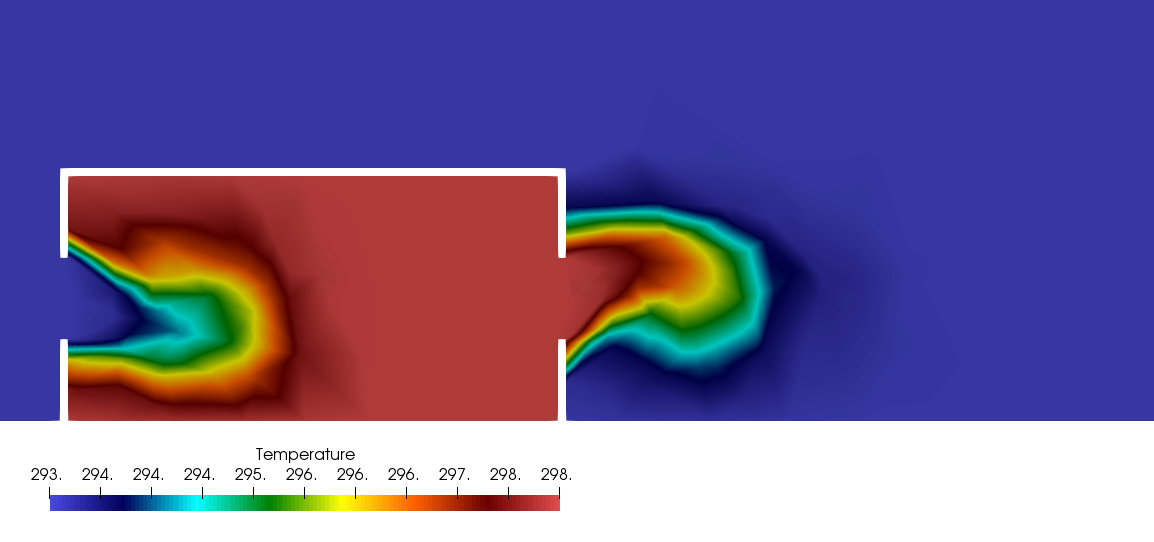}
        \caption{6 s}
        \label{Fig:Case6a_Temp_6sec}
    \end{subfigure}
    \begin{subfigure}{0.42\textwidth}
        \includegraphics[width=\textwidth]{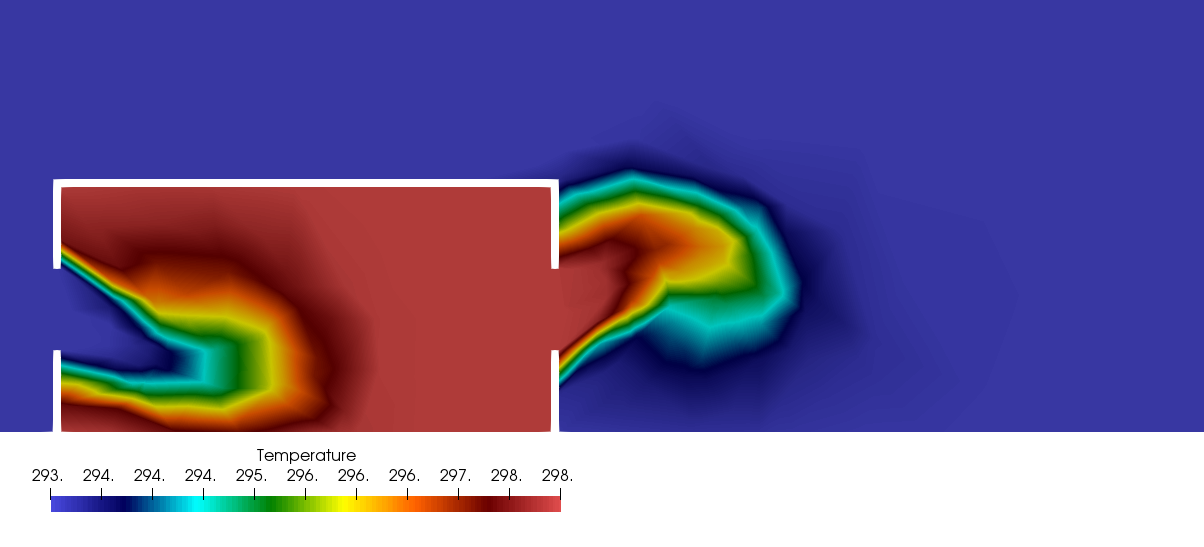}
        \caption{9 s}
        \label{Fig:Case6a_Temp_9sec}
    \end{subfigure}
    \begin{subfigure}{0.38\textwidth}
        \includegraphics[width=\textwidth]{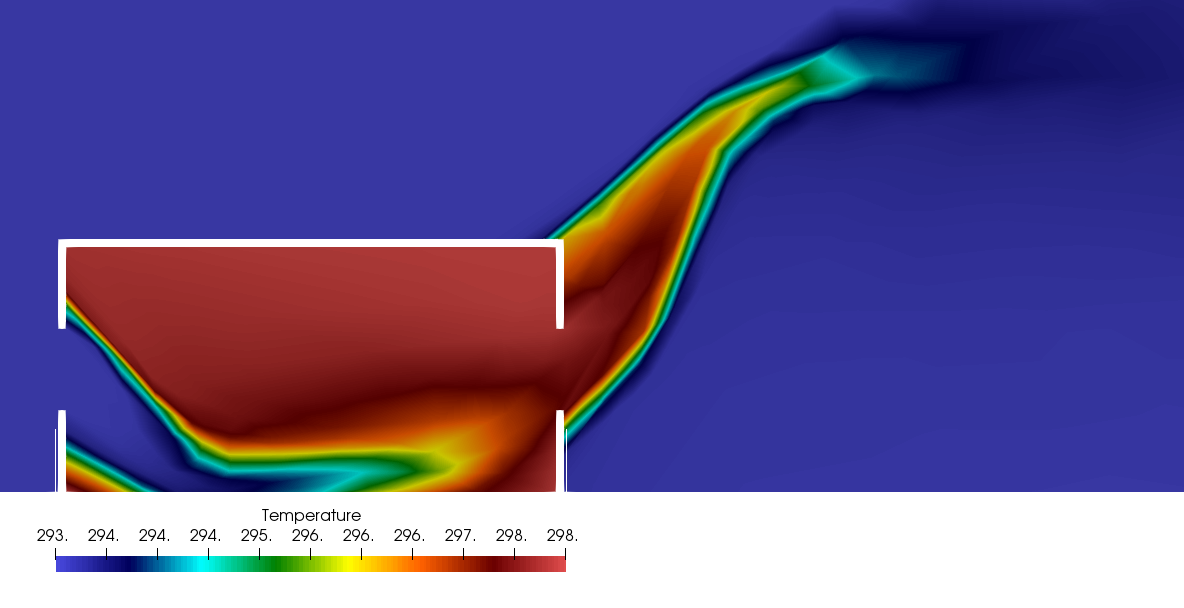}
        \caption{40 s}
        \label{Fig:Case6a_Temp_40sec}
    \end{subfigure}
    \caption{Temperature field at different instant for the simulation \textit{3dBox\_Case6a.flml}. The temperature field only is adapted with an \texttt{error\_bound\_interpolation} equal to 0.5.}
    \label{Fig:Case6a_Temp}
\end{figure}

\begin{figure}
    \centering    
    \begin{subfigure}{0.38\textwidth}
        \includegraphics[width=\textwidth]{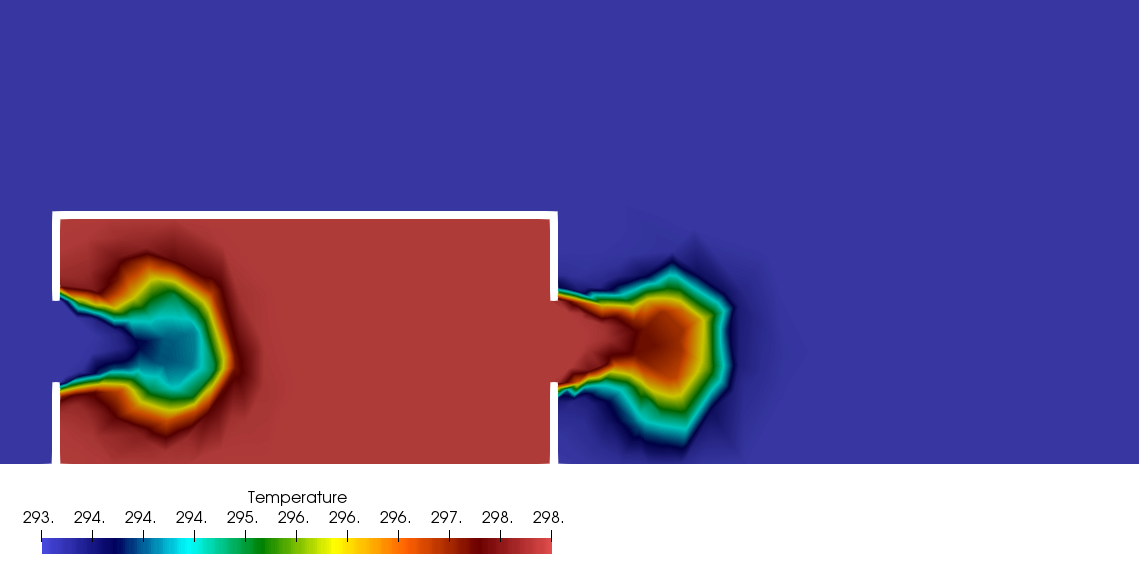}
        \caption{2 s}
        \label{Fig:Case6b_Temp_2sec}
    \end{subfigure}
    \begin{subfigure}{0.4\textwidth}
        \includegraphics[width=\textwidth]{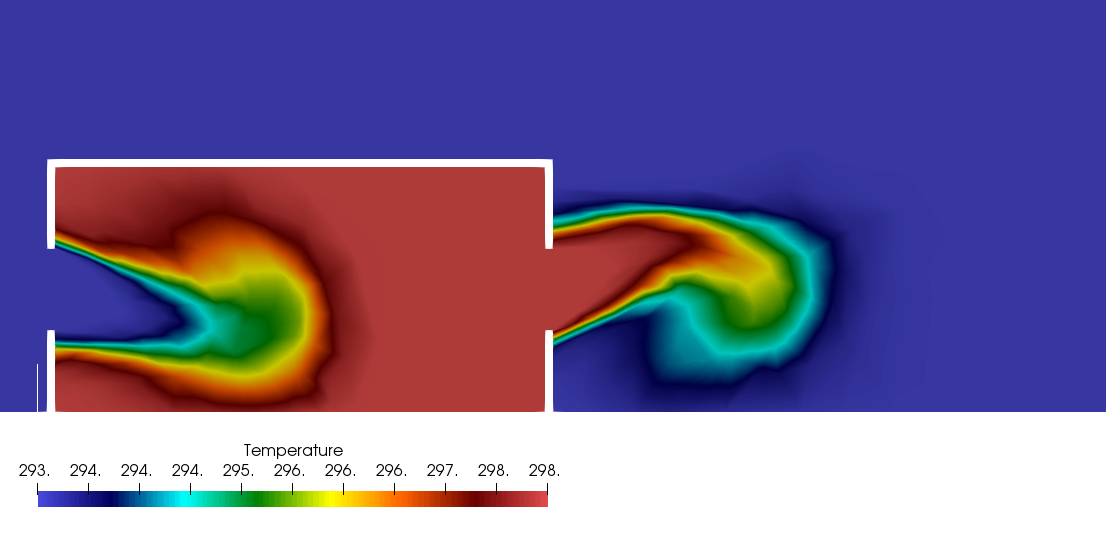}
        \caption{6 sec}
        \label{Fig:Case6b_Temp_6sec}
    \end{subfigure}
    \begin{subfigure}{0.4\textwidth}
        \includegraphics[width=\textwidth]{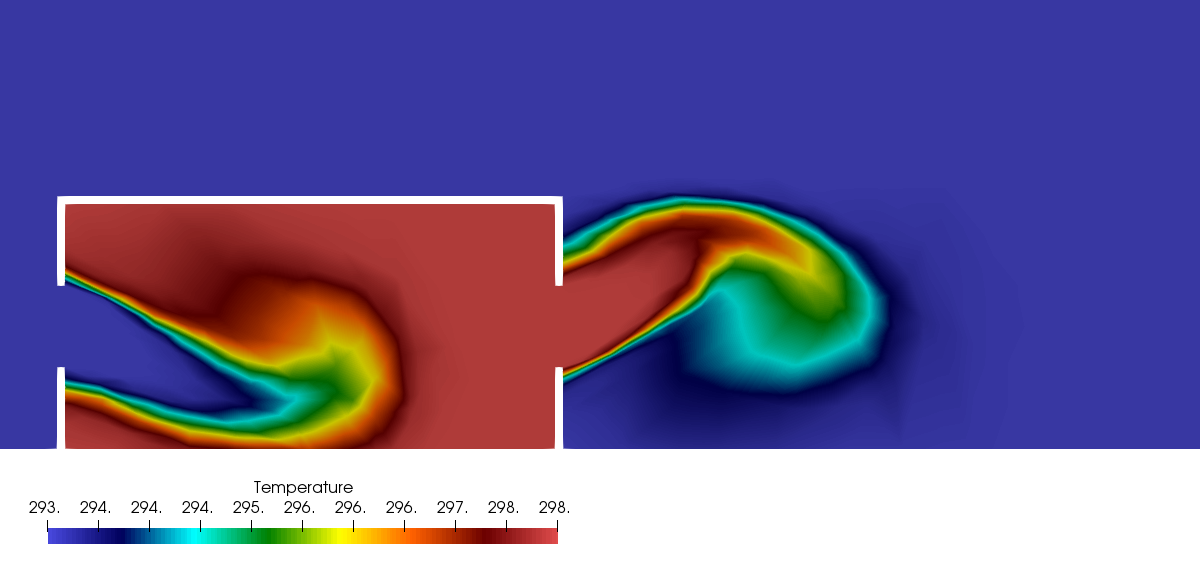}
        \caption{9 sec}
        \label{Fig:Case6b_Temp_9sec}
    \end{subfigure}
    \begin{subfigure}{0.4\textwidth}
        \includegraphics[width=\textwidth]{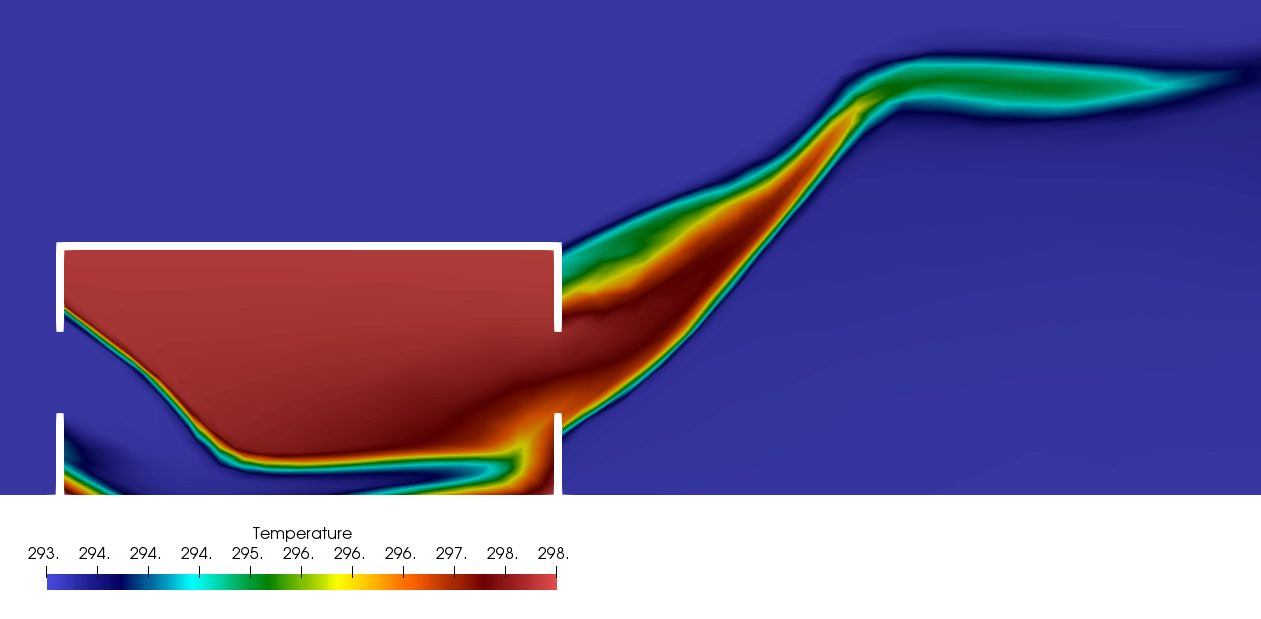}
        \caption{40 sec}
        \label{Fig:Case6b_Temp_40sec}
    \end{subfigure}
    \caption{Temperature field at different instant for the simulation \textit{3dBox\_Case6b.flml}. The temperature field only is adapted with an \texttt{error\_bound\_interpolation} equal to 0.3.}
    \label{Fig:Case6b_Temp}
\end{figure}

\begin{figure}
    \centering
    \begin{subfigure}{0.4\textwidth}
        \includegraphics[width=\textwidth]{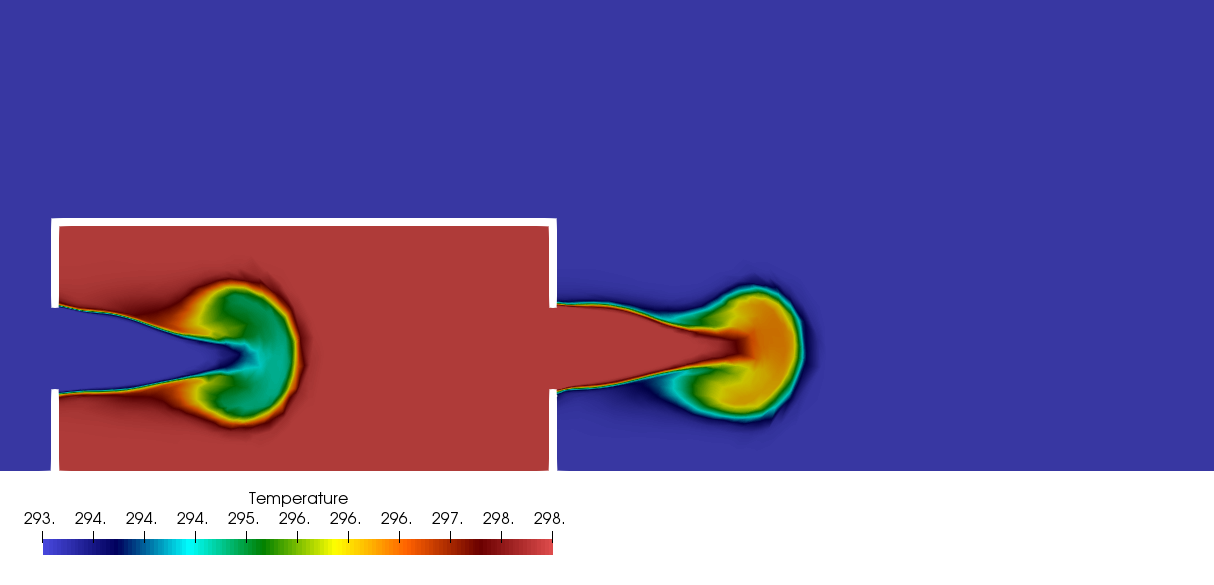}
        \caption{2 s}
        \label{Fig:Case6c_Temp_2sec}
    \end{subfigure}
    \begin{subfigure}{0.39\textwidth}
        \includegraphics[width=\textwidth]{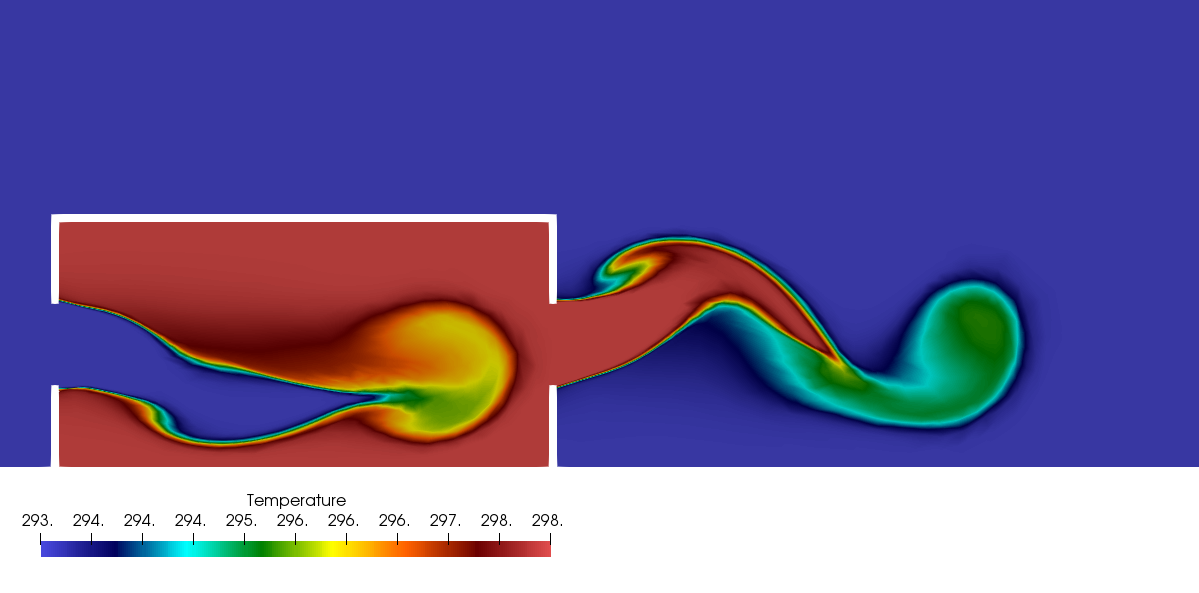}
        \caption{6 s}
        \label{Fig:Case6c_Temp_6sec}
    \end{subfigure}
    \begin{subfigure}{0.38\textwidth}
        \includegraphics[width=\textwidth]{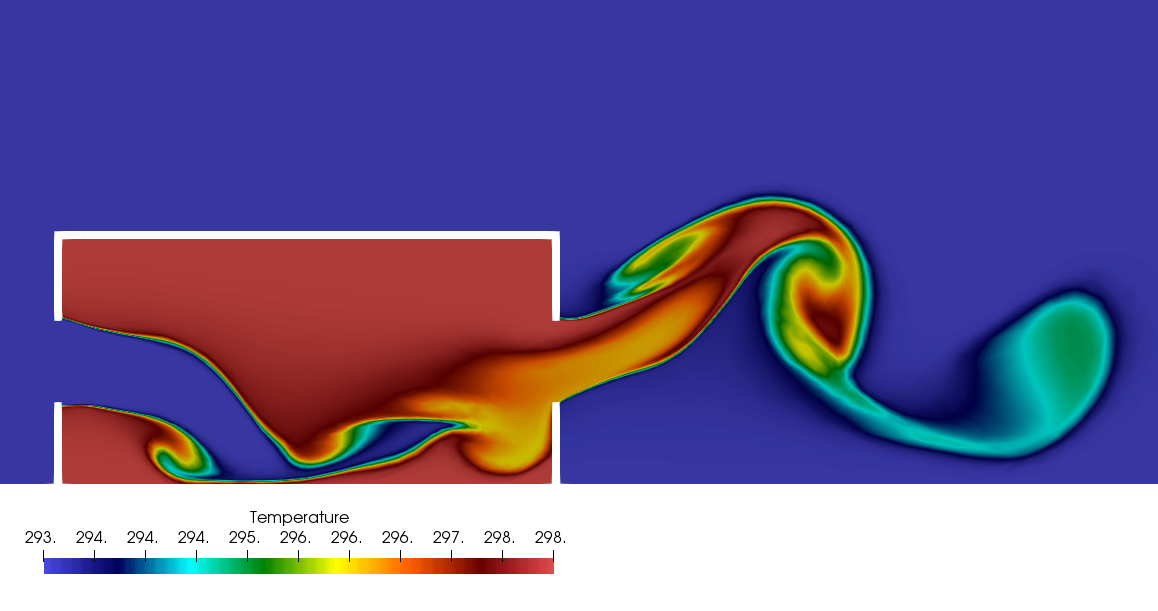}
        \caption{9 s}
        \label{Fig:Case6c_Temp_9sec}
    \end{subfigure}
    \begin{subfigure}{0.4\textwidth}
        \includegraphics[width=\textwidth]{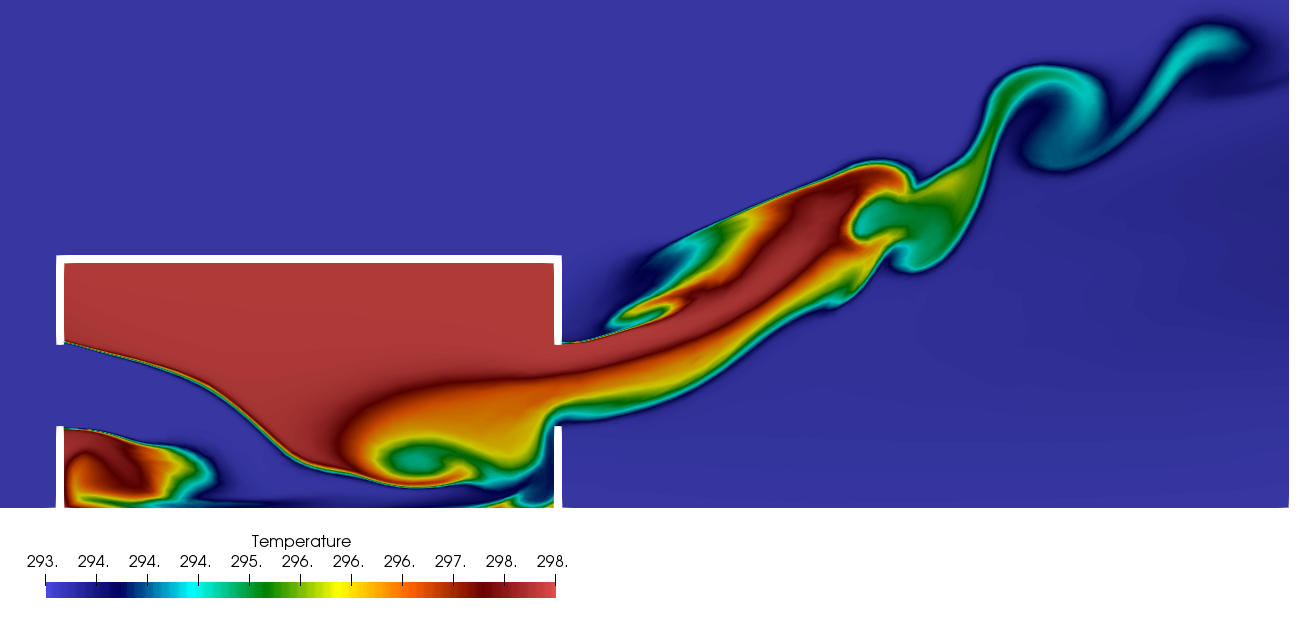}
        \caption{40 s}
        \label{Fig:Case6c_Temp_40sec}
    \end{subfigure}
    \caption{Temperature field at different instant for the simulation \textit{3dBox\_Case6c.flml}. The temperature field only is adapted with an \texttt{error\_bound\_interpolation} equal to 0.1.}
    \label{Fig:Case6c_Temp}
\end{figure}

\begin{figure}
    \centering
    \begin{subfigure}{0.4\textwidth}
        \includegraphics[width=\textwidth]{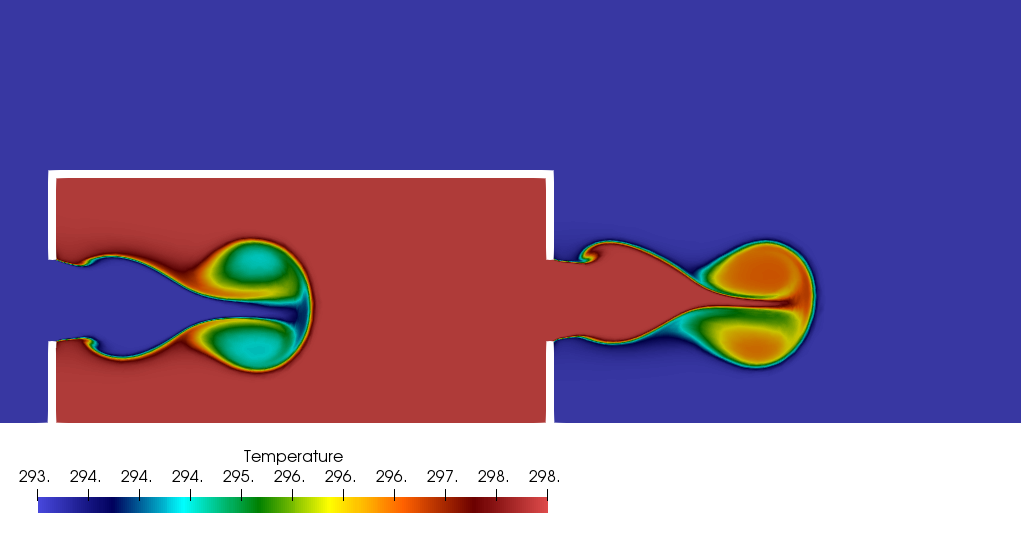}
        \caption{2 s}
        \label{Fig:Case6d_Temp_2sec}
    \end{subfigure}
    \begin{subfigure}{0.42\textwidth}
        \includegraphics[width=\textwidth]{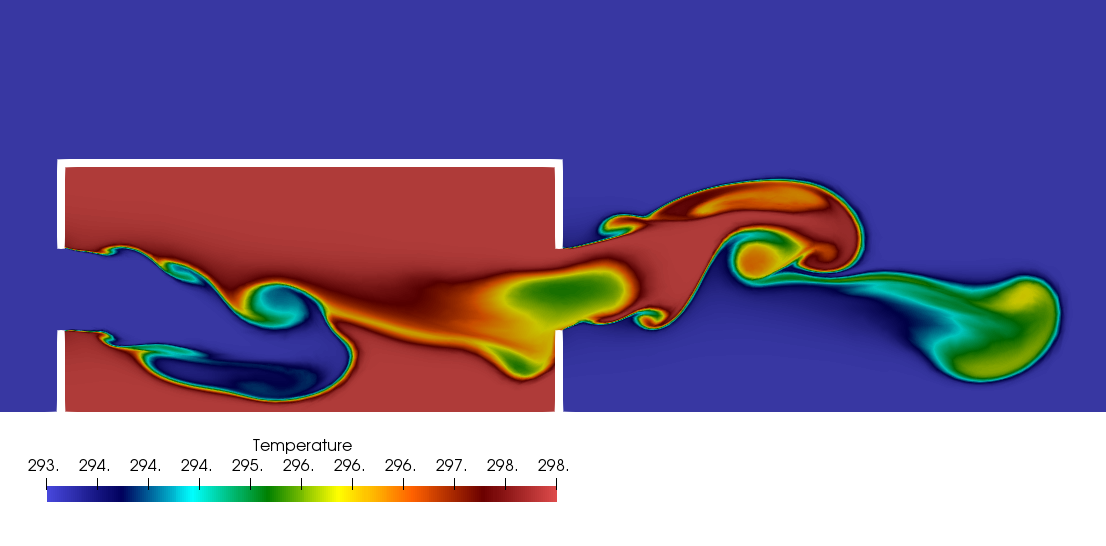}
        \caption{6 s}
        \label{Fig:Case6d_Temp_6sec}
    \end{subfigure}
    \begin{subfigure}{0.42\textwidth}
        \includegraphics[width=\textwidth]{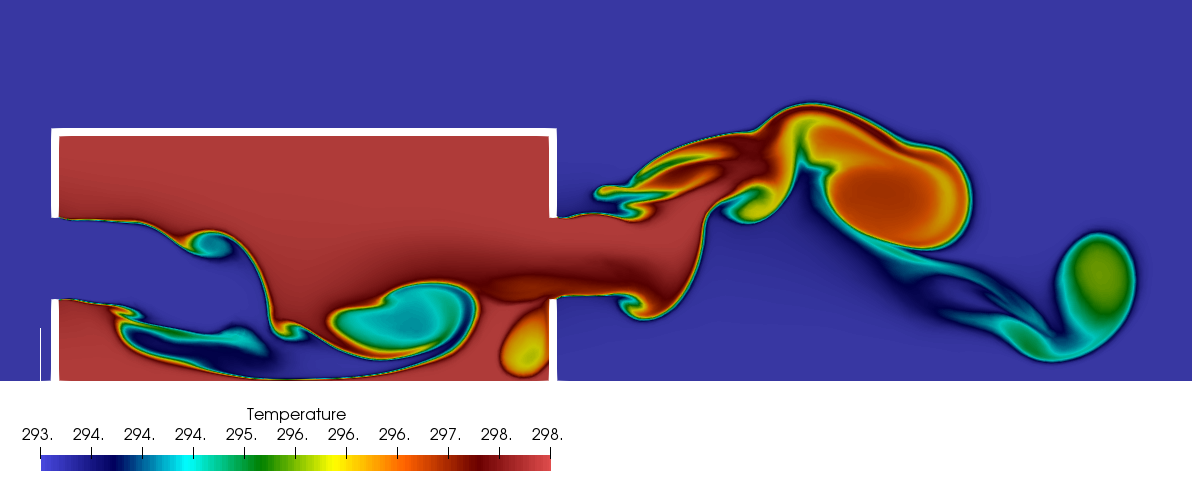}
        \caption{9 s}
        \label{Fig:Case6d_Temp_9sec}
    \end{subfigure}
    \begin{subfigure}{0.4\textwidth}
        \includegraphics[width=\textwidth]{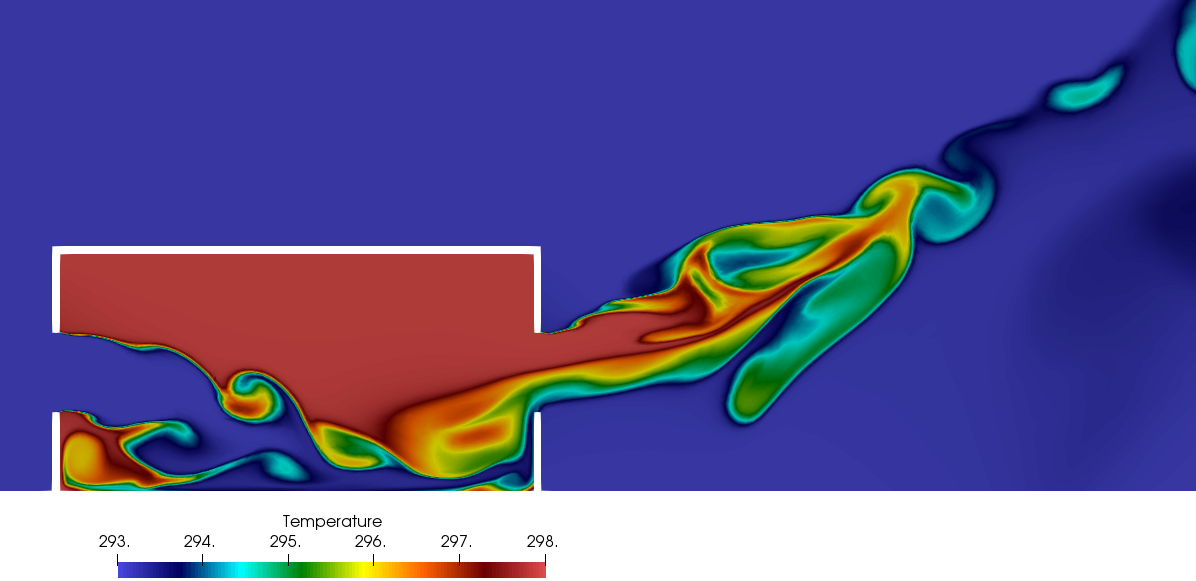}
        \caption{30 s}
        \label{Fig:Case6d_Temp_40sec}
    \end{subfigure}
    \caption{Temperature field at different instant for the simulation \textit{3dBox\_Case6d.flml}. The temperature field only is adapted with an \texttt{error\_bound\_interpolation} equal to 0.05.}
    \label{Fig:Case6d_Temp}
\end{figure}

\begin{figure}
    \centering
    \begin{subfigure}{0.4\textwidth}
        \includegraphics[width=\textwidth]{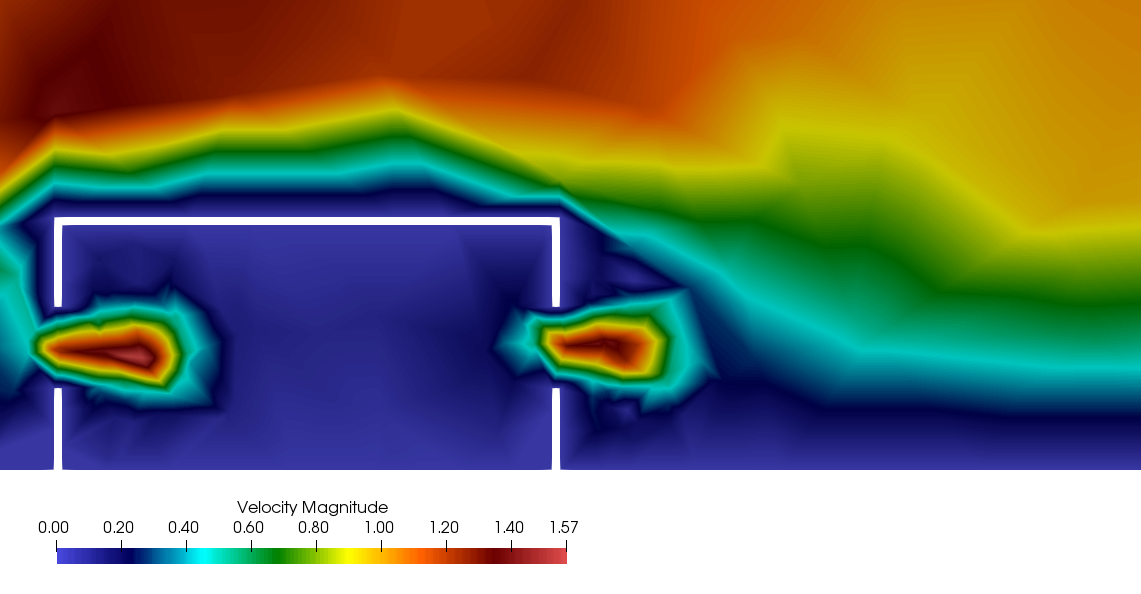}
        \caption{2 s}
        \label{Fig:Case6a_Vel_2sec}
    \end{subfigure}
    \begin{subfigure}{0.39\textwidth}
        \includegraphics[width=\textwidth]{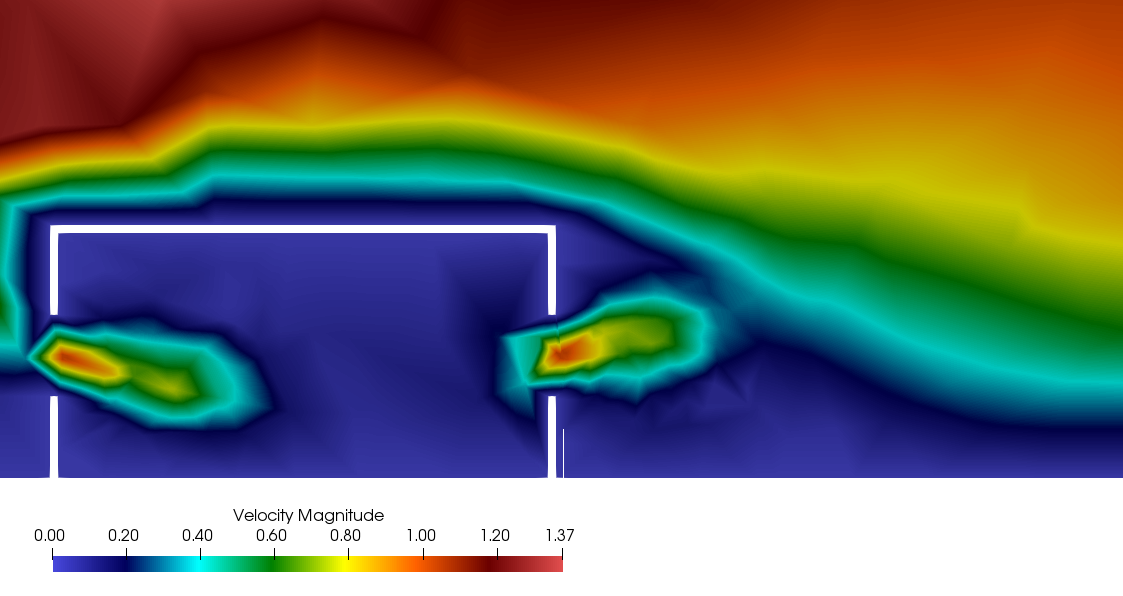}
        \caption{6 s}
        \label{Fig:Case6a_Vel_6sec}
    \end{subfigure}
    \begin{subfigure}{0.38\textwidth}
        \includegraphics[width=\textwidth]{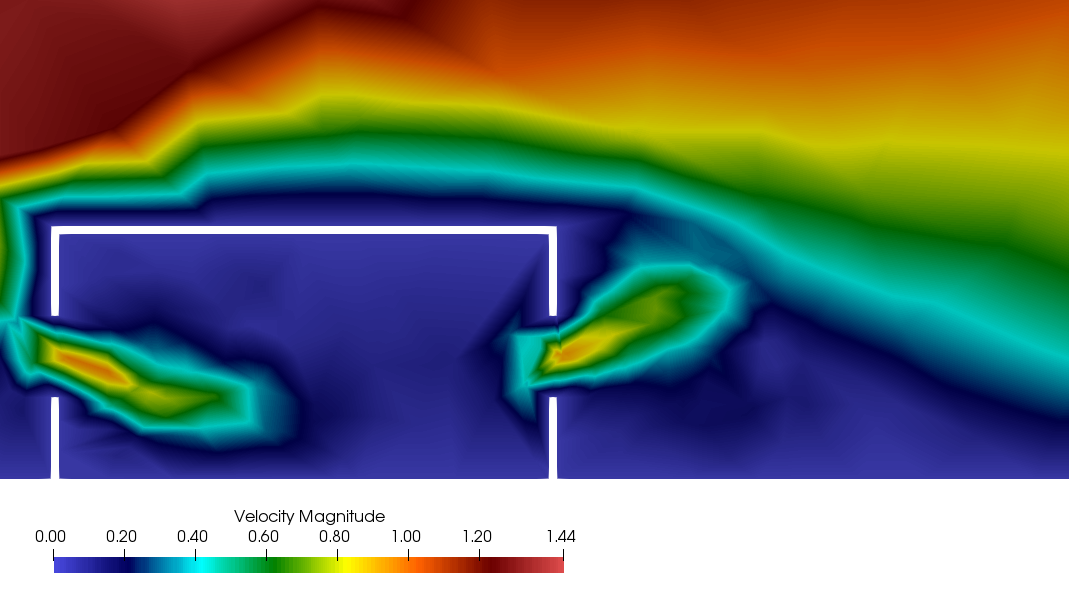}
        \caption{9 s}
        \label{Fig:Case6a_Vel_9sec}
    \end{subfigure}
    \begin{subfigure}{0.4\textwidth}
        \includegraphics[width=\textwidth]{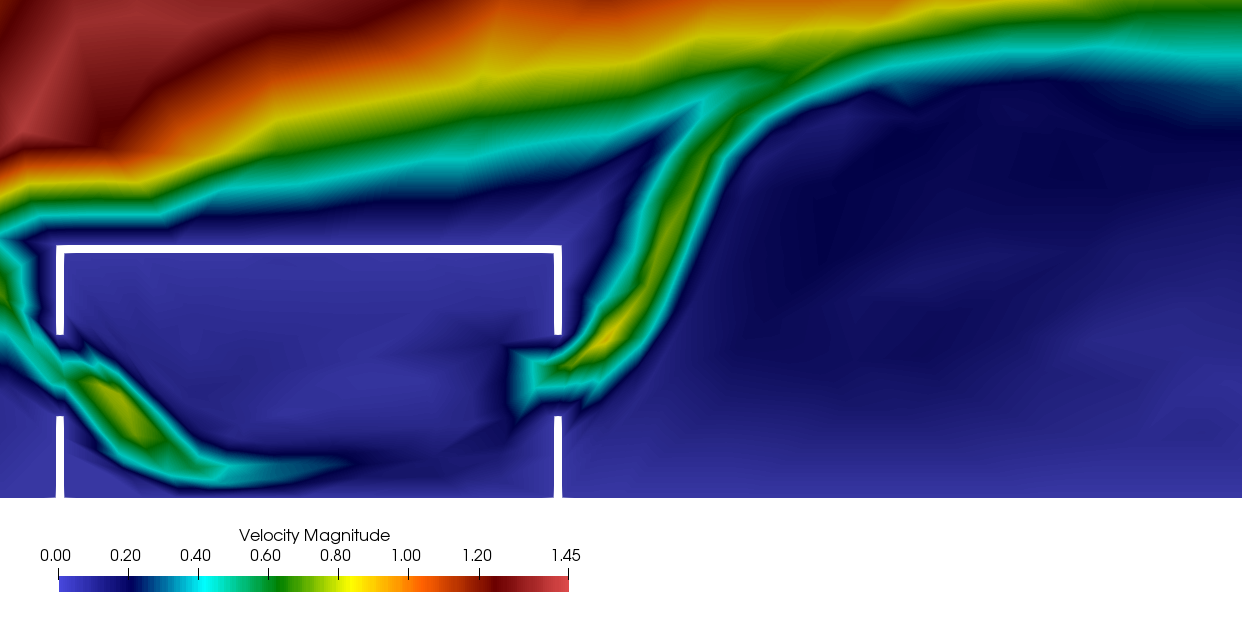}
        \caption{40 s}
        \label{Fig:Case6a_Vel_40sec}
    \end{subfigure}
    \caption{Velocity field at different instant for the simulation \textit{3dBox\_Case6a.flml}. The temperature field only is adapted with an \texttt{error\_bound\_interpolation} equal to 0.5.}
    \label{Fig:Case6a_Vel}
\end{figure}

\begin{figure}
    \centering    
    \begin{subfigure}{0.4\textwidth}
        \includegraphics[width=\textwidth]{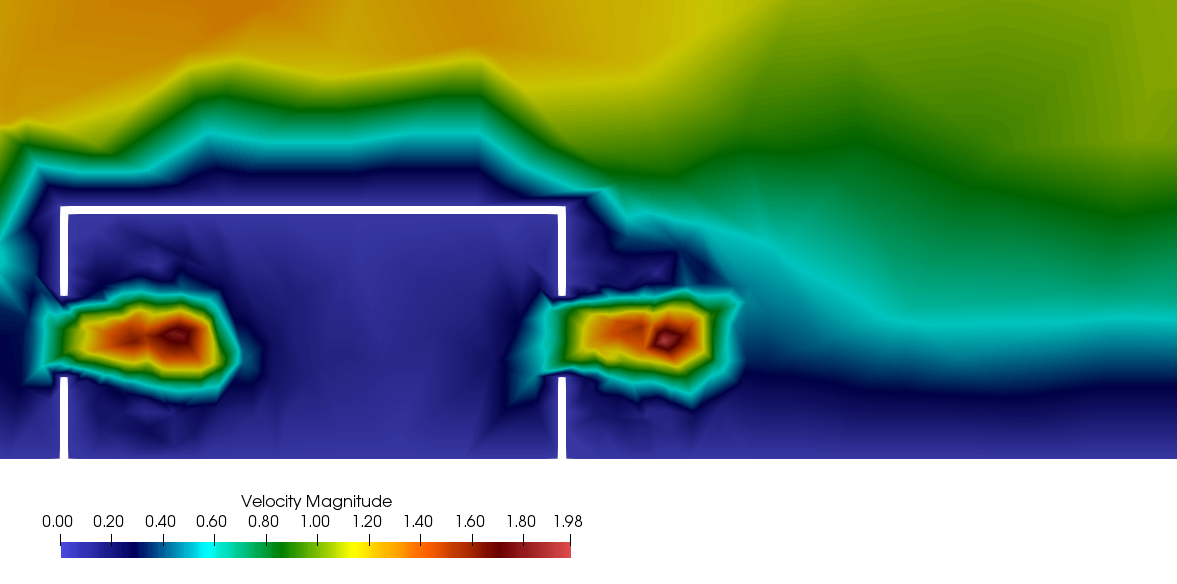}
        \caption{2 s}
        \label{Fig:Case6b_Vel_2sec}
    \end{subfigure}
    \begin{subfigure}{0.39\textwidth}
        \includegraphics[width=\textwidth]{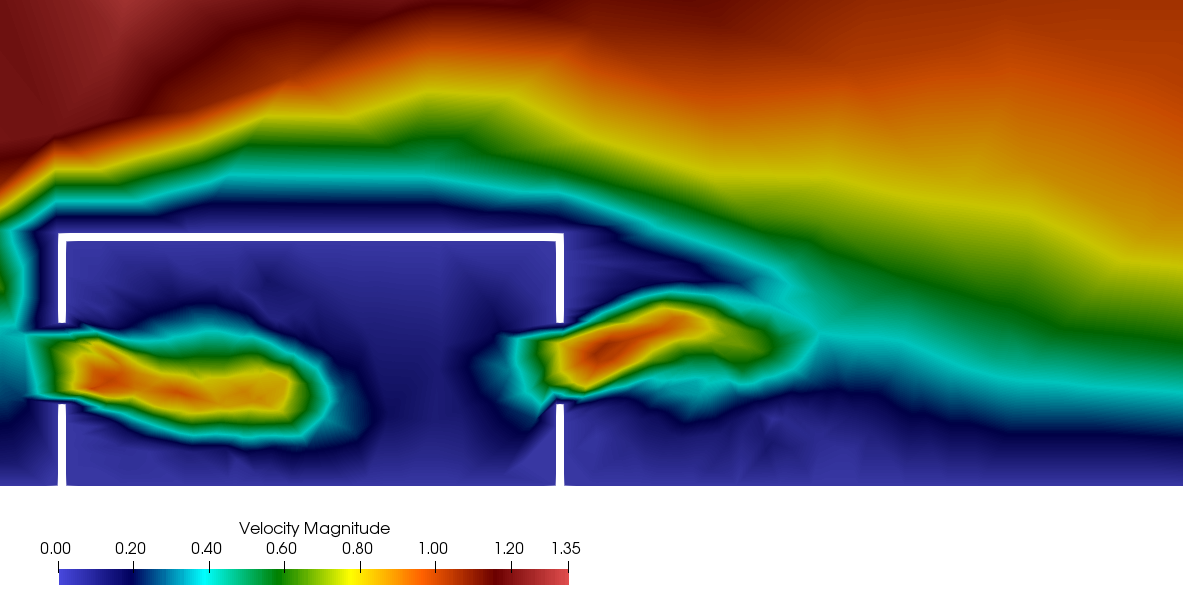}
        \caption{6 s}
        \label{Fig:Case6b_Vel_6sec}
    \end{subfigure}
    \begin{subfigure}{0.4\textwidth}
        \includegraphics[width=\textwidth]{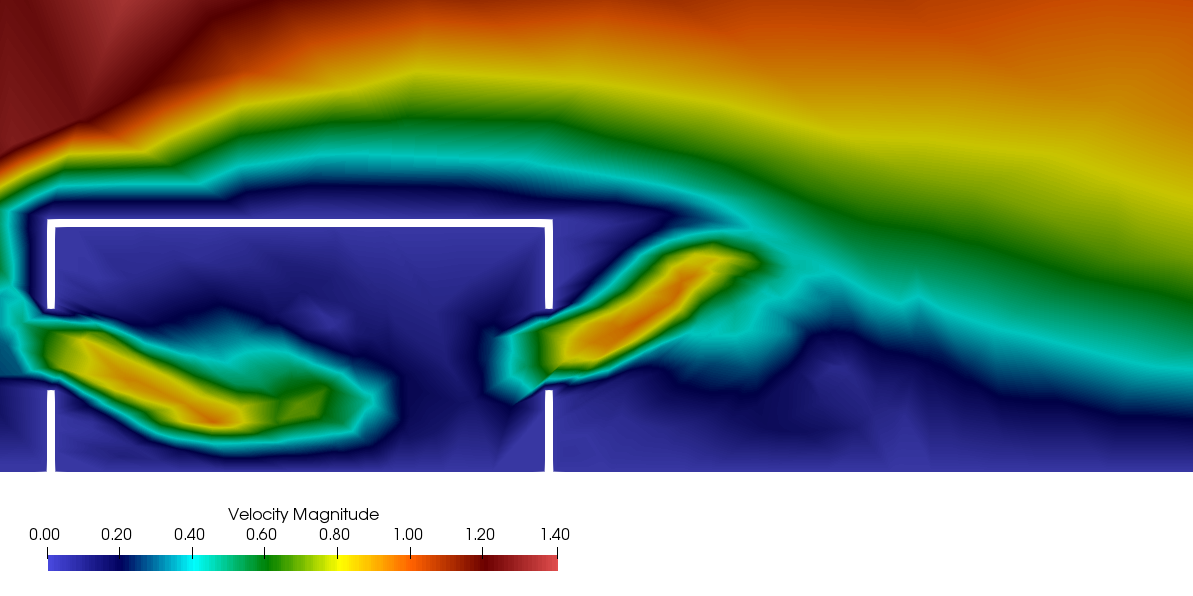}
        \caption{9 s}
        \label{Fig:Case6b_Vel_9sec}
    \end{subfigure}
    \begin{subfigure}{0.4\textwidth}
        \includegraphics[width=\textwidth]{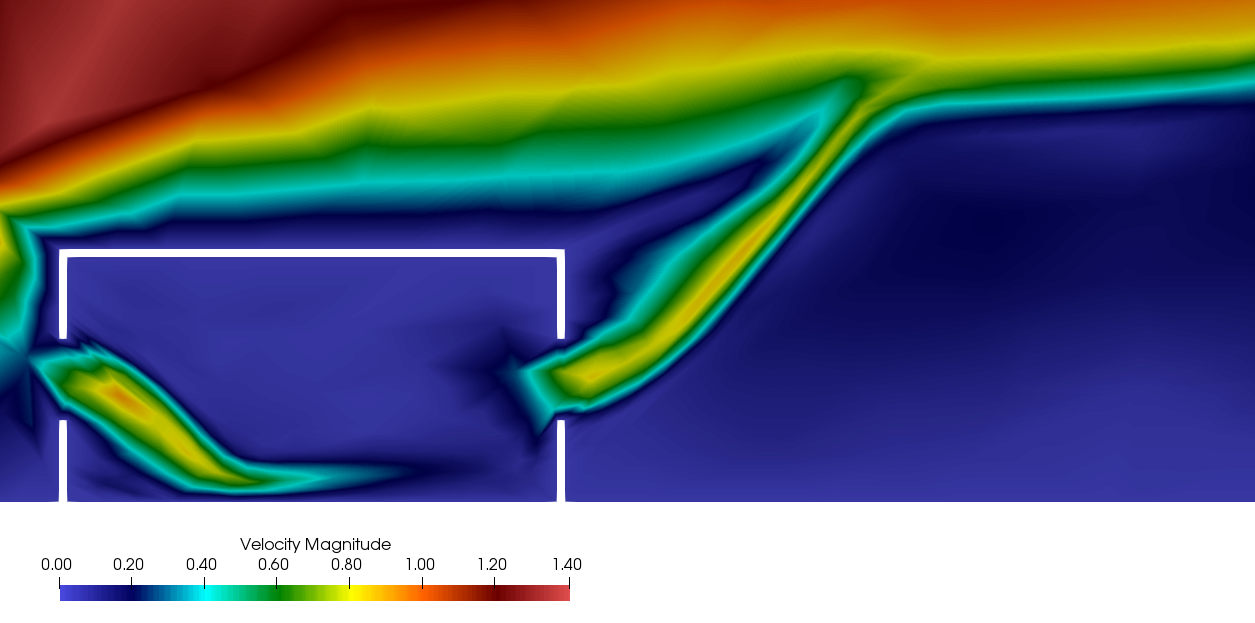}
        \caption{40 s}
        \label{Fig:Case6b_Vel_40sec}
    \end{subfigure}
    \caption{Velocity field at different instant for the simulation \textit{3dBox\_Case6b.flml}. The temperature field only is adapted with an \texttt{error\_bound\_interpolation} equal to 0.3.}
    \label{Fig:Case6b_Vel}
\end{figure}

\begin{figure}
    \centering
    \begin{subfigure}{0.4\textwidth}
        \includegraphics[width=\textwidth]{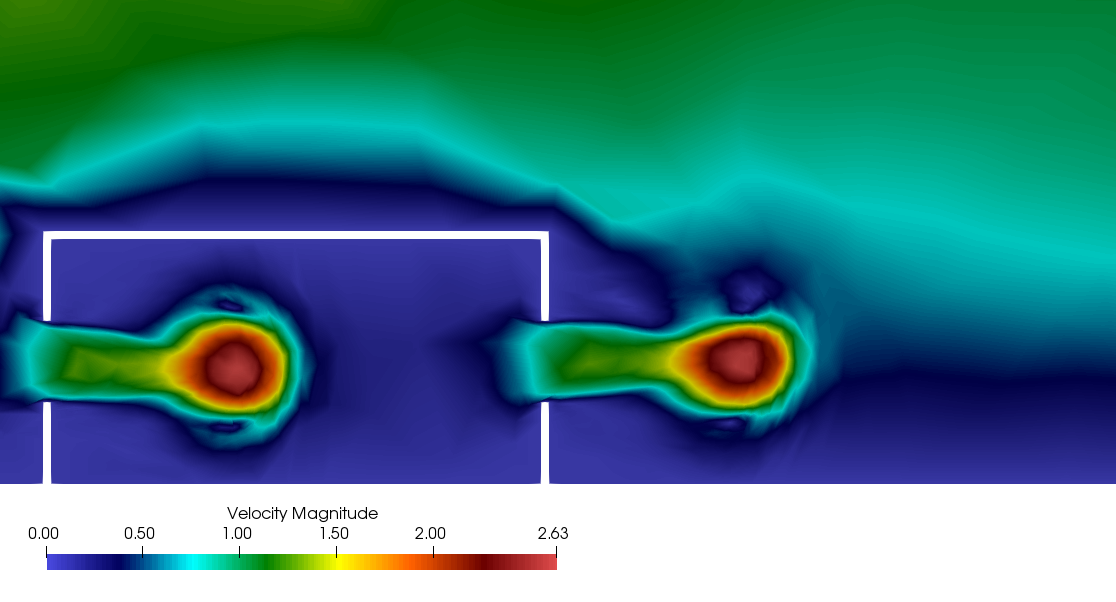}
        \caption{2 s}
        \label{Fig:Case6c_Vel_2sec}
    \end{subfigure}
    \begin{subfigure}{0.39\textwidth}
        \includegraphics[width=\textwidth]{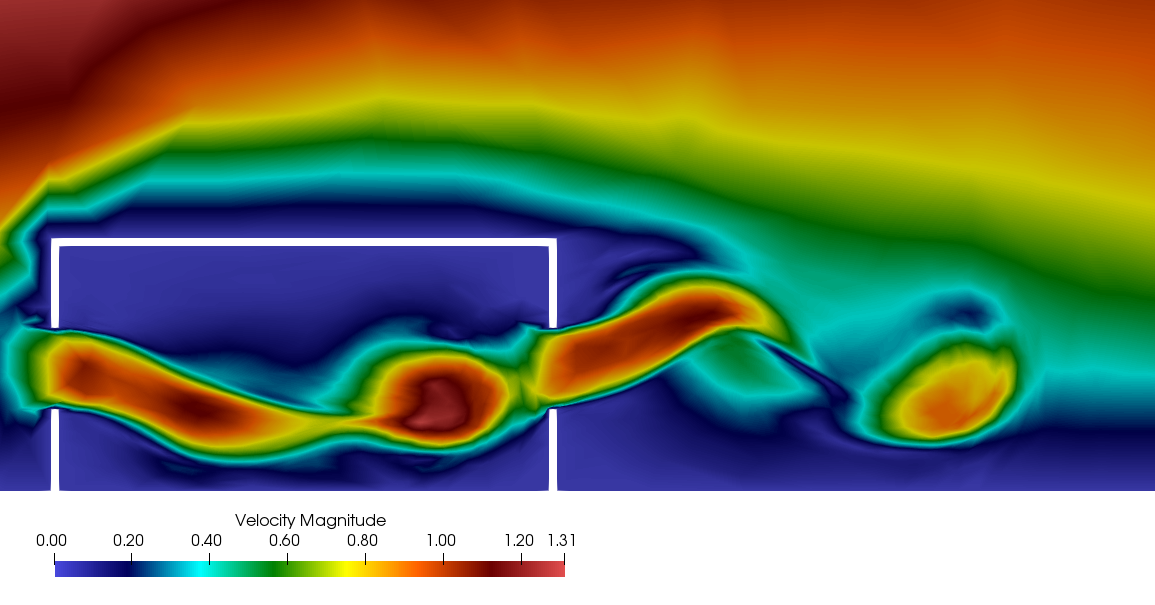}
        \caption{6 s}
        \label{Fig:Case6c_Vel_6sec}
    \end{subfigure}
    \begin{subfigure}{0.39\textwidth}
        \includegraphics[width=\textwidth]{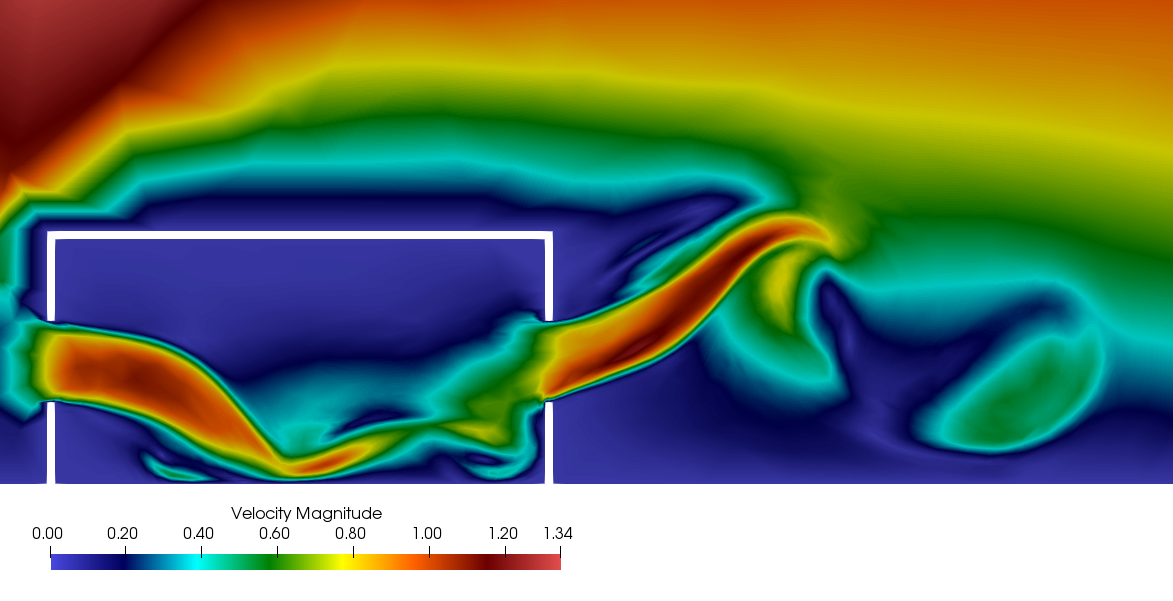}
        \caption{9 s}
        \label{Fig:Case6c_Vel_9sec}
    \end{subfigure}
    \begin{subfigure}{0.4\textwidth}
        \includegraphics[width=\textwidth]{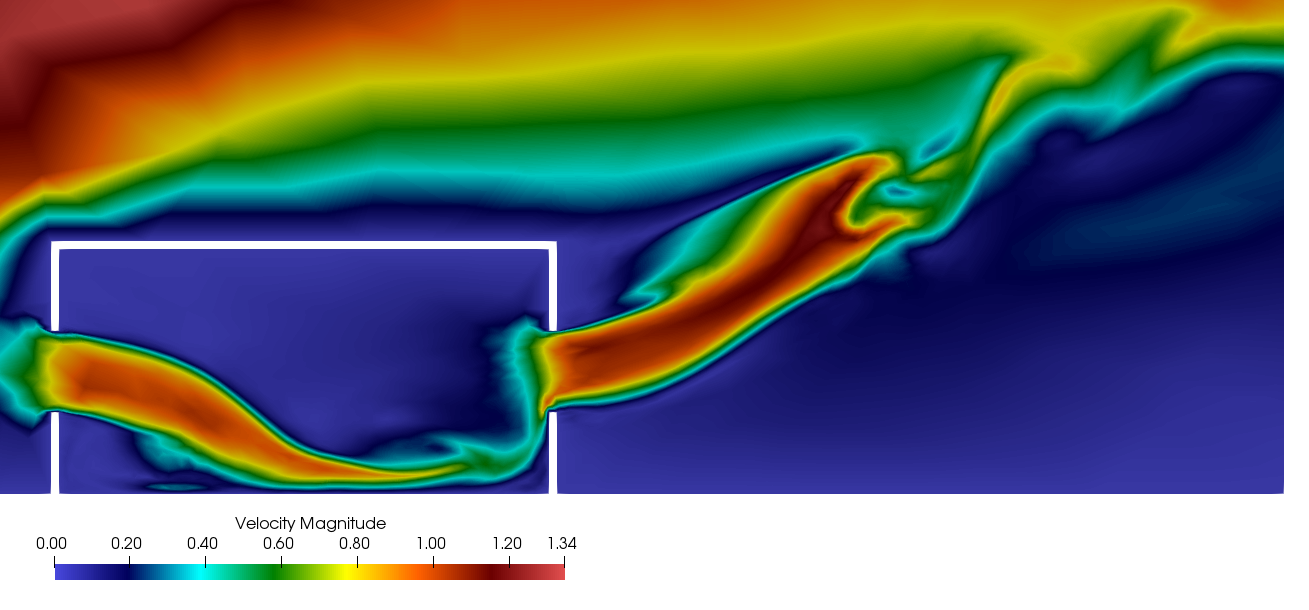}
        \caption{40 s}
        \label{Fig:Case6c_Vel_40sec}
    \end{subfigure}
    \caption{Velocity field at different instant for the simulation \textit{3dBox\_Case6c.flml}. The temperature field only is adapted with an \texttt{error\_bound\_interpolation} equal to 0.1.}
    \label{Fig:Case6c_Vel}
\end{figure}

\begin{figure}
    \centering
    \begin{subfigure}{0.34\textwidth}
        \includegraphics[width=\textwidth]{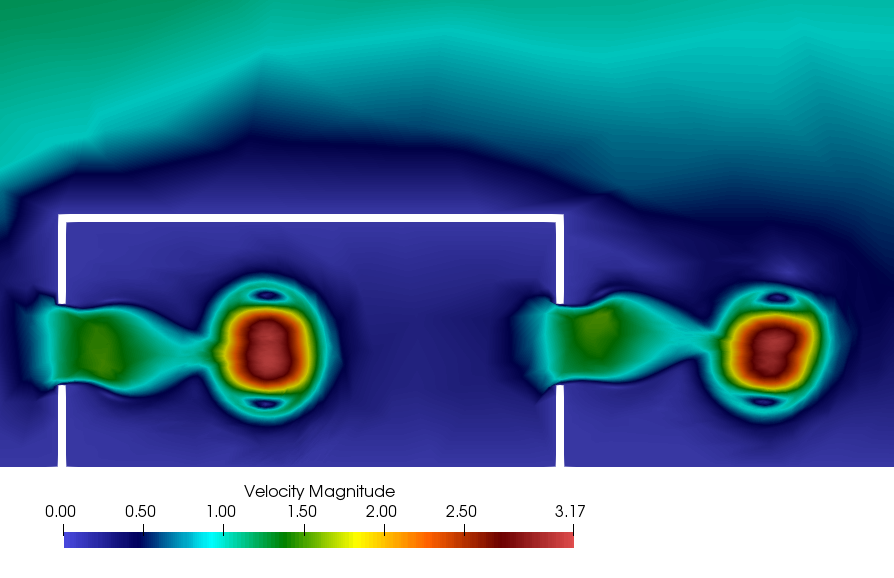}
        \caption{2 s}
        \label{Fig:Case6d_Vel_2sec}
    \end{subfigure}
    \begin{subfigure}{0.43\textwidth}
        \includegraphics[width=\textwidth]{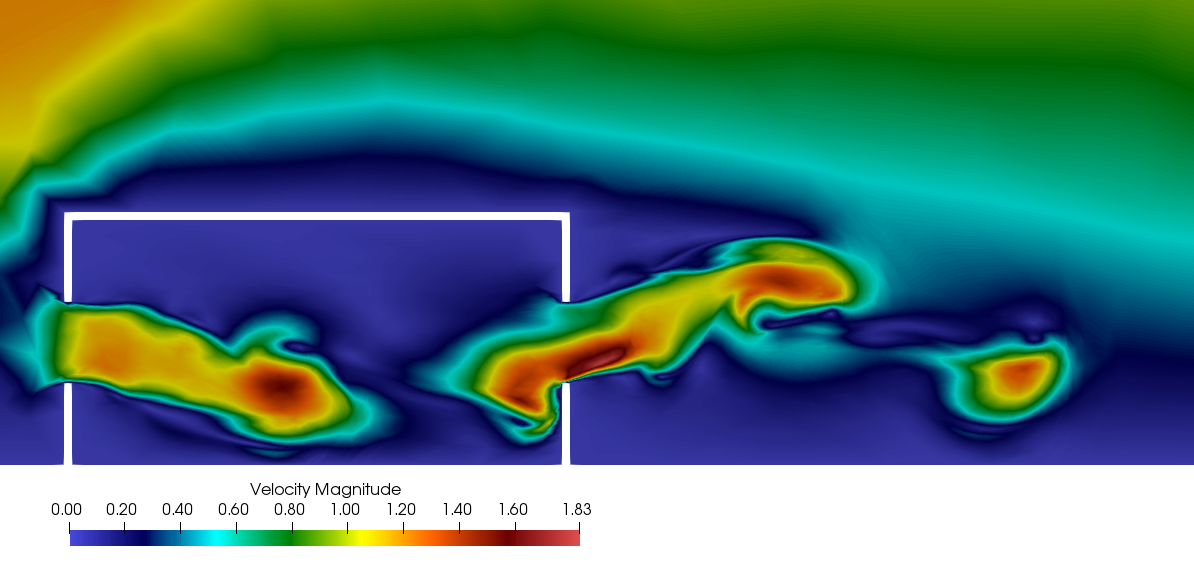}
        \caption{6 s}
        \label{Fig:Case6d_Vel_6sec}
    \end{subfigure}
    \begin{subfigure}{0.4\textwidth}
        \includegraphics[width=\textwidth]{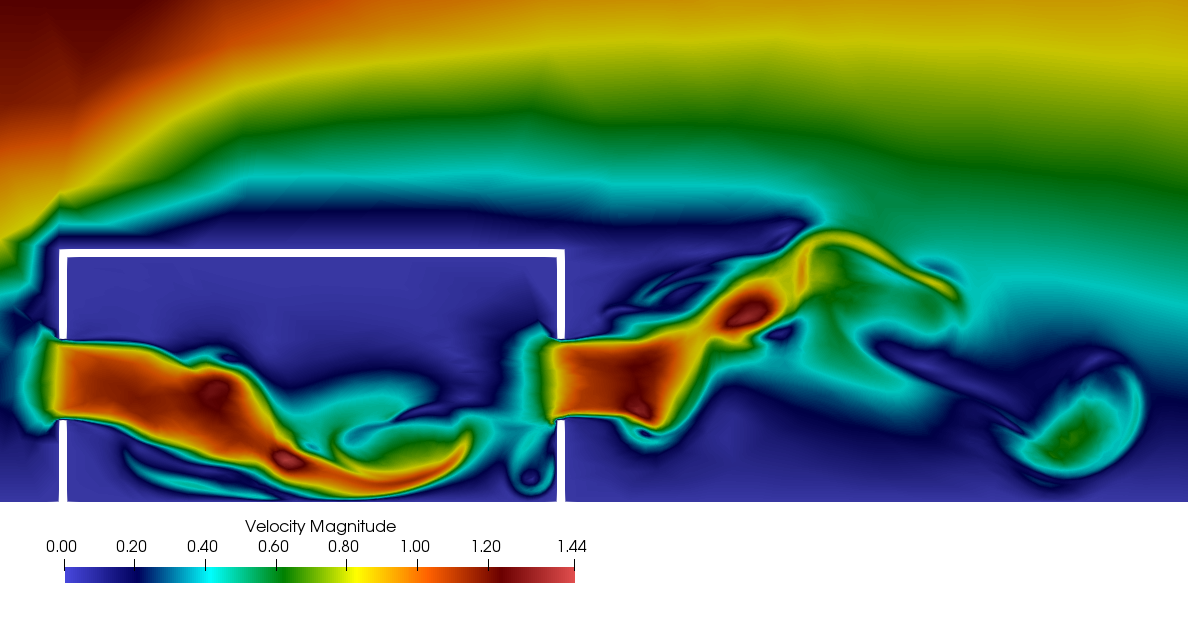}
        \caption{9 s}
        \label{Fig:Case6d_Vel_9sec}
    \end{subfigure}
    \begin{subfigure}{0.4\textwidth}
        \includegraphics[width=\textwidth]{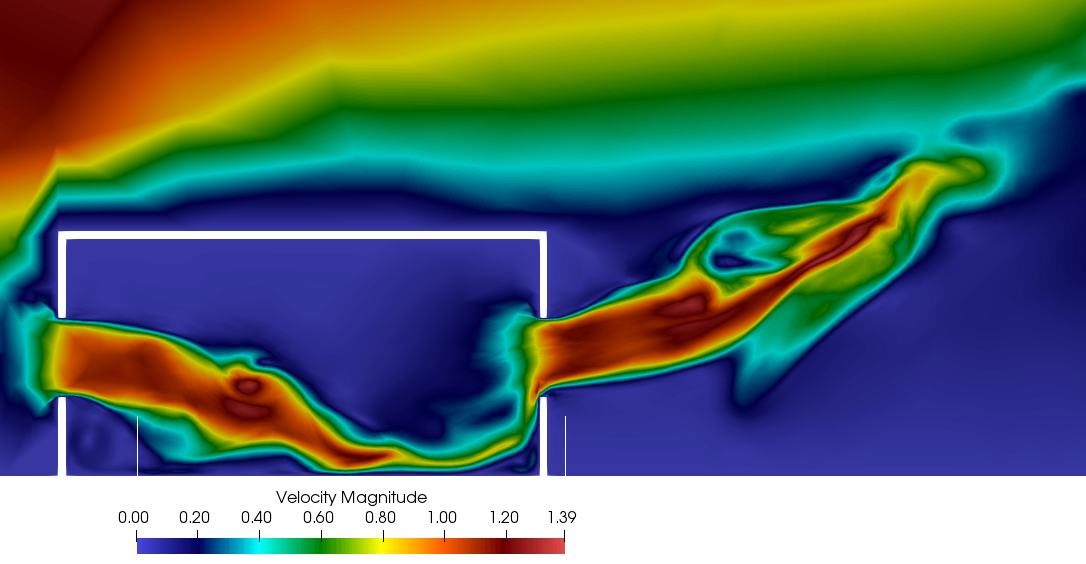}
        \caption{30 s}
        \label{Fig:Case6d_Vel_40sec}
    \end{subfigure}
    \caption{Velocity field at different instant for the simulation \textit{3dBox\_Case6d.flml}. The temperature field only is adapted with an \texttt{error\_bound\_interpolation} equal to 0.05.}
    \label{Fig:Case6d_Vel}
\end{figure}

Snapshots of the meshes are shown in Figure~\ref{Fig:Case6a_Mesh}, Figure~\ref{Fig:Case6b_Mesh}, Figure~\ref{Fig:Case6c_Mesh} and Figure~\ref{Fig:Case6d_Mesh}. Snapshots of the temperature field are shown in Figure~\ref{Fig:Case6a_Temp}, Figure~\ref{Fig:Case6b_Temp}, Figure~\ref{Fig:Case6c_Temp} and Figure~\ref{Fig:Case6d_Temp}. Snapshots of the velocity field are shown in Figure~\ref{Fig:Case6a_Vel}, Figure~\ref{Fig:Case6b_Vel}, Figure~\ref{Fig:Case6c_Vel} and Figure~\ref{Fig:Case6d_Vel}. Go to Chapter~\ref{Sec:PostProcessing} to learn how to visualise the results using \textbf{ParaView}.

\noindent For a given \texttt{error\_bound\_interpolation}, as shown in Figure~\ref{Fig:Case6a_Mesh}, Figure~\ref{Fig:Case6b_Mesh}, Figure~\ref{Fig:Case6c_Mesh} and Figure~\ref{Fig:Case6d_Mesh} the mesh is adapting based on the temperature field, i.e. mainly within the box and at the outlet of the box. Indeed, decreasing the value of the \texttt{error\_bound\_interpolation} results in finer mesh in those regions. However, it is important to mention that the computational time is by consequence larger, see Section~\ref{Sec:ComputationTime}. Choosing an \texttt{error\_bound\_interpolation} higher to 0.3 seems to lead to poor mesh quality (in that particular case). Even with small \texttt{error\_bound\_} \texttt{interpolation} and if the temperature field is well resolved, the velocity field is not properly captured due to a poor mesh quality, especially around the box.

\subsection{Adaptation based on the velocity field}\label{Sec:AdaptVel}
In this section, the mesh adaptivity process is prescribed based on the velocity field only.

\subsubsection{Mesh adaptivity options}
Even if the initial inlet velocity is prescribed equal to 1 m/s, it is not recommended to use 1 m/s as the upper bound to determine the \texttt{error\_bound\_interpolation}. Indeed, the velocity magnitude in the domain will probably be higher than this value and estimate the \texttt{error\_bound\_interpolation} based on 1 m/s will lead to a too small value as a first guess. Therefore, based on the run performed in examples \textit{3dBox\_Case6*.flml}, the velocity range seems to be between 0 m/s and approximately 5 m/s, i.e. a difference of 5 m/s.  The \texttt{error\_bound\_interpolation} is, in a first run, set up at 10$\%$ of this velocity range: the value $5\times10\%=0.5$ is used in example \textit{3dBox\_Case7a.flml}. Then the \texttt{error\_bound\_interpolation} is progressively decreased to values equal to 0.25 (5$\%$ of the velocity range) in \textit{3dBox\_Case7b.flml}, 0.15 (3$\%$ of the velocity range) in \textit{3dBox\_Case7c.flml} and 0.1 (2$\%$ of the velocity range) in \textit{3dBox\_Case7d.flml}.

\noindent These examples can be run using the commands: 
\begin{Terminal}[]
ä\colorbox{davysgrey}{
\parbox{435pt}{
\color{applegreen} \textbf{user@mypc}\color{white}\textbf{:}\color{codeblue}$\sim$
\color{white}\$ <<FluiditySourcePath>>/bin/fluidity -l -v3 3dBox\_Case7a.flml \&
\newline
\color{applegreen} \textbf{user@mypc}\color{white}\textbf{:}\color{codeblue}$\sim$
\color{white}\$ <<FluiditySourcePath>>/bin/fluidity -l -v3 3dBox\_Case7b.flml \&
\newline
\color{applegreen} \textbf{user@mypc}\color{white}\textbf{:}\color{codeblue}$\sim$
\color{white}\$ <<FluiditySourcePath>>/bin/fluidity -l -v3 3dBox\_Case7c.flml \&
\newline
\color{applegreen} \textbf{user@mypc}\color{white}\textbf{:}\color{codeblue}$\sim$
\color{white}\$ <<FluiditySourcePath>>/bin/fluidity -l -v3 3dBox\_Case7d.flml \&
}}
\end{Terminal}

\subsubsection{Results and discussion}
\begin{figure}
    \centering
    \begin{subfigure}{0.39\textwidth}
        \includegraphics[width=\textwidth]{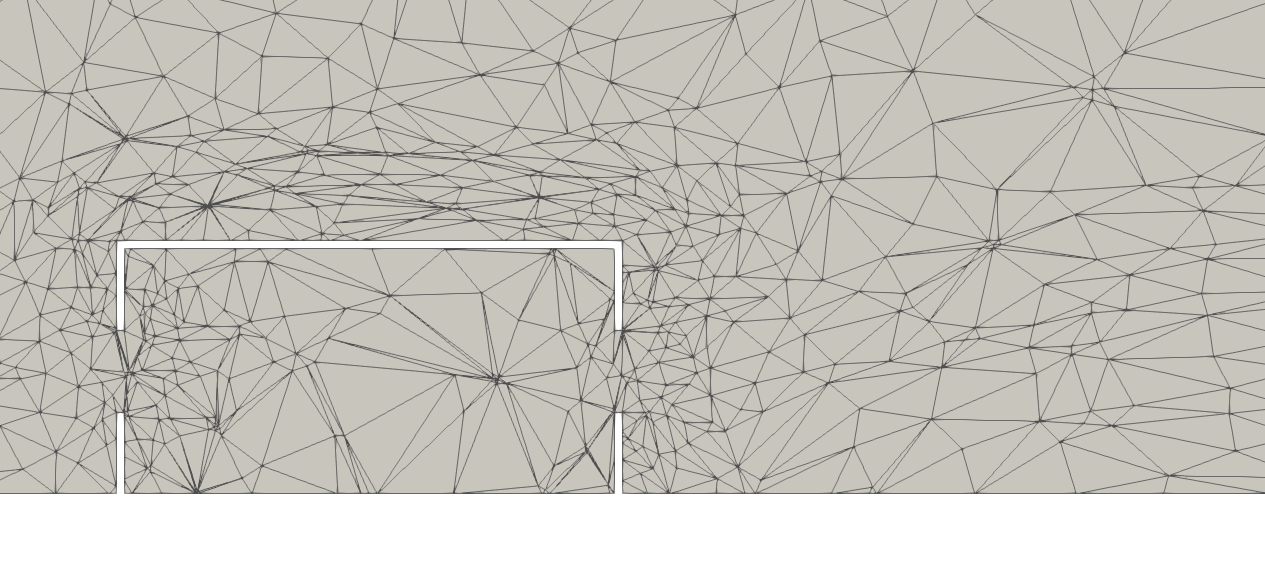}
        \caption{2 s}
        \label{Fig:Case7a_Mesh_2sec}
    \end{subfigure}
    \begin{subfigure}{0.4\textwidth}
        \includegraphics[width=\textwidth]{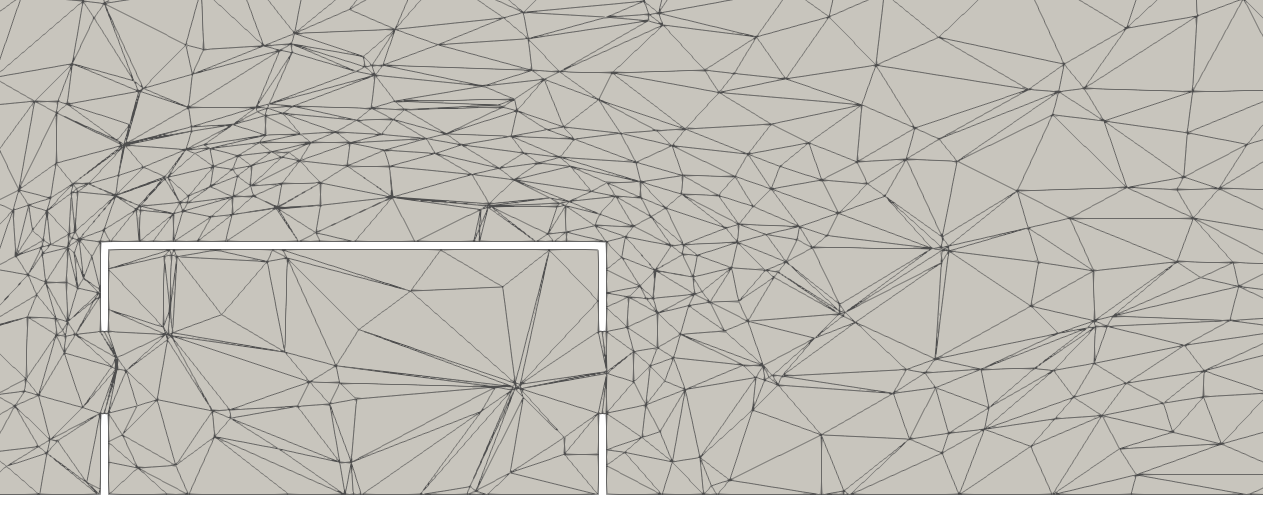}
        \caption{6 s}
        \label{Fig:Case7a_Mesh_6sec}
    \end{subfigure}
    \begin{subfigure}{0.4\textwidth}
        \includegraphics[width=\textwidth]{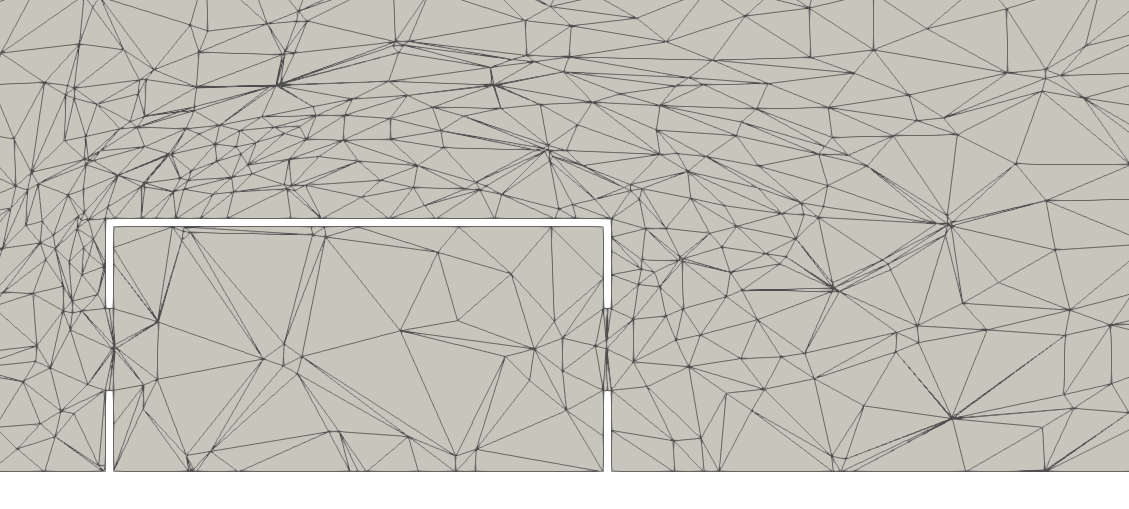}
        \caption{9 s}
        \label{Fig:Case7a_Mesh_9sec}
    \end{subfigure}
    \begin{subfigure}{0.4\textwidth}
        \includegraphics[width=\textwidth]{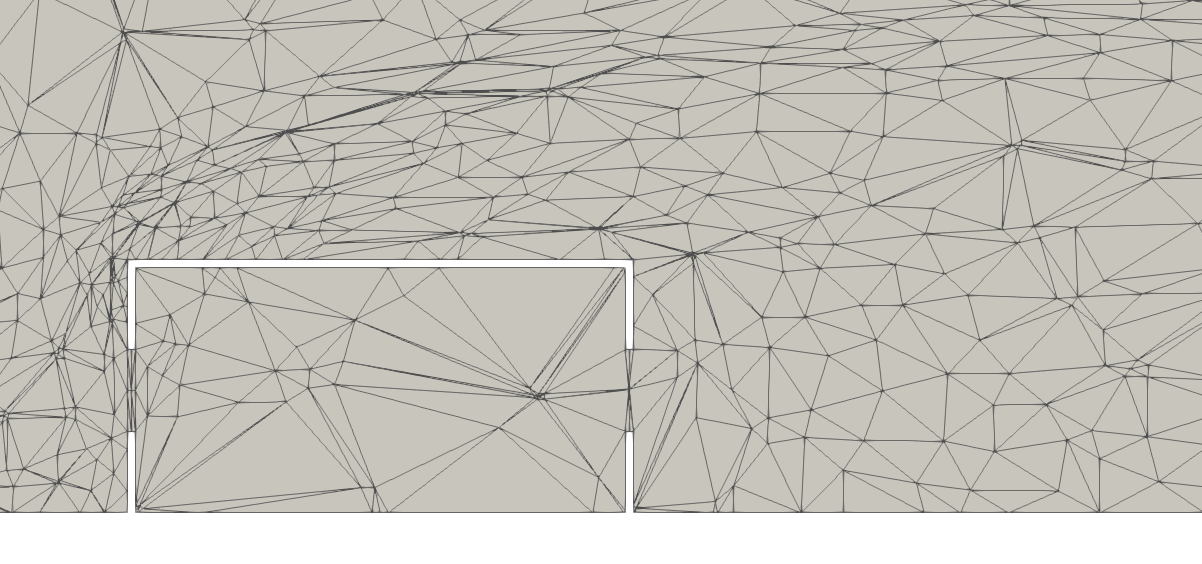}
        \caption{40 s}
        \label{Fig:Case7a_Mesh_40sec}
    \end{subfigure}
    \caption{Meshes at different instant for example \textit{3dBox\_Case7a.flml}. The velocity field only is adapted with an \texttt{error\_bound\_interpolation} equal to 0.5.}
    \label{Fig:Case7a_Mesh}
\end{figure}

\begin{figure}
    \centering    
    \begin{subfigure}{0.42\textwidth}
        \includegraphics[width=\textwidth]{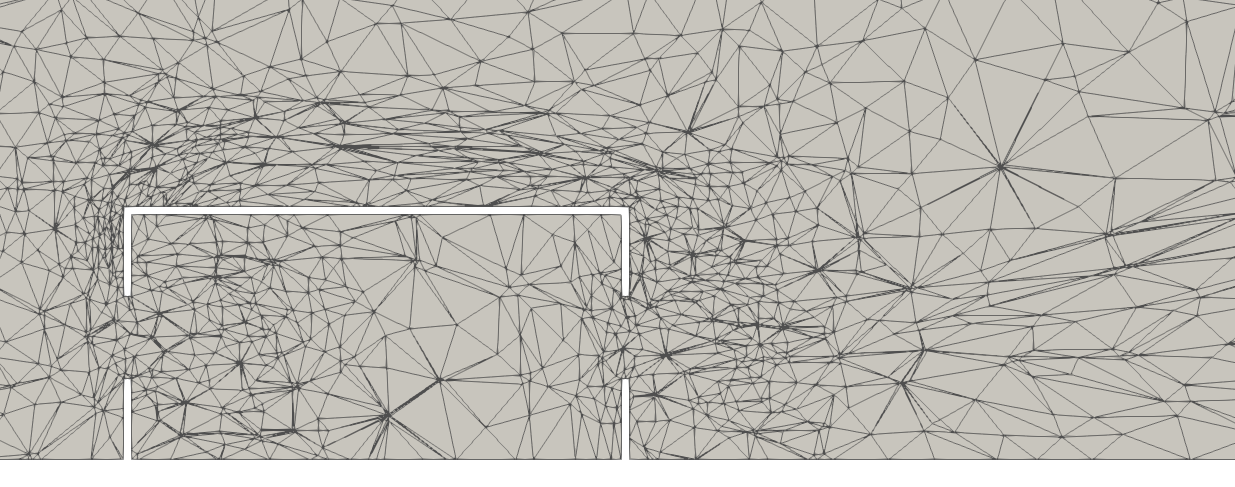}
        \caption{2 s}
        \label{Fig:Case7b_Mesh_2sec}
    \end{subfigure}
    \begin{subfigure}{0.39\textwidth}
        \includegraphics[width=\textwidth]{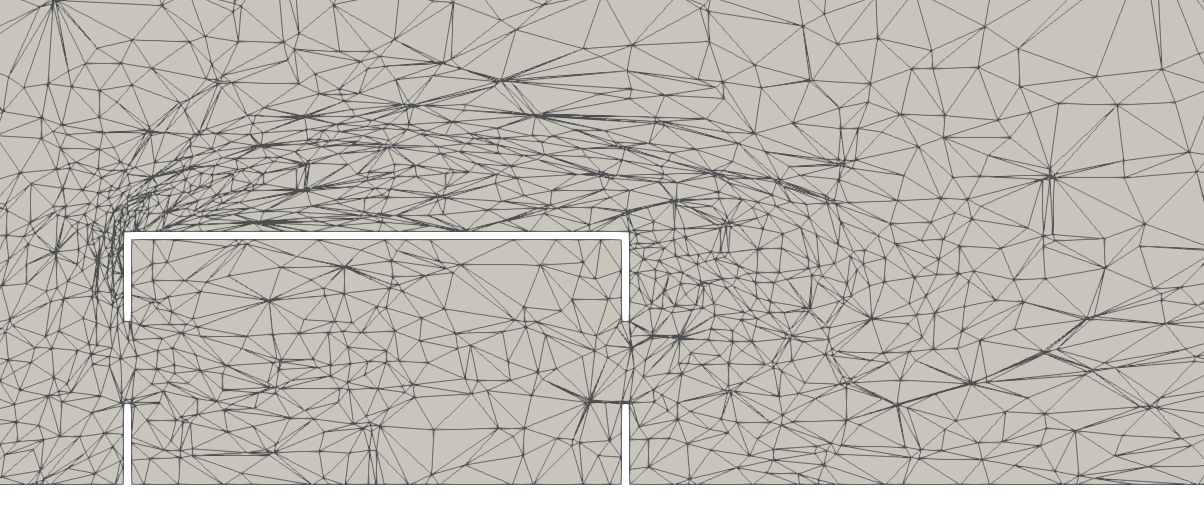}
        \caption{6 s}
        \label{Fig:Case7b_Mesh_6sec}
    \end{subfigure}
    \begin{subfigure}{0.42\textwidth}
        \includegraphics[width=\textwidth]{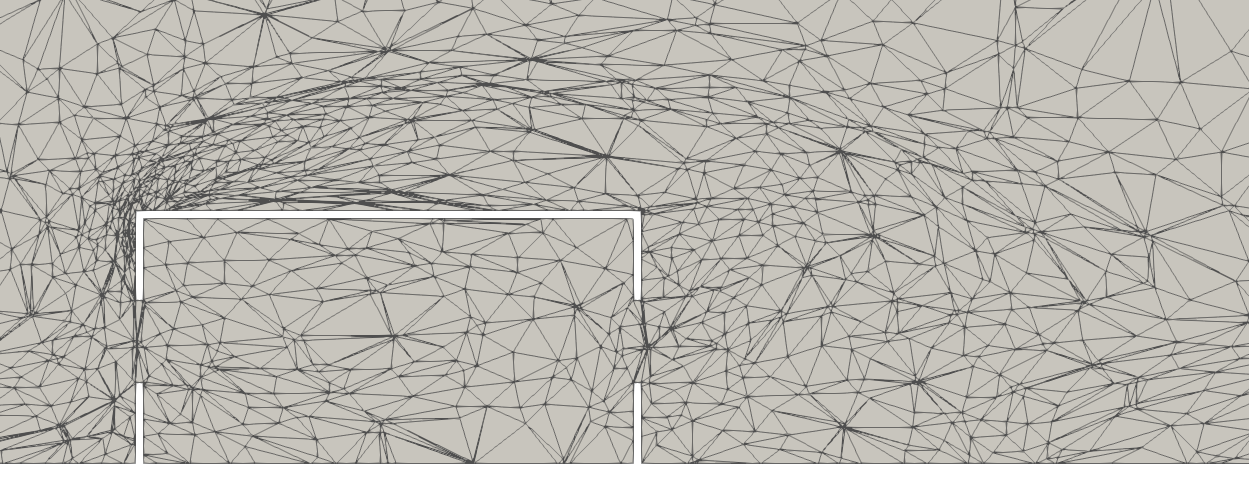}
        \caption{9 s}
        \label{Fig:Case7b_Mesh_9sec}
    \end{subfigure}
    \begin{subfigure}{0.39\textwidth}
        \includegraphics[width=\textwidth]{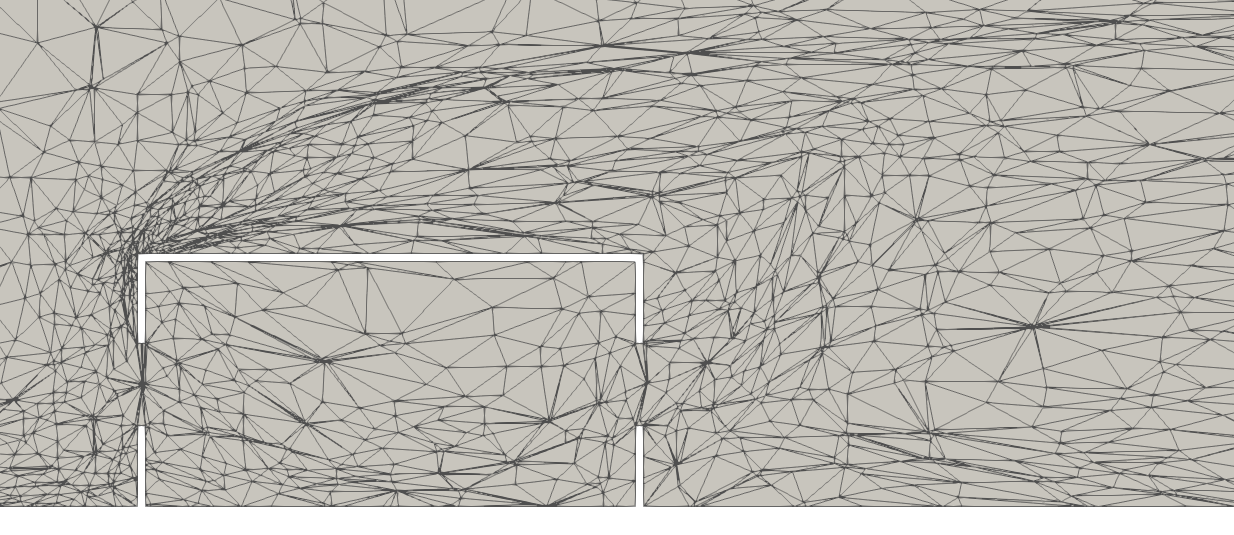}
        \caption{40 s}
        \label{Fig:Case7b_Mesh_40sec}
    \end{subfigure}
    \caption{Meshes at different instant for example \textit{3dBox\_Case7b.flml}. The velocity field only is adapted with an \texttt{error\_bound\_interpolation} equal to 0.25.}
    \label{Fig:Case7b_Mesh}
\end{figure}

\begin{figure}
    \centering
    \begin{subfigure}{0.4\textwidth}
        \includegraphics[width=\textwidth]{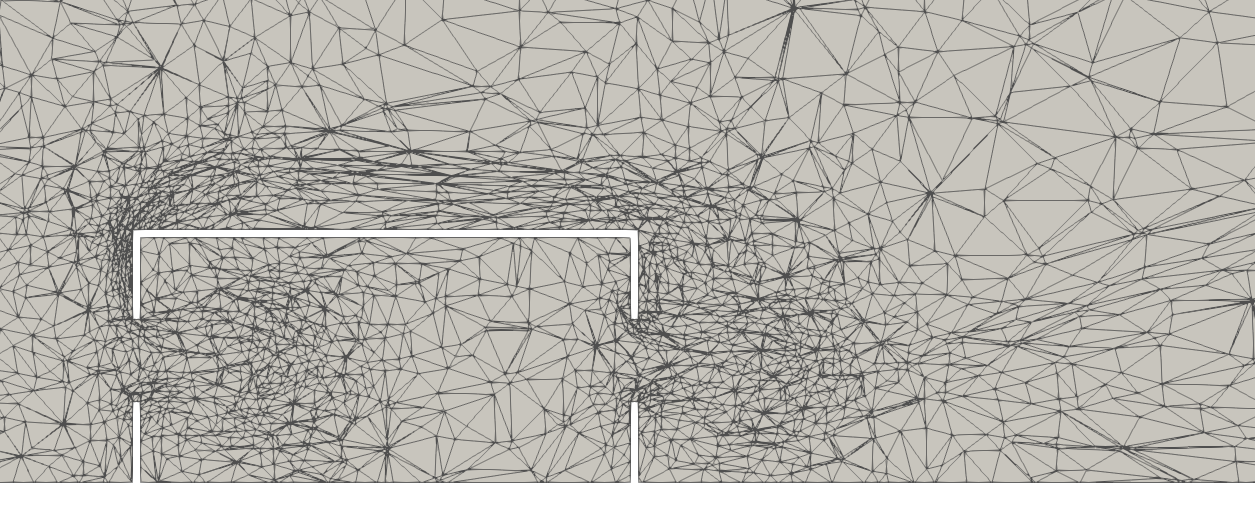}
        \caption{2 s}
        \label{Fig:Case7c_Mesh_2sec}
    \end{subfigure}
    \begin{subfigure}{0.4\textwidth}
        \includegraphics[width=\textwidth]{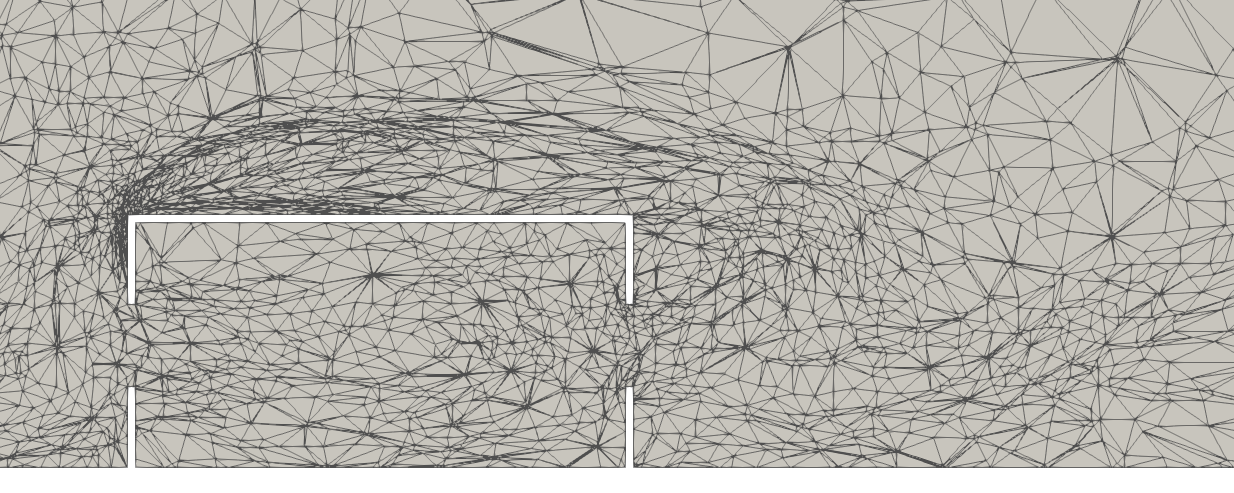}
        \caption{6 s}
        \label{Fig:Case7c_Mesh_6sec}
    \end{subfigure}
    \begin{subfigure}{0.4\textwidth}
        \includegraphics[width=\textwidth]{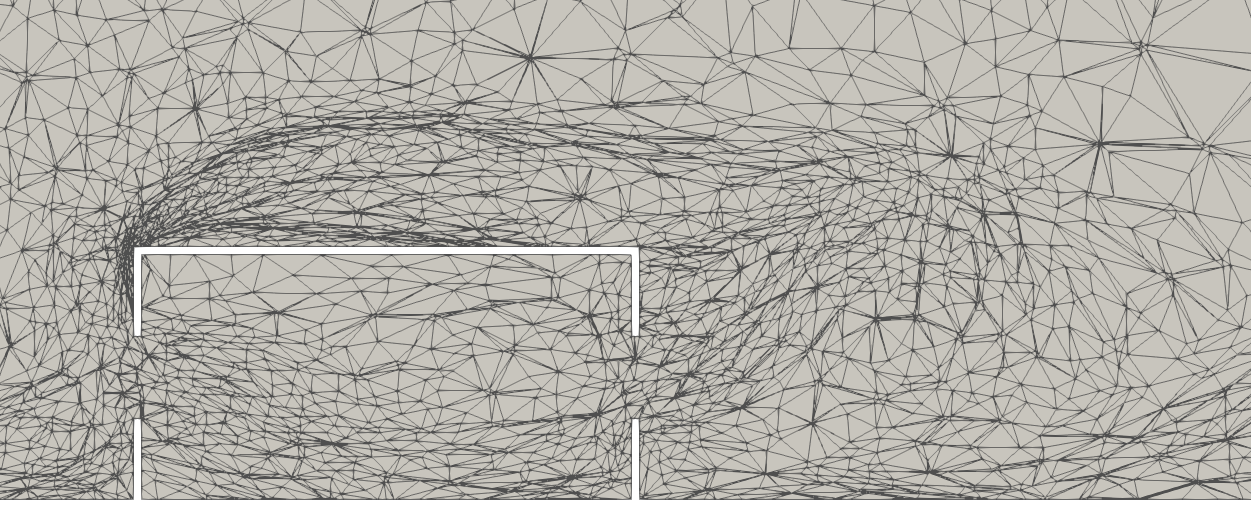}
        \caption{9 s}
        \label{Fig:Case7c_Mesh_9sec}
    \end{subfigure}
    \begin{subfigure}{0.39\textwidth}
        \includegraphics[width=\textwidth]{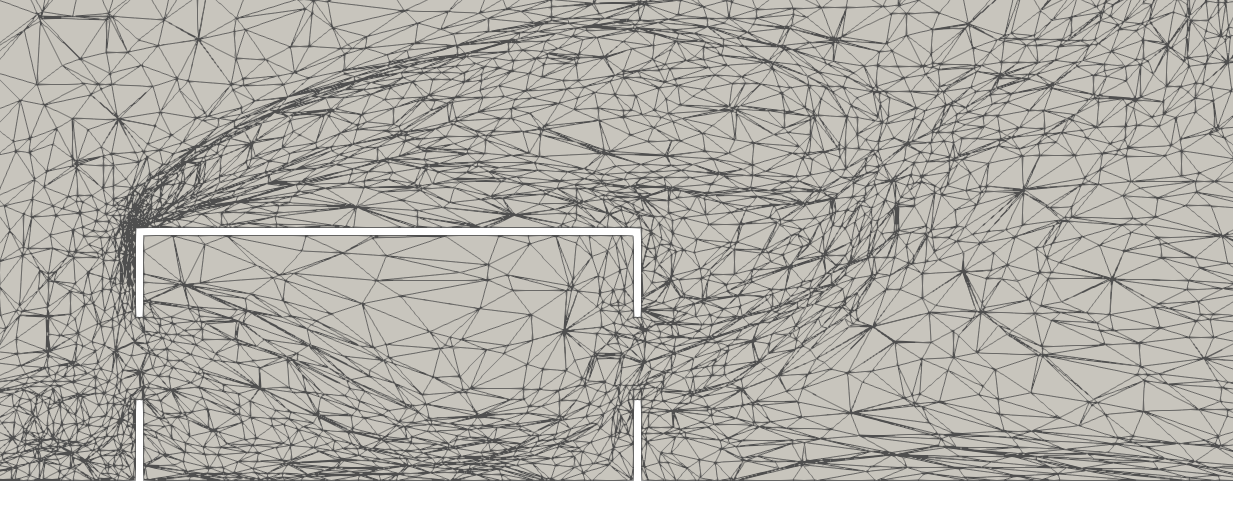}
        \caption{40 s}
        \label{Fig:Case7c_Mesh_40sec}
    \end{subfigure}
    \caption{Meshes at different instant for example \textit{3dBox\_Case7c.flml}. The velocity field only is adapted with an \texttt{error\_bound\_interpolation} equal to 0.15.}
    \label{Fig:Case7c_Mesh}
\end{figure}

\begin{figure}
    \centering
    \begin{subfigure}{0.4\textwidth}
        \includegraphics[width=\textwidth]{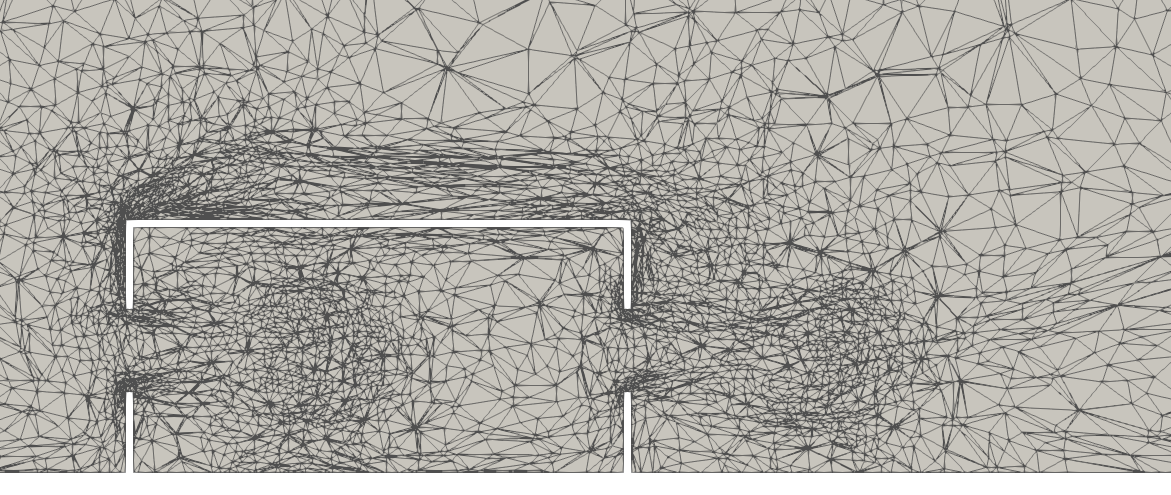}
        \caption{2 s}
        \label{Fig:Case7d_Mesh_2sec}
    \end{subfigure}
    \begin{subfigure}{0.4\textwidth}
        \includegraphics[width=\textwidth]{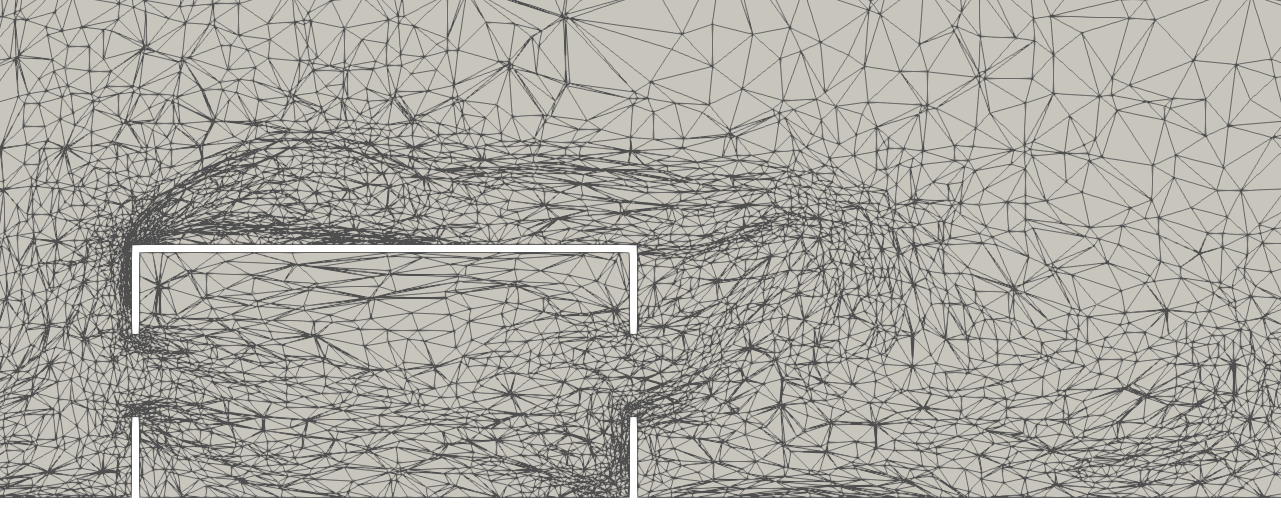}
        \caption{6 s}
        \label{Fig:Case7d_Mesh_6sec}
    \end{subfigure}
    \begin{subfigure}{0.4\textwidth}
        \includegraphics[width=\textwidth]{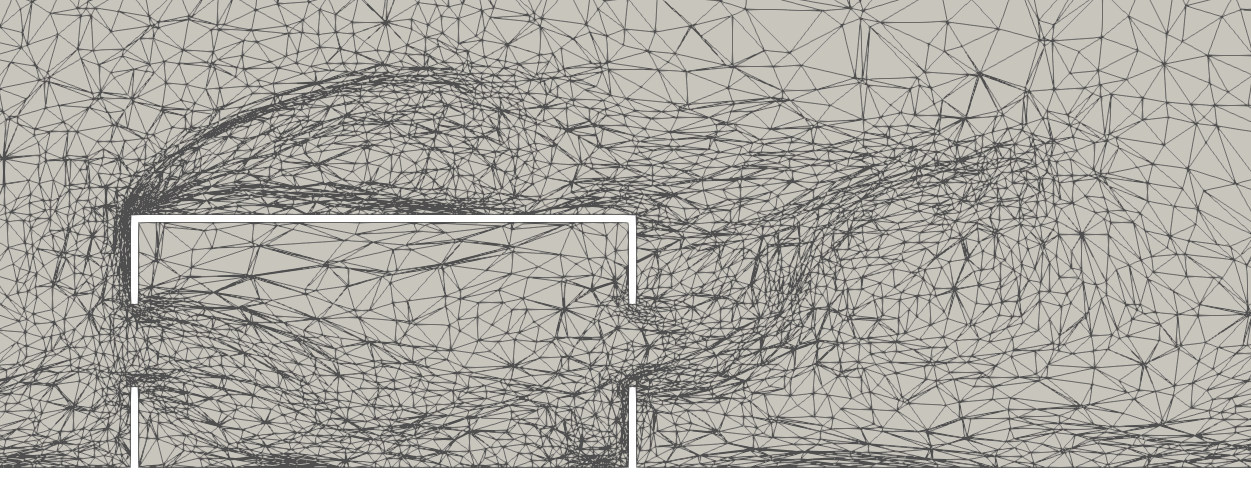}
        \caption{9 s}
        \label{Fig:Case7d_Mesh_9sec}
    \end{subfigure}
    \begin{subfigure}{0.4\textwidth}
        \includegraphics[width=\textwidth]{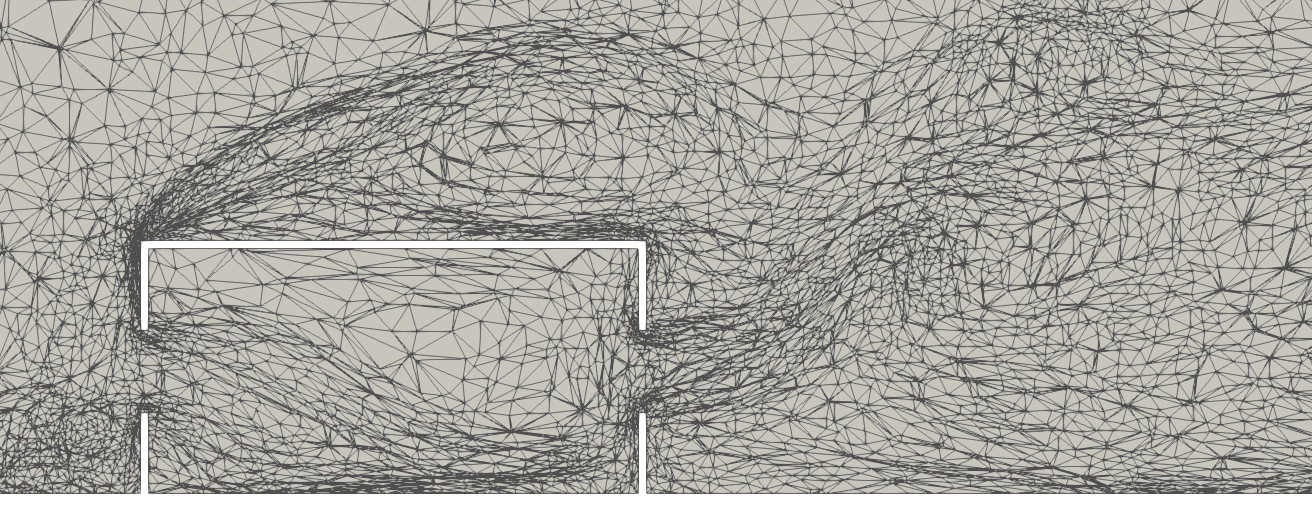}
        \caption{40 s}
        \label{Fig:Case7d_Mesh_40sec}
    \end{subfigure}
    \caption{Meshes at different instant for example \textit{3dBox\_Case7d.flml}. The velocity field only is adapted with an \texttt{error\_bound\_interpolation} equal to 0.1.}
    \label{Fig:Case7d_Mesh}
\end{figure}

\begin{figure}
    \centering
    \begin{subfigure}{0.39\textwidth}
        \includegraphics[width=\textwidth]{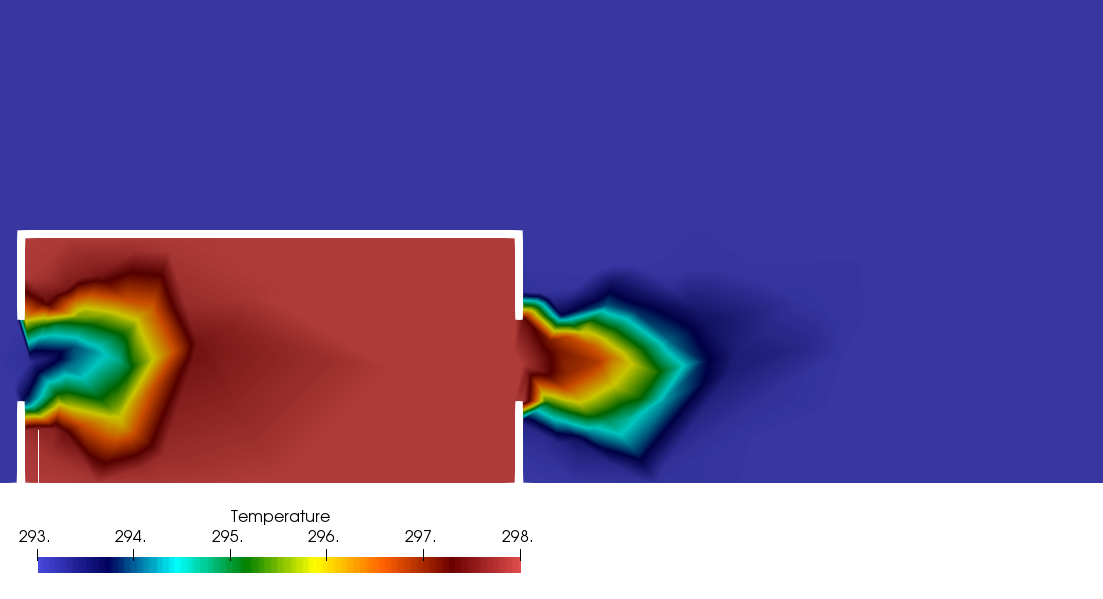}
        \caption{2 s}
        \label{Fig:Case7a_Temp_2sec}
    \end{subfigure}
    \begin{subfigure}{0.33\textwidth}
        \includegraphics[width=\textwidth]{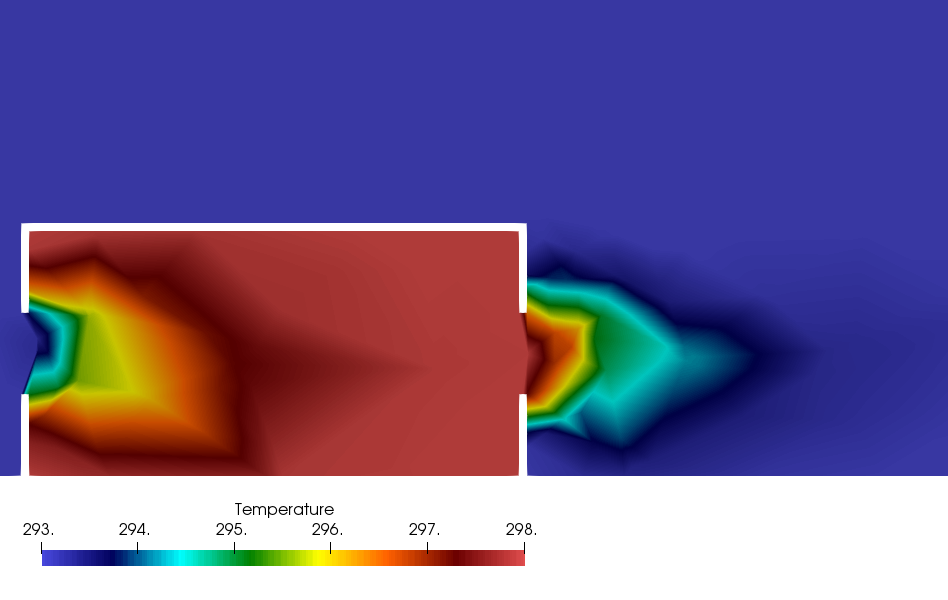}
        \caption{6 s}
        \label{Fig:Case7a_Temp_6sec}
    \end{subfigure}
    \begin{subfigure}{0.39\textwidth}
        \includegraphics[width=\textwidth]{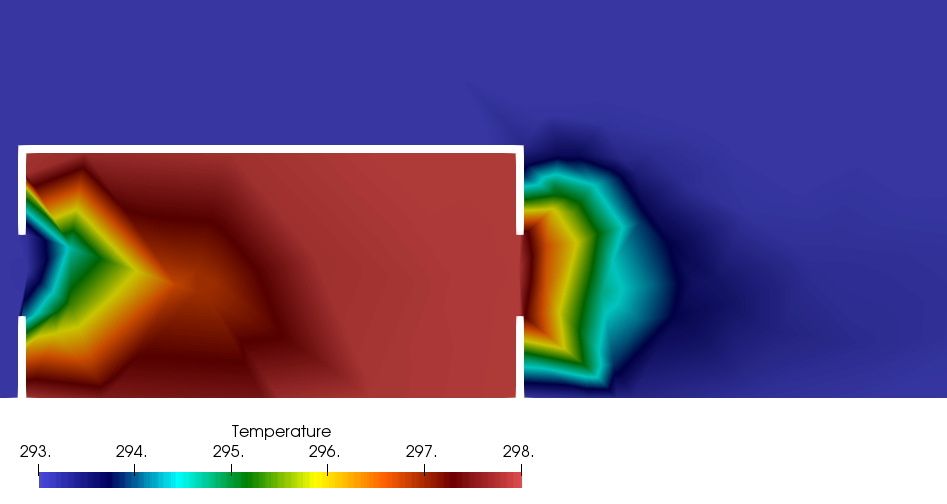}
        \caption{9 s}
        \label{Fig:Case7a_Temp_9sec}
    \end{subfigure}
    \begin{subfigure}{0.37\textwidth}
        \includegraphics[width=\textwidth]{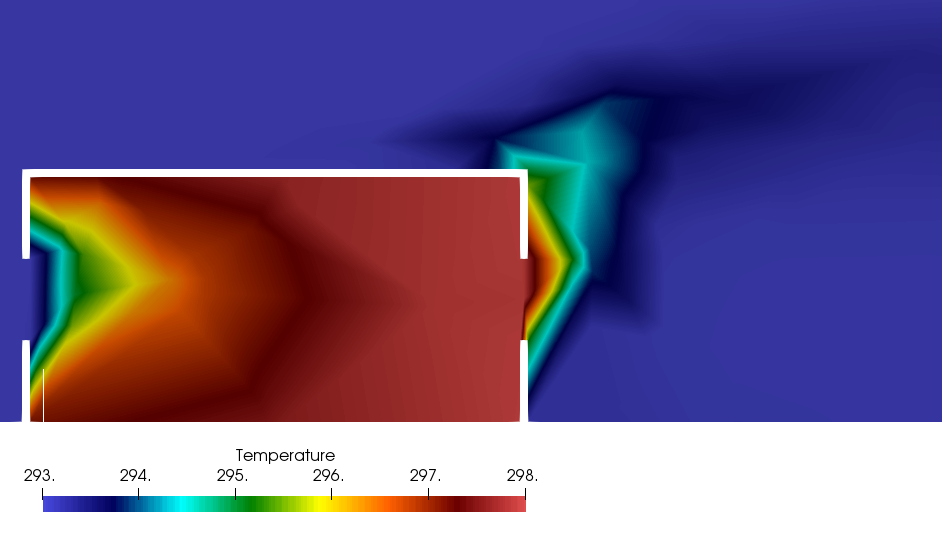}
        \caption{40 s}
        \label{Fig:Case7a_Temp_40sec}
    \end{subfigure}
    \caption{Temperature field at different instant for example \textit{3dBox\_Case7a.flml}. The velocity field only is adapted with an \texttt{error\_bound\_interpolation} equal to 0.5.}
    \label{Fig:Case7a_Temp}
\end{figure}

\begin{figure}
    \centering    
    \begin{subfigure}{0.39\textwidth}
        \includegraphics[width=\textwidth]{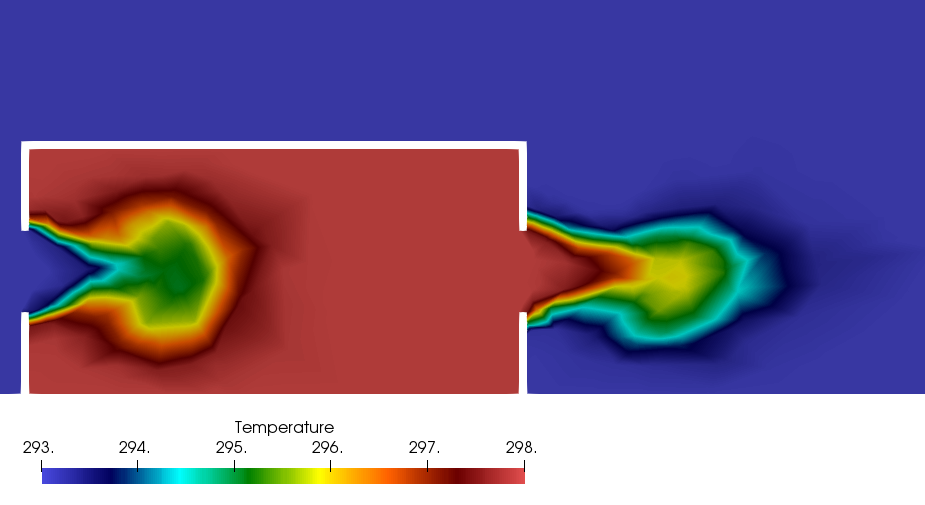}
        \caption{2 s}
        \label{Fig:Case7b_Temp_2sec}
    \end{subfigure}
    \begin{subfigure}{0.4\textwidth}
        \includegraphics[width=\textwidth]{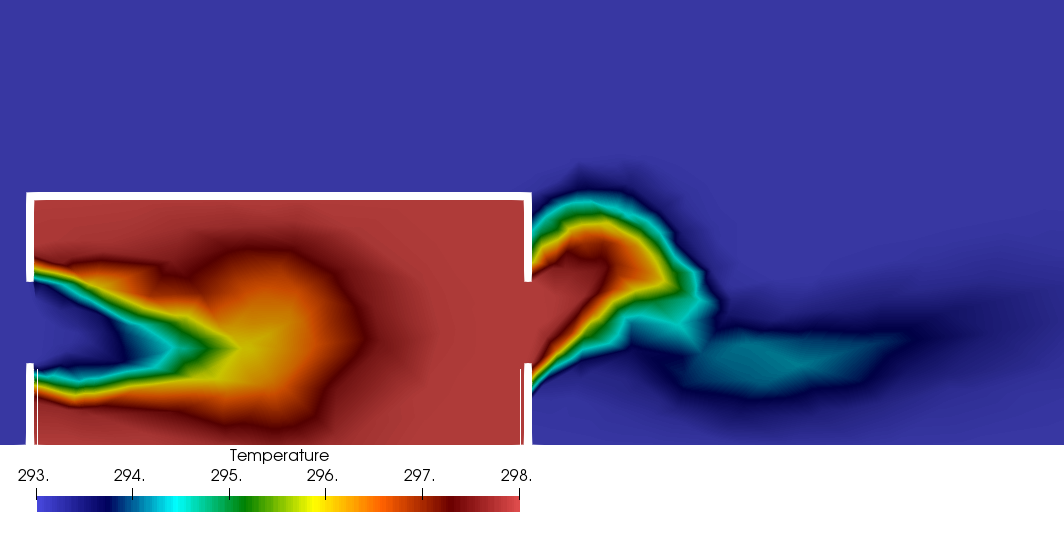}
        \caption{6 s}
        \label{Fig:Case7b_Temp_6sec}
    \end{subfigure}
    \begin{subfigure}{0.4\textwidth}
        \includegraphics[width=\textwidth]{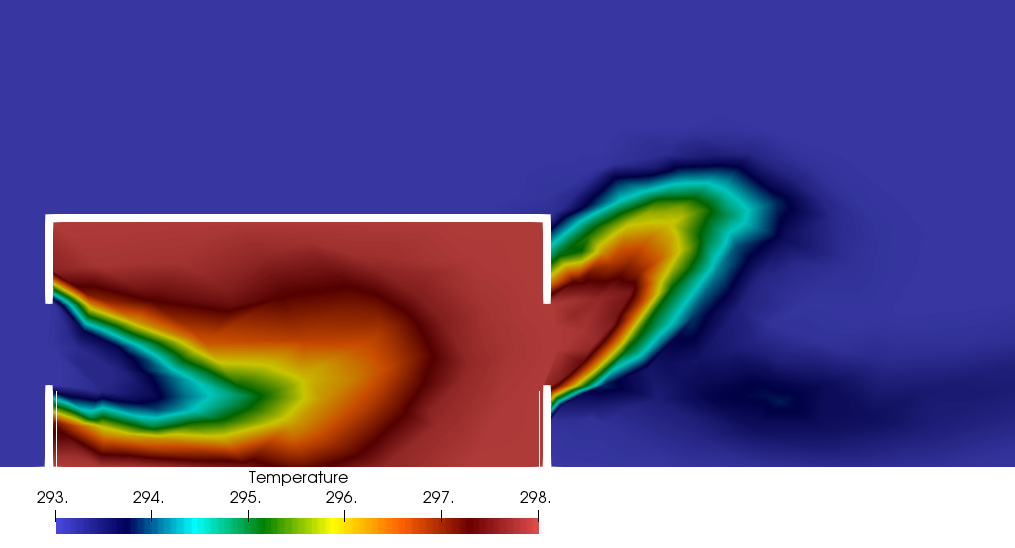}
        \caption{9 s}
        \label{Fig:Case7b_Temp_9sec}
    \end{subfigure}
    \begin{subfigure}{0.4\textwidth}
        \includegraphics[width=\textwidth]{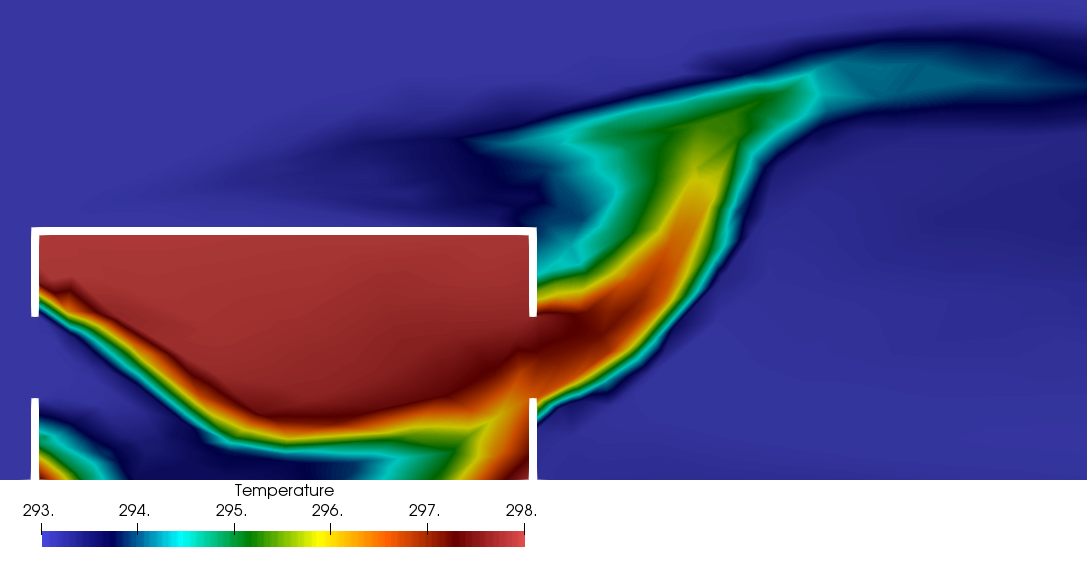}
        \caption{40 s}
        \label{Fig:Case7b_Temp_40sec}
    \end{subfigure}
    \caption{Temperature field at different instant for example \textit{3dBox\_Case7b.flml}. The velocity field only is adapted with an \texttt{error\_bound\_interpolation} equal to 0.25.}
    \label{Fig:Case7b_Temp}
\end{figure}

\begin{figure}
    \centering
    \begin{subfigure}{0.37\textwidth}
        \includegraphics[width=\textwidth]{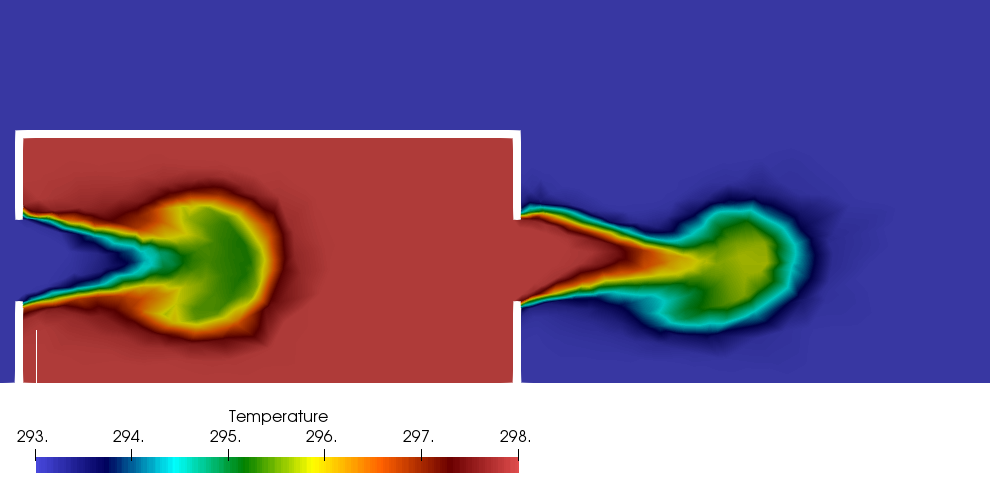}
        \caption{2 s}
        \label{Fig:Case7c_Temp_2sec}
    \end{subfigure}
    \begin{subfigure}{0.4\textwidth}
        \includegraphics[width=\textwidth]{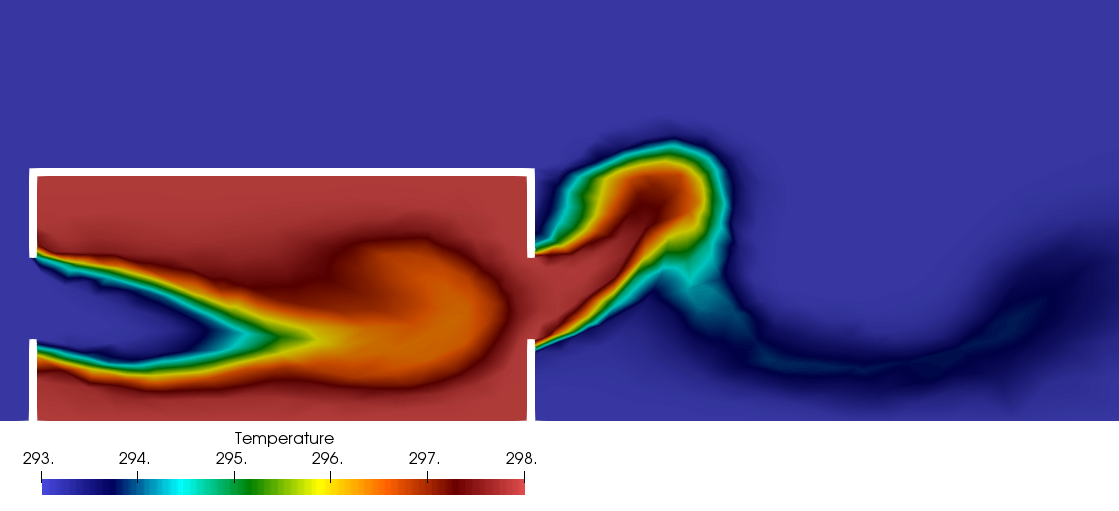}
        \caption{6 s}
        \label{Fig:Case7c_Temp_6sec}
    \end{subfigure}
    \begin{subfigure}{0.4\textwidth}
        \includegraphics[width=\textwidth]{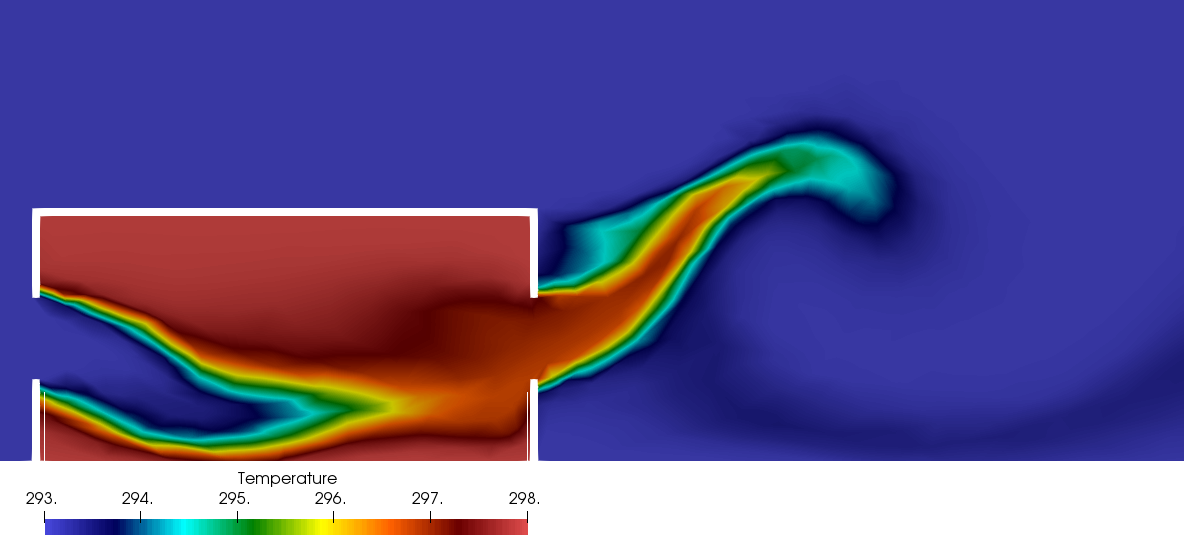}
        \caption{9 s}
        \label{Fig:Case7c_Temp_9sec}
    \end{subfigure}
    \begin{subfigure}{0.38\textwidth}
        \includegraphics[width=\textwidth]{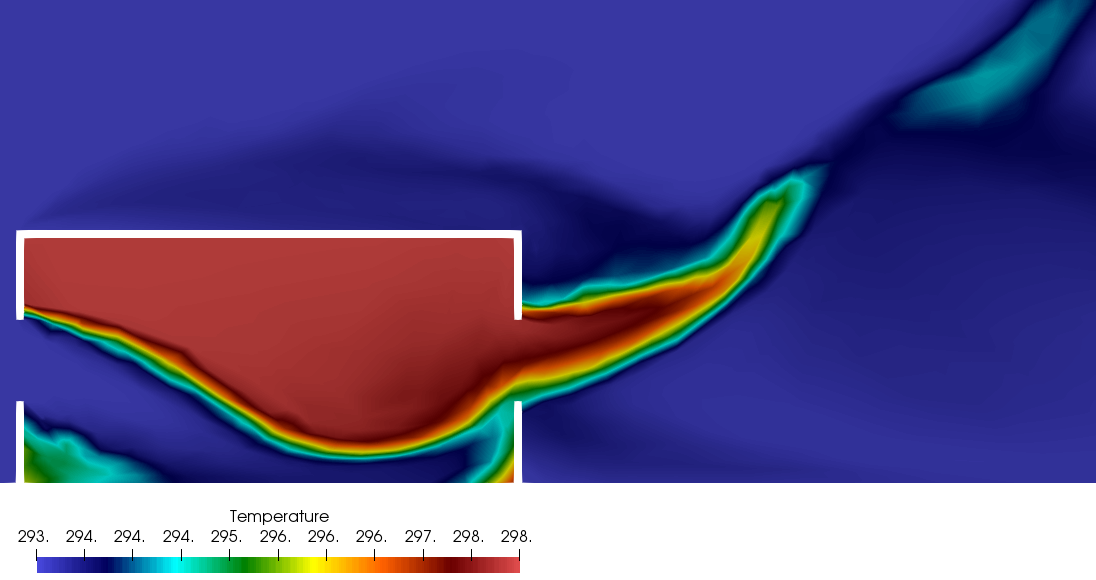}
        \caption{40 s}
        \label{Fig:Case7c_Temp_40sec}
    \end{subfigure}
    \caption{Temperature field at different instant for the simulation \textit{3dBox\_Case7c.flml}. The velocity field only is adapted with an \texttt{error\_bound\_interpolation} equal to 0.15.}
    \label{Fig:Case7c_Temp}
\end{figure}

\begin{figure}
    \centering
    \begin{subfigure}{0.36\textwidth}
        \includegraphics[width=\textwidth]{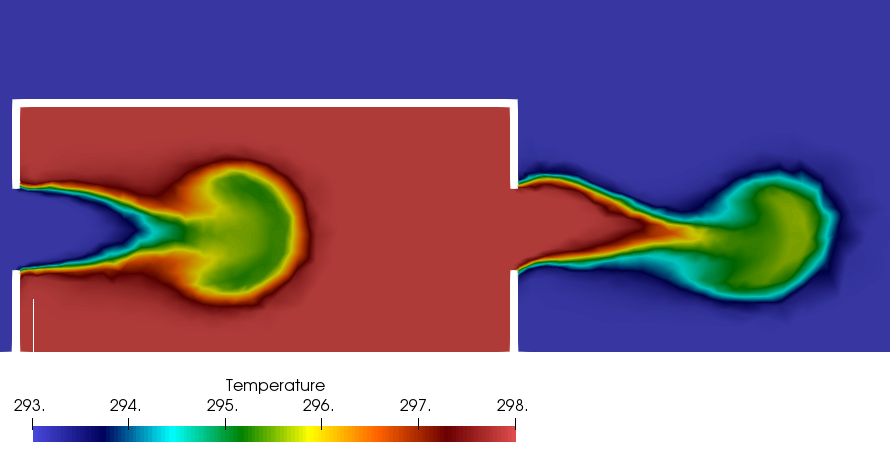}
        \caption{2 s}
        \label{Fig:Case7d_Temp_2sec}
    \end{subfigure}
    \begin{subfigure}{0.42\textwidth}
        \includegraphics[width=\textwidth]{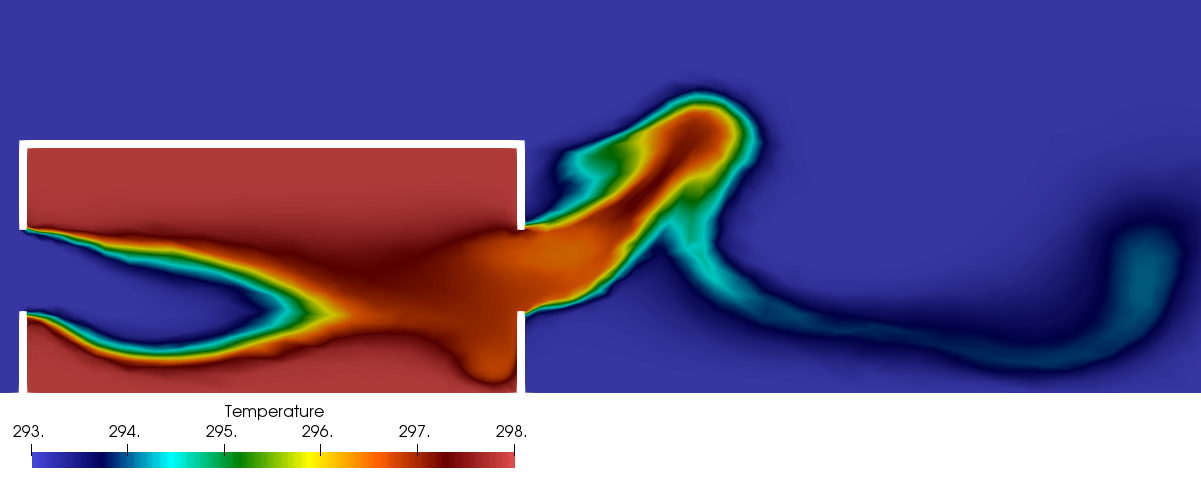}
        \caption{6 s}
        \label{Fig:Case7d_Temp_6sec}
    \end{subfigure}
    \begin{subfigure}{0.4\textwidth}
        \includegraphics[width=\textwidth]{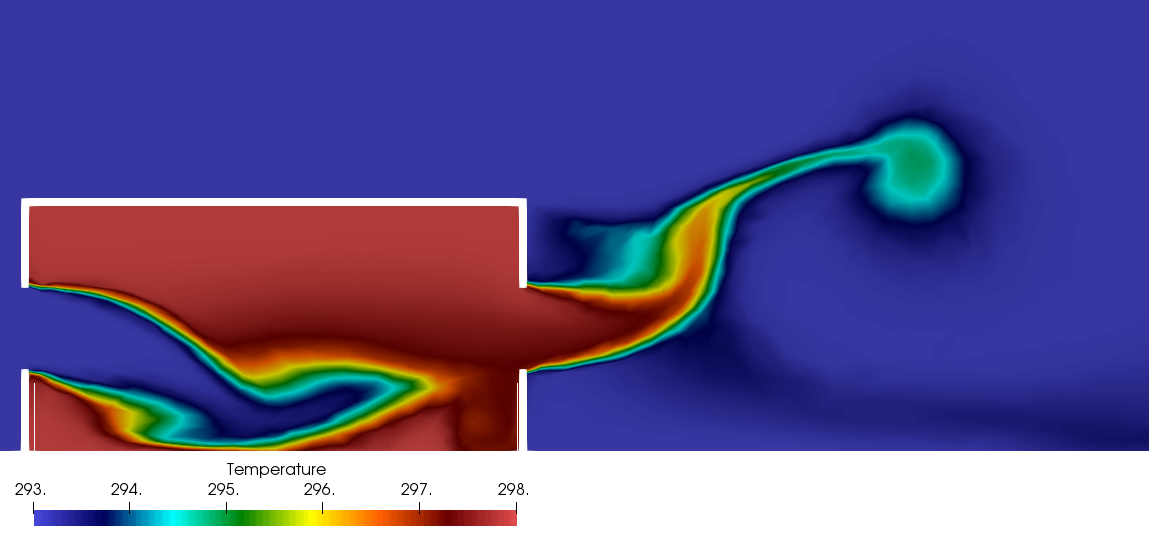}
        \caption{9 s}
        \label{Fig:Case7d_Temp_9sec}
    \end{subfigure}
    \begin{subfigure}{0.4\textwidth}
        \includegraphics[width=\textwidth]{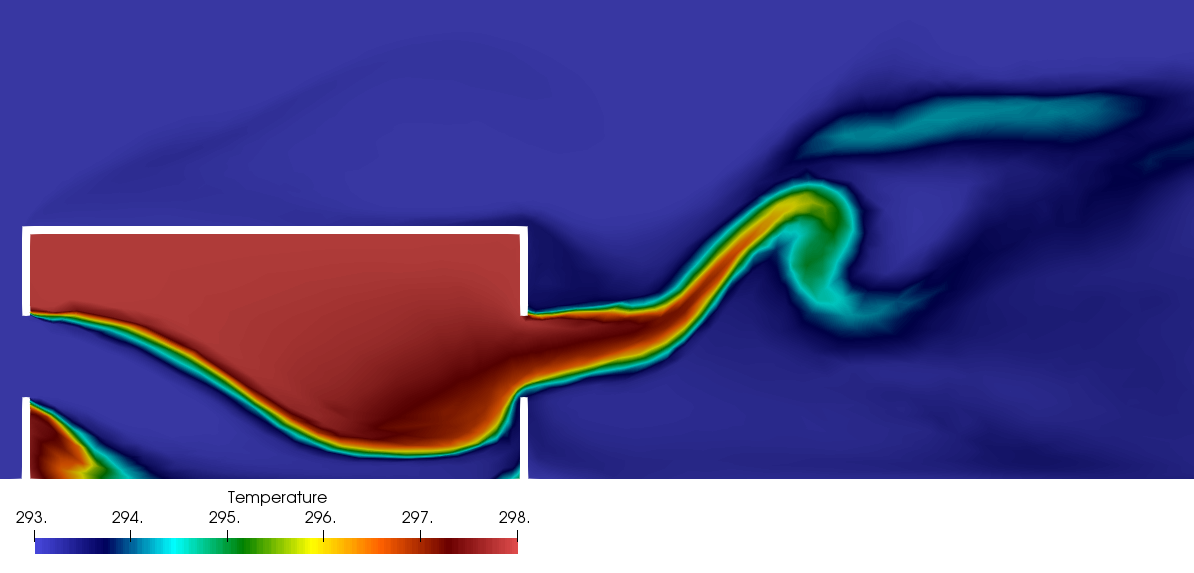}
        \caption{40 s}
        \label{Fig:Case7d_Temp_40sec}
    \end{subfigure}
    \caption{Temperature field at different instant for the simulation \textit{3dBox\_Case7d.flml}. The velocity field only is adapted with an \texttt{error\_bound\_interpolation} equal to 0.1.}
    \label{Fig:Case7d_Temp}
\end{figure}

\begin{figure}
    \centering
    \begin{subfigure}{0.38\textwidth}
        \includegraphics[width=\textwidth]{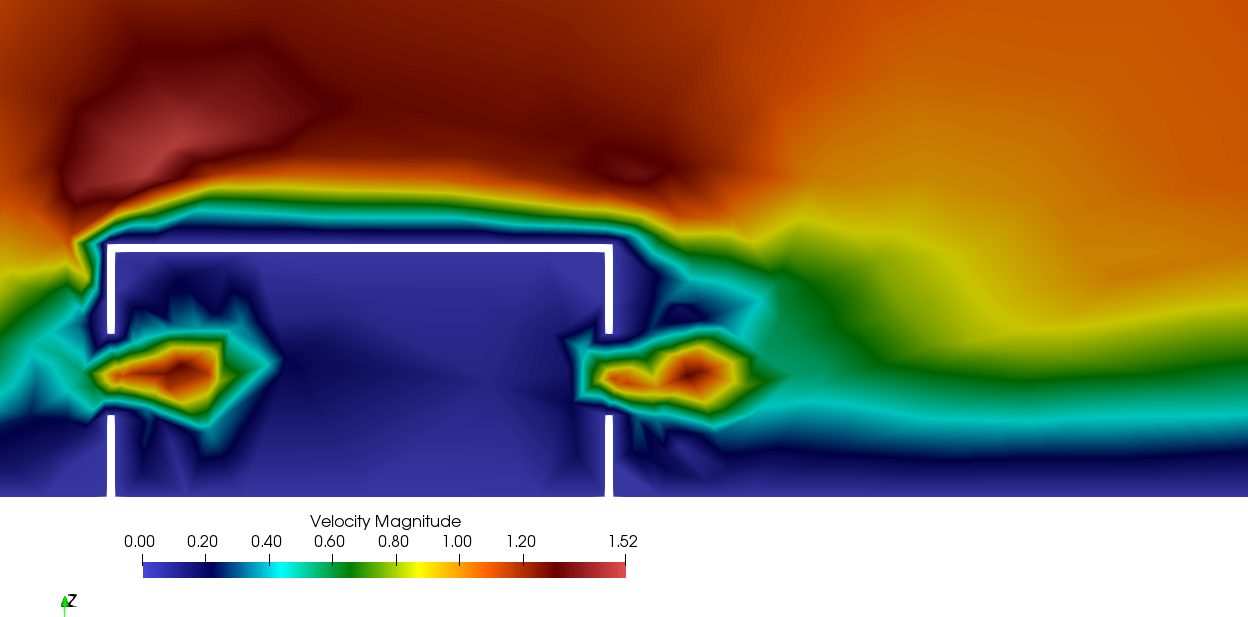}
        \caption{2 s}
        \label{Fig:Case7a_Vel_2sec}
    \end{subfigure}
    \begin{subfigure}{0.4\textwidth}
        \includegraphics[width=\textwidth]{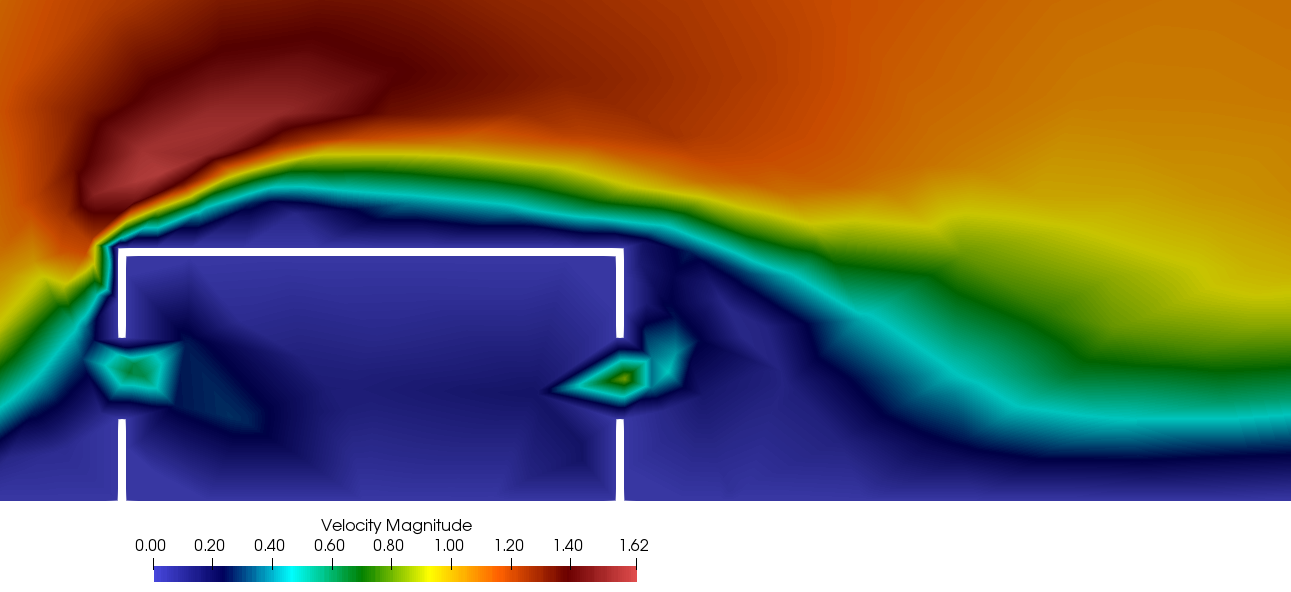}
        \caption{6 s}
        \label{Fig:Case7a_Vel_6sec}
    \end{subfigure}
    \begin{subfigure}{0.4\textwidth}
        \includegraphics[width=\textwidth]{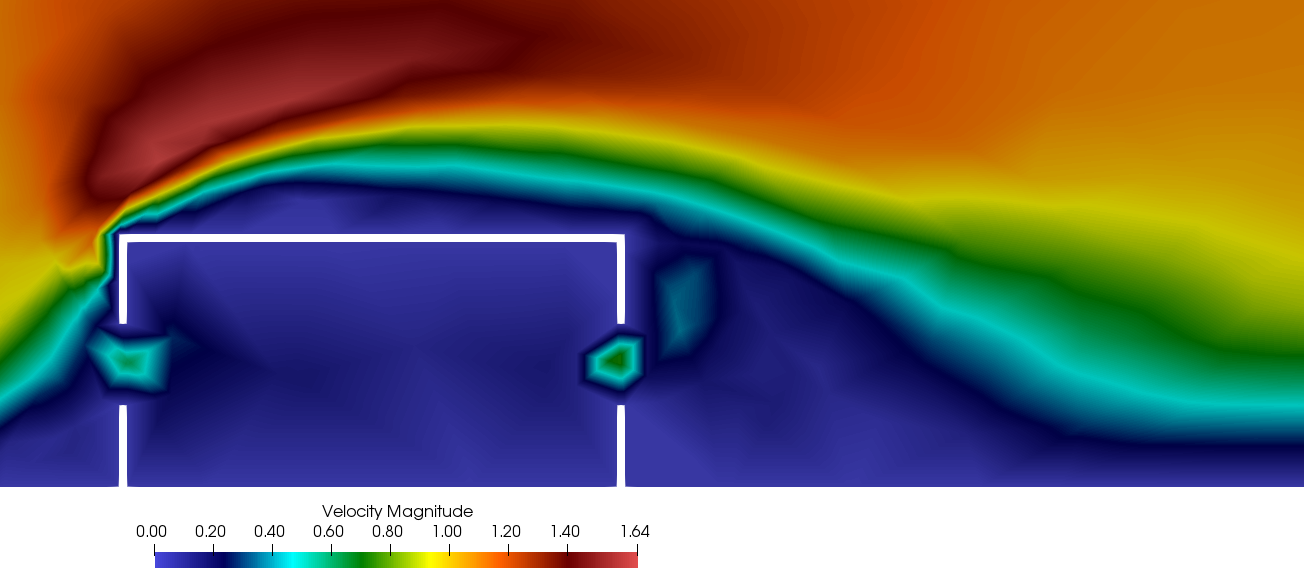}
        \caption{9 s}
        \label{Fig:Case7a_Vel_9sec}
    \end{subfigure}
    \begin{subfigure}{0.4\textwidth}
        \includegraphics[width=\textwidth]{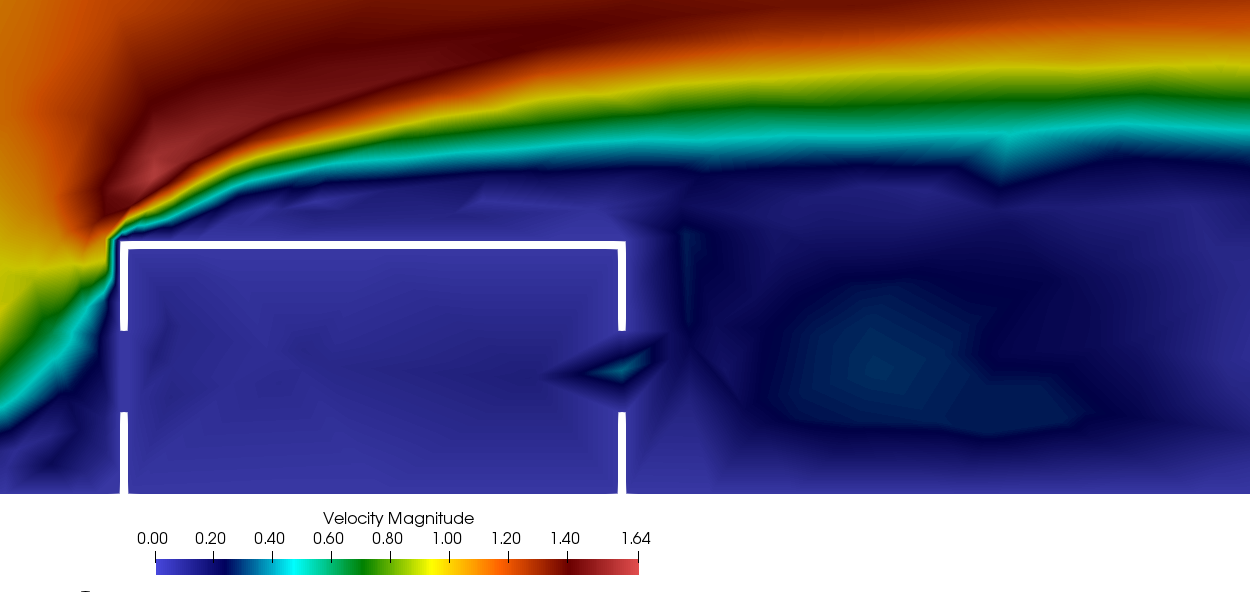}
        \caption{40 s}
        \label{Fig:Case7a_Vel_40sec}
    \end{subfigure}
    \caption{Velocity field at different instant for the simulation \textit{3dBox\_Case7a.flml}. The velocity field only is adapted with an \texttt{error\_bound\_interpolation} equal to 0.5.}
    \label{Fig:Case7a_Vel}
\end{figure}

\begin{figure}
    \centering    
    \begin{subfigure}{0.4\textwidth}
        \includegraphics[width=\textwidth]{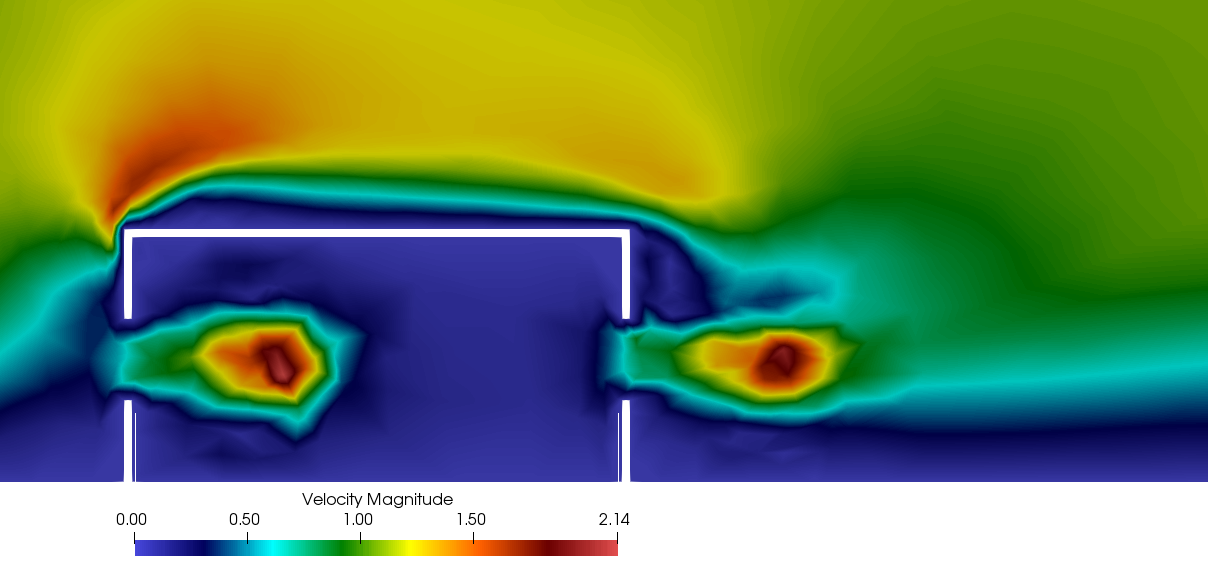}
        \caption{2 s}
        \label{Fig:Case7b_Vel_2sec}
    \end{subfigure}
    \begin{subfigure}{0.4\textwidth}
        \includegraphics[width=\textwidth]{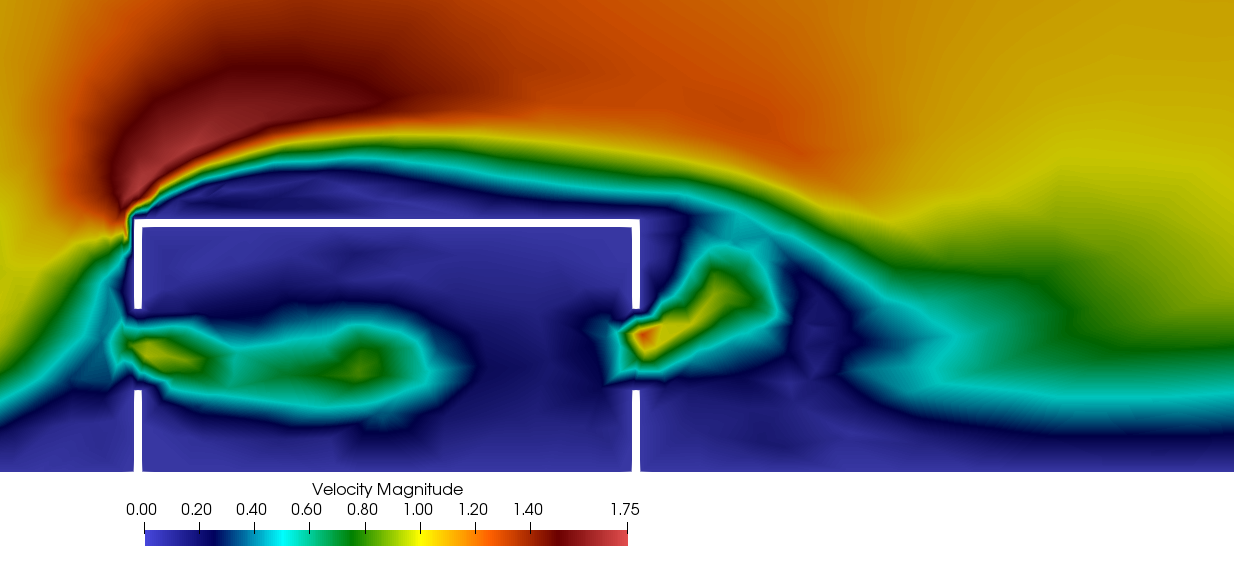}
        \caption{6 s}
        \label{Fig:Case7b_Vel_6sec}
    \end{subfigure}
    \begin{subfigure}{0.4\textwidth}
        \includegraphics[width=\textwidth]{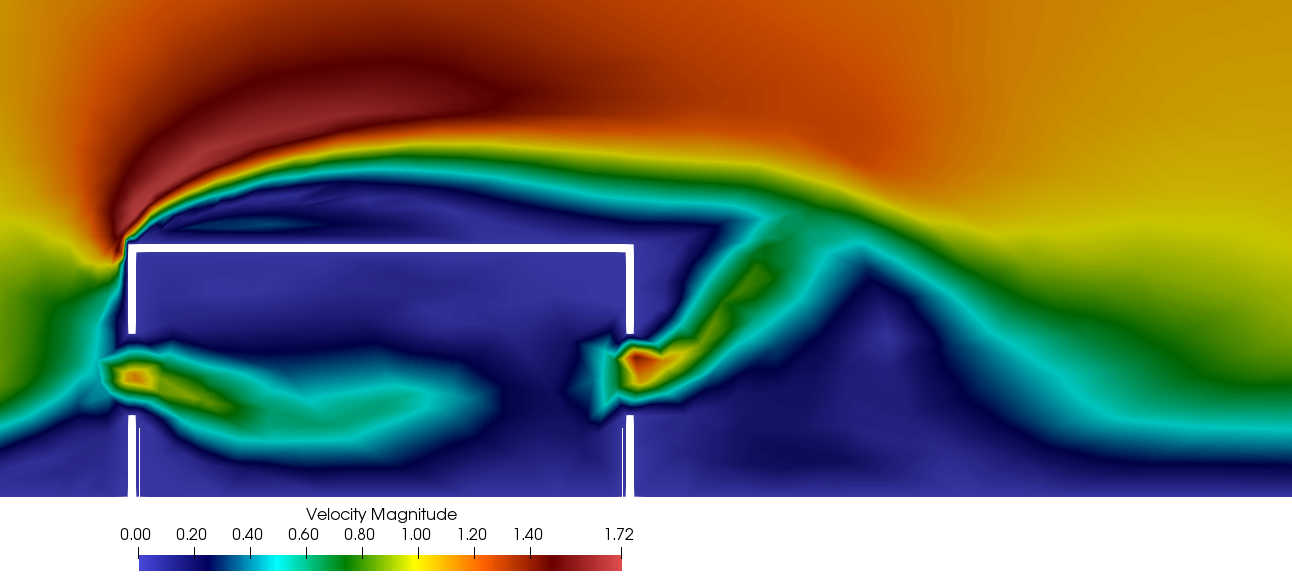}
        \caption{9 s}
        \label{Fig:Case7b_Vel_9sec}
    \end{subfigure}
    \begin{subfigure}{0.4\textwidth}
        \includegraphics[width=\textwidth]{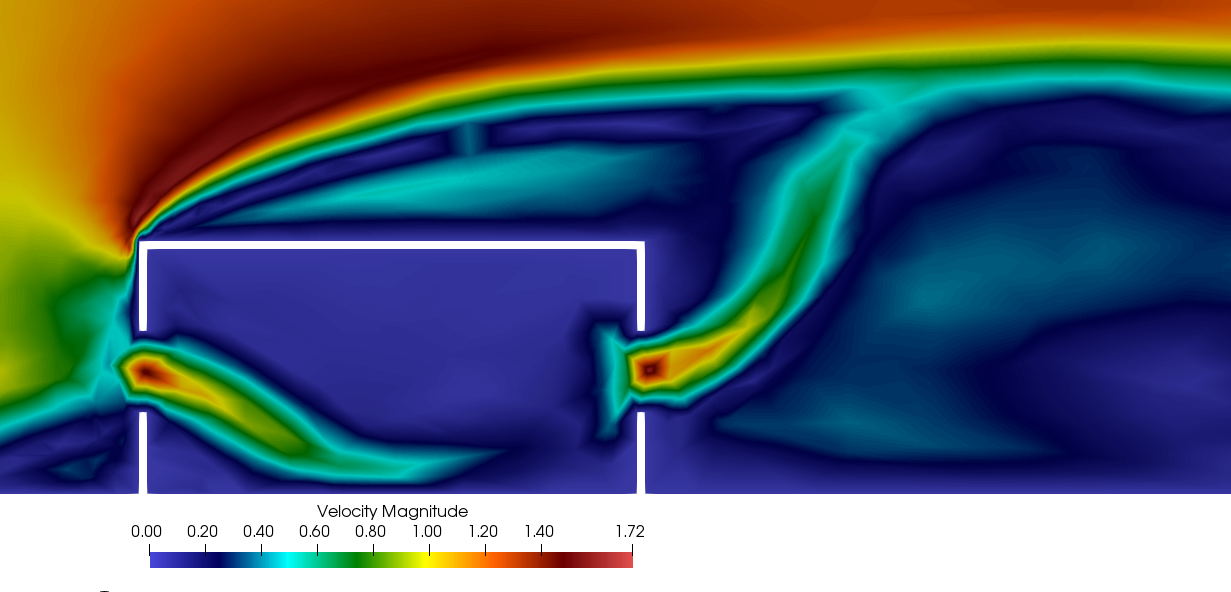}
        \caption{40 s}
        \label{Fig:Case7b_Vel_40sec}
    \end{subfigure}
    \caption{Velocity field at different instant for the simulation \textit{3dBox\_Case7b.flml}. The velocity field only is adapted with an \texttt{error\_bound\_interpolation} equal to 0.25.}
    \label{Fig:Case7b_Vel}
\end{figure}

\begin{figure}
    \centering
    \begin{subfigure}{0.4\textwidth}
        \includegraphics[width=\textwidth]{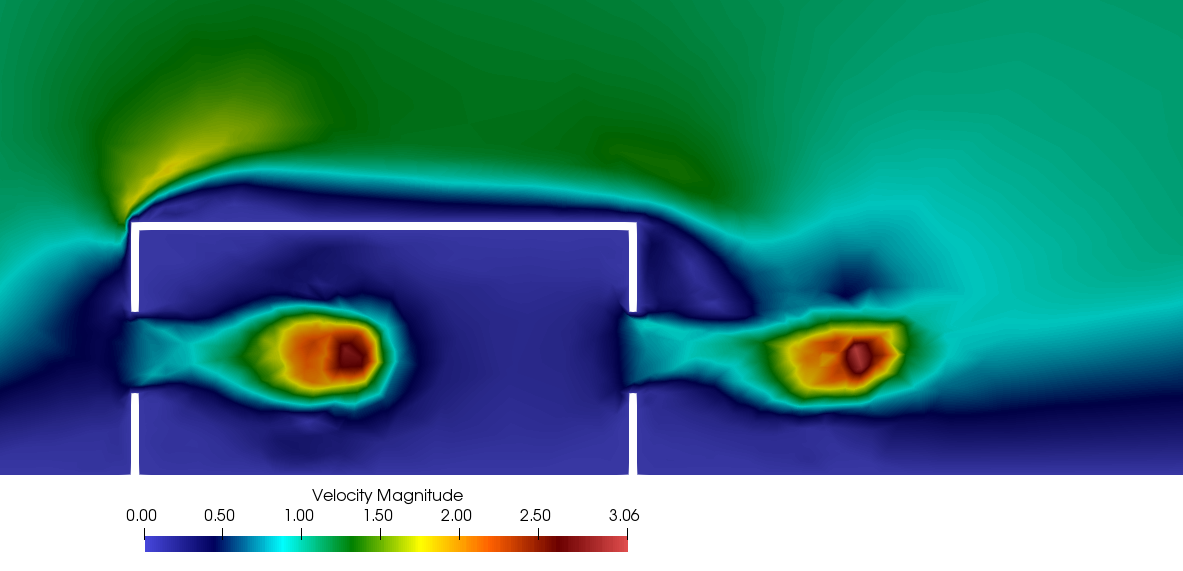}
        \caption{2 s}
        \label{Fig:Case7c_Vel_2sec}
    \end{subfigure}
    \begin{subfigure}{0.4\textwidth}
        \includegraphics[width=\textwidth]{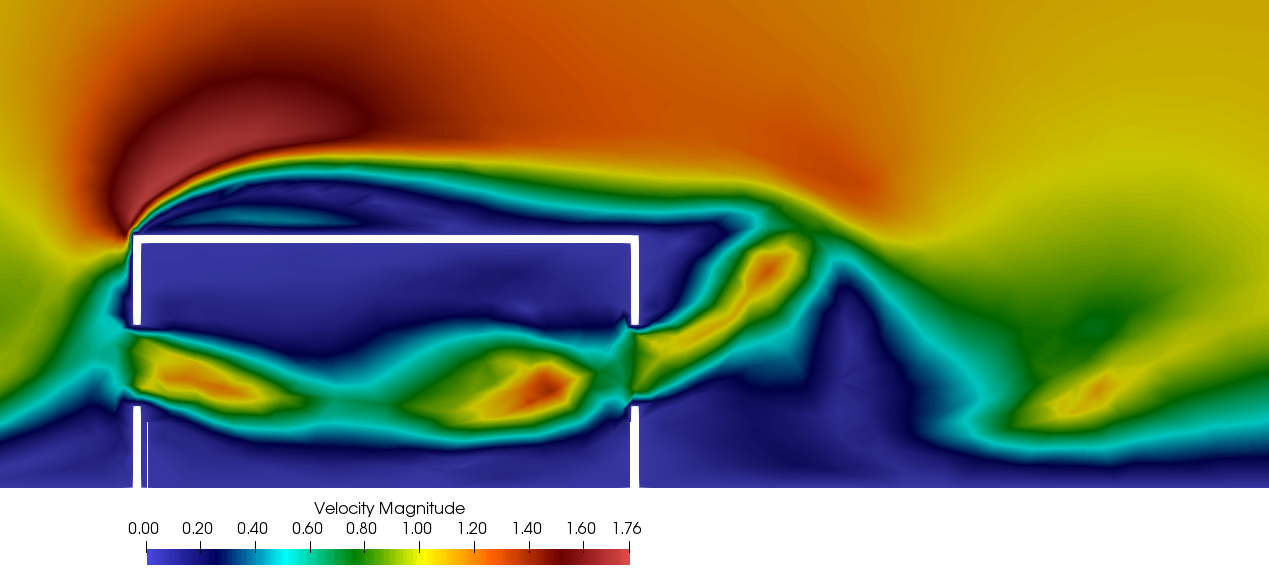}
        \caption{6 s}
        \label{Fig:Case7c_Vel_6sec}
    \end{subfigure}
    \begin{subfigure}{0.4\textwidth}
        \includegraphics[width=\textwidth]{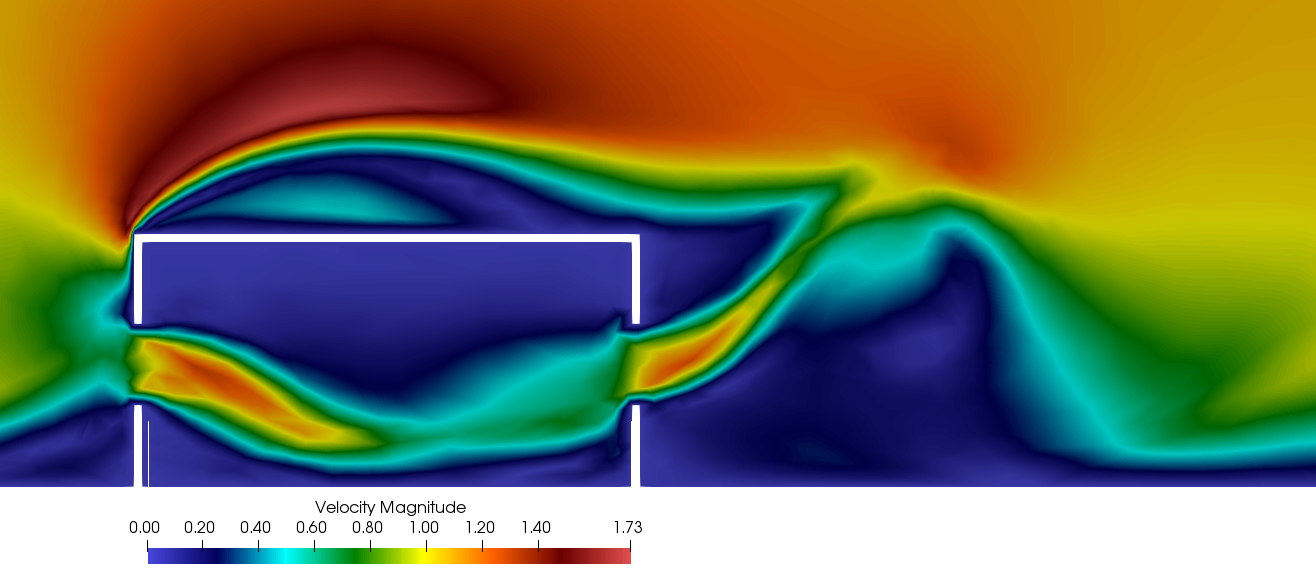}
        \caption{9 s}
        \label{Fig:Case7c_Vel_9sec}
    \end{subfigure}
    \begin{subfigure}{0.4\textwidth}
        \includegraphics[width=\textwidth]{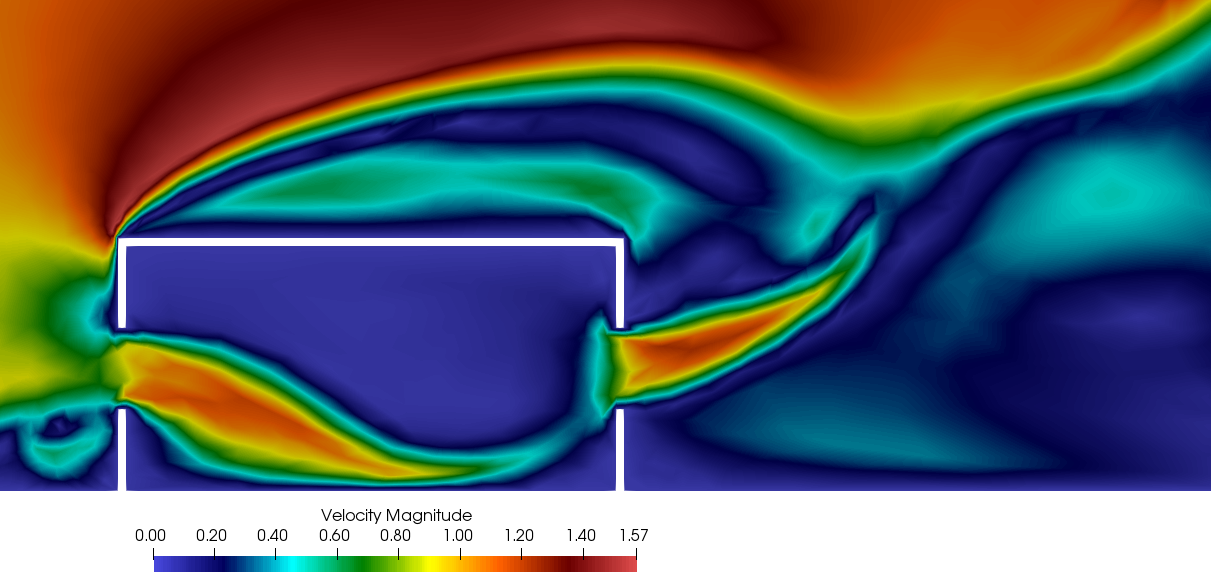}
        \caption{40 s}
        \label{Fig:Case7c_Vel_40sec}
    \end{subfigure}
    \caption{Velocity field at different instant for the simulation \textit{3dBox\_Case7c.flml}. The velocity field only is adapted with an \texttt{error\_bound\_interpolation} equal to 0.15.}
    \label{Fig:Case7c_Vel}
\end{figure}

\begin{figure}
    \centering
    \begin{subfigure}{0.39\textwidth}
        \includegraphics[width=\textwidth]{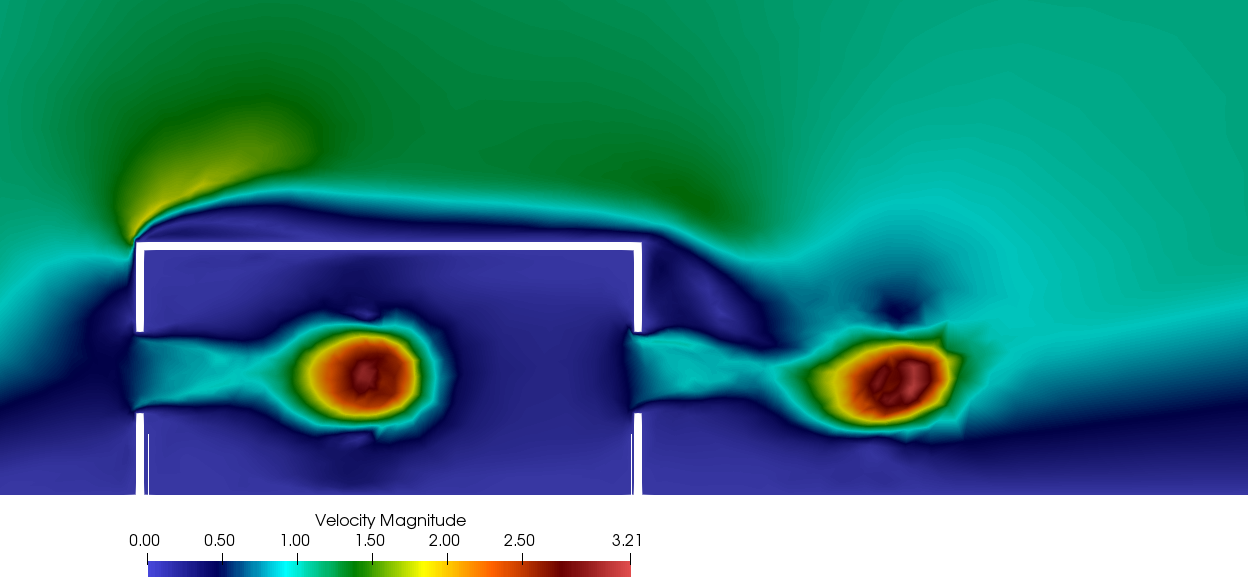}
        \caption{2 s}
        \label{Fig:Case7d_Vel_2sec}
    \end{subfigure}
    \begin{subfigure}{0.4\textwidth}
        \includegraphics[width=\textwidth]{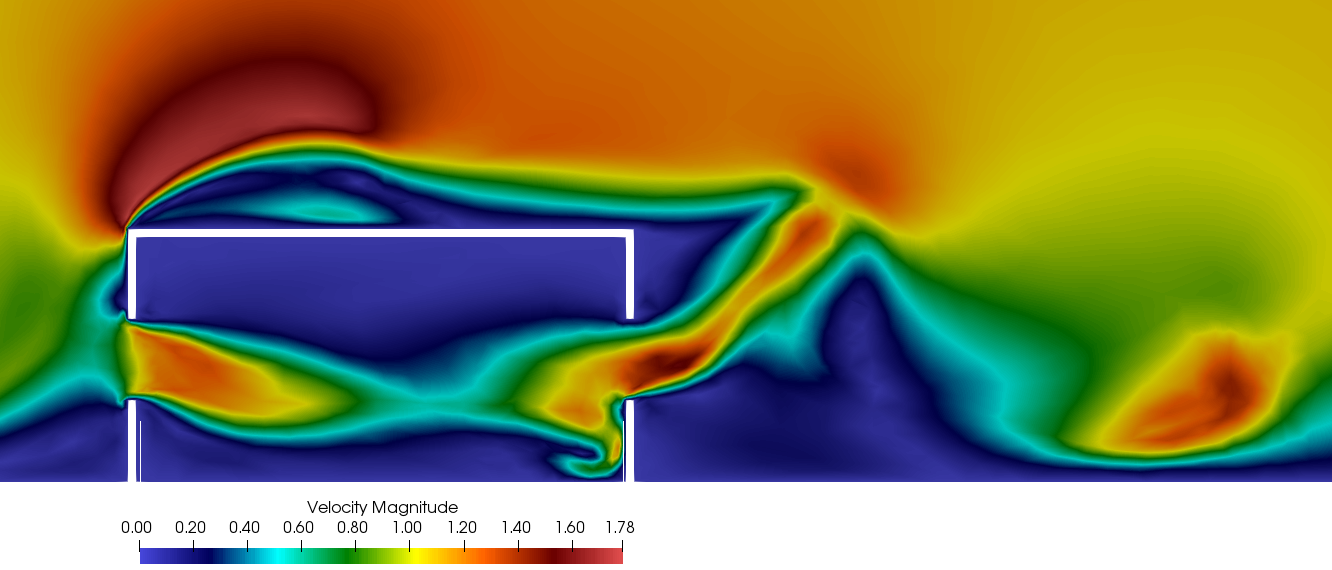}
        \caption{6 s}
        \label{Fig:Case7d_Vel_6sec}
    \end{subfigure}
    \begin{subfigure}{0.4\textwidth}
        \includegraphics[width=\textwidth]{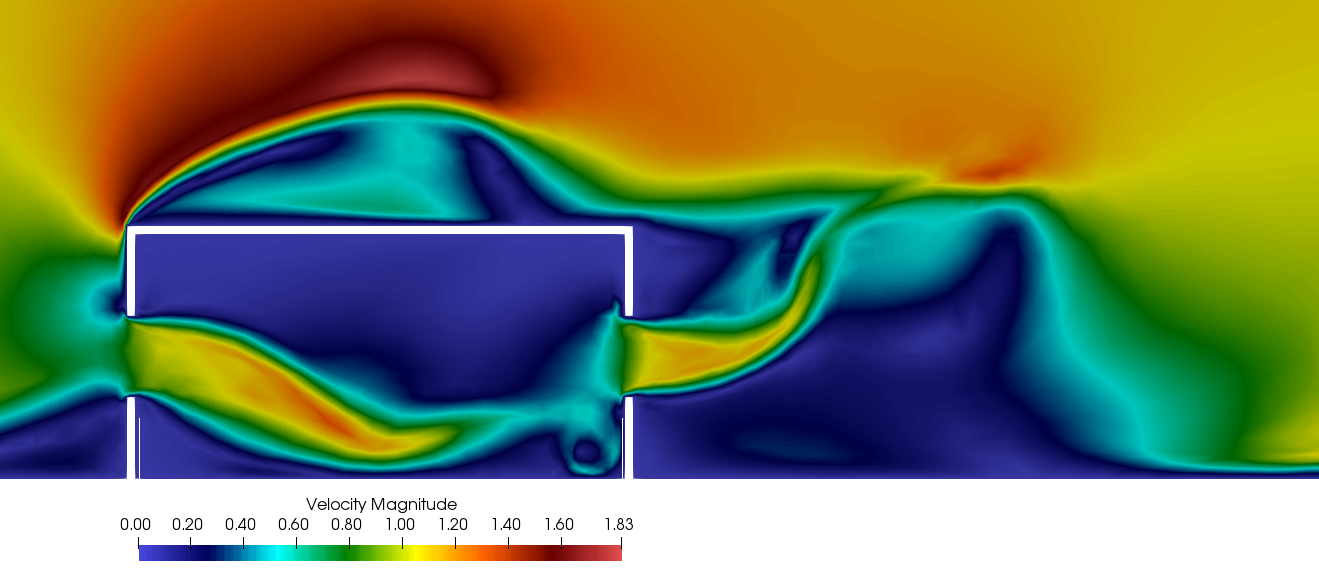}
        \caption{9 s}
        \label{Fig:Case7d_Vel_9sec}
    \end{subfigure}
    \begin{subfigure}{0.38\textwidth}
        \includegraphics[width=\textwidth]{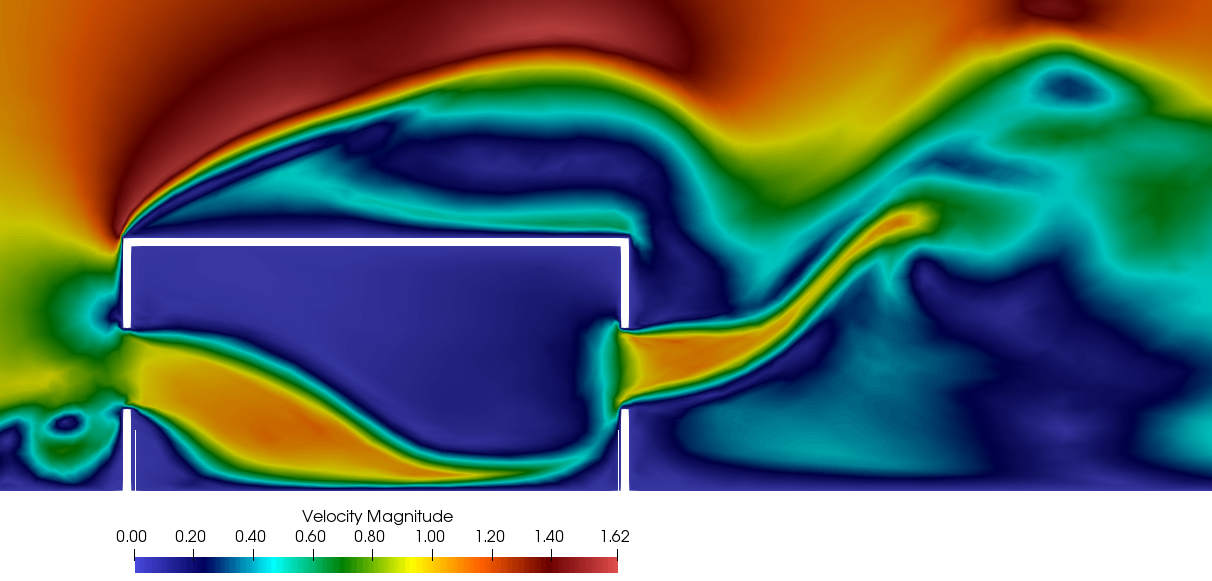}
        \caption{40 s}
        \label{Fig:Case7d_Vel_40sec}
    \end{subfigure}
    \caption{Velocity field at different instant for the simulation \textit{3dBox\_Case7d.flml}. The velocity field only is adapted with an \texttt{error\_bound\_interpolation} equal to 0.1.}
    \label{Fig:Case7d_Vel}
\end{figure}

Snapshots of the meshes are shown in Figure~\ref{Fig:Case7a_Mesh}, Figure~\ref{Fig:Case7b_Mesh}, Figure~\ref{Fig:Case7c_Mesh} and Figure~\ref{Fig:Case7d_Mesh}. Snapshots of the temperature field are shown in Figure~\ref{Fig:Case7a_Temp}, Figure~\ref{Fig:Case7b_Temp}, Figure~\ref{Fig:Case7c_Temp} and Figure~\ref{Fig:Case7d_Temp}. Snapshots of the velocity field are shown in Figure~\ref{Fig:Case7a_Vel}, Figure~\ref{Fig:Case7b_Vel}, Figure~\ref{Fig:Case7c_Vel} and Figure~\ref{Fig:Case7d_Vel}. Go to Chapter~\ref{Sec:PostProcessing} to learn how to visualise the results using \textbf{ParaView}.

\noindent For a given \texttt{error\_bound\_interpolation}, as shown in Figure~\ref{Fig:Case7a_Mesh}, Figure~\ref{Fig:Case7b_Mesh}, Figure~\ref{Fig:Case7c_Mesh} and Figure~\ref{Fig:Case7d_Mesh} the mesh is adapting based on the velocity field, i.e. mainly at the openings and around the exterior surfaces of the box. Indeed, decreasing the value of the \texttt{error\_bound\_interpolation} results in finer mesh in those regions. Choosing an \texttt{error\_bound\_interpolation} higher to 0.15 seems to lead to poor mesh quality (in that particular case). Even with small \texttt{error\_bound\_interpolation} and if the velocity field is well resolved, the temperature field is not properly captured due to a poor mesh quality, especially within the box.

\subsection{Adaptation based on the velocity and the temperature fields}\label{Sec:AdaptVelTemp}
In this section, the mesh adaptivity process is prescribed based on both the velocity field and the temperature field.

\subsubsection{Mesh adaptivity options}
Based on the run performed in examples \textit{3dBox\_Case6*.flml} and \textit{3dBox\_Case7*.flml}, the \texttt{error\_bound\_interpolation} values for the temperature and the velocity are chosen to be both equal to 0.15 in example \textit{3dBox\_Case8.flml}. Note that using the same value is a coincidence, and these values can be of course different. These values were chosen to capture properly both the temperature and the velocity fields, while keeping an acceptable computational time for the purpose of this manual. It is therefore recommended to use 0.1 for both field to fully capture the dynamics.

\noindent This example can be run using the command:
\begin{Terminal}[]
ä\colorbox{davysgrey}{
\parbox{435pt}{
\color{applegreen} \textbf{user@mypc}\color{white}\textbf{:}\color{codeblue}$\sim$
\color{white}\$ <<FluiditySourcePath>>/bin/fluidity -l -v3 3dBox\_Case8.flml \&
}}
\end{Terminal}

\subsubsection{Results and discussion}
\begin{figure}
    \centering
    \begin{subfigure}{0.4\textwidth}
        \includegraphics[width=\textwidth]{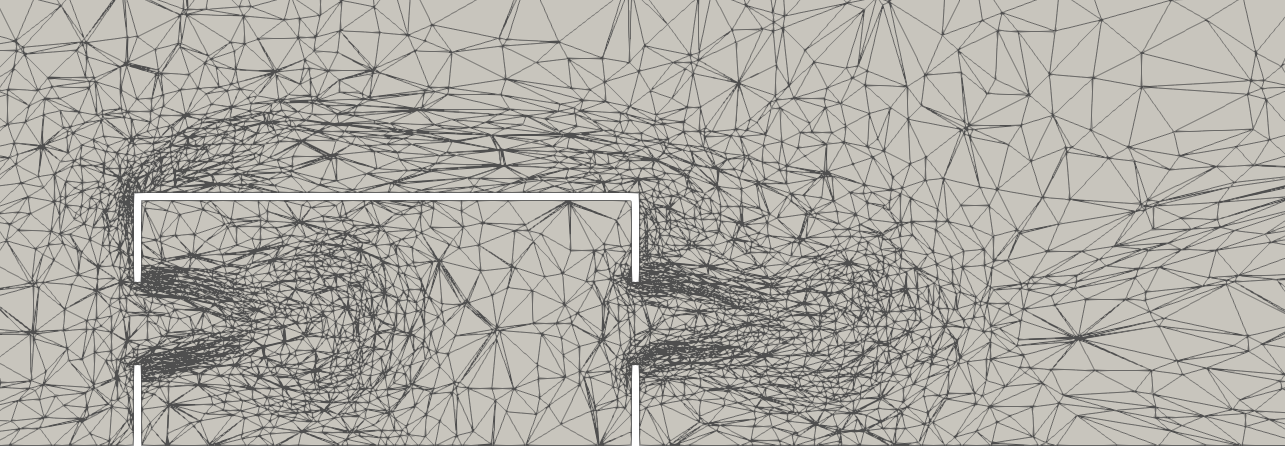}
        \caption{2 s}
        \label{Fig:Case8_Mesh_2sec}
    \end{subfigure}
    \begin{subfigure}{0.4\textwidth}
        \includegraphics[width=\textwidth]{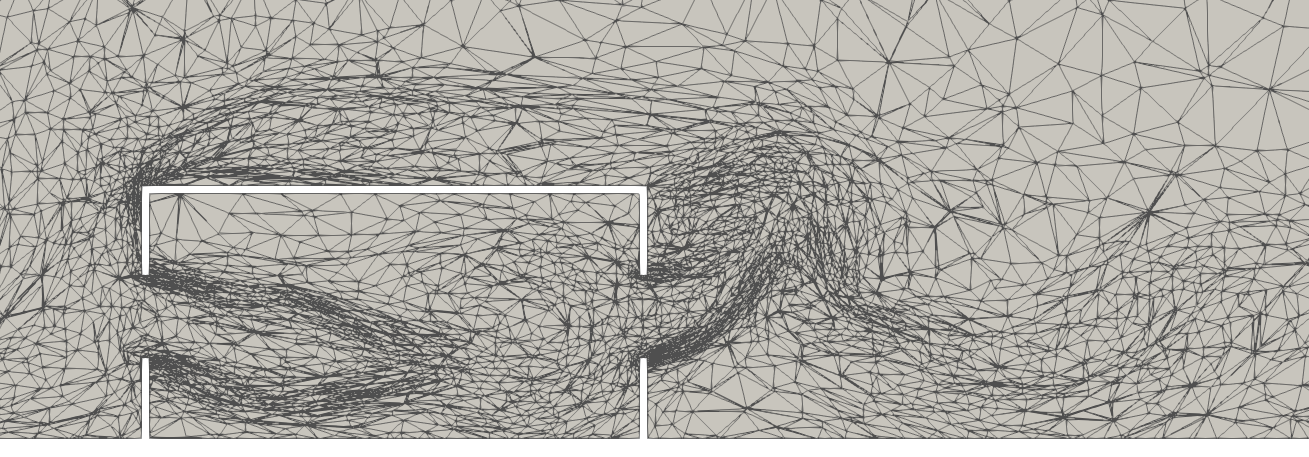}
        \caption{6 s}
        \label{Fig:Case8_Mesh_6sec}
    \end{subfigure}
    \begin{subfigure}{0.4\textwidth}
        \includegraphics[width=\textwidth]{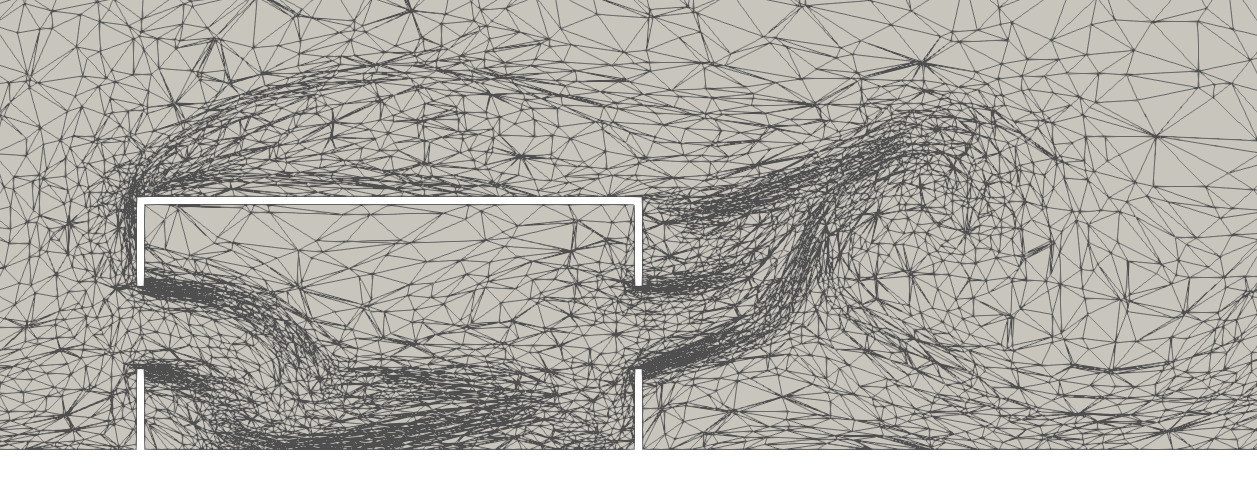}
        \caption{9 s}
        \label{Fig:Case8_Mesh_9sec}
    \end{subfigure}
    \begin{subfigure}{0.4\textwidth}
        \includegraphics[width=\textwidth]{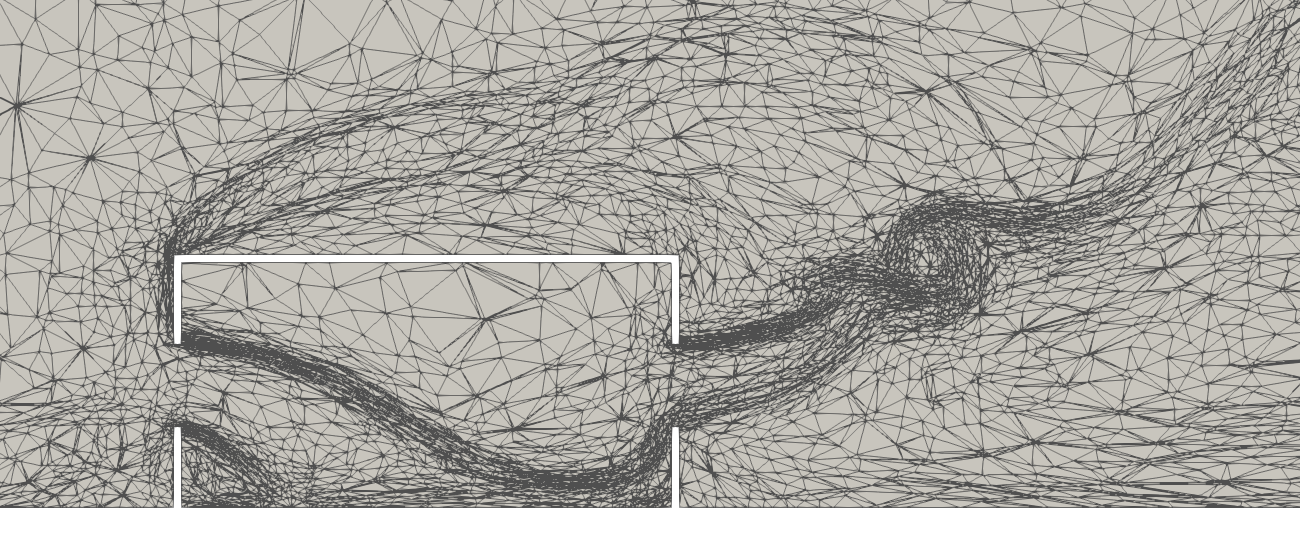}
        \caption{40 s}
        \label{Fig:Case8_Mesh_40sec}
    \end{subfigure}
    \caption{Meshes at different instant for the simulation \textit{3dBox\_Case8.flml}. Temperature and velocity fields are both adapted with \texttt{error\_bound\_interpolation} values equal to 0.15.}
    \label{Fig:Case8_Mesh}
\end{figure}

\begin{figure}
    \centering
    \begin{subfigure}{0.35\textwidth}
        \includegraphics[width=\textwidth]{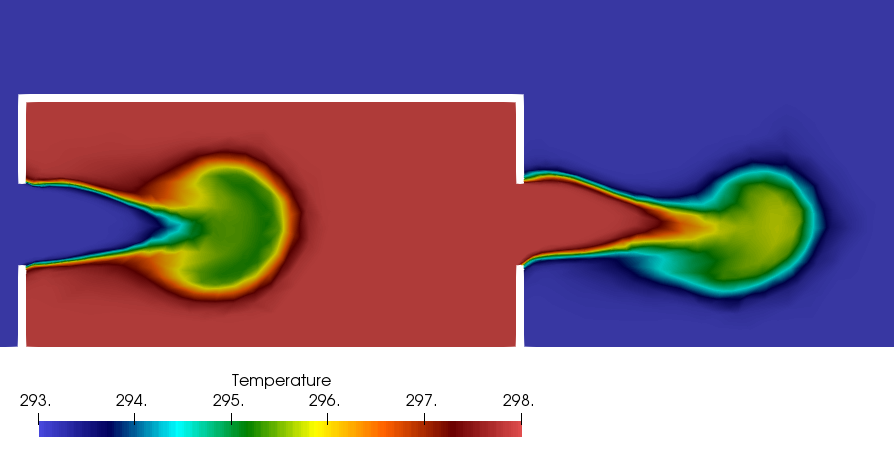}
        \caption{2 s}
        \label{Fig:Case8_Temp_2sec}
    \end{subfigure}
    \begin{subfigure}{0.42\textwidth}
        \includegraphics[width=\textwidth]{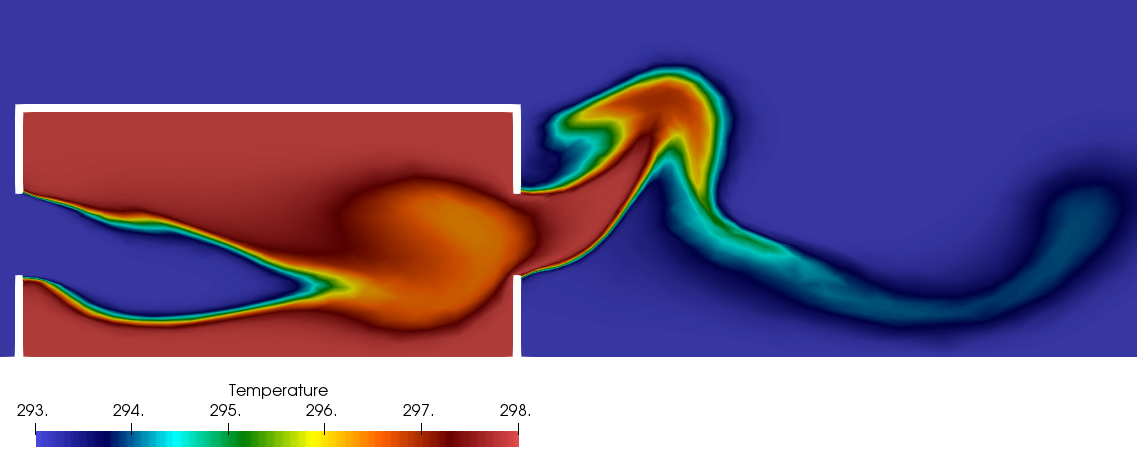}
        \caption{6 s}
        \label{Fig:Case8_Temp_6sec}
    \end{subfigure}
    \begin{subfigure}{0.4\textwidth}
        \includegraphics[width=\textwidth]{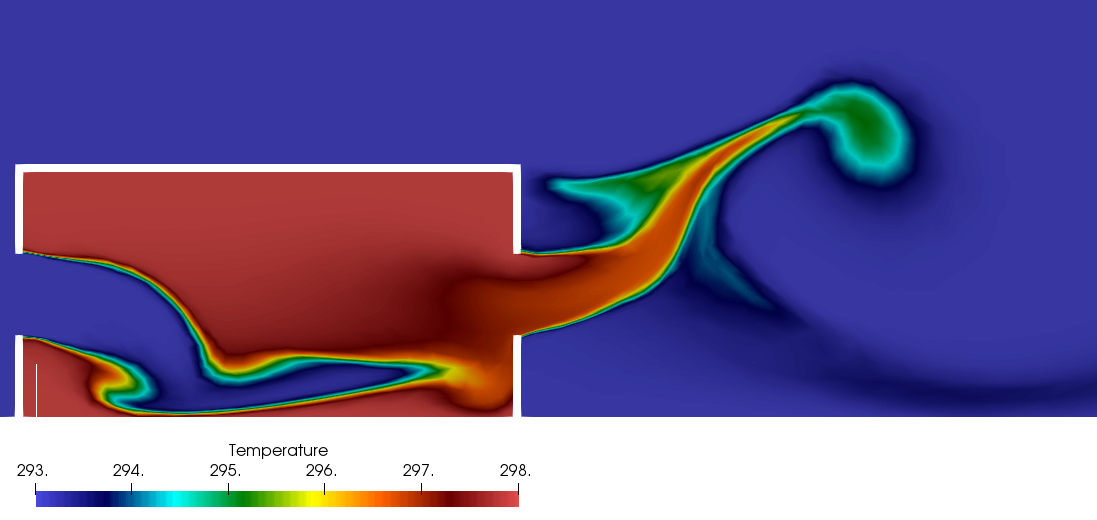}
        \caption{9 s}
        \label{Fig:Case8_Temp_9sec}
    \end{subfigure}
    \begin{subfigure}{0.4\textwidth}
        \includegraphics[width=\textwidth]{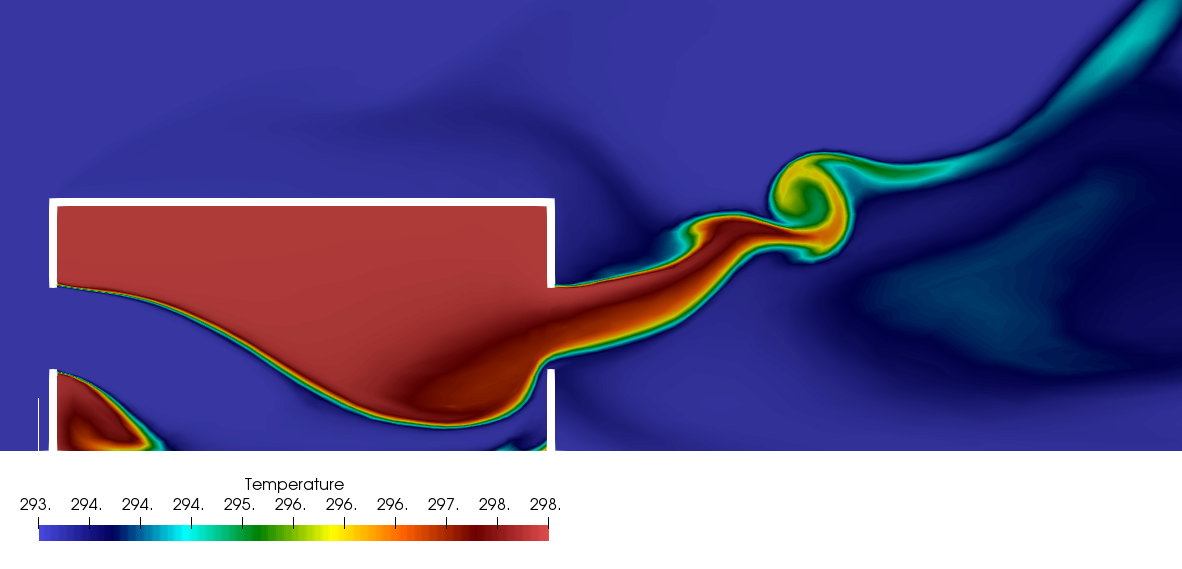}
        \caption{40 s}
        \label{Fig:Case8_Temp_40sec}
    \end{subfigure}
    \caption{Temperature field at different instant for example \textit{3dBox\_Case8.flml}. Temperature and velocity fields are both adapted with \texttt{error\_bound\_interpolation} values equal to 0.15.}
    \label{Fig:Case8_Temp}
\end{figure}

\begin{figure}
    \centering
    \begin{subfigure}{0.38\textwidth}
        \includegraphics[width=\textwidth]{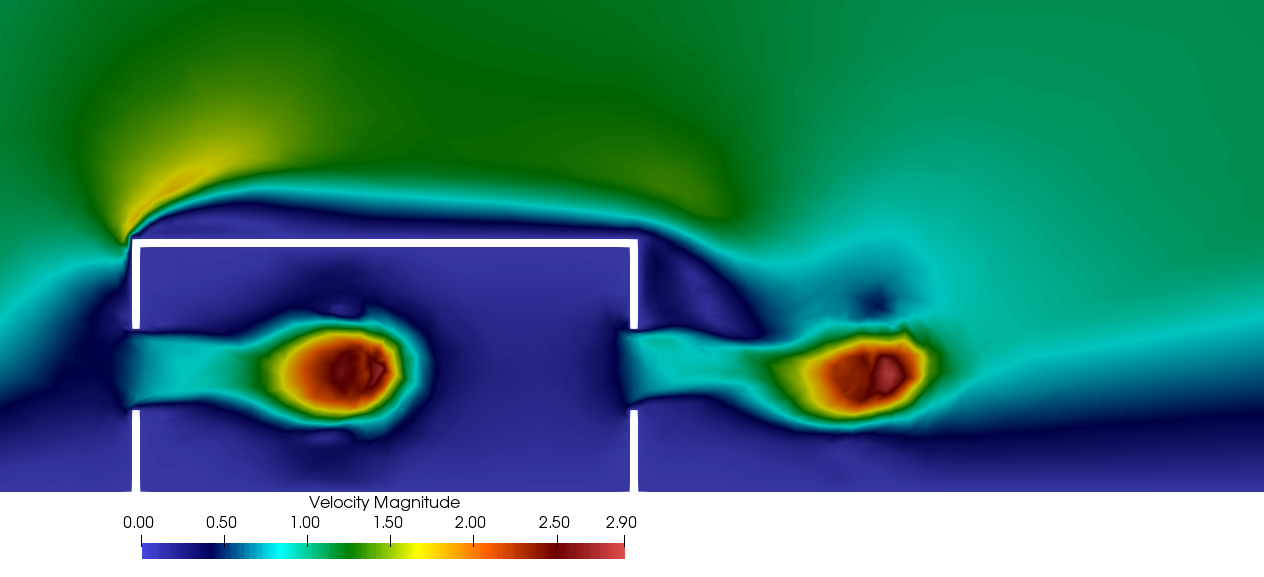}
        \caption{2 s}
        \label{Fig:Case8_Vel_2sec}
    \end{subfigure}
    \begin{subfigure}{0.4\textwidth}
        \includegraphics[width=\textwidth]{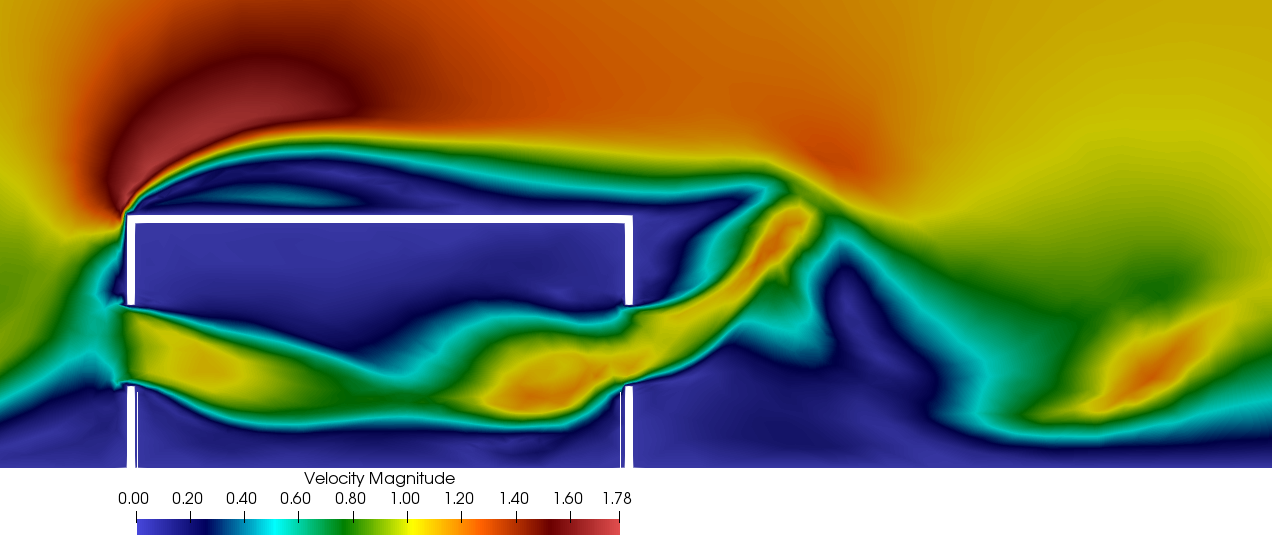}
        \caption{6 s}
        \label{Fig:Case8_Vel_6sec}
    \end{subfigure}
    \begin{subfigure}{0.4\textwidth}
        \includegraphics[width=\textwidth]{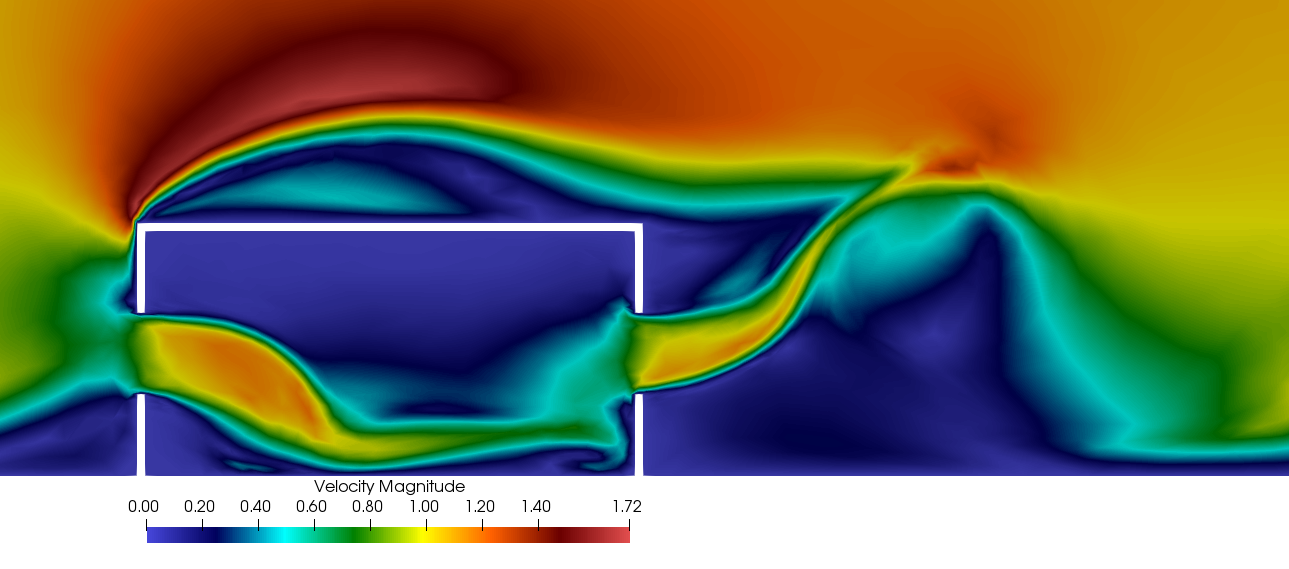}
        \caption{9 s}
        \label{Fig:Case8_Vel_9sec}
    \end{subfigure}
    \begin{subfigure}{0.4\textwidth}
        \includegraphics[width=\textwidth]{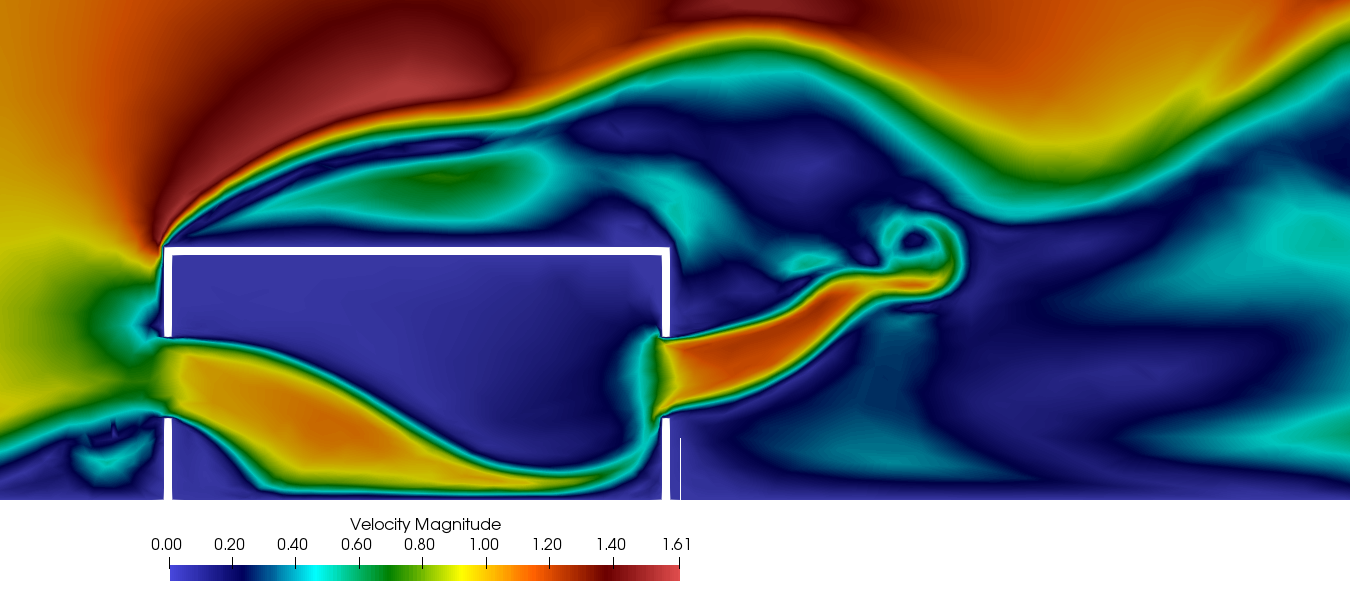}
        \caption{40 s}
        \label{Fig:Case8_Vel_40sec}
    \end{subfigure}
    \caption{Velocity field at different instant for example \textit{3dBox\_Case8.flml}. Temperature and velocity fields are both adapted with \texttt{error\_bound\_interpolation} values equal to 0.15.}
    \label{Fig:Case8_Vel}
\end{figure}

Snapshots of the meshes are shown in Figure~\ref{Fig:Case8_Mesh}. Snapshots of the temperature field are shown in Figure~\ref{Fig:Case8_Temp}. Snapshots of the velocity field are shown in Figure~\ref{Fig:Case8_Vel}. Go to Chapter~\ref{Sec:PostProcessing} to learn how to visualise the results using \textbf{ParaView}.

\noindent As shown in Figure~\ref{Fig:Case8_Mesh} the mesh is well-adapted based on the velocity field, i.e. at the openings and around the exterior surfaces of the box and based on the temperature field, i.e. within the box. In other words, the mesh is not only adapted in the interior or the exterior of the box but in both regions, thus capturing the full dynamics of the velocity field and the temperature field.

\subsection{Computation time}\label{Sec:ComputationTime}
The computation time for all the simulations is reported in Table~\ref{Tab:ComputationTime}, highlighting that refining the mesh, i.e. decreasing the \texttt{error\_bound\_interpolation}, drastically increases the computational time. Therefore, a trade-off has to be made between the accuracy desired and an acceptable computational time. 

\begin{table}
    \centering
    \begin{tabular}{ |p{1.1cm}||p{3cm}|p{3.0cm}|p{3cm}|p{3cm}|  }
         \hline
         \textbf{Case \newline Nbr} & \textbf{Simulation \newline  time: 2 s} & \textbf{Simulation \newline time: 6 s} & \bf{Simulation \newline time: 9 s} & \bf{Simulation \newline time: 40 s} \\
         \hline
         6a & 00H 01min 05s & 00H 01min 33s & 00H 01min 53s & 00H 02min 08s \\\hline
         6b & 00H 01min 35s & 00H 02min 59s & 00H 04min 27s & 00H 32min 29s \\\hline
         6c & 00H 25min 12s & 01H 24min 33s & 02H 11min 44s & 16H 51min 14s \\\hline
         6d & 03H 43min 19s & 10H 28min 17s & 19H 19min 28s & Too long... \\\hline
         7a & 00H 02min 39s & 00H 03min 38s & 00H 04min 12s & 00H 09min 56s \\\hline
         7b & 00H 07min 37s & 00H 11min 42s & 00H 16min 45s & 01H 21min 22s \\\hline
         7c & 00H 25min 37s & 00H 58min 26s & 01H 34min 29s & 08H 25min 53s \\\hline
         7d & 01H 04min 17s & 02H 35min 45s & 03H 56min 29s & 31H 04min 48s \\\hline
         8  & 00H 32min 27s & 01H 10min 05s & 01H 50min 02s & 11H 08min 00s \\\hline
    \end{tabular}
    \caption{\label{Tab:ComputationTime} Computation time for the simulations with mesh adaptivity at different instants.}
\end{table}

\section{Ensuring enough resolution in specific region}\label{Sec:AdaptPython}
\subsection{Python script to refine zone}
\subsubsection{Boundary conditions}
For the cases \textit{3dBox\_Case9a.flml} and \textit{3dBox\_Case9b.flml}, the initial and the ambient temperatures are set to 293 K and $u$ is set to $0$ m/s. The heat flux applied at the source $\phi_{source}$ is equal to 1000 W/m\textsuperscript{2}, hence $\phi_{fluidity}$, set up in \textbf{Diamond}, is equal to 0.8163 according to equation~\ref{Eq:FluxFluidity}. The surface of the heat source is $0.2 \times 0.2$ m\textsuperscript{2}: the flux applied at the source is then equal to 40 W. This example is similar to \textit{3dBox\_Case2b.flml} which is set up without mesh adaptivity, i.e on a fixed mesh.

\subsubsection{CFL number}
To avoid any crash of the simulations, the time step is now also adaptive using a CFL condition as described in Section~\ref{Sec:CFLNumber}, with the CFL number taken as 2 in the following example. This value can be increased later on to speed up the simulations. However, it is recommended to do a sensibility analysis of the results as a function of the CFL number to ensure that no important information is missed.
Moreover, a maximum time step is prescribed equal to 1 s: at the beginning of the simulation, the velocity within the domain is almost equal to 0 m/s. As a consequence, based on the CFL number, the computed time step will be unreasonably very large (see equation~\ref{Eq:CFLNumber}) causing the simulation to crash.

\subsubsection{General mesh adaptivity options}
In the following sections, mesh adaptivity will be performed based on the temperature and the velocity fields. The maximum number of nodes is set to $200,000$ and the mesh is adapted every 10 time steps. 

\noindent Based on the results of \textit{3dBox\_Case2b.flml}, the velocity is estimated to be approximately between 0 m/s and 0.45 m/s. The velocity \texttt{error\_bound\_interpolation} is taken to be equal to 10\% of the velocity range, i.e equal to 0.045.
Based on the results of \textit{3dBox\_Case2b.flml}, the temperature is estimated to be approximately between 293 K and 333 K. However, the highest temperatures are located near the source only: the rest of the inner box is much cooler, i.e. between 293 K and approximately 300 K. The choice is then made not to take into account the entire temperature range, but only the one of interest, i.e. between 293 K and 300 K. The temperature \texttt{error\_bound\_interpolation} is taken to be equal to 10\% of the interesting temperature range, i.e of 7 K. Therefore, the \texttt{error\_bound\_interpolation} is set to 0.7.

\subsubsection{Minimum and maximum edge lengths}
The characteristic length $L$ of the domain is taken to be the windows opening, i.e. $1 m$: the minimum edge length is set up to $L/100$, i.e. 0.01 m. The height $H$ of the domain is 21 m: the maximum edge length is set up to $H/10$, i.e 2.1 m.

\begin{figure}
    \centering
    \begin{subfigure}{0.45\textwidth}
        \includegraphics[width=\textwidth]{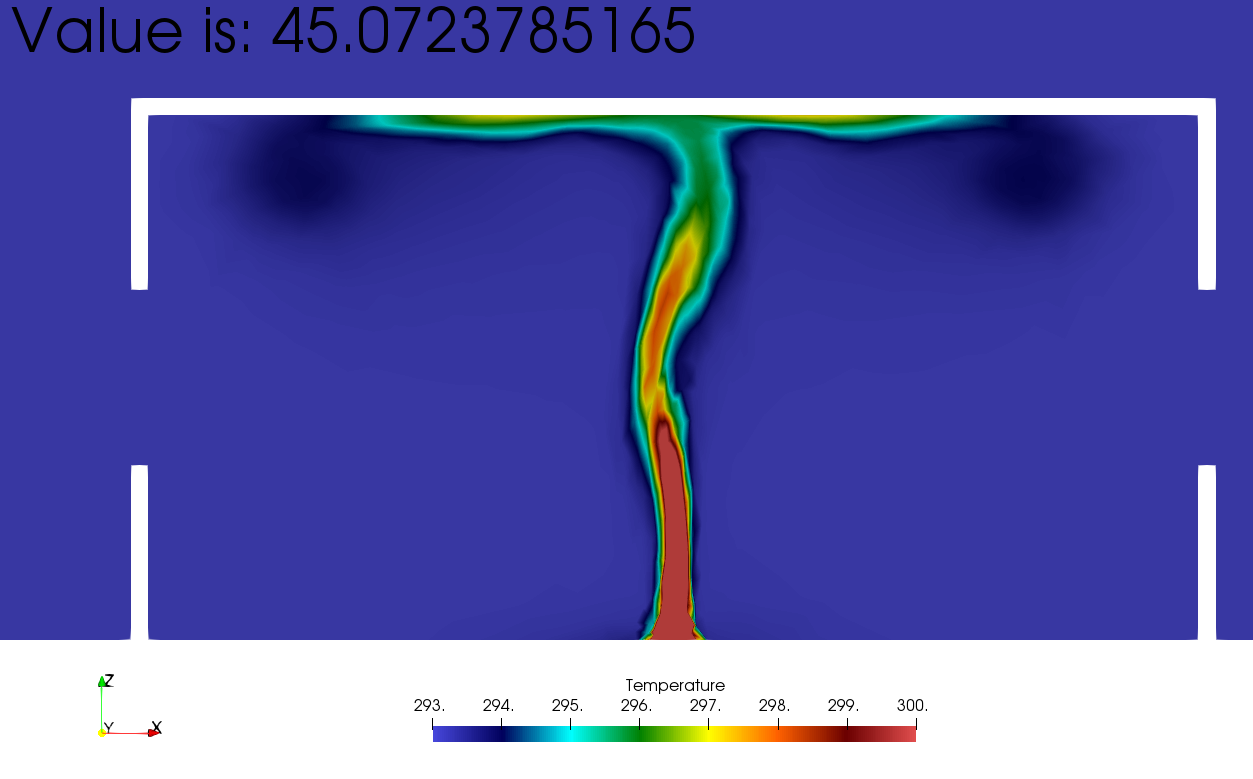}
        \caption{}
        \label{Fig:Case9a_Temp}
    \end{subfigure}
    \begin{subfigure}{0.47\textwidth}
        \includegraphics[width=\textwidth]{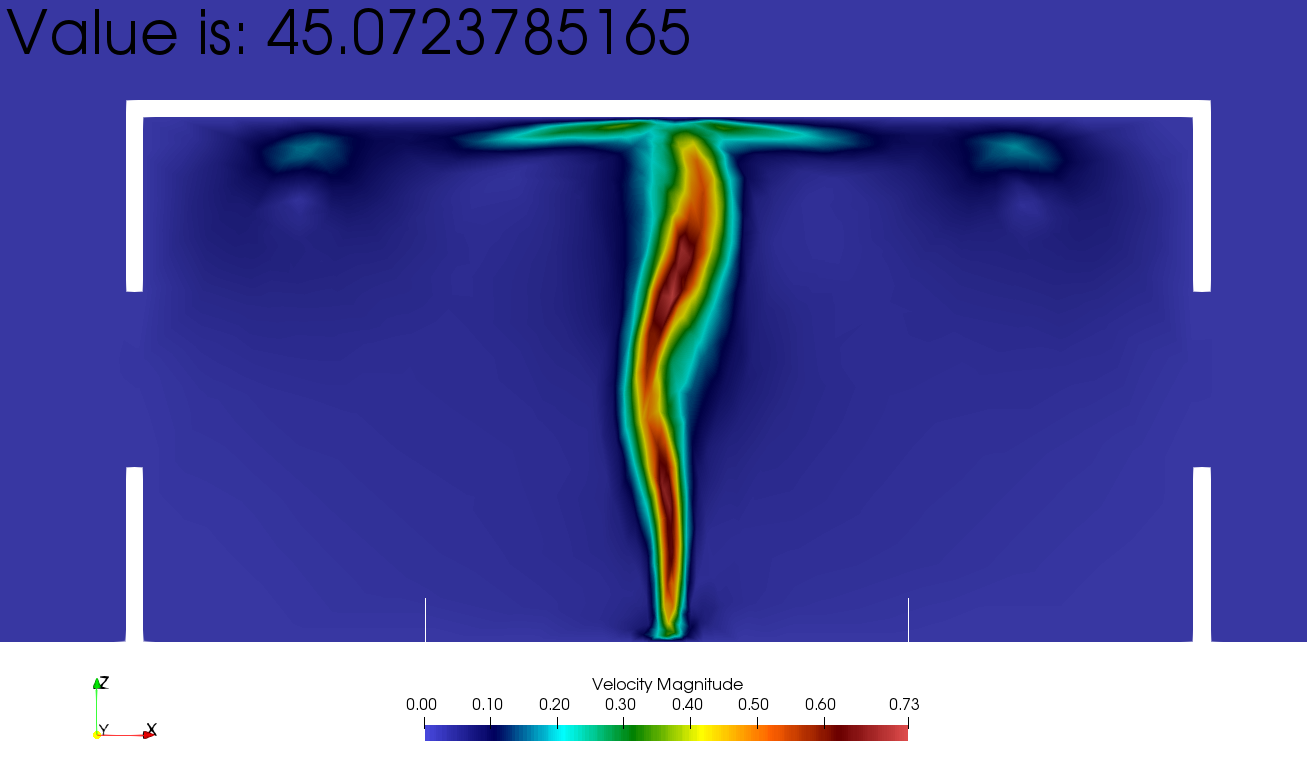}
        \caption{}
        \label{Fig:Case9a_Vel}
    \end{subfigure}
    \caption{(a) Temperature and (b) Velocity fields at 45 s for the simulation \textit{3dBox\_Case9a.flml}. The temperature and the velocity fields are both adapted with \texttt{error\_bound\_interpolation} values equal to 0.7 and 0.045, respectively.}
    \label{Fig:Case9a_Results}
\end{figure}

\begin{figure}
    \centering
    \begin{subfigure}{0.22\textwidth}
        \includegraphics[width=\textwidth]{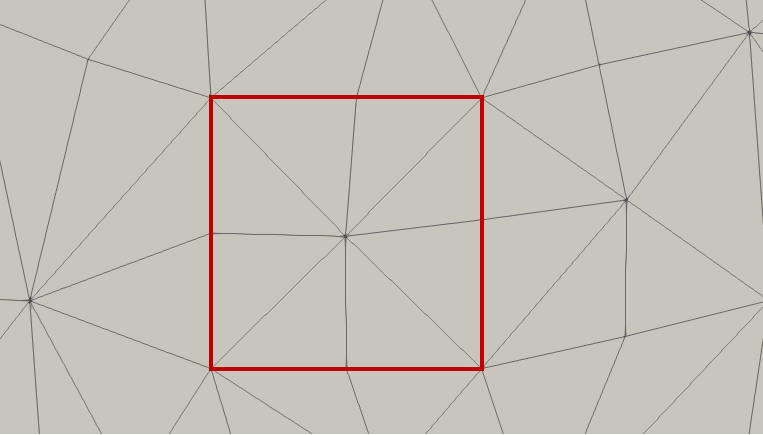}
        \caption{20 sec}
        \label{Fig:Case9a_SourceMesh_20s}
    \end{subfigure}
    \begin{subfigure}{0.2\textwidth}
        \includegraphics[width=\textwidth]{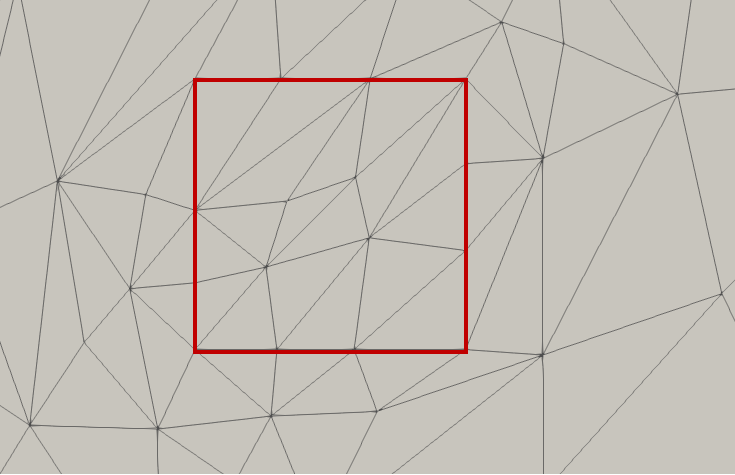}
        \caption{45 sec}
        \label{Fig:Case9a_SourceMesh_45s}
    \end{subfigure}
    \begin{subfigure}{0.2\textwidth}
        \includegraphics[width=\textwidth]{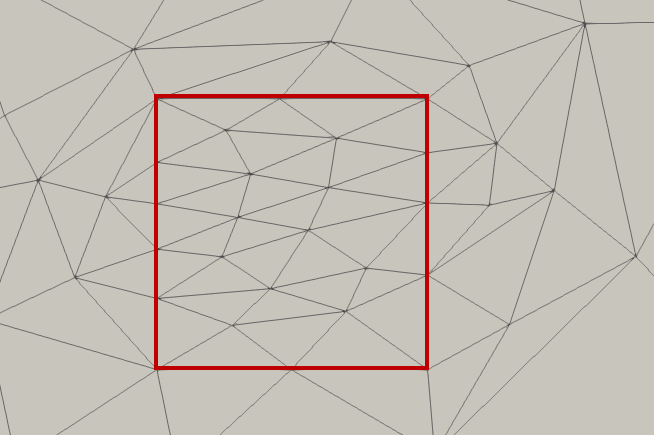}
        \caption{70 sec}
        \label{Fig:Case9a_SourceMesh_70s}
    \end{subfigure}
    \begin{subfigure}{0.2\textwidth}
        \includegraphics[width=\textwidth]{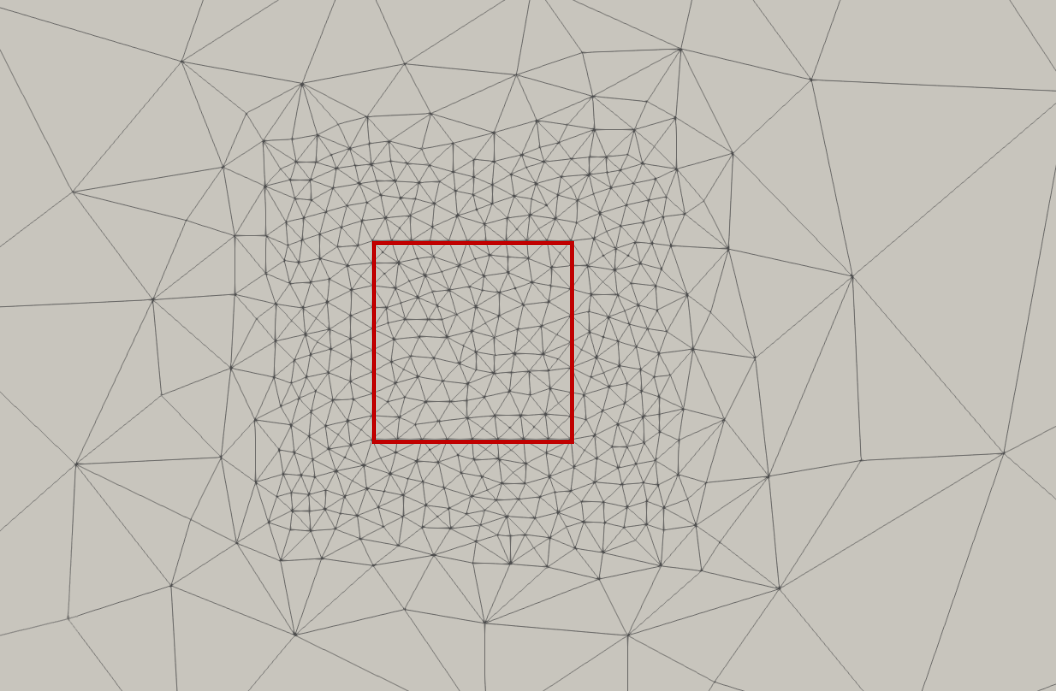}
        \caption{}
        \label{Fig:Case9b_SourceMesh}
    \end{subfigure}
    \caption{Meshes at the source location (a), (b) and (c) at different instant and for the simulation \textit{3dBox\_Case9a.flml} when no constraint is imposed on the mesh at the source and (d) meshes at the source location for the simulation \textit{3dBox\_Case9b.flml} when refinement is enforced around the source. The red squares depict the location of the source.}
    \label{Fig:Case9a_Meshes}
\end{figure}

\noindent As shown in Figure~\ref{Fig:Case9a_Results}, these \texttt{error\_bound\_interpolation} values gives acceptable results as a first run. However, zooming on the source,  Figure~\ref{Fig:Case9a_Meshes} shows that the region near the source is not well resolved. It is commonly assumed that at least 10-20 elements should define a source, which is clearly not the case in \textit{3dBox\_Case9a.flml}.

\noindent Reducing the \texttt{error\_bound\_interpolation} value for the temperature will have a tendency to overly increase the resolution near the source at the expense of the rest of the plume. One shortcut, to ensure a well resolved source, is to use a python script to vary the specified minimum and the maximum edge lengths over the domain, constraining finer resolutions over areas of interest, in this case the source. It is worth noting that this technique can also be used to ensure refinement near openings for example. In example \textit{3dBox\_Case9b.flml}, the python scripts in Code~\ref{Lst:PythonMinEdge} and Code~\ref{Lst:PythonMaxEdge} are used to set up the minimum and the maximum edge lengths. The refined region of the mesh near the source is twice as large as the source itself and at least 20 elements are forced in that region. The mesh obtained near the source is shown in Figure~\ref{Fig:Case9b_SourceMesh}.

\noindent These two python scripts can be used for a square/cubic source. For a circular/spherical/ellipsoidal source, a python script similar to the one in Code~\ref{Lst:PythonEdgeEllipsoid} can be used.

\begin{Code}[language=python, caption={Python script to prescribed different minimum edge length in different region used in \textit{3dBox\_Case9b.flml}.}, label={Lst:PythonMinEdge}]
def val(X, t):
  # Function code
  # Geometry variables
  x_source = 9.0 # x-position of the center of the source
  y_source = 9.0 # y-position of the center of the source
  z_source = 0.0 # z-position of the center of the source
  
  lx = 0.4 # x dimension of the refinement region (twice the source)
  ly = 0.4 # y dimension of the refinement region (twice the source)
  lz = 0.2 # z dimension of the refinement region
  
  xmin = x_source - (lx/2.)
  xmax = x_source + (lx/2.)
  ymin = y_source - (ly/2.)
  ymax = y_source + (ly/2.)
  zmin = z_source - (lz/2.)
  zmax = z_source + (lz/2.)
  
  # Values of the edge length
  s_domain = 0.01   # L/100, L: windows opening
  s_source = 0.002  # lx/100
  
  val = [[s_domain,0.0,0.0],[0.0,s_domain,0.0],[0.0,0.0,s_domain]]
  
  if (X[2] >= zmin) and (X[2] <= zmax):
    if (X[1] >= ymin) and (X[1] <= ymax):
      if (X[0] >= xmin) and (X[0] <= xmax):
        val = [[s_source,0.0,0.0],[0.0,s_source,0.0],[0.0,0.0,s_source]]

  return val #Return value
\end{Code}

\begin{Code}[language=python, caption={Python script to prescribed different maximum edge length in different region used in \textit{3dBox\_Case9b.flml}.}, label={Lst:PythonMaxEdge}]
def val(X, t):
  # Function code
  # Geometry variables
  x_source = 9.0 # x-position of the center of the source
  y_source = 9.0 # y-position of the center of the source
  z_source = 0.0 # z-position of the center of the source
  
  lx = 0.4 # x dimension of the refinement region (twice the source)
  ly = 0.4 # y dimension of the refinement region (twice the source)
  lz = 0.2 # z dimension of the refinement region
  
  xmin = x_source - (lx/2.)
  xmax = x_source + (lx/2.)
  ymin = y_source - (ly/2.)
  ymax = y_source + (ly/2.)
  zmin = z_source - (lz/2.)
  zmax = z_source + (lz/2.)
  
  # Values of the edge length
  b_domain = 2.1   # H/10, H: domain height
  b_source = 0.02  # lx/20
  
  val = [[b_domain,0.0,0.0],[0.0,b_domain,0.0],[0.0,0.0,b_domain]]
  
  if (X[2] >= zmin) and (X[2] <= zmax):
    if (X[1] >= ymin) and (X[1] <= ymax):
      if (X[0] >= xmin) and (X[0] <= xmax):
        val = [[b_source,0.0,0.0],[0.0,b_source,0.0],[0.0,0.0,b_source]]

  return val #Return value
\end{Code}

\begin{Code}[language=python, caption={Python script to prescribed different edge length in different region in the case of an ellipsoidal source.}, label={Lst:PythonEdgeEllipsoid}]
def val(X, t):
  # Function code
  # Geometry variables
  x_source = 9.0 # x-position of the center of the source
  y_source = 9.0 # y-position of the center of the source
  z_source = 0.0 # z-position of the center of the source
  
  dx = 0.4 # x radius of the refinement region (twice the source)
  dy = 0.4 # y radius of the refinement region (twice the source)
  dz = 0.2 # z radius of the refinement region

  # Values of the edge length
  b_domain = 2.1   # H/10, H: domain height
  b_source = 0.02  # lx/20
  
  val = [[b_domain,0.0,0.0],[0.0,b_domain,0.0],[0.0,0.0,b_domain]]
  
  if(((X[0]-x_source)**2/dx**2)+((X[1]-y_source)**2/dy**2)+((X[2]-z_source)**2/dz**2)<=1.0): 
        val = [[b_source,0.0,0.0],[0.0,b_source,0.0],[0.0,0.0,b_source]]

  return val #Return value
\end{Code}

\noindent These examples can be run using the commands:
\begin{Terminal}[]
ä\colorbox{davysgrey}{
\parbox{435pt}{
\color{applegreen} \textbf{user@mypc}\color{white}\textbf{:}\color{codeblue}$\sim$
\color{white}\$ <<FluiditySourcePath>>/bin/fluidity -l -v3 3dBox\_Case9a.flml \&
\newline
\color{applegreen} \textbf{user@mypc}\color{white}\textbf{:}\color{codeblue}$\sim$
\color{white}\$ <<FluiditySourcePath>>/bin/fluidity -l -v3 3dBox\_Case9b.flml \&
}}
\end{Terminal}

\noindent Go to Chapter~\ref{Sec:PostProcessing} to learn how to visualise the results using \textbf{ParaView}.

\subsection{Locking nodes}
\begin{figure}
    \centering
    \begin{subfigure}{0.4\textwidth}
        \includegraphics[width=\textwidth]{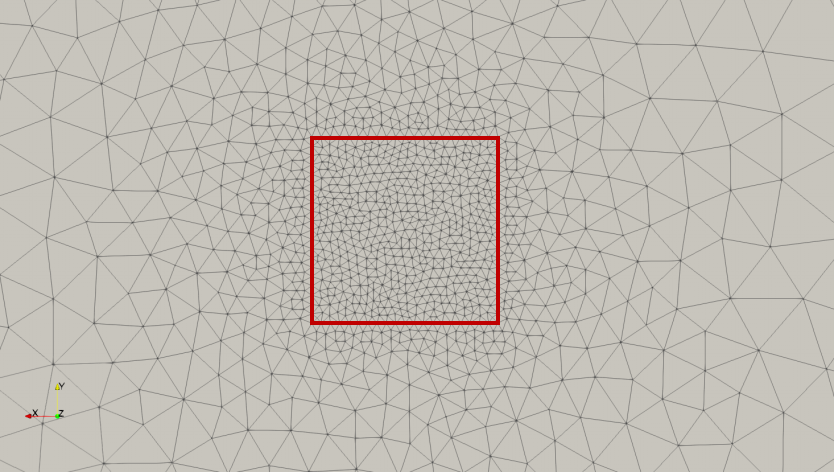}
        \caption{Initial mesh from \textit{Box.msh}}
        \label{Fig:Case9c_MeshInit}
    \end{subfigure}
    \begin{subfigure}{0.4\textwidth}
        \includegraphics[width=\textwidth]{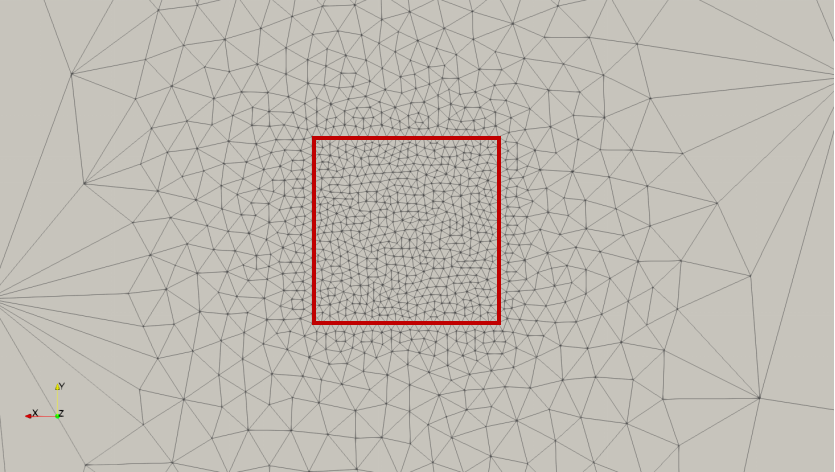}
        \caption{Mesh after the 1\textsuperscript{st} adaptation}
        \label{Fig:Case9c_MeshFirstAdapt}
    \end{subfigure}
    \caption{Meshes at the source location fro example \textit{3dBox\_Case9c.flml}. (a) Initial mesh generated by \textbf{GMSH} from file \textit{Box.geo}. (b) Mesh obtained after the first adaptation using mesh adaptivity in \textbf{Fluidity}. The red squares depict the location of the source.}
    \label{Fig:Case9c_Meshes}
\end{figure}

\begin{figure}
    \begin{center}
        \includegraphics[scale=0.2]{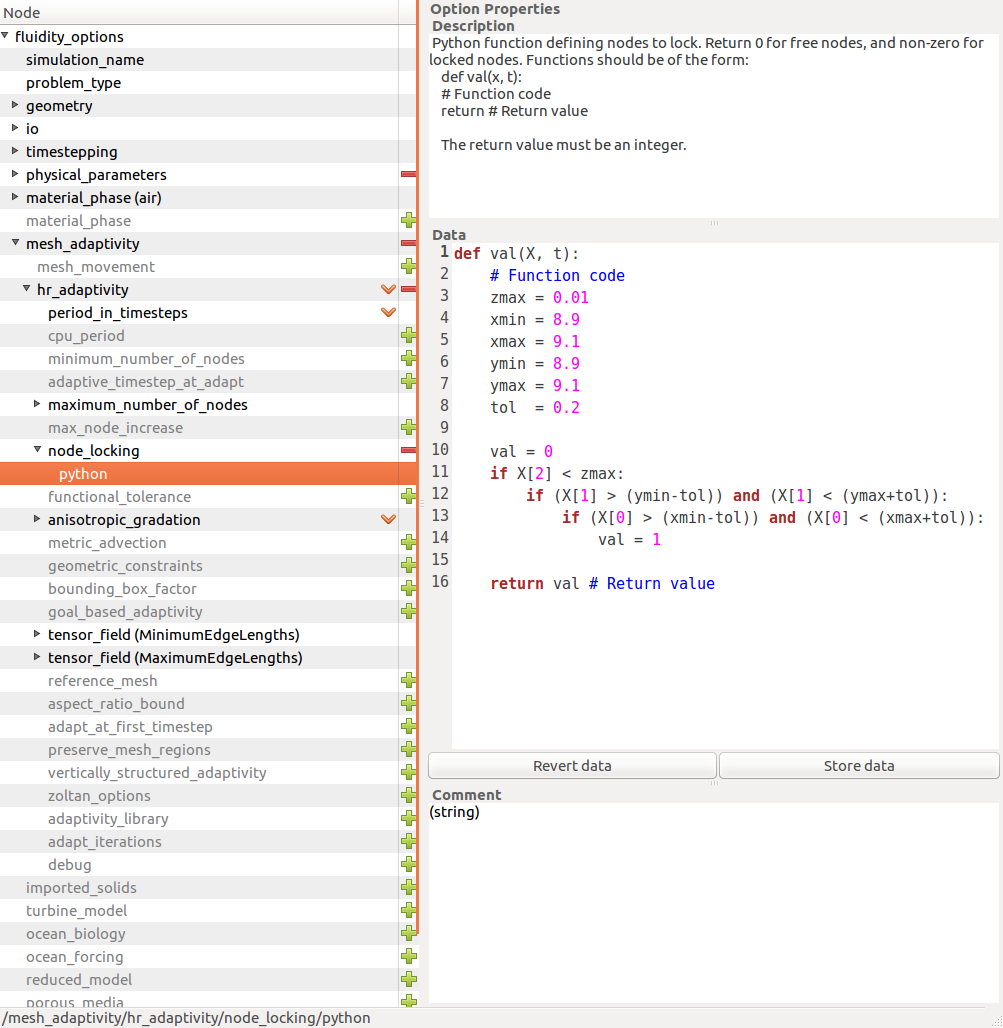}
        \caption{Options to lock nodes and prevent mesh adaptation in specific region in \textbf{Diamond}.}
        \label{Fig:NodeLock}
    \end{center}
\end{figure}

Set-up of example \textit{3dBox\_Case9c.flml} is the same than example \textit{3dBox\_Case9a.flml}. In example \textit{3dBox\_Case9c.flml}, the \textit{Box.geo} and by consequence \textit{Box.msh} files were modified to initially assign an higher resolution near the source as shown in Figure~\ref{Fig:Case9c_MeshInit}. In the mesh adaptivity option, the option \texttt{node\_locking} (Figure~\ref{Fig:NodeLock}) is turned on to specify that nodes around the source are locked using Code~\ref{Lst:PythonNodeLock}. This option prevent the mesh to be adapted where specified and thus allow to ensure that the desire resolution in specific regions are kept. Once the simulation is running, it can be observed that the mesh is preserved at the source location after the adaptation process (Figure~\ref{Fig:Case9c_MeshFirstAdapt}). However, this option can sometimes lead to weird meshes.

\noindent This example can be run using the command:
\begin{Terminal}[]
ä\colorbox{davysgrey}{
\parbox{435pt}{
\color{applegreen} \textbf{user@mypc}\color{white}\textbf{:}\color{codeblue}$\sim$
\color{white}\$ <<FluiditySourcePath>>/bin/fluidity -l -v3 3dBox\_Case9c.flml \&
}}
\end{Terminal}

\noindent Go to Chapter~\ref{Sec:PostProcessing} to learn how to visualise the results using \textbf{ParaView}.

\begin{Code}[language=python, caption={Python script to lock nodes and prevent mesh adaptation near the source.}, label={Lst:PythonNodeLock}]
def val(X, t):
    # Function code
    zmax = 0.01
    xmin = 8.9
    xmax = 9.1 
    ymin = 8.9
    ymax = 9.1
    tol  = 0.2
    
    val = 0
    if X[2] < zmax:
        if (X[1] > (ymin-tol)) and (X[1] < (ymax+tol)):
            if (X[0] > (xmin-tol)) and (X[0] < (xmax+tol)):
                val = 1
    
    return val # Return value
\end{Code}

\section{Advection of the mesh}\label{Sec:AdaptAdvec}
Metric advection is a technique that uses the current flow velocity to advect the metric forward in
time over the period until the next mesh adapt. With metric advection, mesh resolution is pushed ahead of the flow such that, between mesh adapts, the dynamics of interest are less likely to propagate out of the region of higher resolution. This leads to a larger area that requires refinement and, therefore, an increase in the number of nodes. However, metric advection can allow the frequency of adapt to be reduced whilst maintaining a good representation of the dynamics.

\begin{figure}
    \begin{center}
        \includegraphics[scale=0.2]{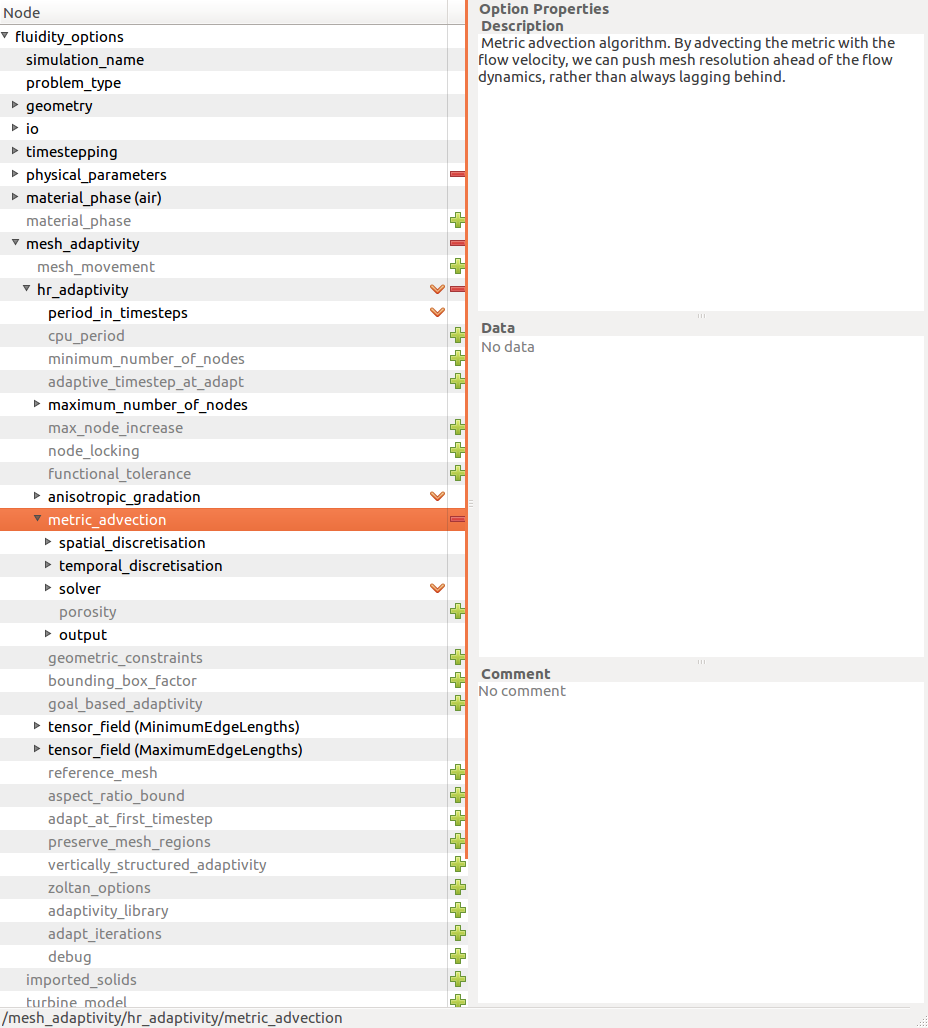}
        \caption{Options to advect the mesh in \textbf{Diamond}.}
        \label{Fig:AdaptAdvect}
    \end{center}
\end{figure}

\noindent The options for metric advection can be found under the \texttt{mesh\_adaptivity} options as shown in Figure~\ref{Fig:AdaptAdvect}. The advection equation is discretised with a control volume method. For spatial discretisation, a first order upwind scheme and non-conservative (\texttt{conservative\_advection}=0) form are generally recommended. For temporal discretisation a semi-implicit discretisation in time is recommended, i.e \texttt{theta}=0.5. The time step is controlled by the choice of CFL number under the option \texttt{temporal\_} \texttt{discretisation/maximum\_courant\_number\_per\_subcycle} and the value set should be the same as the one specified under the option \texttt{timestepping} \texttt{/adaptive\_timestep/} \texttt{requested\_cfl}.

\begin{figure}
    \centering
    \begin{subfigure}{0.35\textwidth}
        \includegraphics[width=\textwidth]{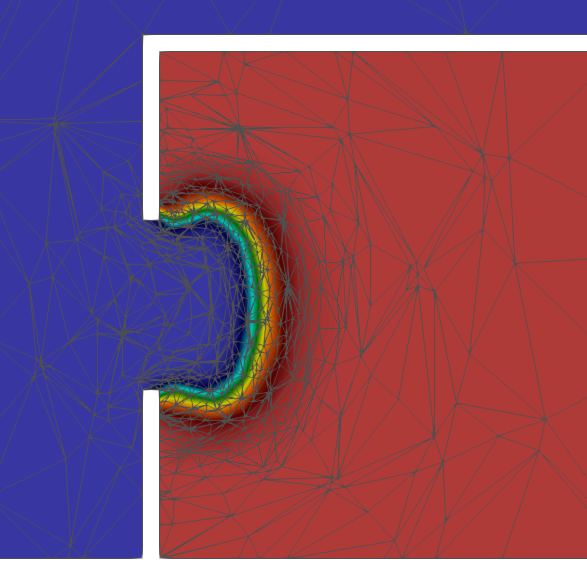}
        \caption{After the 2\textsuperscript{nd} adaptation}
        \label{Fig:Case6c_2ndAdapt}
    \end{subfigure}
    \begin{subfigure}{0.35\textwidth}
        \includegraphics[width=\textwidth]{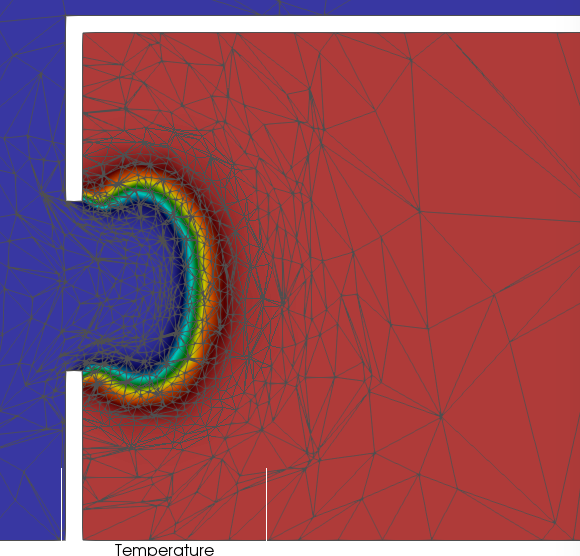}
        \caption{After the 2\textsuperscript{nd} adaptation}
        \label{Fig:Case10_2ndAdapt}
    \end{subfigure}
    \begin{subfigure}{0.32\textwidth}
        \includegraphics[width=\textwidth]{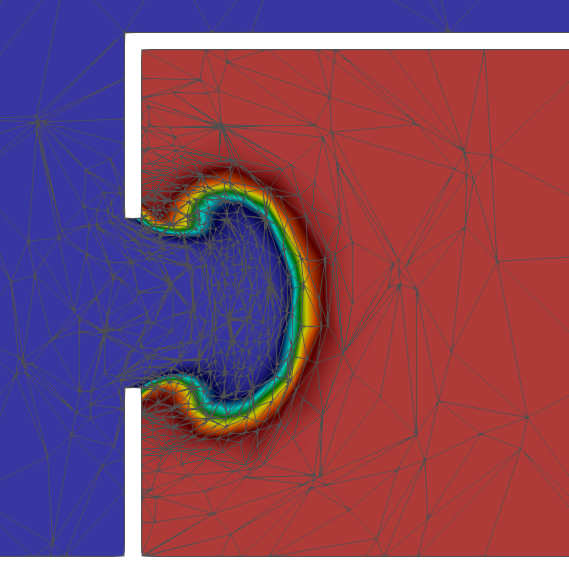}
        \caption{Before the 3\textsuperscript{rd} adaptation}
        \label{Fig:Case6c_Before3rdAdapt}
    \end{subfigure}
    \begin{subfigure}{0.35\textwidth}
        \includegraphics[width=\textwidth]{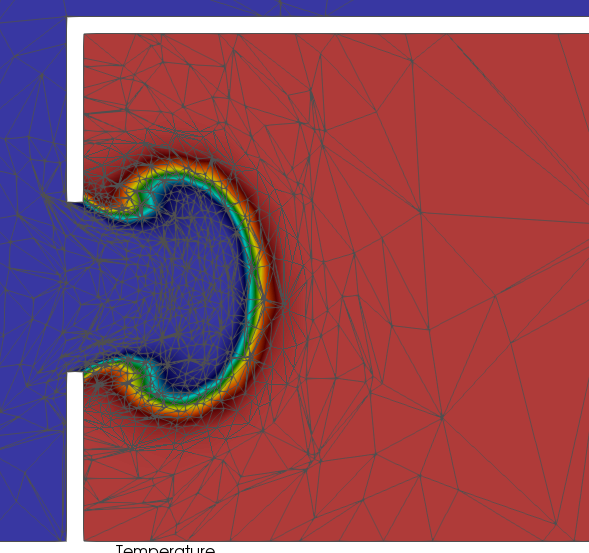}
        \caption{Before 3\textsuperscript{rd} adaptation}
        \label{Fig:Case10_Before3rdAdapt}
    \end{subfigure}
    \caption{Temperature field and meshes for (a), (c) \textit{3dBox\_Case6c.flml} when the mesh is not advected and (b), (d) \textit{3dBox\_Case10.flml} when the mesh is advected.}
    \label{Fig:Case10_Advec_Adapt}
\end{figure}

\begin{figure}
    \centering
    \begin{subfigure}{0.36\textwidth}
        \includegraphics[width=\textwidth]{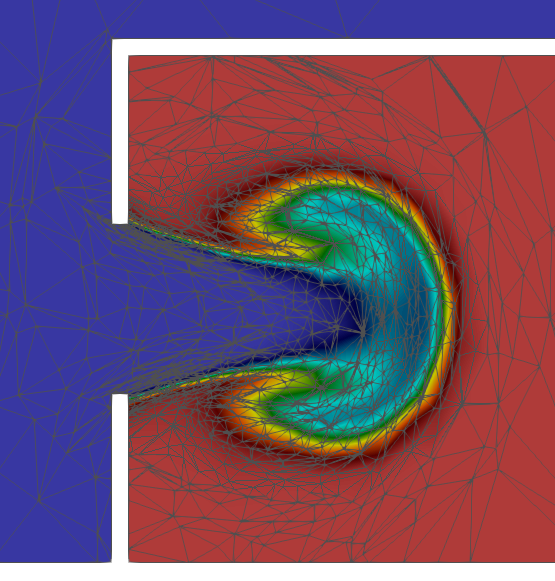}
        \caption{\textit{3dBox\_Case6c.flml}}
        \label{Fig:Case6c_1sec}
    \end{subfigure}
    \begin{subfigure}{0.39\textwidth}
        \includegraphics[width=\textwidth]{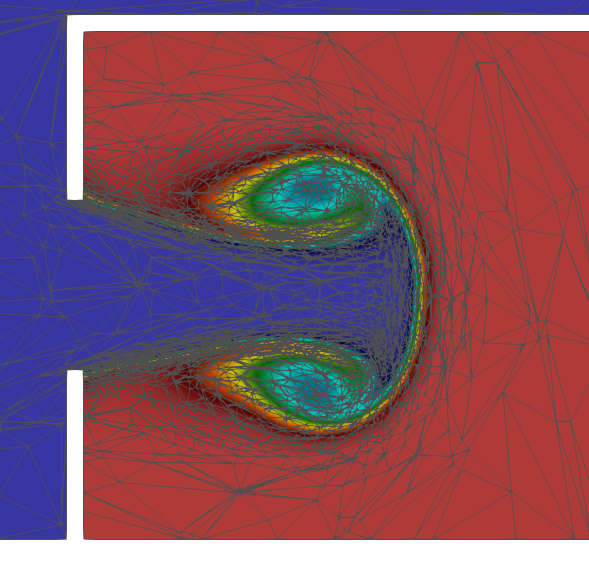}
        \caption{\textit{3dBox\_Case10.flml}}
        \label{Fig:Case10_1sec}
    \end{subfigure}
    \caption{Temperature field and meshes for (a) \textit{3dBox\_Case6c.flml} when the mesh is not advected and (b) \textit{3dBox\_Case10.flml} when the mesh is advected at $t=1 $ s.}
    \label{Fig:Case10_Advec_1sec}
\end{figure}

\noindent The case \textit{3dBox\_Case10.flml} is similar to the case \textit{3dBox\_Case6c.flml} with the mesh advection turned on. Difference between the meshes generated at different times for both cases is shown in Figure~\ref{Fig:Case10_Advec_Adapt} and Figure~\ref{Fig:Case10_Advec_1sec}. Figure~\ref{Fig:Case10_Advec_Adapt} shows the meshes and the temperature field just after the second adaptation and just before the third adaptation. The region where the mesh is refined in Figure~\ref{Fig:Case6c_2ndAdapt} is clearly much smaller than in Figure~\ref{Fig:Case10_2ndAdapt}. Therefore, just before the next adaptation, the temperature starts to propagate out of the region with higher resolution when mesh advection is not used, as shown in Figure~\ref{Fig:Case6c_Before3rdAdapt}. However, the temperature is maintained in the finer region when using the mesh advection option as shown in Figure~\ref{Fig:Case10_Before3rdAdapt}.

\noindent Figure~\ref{Fig:Case10_Advec_1sec} shows the temperature field at 1 s. As the advection of the mesh allows the field of interest to remain in the fine region, there should be less artificial diffusion due to the presence of coarse mesh elements as shown in Figure~\ref{Fig:Case10_Advec_1sec}.

\section{Common errors}
All errors are listed in the \texttt{fluidity-err.0} file, including those linked to mesh adaptivity. Some might only be warnings and if the simulation is still running they can be safely ignored.  However, if something went wrong during the adaptation process, the simulation will automatically crash. This is primarily caused by the user pushing the adaptation parameters too far... Finally, the user should be reminded that it is not recommended to adapt the pressure field.
    \chapter{Other fields}

\section{Case set-up}
Example \textit{3dBox\_Case11.flml} is based on example \textit{3dBox\_Case8.flml}, apart from the following properties that were changed to speed-up the simulation:
\begin{itemize}
    \item CFL number is now equal to 5.
    \item Mesh adaptivity options:
    \begin{itemize}
        \item Temperature: \texttt{error\_bound\_interpolation}=0.2
        \item Velocity: \texttt{error\_bound\_interpolation}=0.17
        \item The mesh is advected.
    \end{itemize}
\end{itemize}

\noindent The mesh is also refined at the openings of the box, using a python script in the maximum edge length field under the adaptivity option, to be sure that the flow is well-captured (Figure~\ref{Fig:TracerFieldRefinement}).

\noindent A number of other fields can be added in \textbf{Diamond} and some are described below. Example \textit{3dBox\_Case11.flml} contains all the interesting fields discussed in this chapter and that the user might want to use. The fields in \textit{3dBox\_Case11.flml} are the following:
\begin{itemize}
    \item Prognostic fields:
    \begin{itemize}
        \item Pressure
        \item Velocity
        \item Temperature
        \item Tracer
    \end{itemize}
    \item Diagnostic fields:
    \begin{itemize}
        \item Density
        \item VelocityAverage: time-average of the velocity
        \item u: $u'$, fluctuation term of the $u$-component of the velocity
        \item v: $v'$, fluctuation term of the $v$-component of the velocity
        \item w: $w'$, fluctuation term of the $w$-component of the velocity
        \item uu: $u'u'$, squared fluctuation term of the $u$-component of the velocity
        \item vv: $v'v'$, squared fluctuation term of the $v$-component of the velocity
        \item ww: $w'w'$, squared fluctuation term of the $w$-component of the velocity
        \item uuAverage: $\overline{u'u'}$, $u$-component of the Reynolds stresses (time-average)
        \item vvAverage: $\overline{v'v'}$, $v$-component of the Reynolds stresses (time-average)
        \item wwAverage: $\overline{w'w'}$, $w$-component of the Reynolds stresses (time-average)
        \item PressureAverage: time-average of the pressure
        \item TemperatureAverage: time-average of the temperature
        \item TracerAverage: time-average of the passive tracer
        \item CFLNumber: CLF number in the mesh
        \item EdgeLength: edge length in the domain
    \end{itemize}
\end{itemize}

\section{Passive tracer}
\begin{figure}
    \centering
    \begin{subfigure}{0.35\textwidth}
        \includegraphics[width=\textwidth]{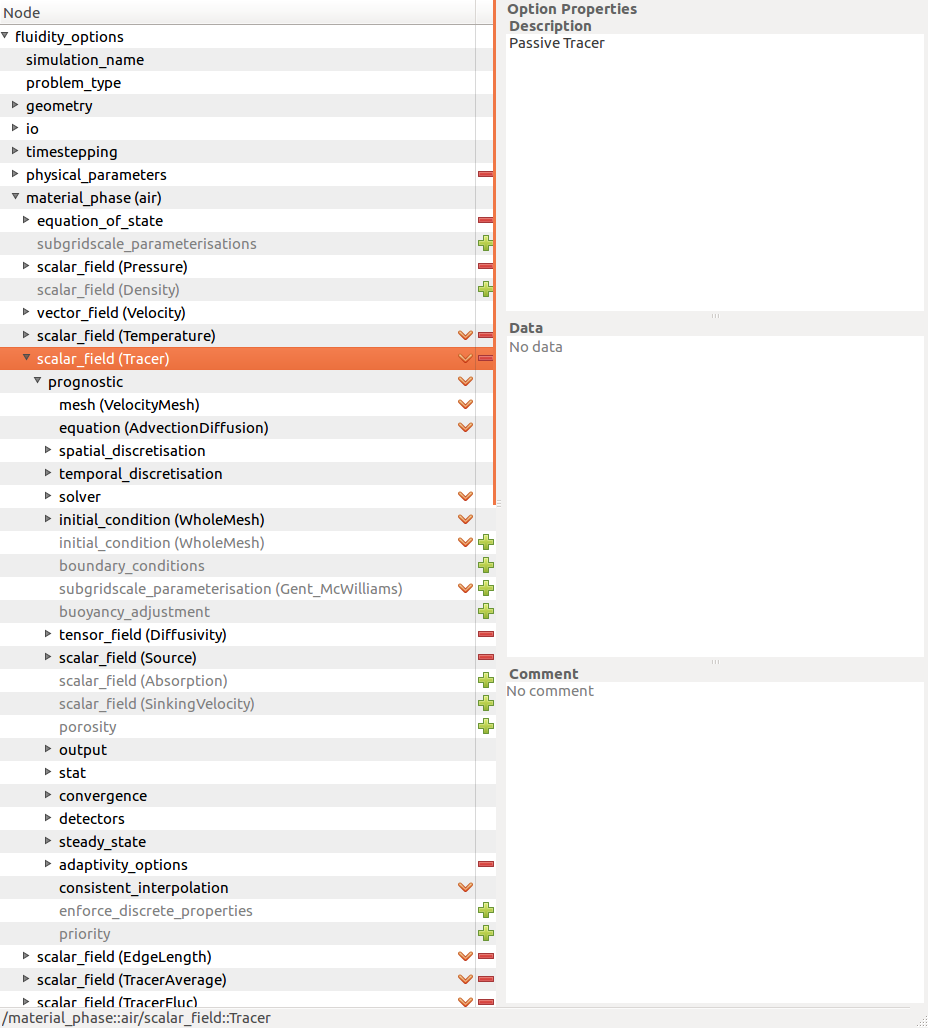}
        \caption{Tracer field}
        \label{Fig:TracerField}
    \end{subfigure}
    \begin{subfigure}{0.37\textwidth}
        \includegraphics[width=\textwidth]{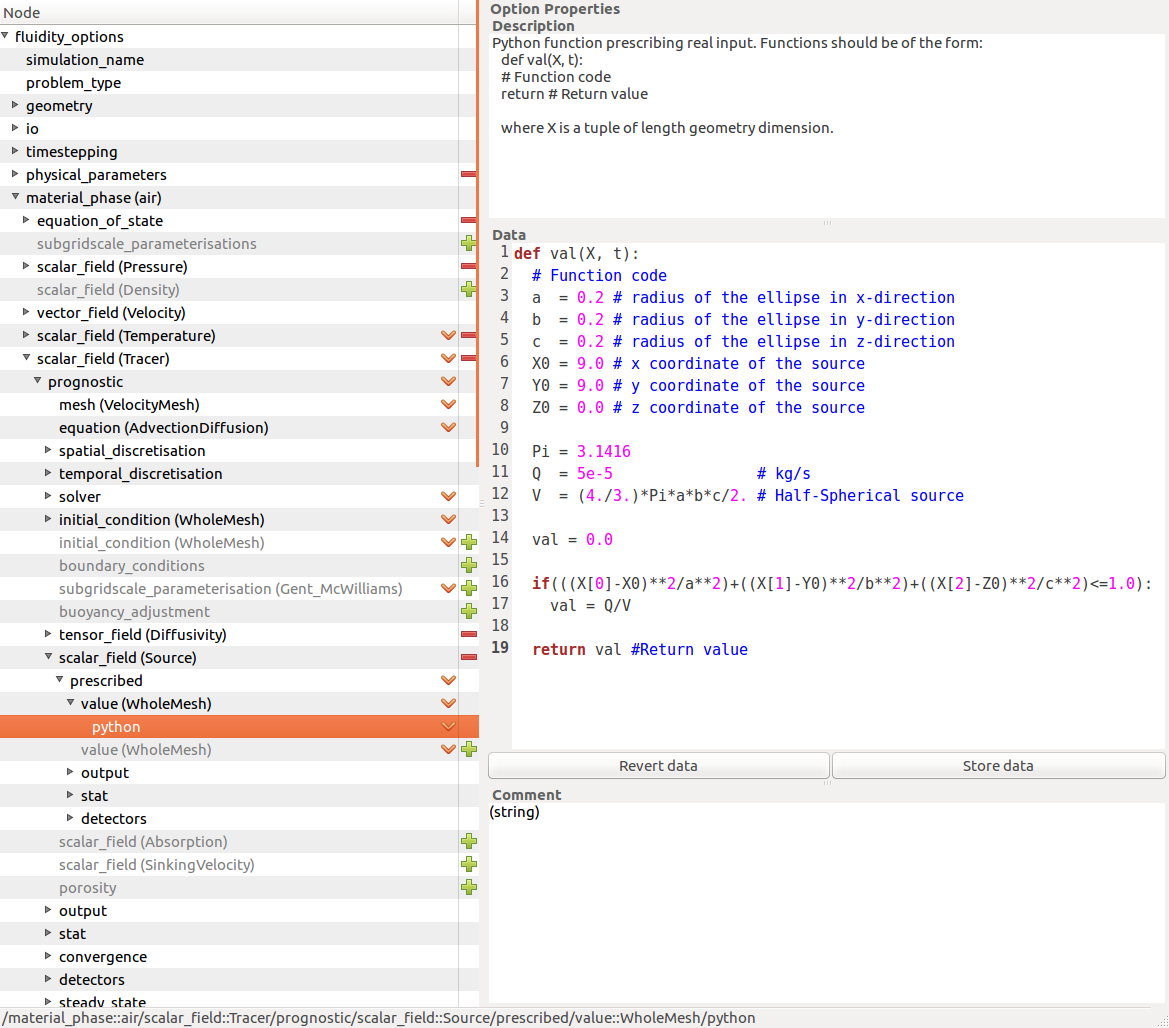}
        \caption{Source of the tracer}
        \label{Fig:TracerFieldSource}
    \end{subfigure}
    \begin{subfigure}{0.5\textwidth}
        \includegraphics[width=\textwidth]{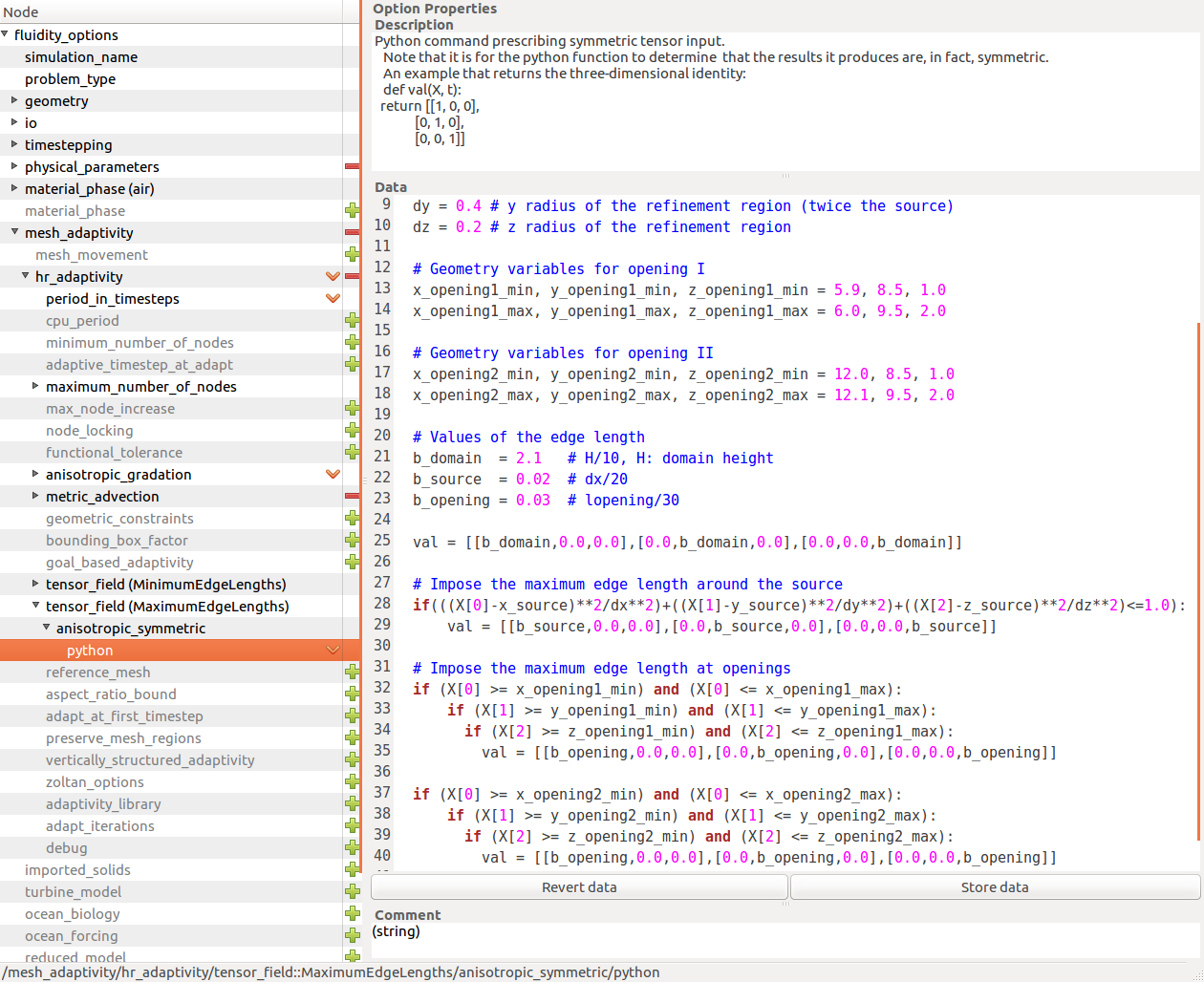}
        \caption{Refinement of the source and openings}
        \label{Fig:TracerFieldRefinement}
    \end{subfigure}
    \caption{(a) Addition of a tracer field in \textbf{Diamond}. (b) The tracer has a source located in the middle of the box near the ground. (c) The mesh is refinement using a python script to ensure that the source and flow through openings are well-resolved.}
    \label{Fig:TracerDiamond}
\end{figure}

A passive tracer field can notably be added as shown in Figure~\ref{Fig:TracerDiamond}. In example \textit{3dBox\_Case11.flml}, the passive tracer has a source located in the middle of the box. Let's assume that the passive tracer is $NO_{2}$, the diffusion coefficient of $NO_{2}$ in air is equal to $1.54\times10^{-5} $ m\textsuperscript{2}/s. A source, located in the middle of the box, is assumed to release $NO_{2}$ with a mass flow rate equal to $5\times10^{-5} $ kg/s. The volume of the source is assumed to be half a sphere with a radius of $0.2 $ m. Therefore, based on equation~\ref{Eq:TracerEq} and equation~\ref{Eq:SourceConcentrationMass}, the python script in Code~\ref{Lst:PythonSource} is assigned to the source scalar field in the Tracer field. Moreover, the mesh in the vicinity of the source is forced to be refined using python scripts, as previously described.

\begin{Code}[language=python, caption={Python script to prescribe a source term in the advection-diffusion equation.}, label={Lst:PythonSource}]
def val(X, t):
  # Function code
  a  = 0.2 # radius of the ellipse in x-direction
  b  = 0.2 # radius of the ellipse in y-direction
  c  = 0.2 # radius of the ellipse in z-direction
  X0 = 9.0 # x coordinate of the source
  Y0 = 9.0 # y coordinate of the source
  Z0 = 0.0 # z coordinate of the source

  Pi = 3.1416
  Q  = 5e-5                # kg/s
  V  = (4./3.)*Pi*a*b*c/2. # Half-Spherical source
  
  val = 0.0
  
  if(((X[0]-X0)**2/a**2)+((X[1]-Y0)**2/b**2)+((X[2]-Z0)**2/c**2)<=1.0): 
    val = Q/V
    
  return val #Return value
\end{Code}

\section{Diagnostic fields in \textbf{Diamond}} \label{Sec:DiagnosticFields}
Some interesting diagnostic fields can also be added in \textbf{Diamond}. Diagnostic fields are calculated from other fields without solving a partial differential equation. Note that scalar, vector and tensor diagnostic field can be added, but only the interesting scalar and vector ones are discussed here.

\subsection{Density field}
The density field is automatically available in \textbf{Diamond} like the Pressure and the Velocity ones, except it is turned off by default. The user can turn on this field if wanted. 

\subsection{Diagnostic fields}
See Chapter 9 of \textbf{Fluidity} manual~\cite{AMCG2015} for more details.

\noindent As shown in Figure~\ref{Fig:DiagnosticFields}, already implemented diagnostic fields can be found in several locations in \textbf{Diamond}:
\begin{itemize}
    \item \textbf{Internal diagnostic fields:} Directly under the scalar or vector fields: when the user wants to add a new scalar or vector field, it is possible to directly choose already implemented fields from a drop down list  (see Figure~\ref{Fig:Diagnos_Scalar} and Figure~\ref{Fig:Diagnos_Vector}).
    \item \textbf{Diagnostic algorithms:} Add a new scalar or vector field, then open the option tree and choose \texttt{diagnostic} instead of \texttt{prognostic}. Under the \texttt{diagnostic} option, the user will see the \texttt{algorithm}, set by default to \texttt{internal}. Here, the algorithm can be changed into another using the drop down list as shown in Figure~\ref{Fig:DiagnosAlgo_Scalar} and Figure~\ref{Fig:DiagnosAlgo_Vector}.
\end{itemize}

\subsubsection{Internal diagnostic fields}
The scalar field to output the CFL number is for example already implemented as a scalar field as shown in ~\ref{Fig:Diagnos_Scalar}. In the same list, the user can notice that it is also possible to output the grid Reynolds number and the grid Peclet number.

\subsubsection{Diagnostic algorithms}
\begin{figure}
    \centering
    \begin{subfigure}{0.4\textwidth}
        \includegraphics[width=\textwidth]{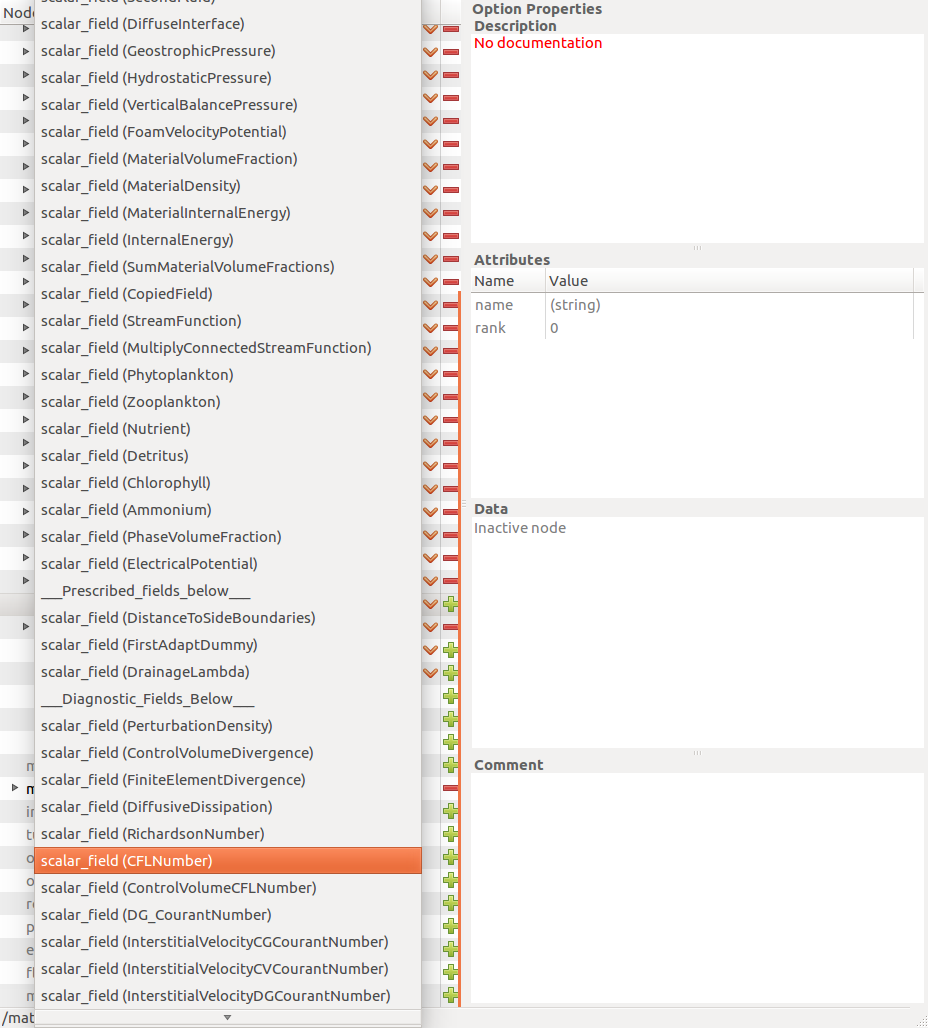}
        \caption{Scalar field}
        \label{Fig:Diagnos_Scalar}
    \end{subfigure}
    \begin{subfigure}{0.4\textwidth}
        \includegraphics[width=\textwidth]{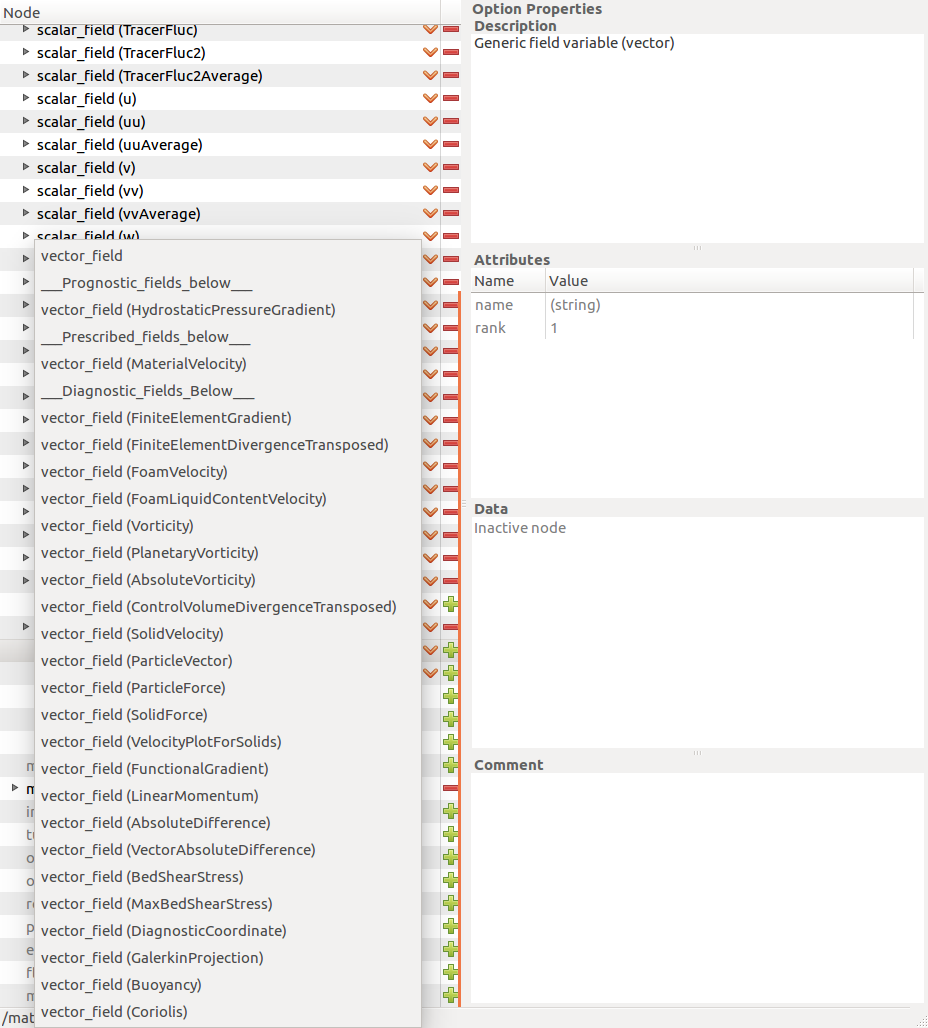}
        \caption{Vector field}
        \label{Fig:Diagnos_Vector}
    \end{subfigure}
    \begin{subfigure}{0.4\textwidth}
        \includegraphics[width=\textwidth]{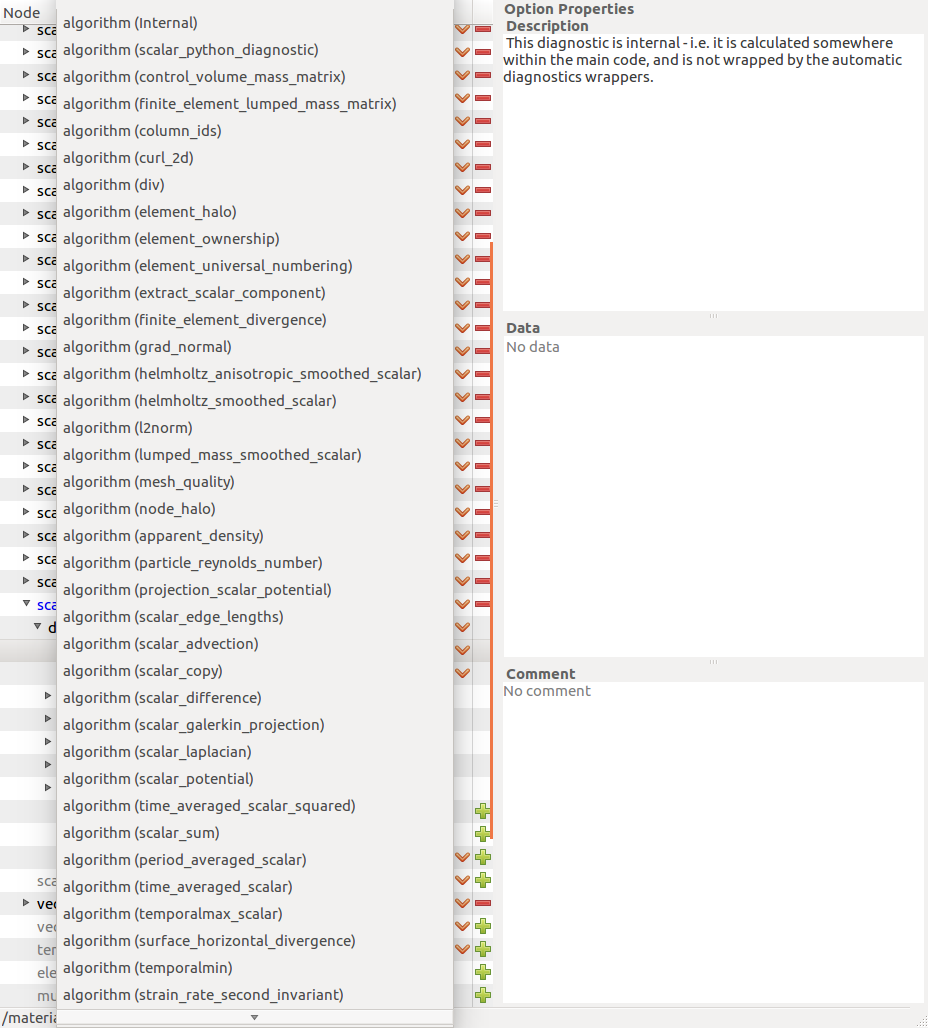}
        \caption{Algorithm scalar field}
        \label{Fig:DiagnosAlgo_Scalar}
    \end{subfigure}
    \begin{subfigure}{0.4\textwidth}
        \includegraphics[width=\textwidth]{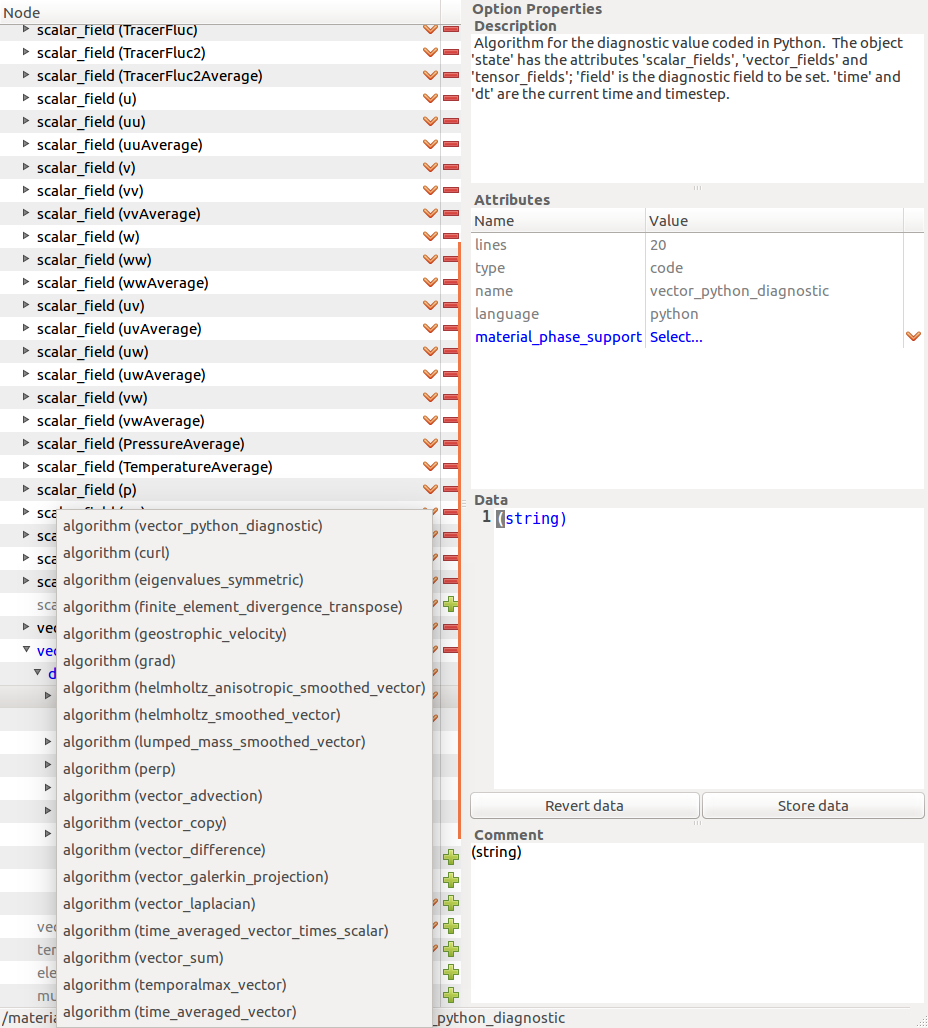}
        \caption{Algorithm vector field}
        \label{Fig:DiagnosAlgo_Vector}
    \end{subfigure}
    \caption{All the different diagnostic fields available in \textbf{Diamond} for the scalar and the vector fields.}
    \label{Fig:DiagnosticFields}
\end{figure}

Once a new scalar or vector filed is added and \texttt{diagnostic} is chosen, the user can notice other interesting diagnostic fields (Figure~\ref{Fig:DiagnosAlgo_Scalar} and Figure~\ref{Fig:DiagnosAlgo_Vector}) such as:
\begin{itemize}
    \item \texttt{time\_averaged\_scalar}: calculates the time-average of a scalar field over the duration of a simulation. A spin-up time can be added: the average will start after the set-up value (value in second).
    \item \texttt{time\_averaged\_vector}: calculates the time-average of a vector field over the duration of a simulation. A spin-up time can be added: the average will start after the set-up value (value in second).
    \item \texttt{scalar\_edge\_lengths}: outputs the edge lengths of the mesh.
    \item \texttt{scalar\_python\_diagnostic}: allows direct access to the internal \textbf{Fluidity} data structures in the computation of a scalar diagnostic field.
    \item \texttt{vector\_python\_diagnostic}: allows direct access to the internal \textbf{Fluidity} data structures in the computation of a vector diagnostic field.
\end{itemize}

\begin{figure}
    \centering
    \begin{subfigure}{0.35\textwidth}
        \includegraphics[width=\textwidth]{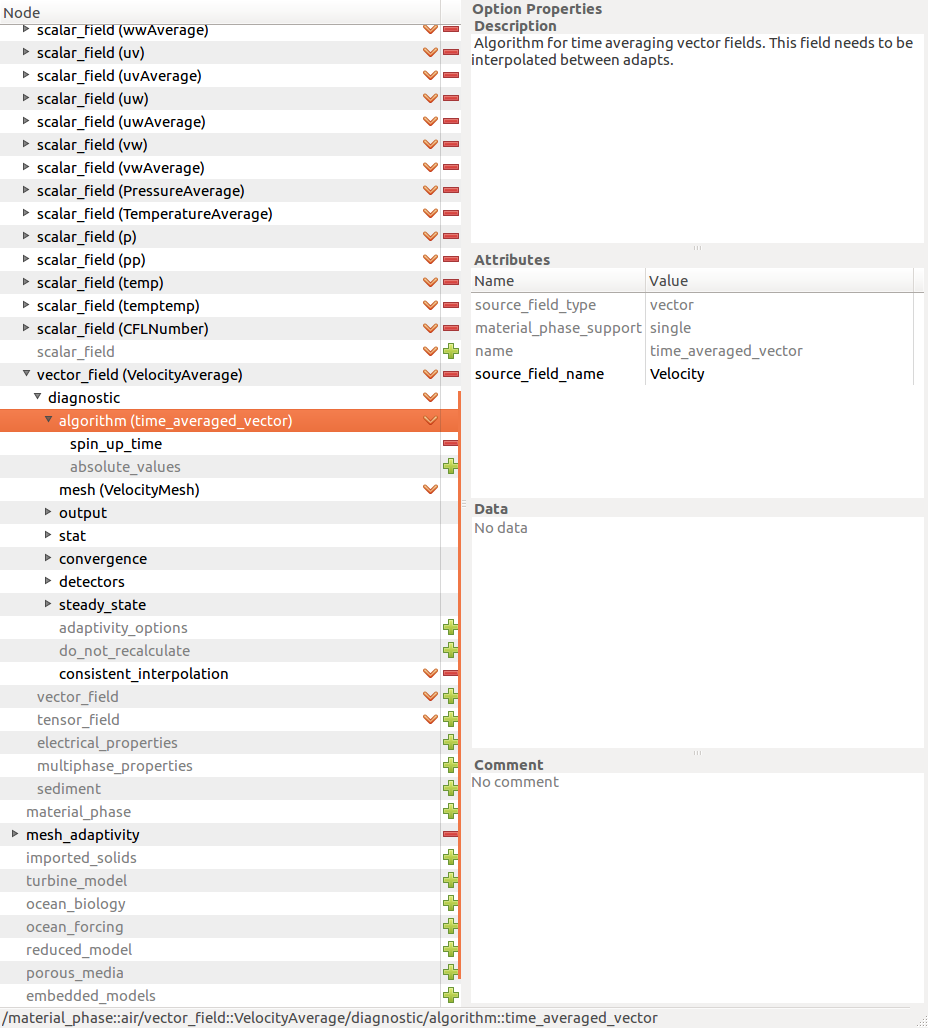} 
        \caption{}
        \label{Fig:DiagnosticFieldVelAverage}
    \end{subfigure}
    \begin{subfigure}{0.4\textwidth}
        \includegraphics[width=\textwidth]{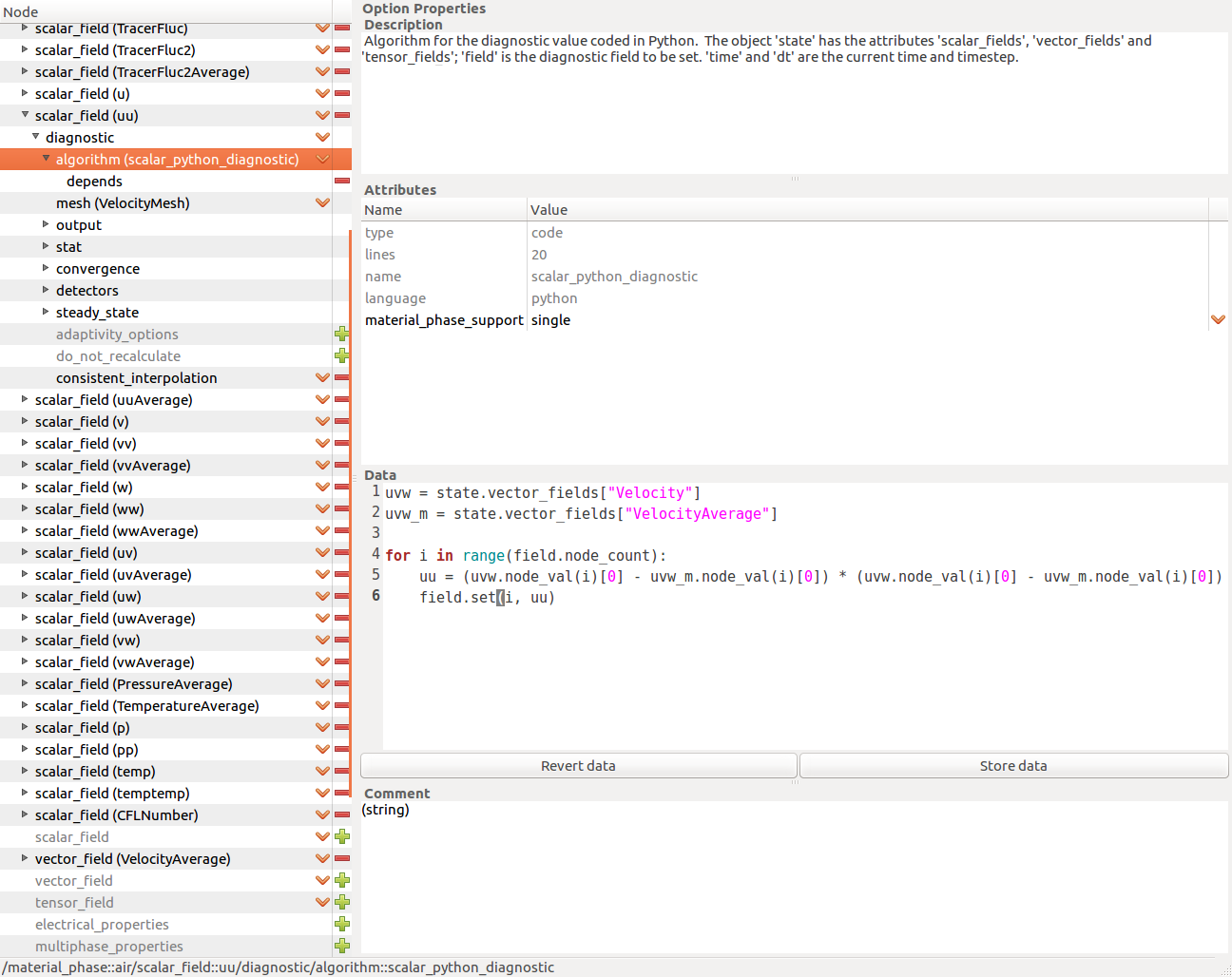}
        \caption{}
        \label{Fig:DiagnosticFielduu}
    \end{subfigure}
    \caption{(a) Time-average of the Velocity field using an internal diagnostic algorithm in \textbf{Diamond}. (b) Use of a python script to compute $u'u'$.}
    \label{Fig:DiagnosticFieldsReynolds}
\end{figure}

\noindent Some diagnostic algorithms need a source field attribute defining the field used to compute the diagnostic (for example the field used to compute the time-average). The name of the field as to be set appropriately, see Figure~\ref{Fig:DiagnosticFieldVelAverage} for example.

\noindent By selecting the \texttt{scalar\_python\_diagnostics} or \texttt{vector\_python\_diagnostic} options, a python script can be used to calculate a new field from existing fields. Under the option \texttt{depends}, the user can specify dependencies manually. Any field specified here will be calculated before the python diagnostic field. An example is shown in Figure \ref{Fig:DiagnosticFielduu} where the instantaneous $u'u'$ component is calculated directly from the velocity field and its time-average. The python code to compute the instantaneous value $u'u'$ is shown in Code~\ref{Lst:Reynoldsdiamond}. Then, applying the \texttt{time\_averaged\_scalar} to this field will give the $\overline{u'u'}$-component of the Reynolds stresses.

\begin{Code}[language=python, caption={Python script to compute $u'u'$ from the velocity and the time-averaged velocity fields.}, label={Lst:Reynoldsdiamond}]
uvw   = state.vector_fields["Velocity"]
uvw_m = state.vector_fields["VelocityAverage"]

for i in range(field.node_count):
    uu = (uvw.node_val(i)[0] - uvw_m.node_val(i)[0]) * (uvw.node_val(i)[0] - uvw_m.node_val(i)[0])
    field.set(i, uu)
\end{Code}

\subsection{Note about the time-averaged field}
Note that there is currently a bug in \textbf{Fluidity} concerning the time-averaged field: when a simulation is run from a checkpoint the value of the previous time-average is not properly read from the checkpointed files implying that the average is not correct. This will later be fixed but for now it is recommended to the user to specify a spin-up time higher than the checkpoint time to start a new time-average. See Section~\ref{Sec:Checkpoint} for more details about checkpointing.\\
For the time averaged fields, when using adaptivity, it is also recommended to use the \texttt{galerkin\_projection} option instead of the \texttt{consistent\_interpolation} one, more details are given in Section~\ref{sec:interpolation}.

    \chapter{Details of options}
\begin{figure}
    \begin{center}
        \includegraphics[scale=0.8]{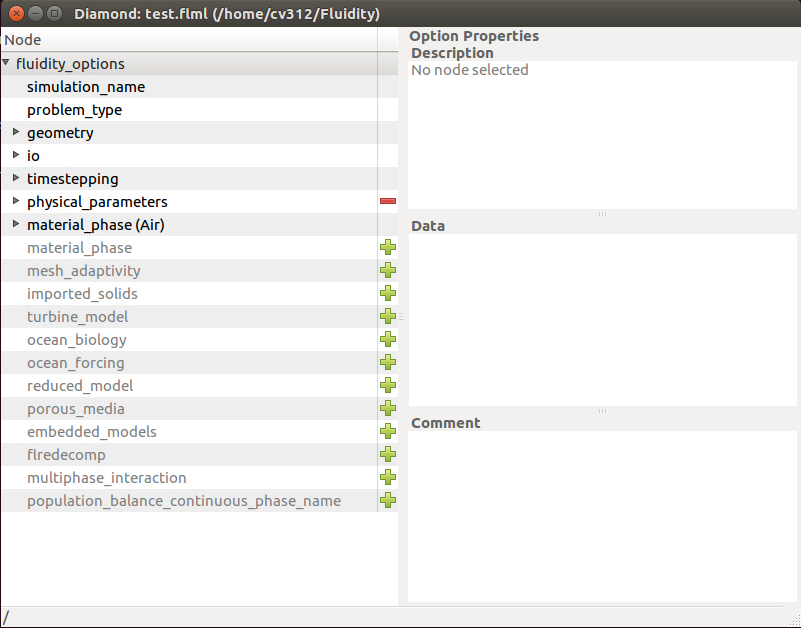}
        \caption{Drop down menu of the options available in \textbf{Diamond}.}
        \label{Fig:fluidity_options}
    \end{center}
\end{figure}

A number of options for the simulation are available to be set up in \textbf{Diamond}. Some are detailed here and most are described in the manual too \cite{AMCG2015}. For simplicity a screen shot of the drop down menu found in \textbf{Diamond} is shown in Figure \ref{Fig:fluidity_options}. The name of the simulation is inputted in \texttt{simulation\_name} and the problem type is defined as \texttt{fluids} for the cases considered here. 

\section{Geometry}
\begin{figure}
    \begin{center}
        \includegraphics[scale=0.3]{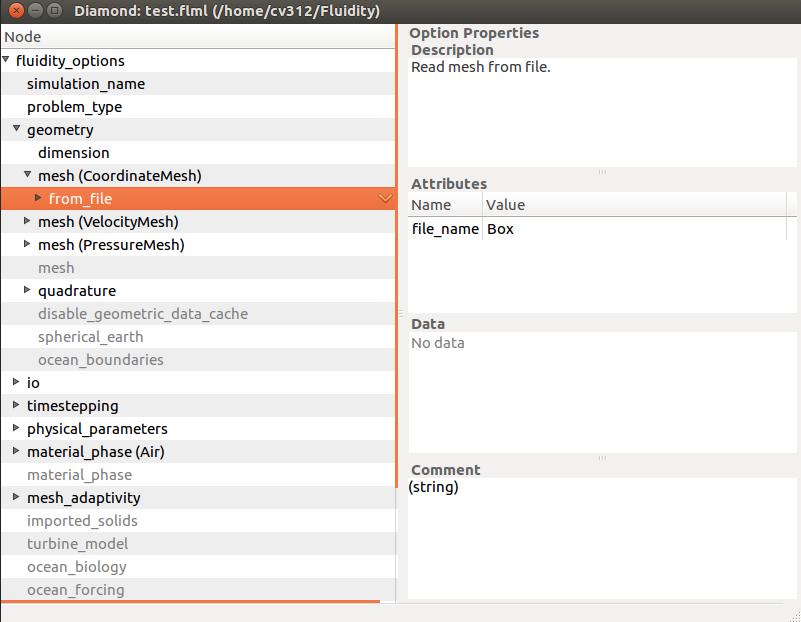} 
        \caption{Geometry options in \textbf{Diamond}.}
        \label{Fig:geometry}
    \end{center}
\end{figure}

The first part of the options defines the geometry used for the simulations as shown in Figure \ref{Fig:geometry}. The file used for the mesh needs to be inputted in \texttt{mesh (CoordinateMesh)} \texttt{/from\_file}. 

\section{Input-Output}
\begin{figure}
    \begin{center}
        \includegraphics[scale=0.6]{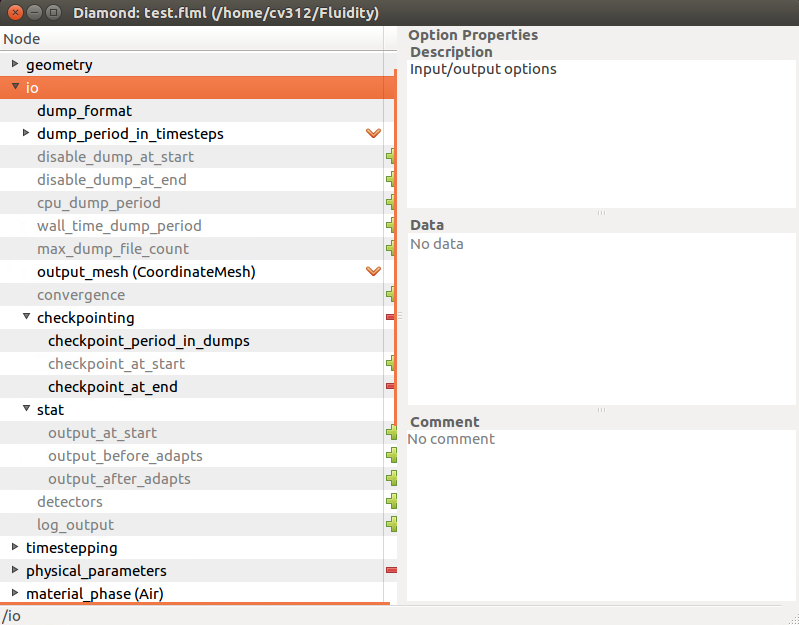} 
        \caption{Input-output options in \textbf{Diamond}.}
        \label{Fig:io}
    \end{center}
\end{figure}

The options for the inputs and outputs are defined in the \texttt{io} menu shown in Figure \ref{Fig:io}.

\noindent In these options, \texttt{dump\_period\_in\_timesteps} defines how often \textbf{ParaView} files (\textit{*.vtu}) should be outputted. This can be defined as a function of time (in second) or time steps. \\
The checkpointing options are also defined in this section. They will determine how often checkpoints are made. Checkpoints will become useful if a simulation needs to be restarted from a particular point in time. See Section~\ref{Sec:Checkpoint} for more details.

\section{Time stepping}
\begin{figure}
    \begin{center}
        \includegraphics[scale=0.6]{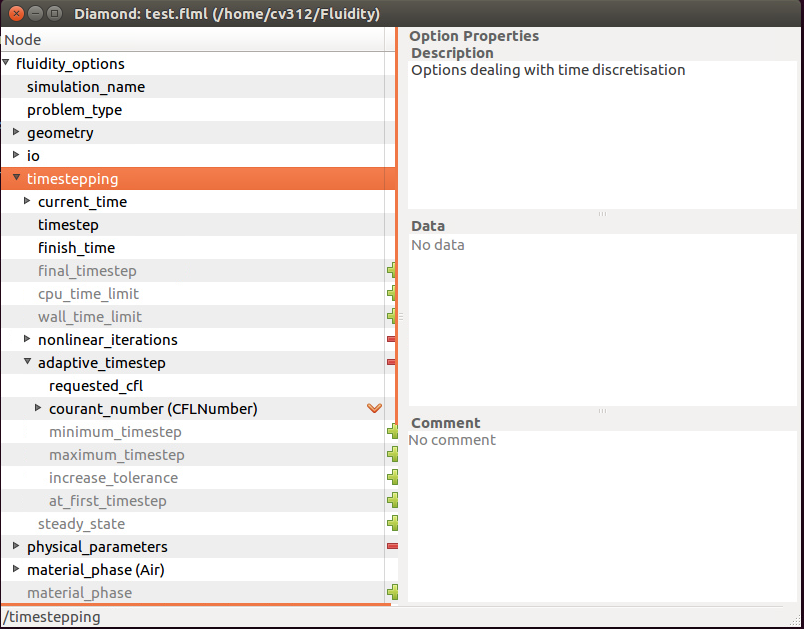} 
        \caption{Time stepping options in \textbf{Diamond}.}
        \label{Fig:timestepping}
    \end{center}
\end{figure}

The options for time stepping are then defined in the \texttt{timestepping} menu shown in Figure \ref{Fig:timestepping}. They comprise the initial, current and final time of the simulation as well as the time step requested, and are all defined in seconds. In the case of an adaptive time step, this is the initial time step. The adaptive time step can be turned on by selecting the option and choosing a CFL number under requested CFL. The non-linear iteration parameter has to be set equal to 2.

\section{Physical parameters}
\begin{figure}
    \begin{center}
        \includegraphics[scale=0.6]{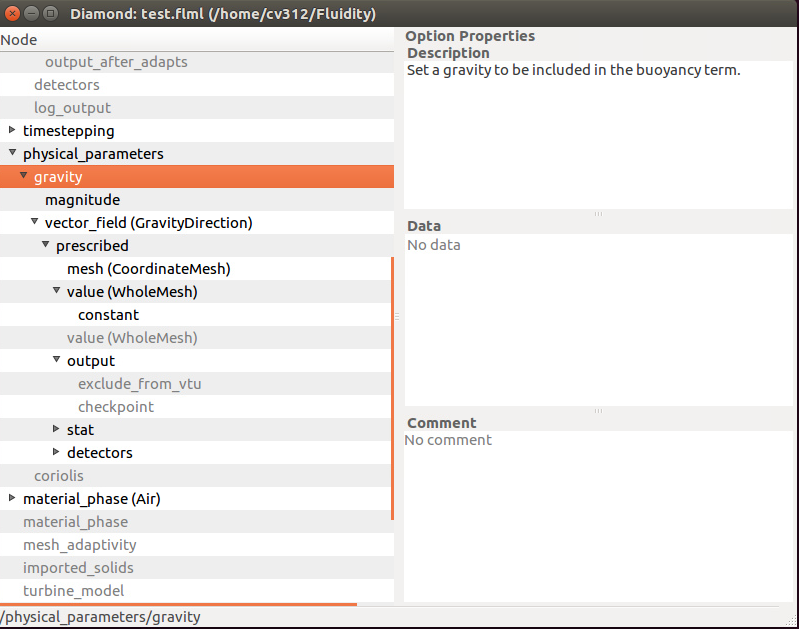} 
        \caption{Physical parameters options in \textbf{Diamond}.}
        \label{Fig:physical_param}
    \end{center}
\end{figure}

The magnitude and direction of gravity is defined in the drop down menu of the physical parameters shown in Figure \ref{Fig:physical_param}. It is valuable to check whether the value of gravity corresponds to the coordinate system used. For example, the magnitude is usually set to 9.81 (if you are not on Mars) and the \texttt{GravityDirection} to $(0,0,-1)$ if the $z$-axis corresponds to the height pointing upwards. 

\section{Material phase}
All the fields used for a given fluid are defined in \texttt{material\_phase/}, this include their initial and boundary conditions. In a case where different materials are used multiple \texttt{material\_phase/} sections will have to be added. In the cases presented here however, only air is present. 

\subsection{Equation of state}
\begin{figure}
    \begin{center}
        \includegraphics[scale=0.6]{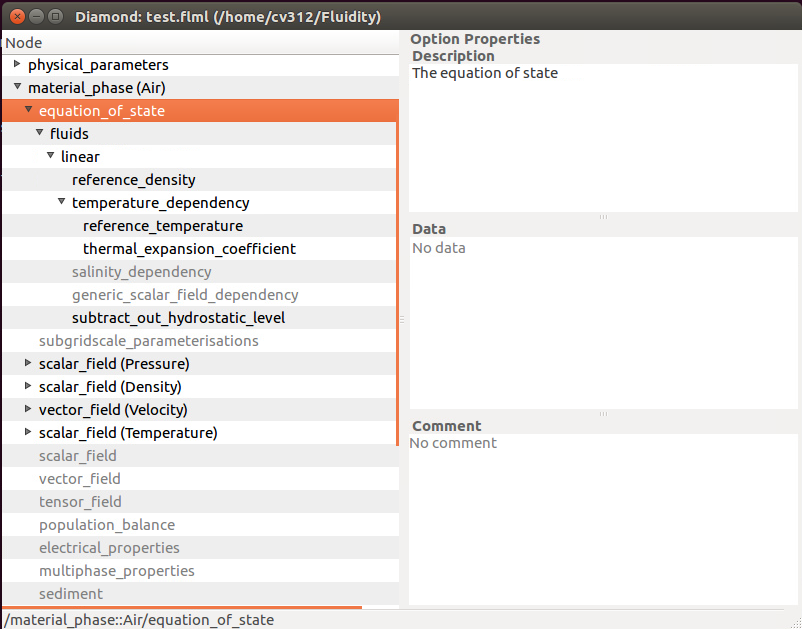} 
        \caption{Equation of state options in \textbf{Diamond}.}
        \label{Fig:eq_of_state}
    \end{center}
\end{figure}

The equation of state options, shown in Figure \ref{Fig:eq_of_state}, define the dependency between temperature and density. This is also where the reference values for temperature and density are defined. The option \texttt{subtract\_out\_hydrostatic\_level} has to be turned on if the Boussinesq approximation is used, which is the case in all the examples presented in this manual.

\subsection{Prognostic fields}
\begin{figure}
    \begin{center}
        \includegraphics[scale=0.6]{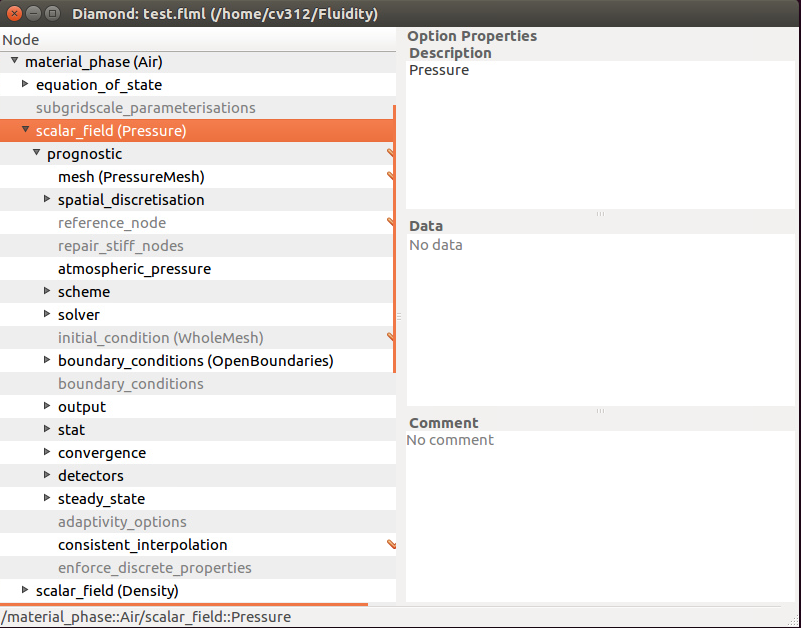} 
        \caption{Prognostic field options in \textbf{Diamond}.}
        \label{Fig:pressure}
    \end{center}
\end{figure}

The prognostic \texttt{scalar\_field} and \texttt{vector\_field} are used to define certain fields such as pressure and velocity as shown in Figure~\ref{Fig:pressure}. 

\noindent This is where boundary and initial conditions are defined as well as how each field is discretised and solved. Output options are also available. Finally \texttt{adaptivity\_options} will allow to define the interpolation error bound used as a weight during the adaptivity process. The following options are chosen for the main prognostic fields.

\noindent For the pressure field: 
\begin{itemize}
    \item Spatial discretisation: \texttt{continuous\_galerkin}.
    \item Scheme: \texttt{poisson\_pressure\_solution} with a \texttt{use\_projection\_method}.
    \item Solver: \texttt{iterative\_method (cg)} with a \texttt{preconditioner (hypre)} and \texttt{hypre\_} \texttt{type (boomeramg)}. Other options are \texttt{relative\_error} and \texttt{max\_iterations} with the suggested values  as well as adding the \texttt{never\_ignore\_solver\_failures}.
\end{itemize}
\noindent For the velocity field: 
\begin{itemize}
    \item Spatial discretisation: \texttt{continuous\_galerkin}, including the options
    \begin{itemize}
        \item \texttt{stabilisation}\texttt{/no\_stabilisation}
        \item \texttt{mass\_terms/lump\_mass\_matrix}
        \item \texttt{stress\_terms/tensor\_form}
        \item \texttt{les\_model/second\_order} (with a Smagorinsky coefficient of 0.1 and a tensor \texttt{length\_scale\_type})
    \end{itemize}
    \item Temporal discretisation: with a \texttt{theta} of 0.5 and \texttt{relaxation} of 0.5.
    \item Solver: \texttt{iterative\_method(gmres)} and \texttt{preconditioner (sor)}. Other options are \texttt{relative\_error} and \texttt{max\_iterations} with the suggested values  as well as adding the \texttt{never\_ignore\_solver\_failures}.
\end{itemize}
\noindent For the temperature field: 
\begin{itemize}
    \item Spatial discretisation: \texttt{control\_volumes} with a \texttt{face\_value(FiniteElement)} of \texttt{limit\_face\_value} specified by a \texttt{limiter(Sweby)}. The \texttt{diffusion\_scheme} is \texttt{ElementGradient} and the \texttt{conservative\_advection} is set to 0. 
    \item Temporal discretisation: \texttt{theta} is 0.5. The \texttt{control\_volumes} option is selected with a \texttt{number\_advection\_iterations} set to 2. 
    \item Solver: \texttt{iterative\_method(gmres)} and \texttt{preconditioner (sor)}. Other options are \texttt{relative\_error} and \texttt{max\_iterations} with the suggested values  as well as adding the \texttt{never\_ignore\_solver\_failures}.
\end{itemize}

\subsection{Diagnostic fields}
\begin{figure}
    \begin{center}
        \includegraphics[scale=0.6]{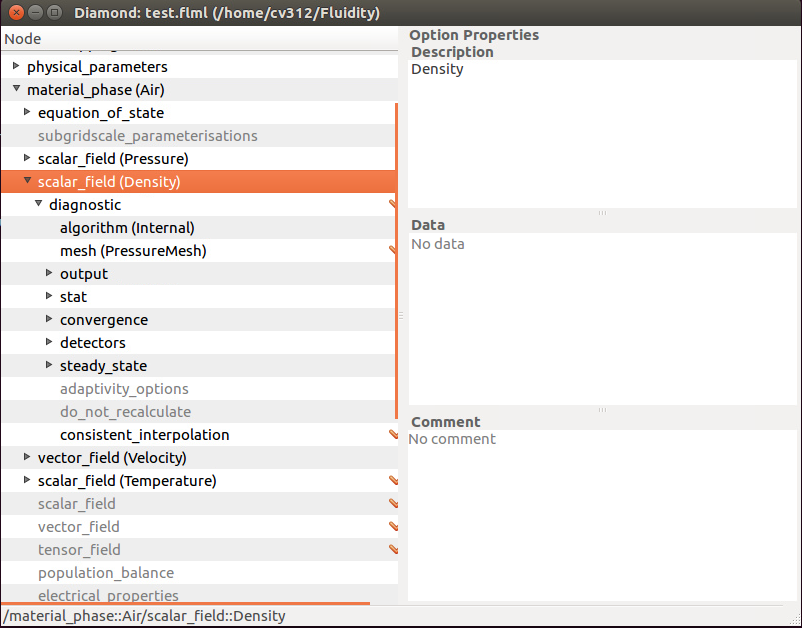} 
        \caption{Diagnostic field options in \textbf{Diamond}.}
        \label{Fig:density}
    \end{center}
\end{figure}

The \texttt{material\_phase} section also includes the calculated diagnostic fields as shown in Figure \ref{Fig:density}. See Section~\ref{Sec:DiagnosticFields} for more details. 

\section{Mesh adaptivity}
\begin{figure}
    \begin{center}
        \includegraphics[scale=0.6]{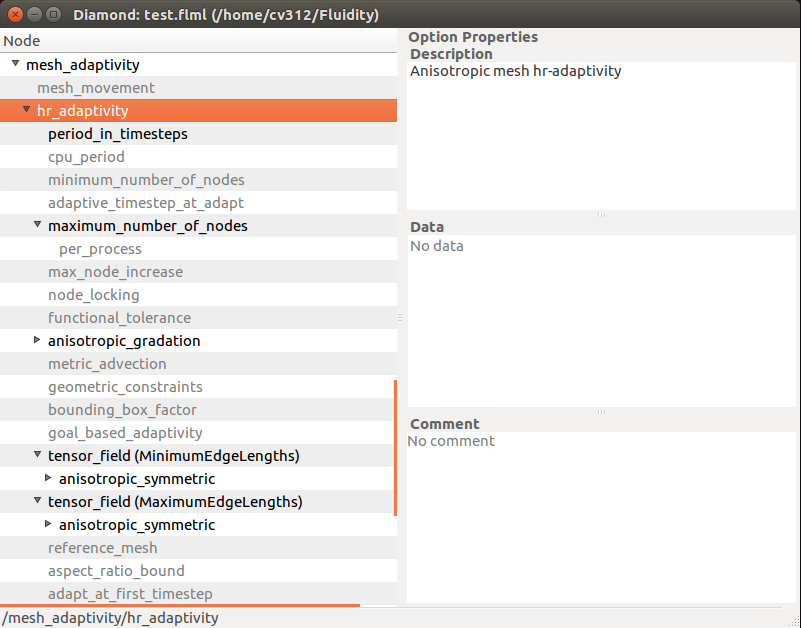} 
        \caption{Mesh adaptivity options in \textbf{Diamond}.}
        \label{Fig:mesh_adaptivity}
    \end{center}
\end{figure}

This section defines the adaptivity process and how it will be conducted. The available options are shown in Figure \ref{Fig:mesh_adaptivity} for the chosen \texttt{hr\_adaptivity} algorithm. 

\noindent In particular, \texttt{period\_in\_timesteps} allows to define how often adaptivity is to happen, \texttt{anistropic\_gradation} how the mesh should be smoothed out, while \texttt{maximum\_number\_} \texttt{of\_nodes} sets an upper bound on the number of nodes the mesh can be decomposed into. As a rule of thumb it is recommended to use at least 50,000 nodes per processor.

\noindent Finally, the minimum and maximum edge lengths can also be defined in this section. In combination with the interpolation error bound they will define how the resolution of the mesh will vary with adaptation. Python scripts can be used to vary the specified lengths over the domain, allowing finer resolutions over areas of interest for instance. More details on the adaptivity process can be found in Chapter~\ref{Sec:MeshAdaptivity}.
    \chapter{Tricks}

\section{Size of the domain}
\subsection{Blockage ratio}
The size of the computational domain has to be properly chosen and it is recommended for the user to read carefully~\cite{Franke2007} for a detailed method. The following summarises the principal rule of thumb and basically paraphrases~\cite{Franke2007}. First, let's define the blockage ratio $r$ as expressed by equation~\ref{Eq:BlockageRatio}:
\begin{equation}
r=\frac{A_{build}}{A_{inlet}}
\label{Eq:BlockageRatio}
\end{equation}

\noindent where $A_{build}$ is the projected area of the building in flow direction and $A_{inlet}$ is the inlet area.

\noindent In the case of a single building: $A_{build}=lH$, where $l$ is the width of the building and $H$ its height. In the case of multiple buildings: $A_{build}=d_{area}H_{max}$, where $d_{area}$ is the average diameter of the domain of interest and $H_{max}$ is the height of the tallest building.

\subsection{Height of the domain}
To prevent artificial acceleration of the flow over the building, the height of the domain has to be chosen as follows.
\begin{itemize}
    \item For a single building: the top of the computational domain should be at least $5H$ above the roof of the building. To eliminate errors due to the size of the computational domain, the blockage ratio is recommended to be lower than 3\% in CFD simulations.
    \item For urban environment with more than one building: the top of the computational domain should be at least $5H_{max}$ above the tallest building.
\end{itemize}
 
\noindent Based on the blockage ratio, smaller and larger values can be used: $4H$ is sufficient for a small blockage, while $10H$ is recommended for a large blockage.
 
\noindent If the simulations are to be compared with wind tunnel experiments, it is recommended to use the wind tunnel's test section geometry for the computational domain. However, if the height of the wind tunnel is much larger than $6H_{max}$, then a lower height of the computational domain can be tested. 

\subsection{Width of the domain}
Once the height of the domain is defined, the width of the domain can be determined using the blockage ratio formula.
\begin{itemize}
    \item For a single building: assuming a domain height of $6H$ the requirement of 3\% blockage leads to a distance of $2.3H$ between the building's side and the lateral boundaries of the computational domain. The published recommendations for this distance are however much larger and using $5H$, leading to a blockage of only 1.5\%, is recommended. 
    \item For urban environment with more than one building: the lateral boundaries of the computational domain can  be placed closer than $5H_{max}$ from the built area.
\end{itemize}

\noindent If the simulations are to be compared with wind tunnel experiments, it is recommended to use the wind tunnel's test section geometry for the computational domain. However, if the lateral walls of the wind tunnel from the built area is much larger than $5H_{max}$, then a smaller width of the computational domain can be tested. 

\subsection{Length of the domain}
The length of the domain has to be divided into two regions:
\begin{itemize}
    \item The upstream region:
    \begin{itemize}
        \item For a single building: a distance of $5H$ between the inflow boundary and the building is recommended.
        \item For urban areas: a distance of $5H$ between the inflow boundary and the building is recommended.
    \end{itemize}
    \item The downstream region:
    \begin{itemize}
        \item For a single building: a distance of again $15H$ behind the building is recommended.
        \item For urban areas: a smaller distance between the outflow boundary and the built area can be used. 
    \end{itemize}
\end{itemize}

\noindent Large values are recommended for the downstream region to avoid flow entering the domain through the outflow boundary. Indeed, this situation should be avoided as it can lead to the divergence of the solver and the crash of the simulation. However the recommended value of $15H$ can lead to too big domain that will increase considerably your computational time. Solutions to tackle this issue are discussed in the next Section~\ref{Sec:AbsorptionTricks}.

\section{Instabilities at the edges of the domain} \label{Sec:AbsorptionTricks}
\begin{figure}
    \centering
    \begin{subfigure}{0.3\textwidth}
        \includegraphics[width=\textwidth]{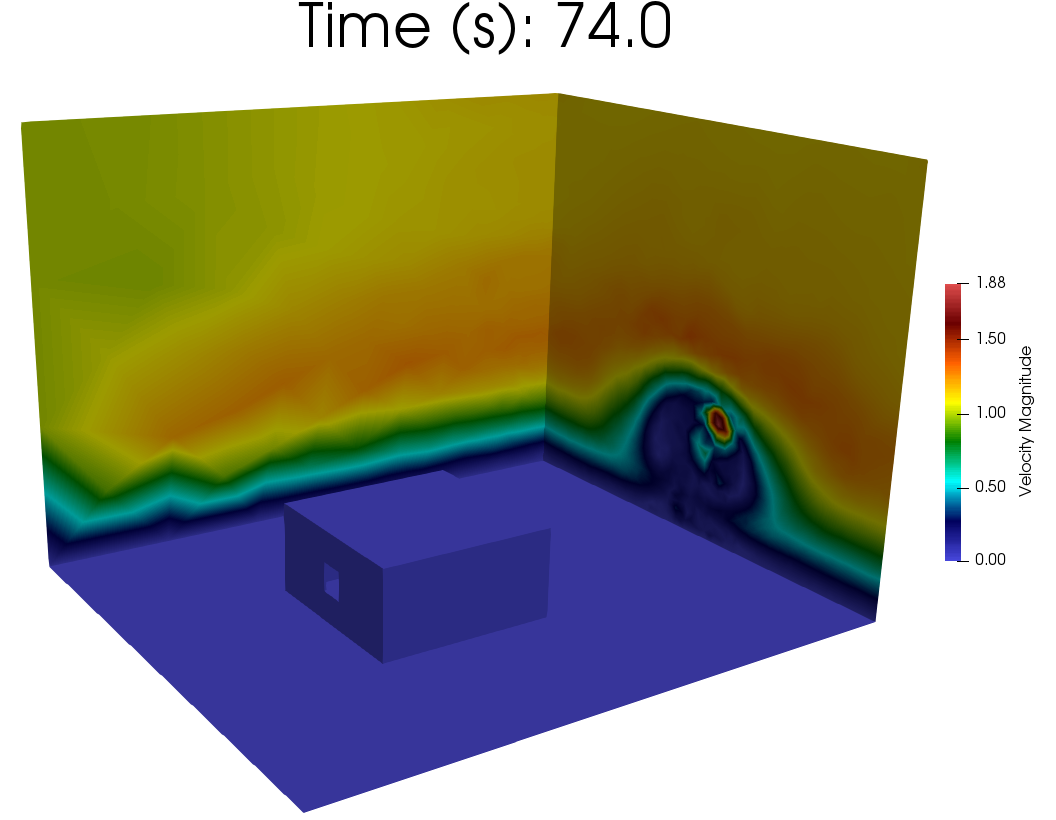}
        \caption{}
        \label{Fig:Case12a_Instabilities74}
    \end{subfigure}
    \begin{subfigure}{0.3\textwidth}
        \includegraphics[width=\textwidth]{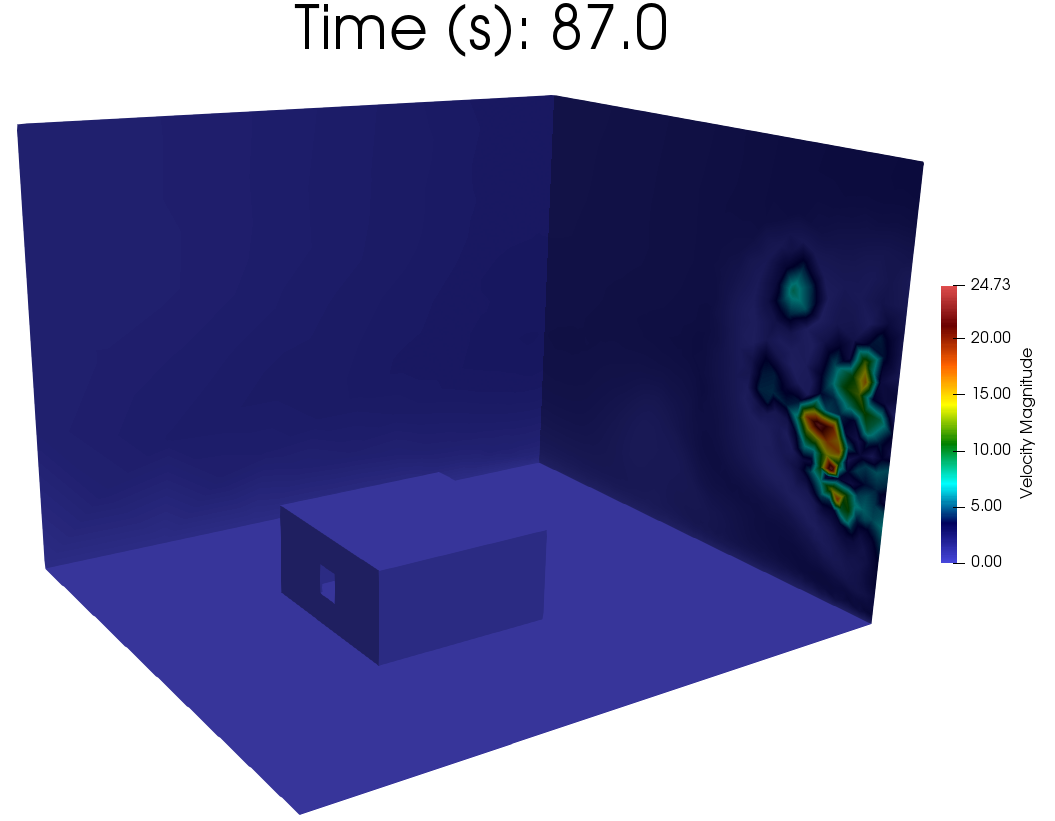}
        \caption{}
        \label{Fig:Case12a_Instabilities87}
    \end{subfigure}
    \begin{subfigure}{0.3\textwidth}
        \includegraphics[width=\textwidth]{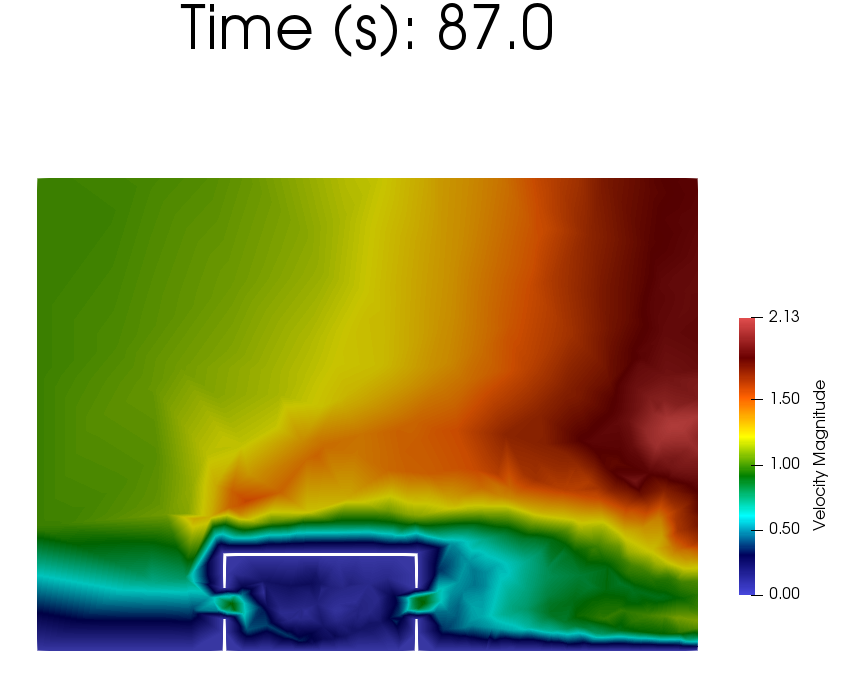}
        \caption{}
        \label{Fig:Case12a_InstabilitiesPlane}
    \end{subfigure}
    \caption{In example \textit{3dBox\_Case12a.flml}, instabilities occurs at the outlet of the domain. The instabilities are characterised by unrealistic high velocity magnitude at the outlet. Screenshots of the velocity magnitudes at (a) 74 seconds; (b) 87 seconds and (c) 87 seconds on a vertical plane.}
    \label{Fig:Case12a}
\end{figure}

Running example \textit{3dBox\_Case12a.flml} long enough, the user can observe that the simulation starts to behave weirdly around 74 seconds and finally crashes at 87 seconds. The user can test this example easily, it takes only 8 minutes to run until crashing. As shown in Figure~\ref{Fig:Case12a}, very high and unrealistic velocity magnitudes appear at the outlet of the domain: this is due to turbulent recirculation at the outlet which causes numerical instabilities until the divergence of the solver. This is a well-known numerical issue which unfortunately occurs often. There are two ways to avoid this issue:
\begin{itemize}
    \item Extend the length of the domain as far as needed. This should be the first option to consider.
    \item Add a `sponge' layer at the end of the domain to artificially dissipate the eddies.
\end{itemize}

\begin{figure}
    \centering
    \begin{subfigure}{0.3\textwidth}
        \includegraphics[width=\textwidth]{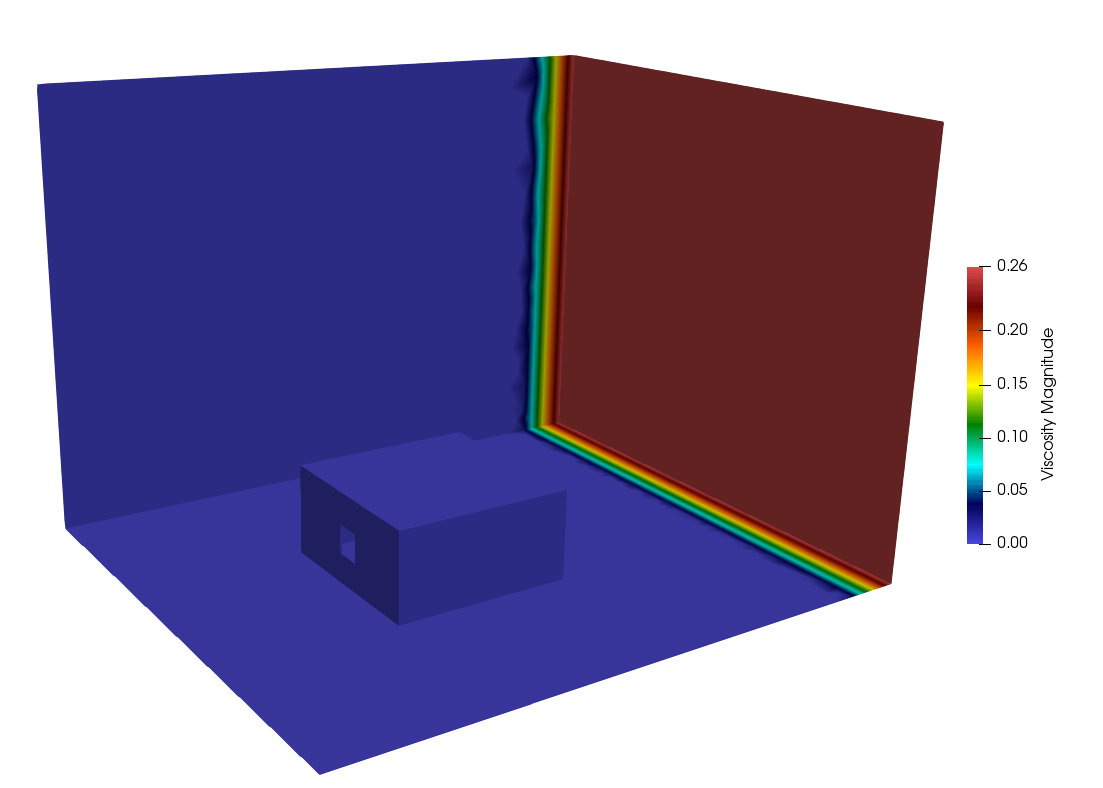}
        \caption{}
        \label{Fig:Case12b_Viscosity}
    \end{subfigure}
    \begin{subfigure}{0.3\textwidth}
        \includegraphics[width=\textwidth]{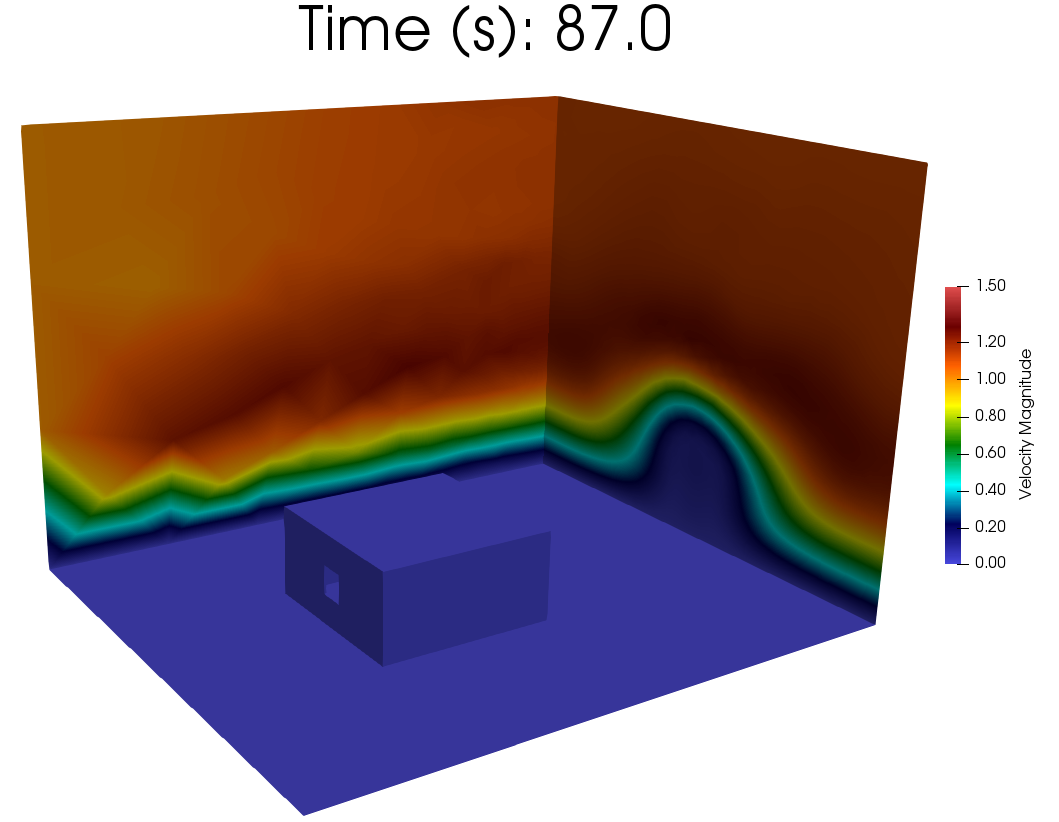}
        \caption{}
        \label{Fig:Case12b_VelocityDomain}
    \end{subfigure}
    \begin{subfigure}{0.3\textwidth}
        \includegraphics[width=\textwidth]{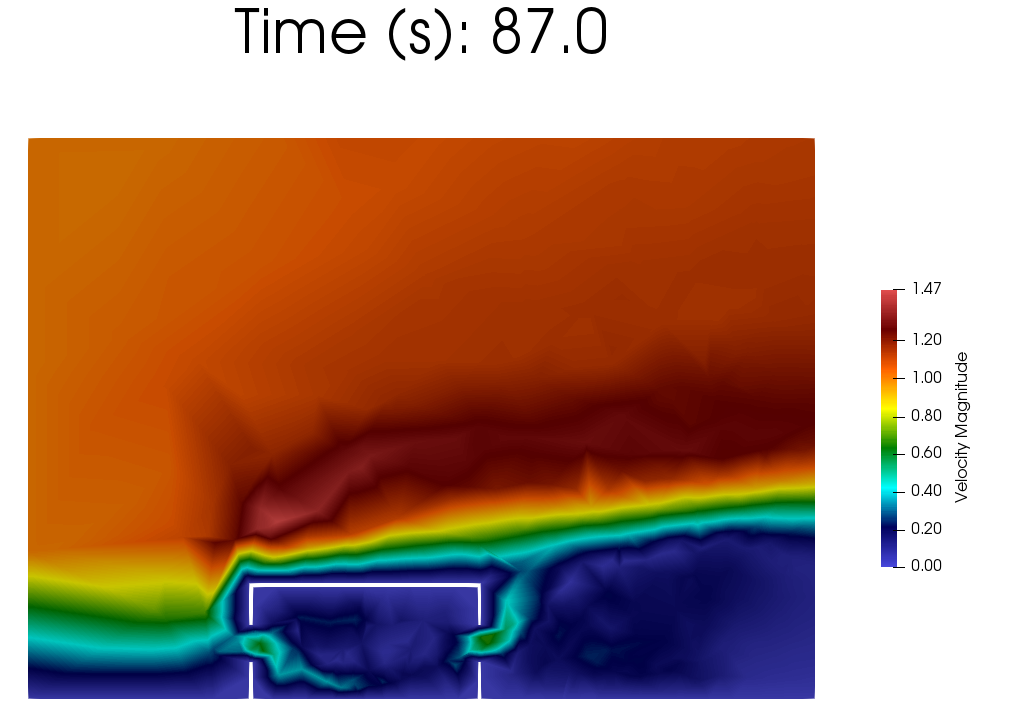}
        \caption{}
        \label{Fig:Case12b_VelocityPlane}
    \end{subfigure}
    \caption{In example \textit{3dBox\_Case12b.flml} a viscosity layer is added at the outlet of the domain to avoid crash of the simulation. (a) Viscosity layer, (b) Velocity magnitude in the domain and (c) Velocity magnitude on a vertical plane.}
    \label{Fig:Case12b}
\end{figure}

\begin{figure}
    \centering
    \begin{subfigure}{0.3\textwidth}
        \includegraphics[width=\textwidth]{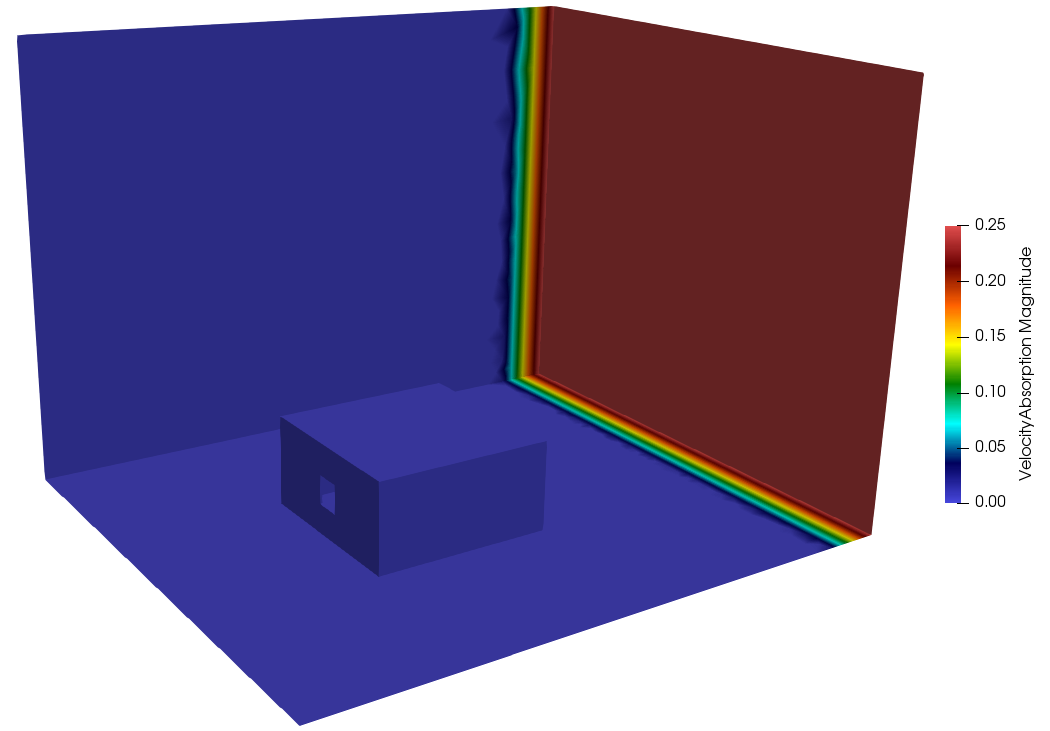}
        \caption{}
        \label{Fig:Case12c_Absorption}
    \end{subfigure}
    \begin{subfigure}{0.3\textwidth}
        \includegraphics[width=\textwidth]{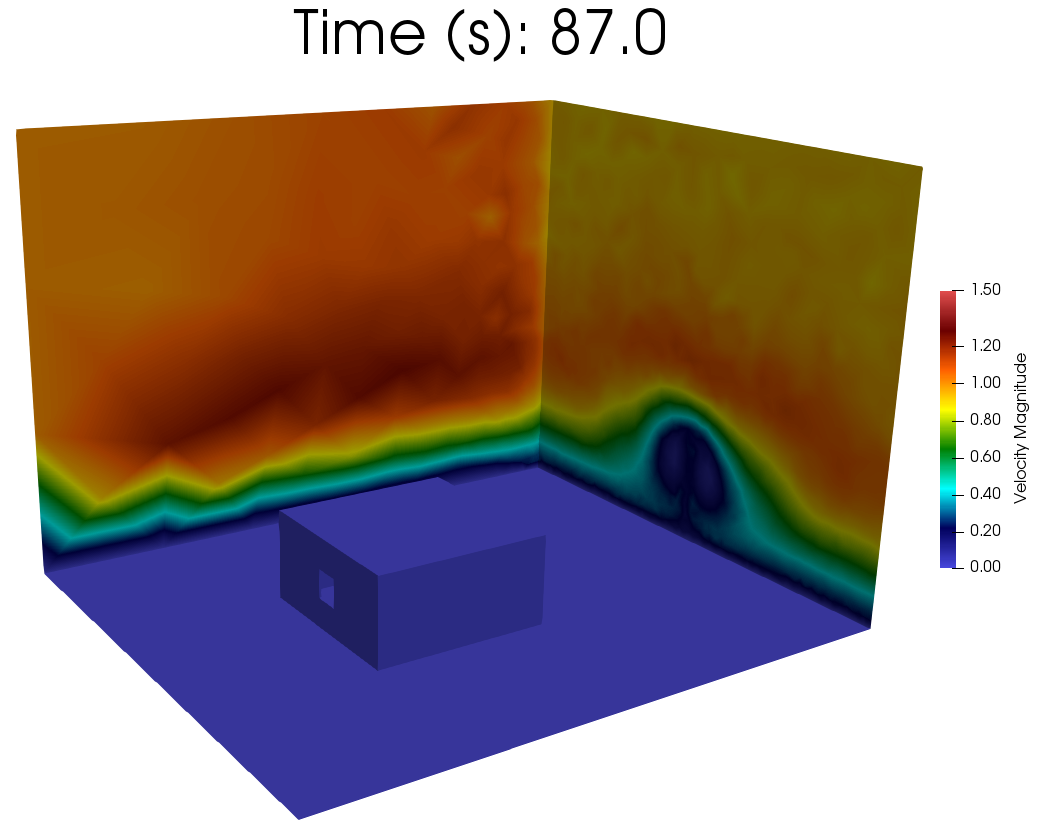}
        \caption{}
        \label{Fig:Case12c_VelocityDomain}
    \end{subfigure}
    \begin{subfigure}{0.3\textwidth}
        \includegraphics[width=\textwidth]{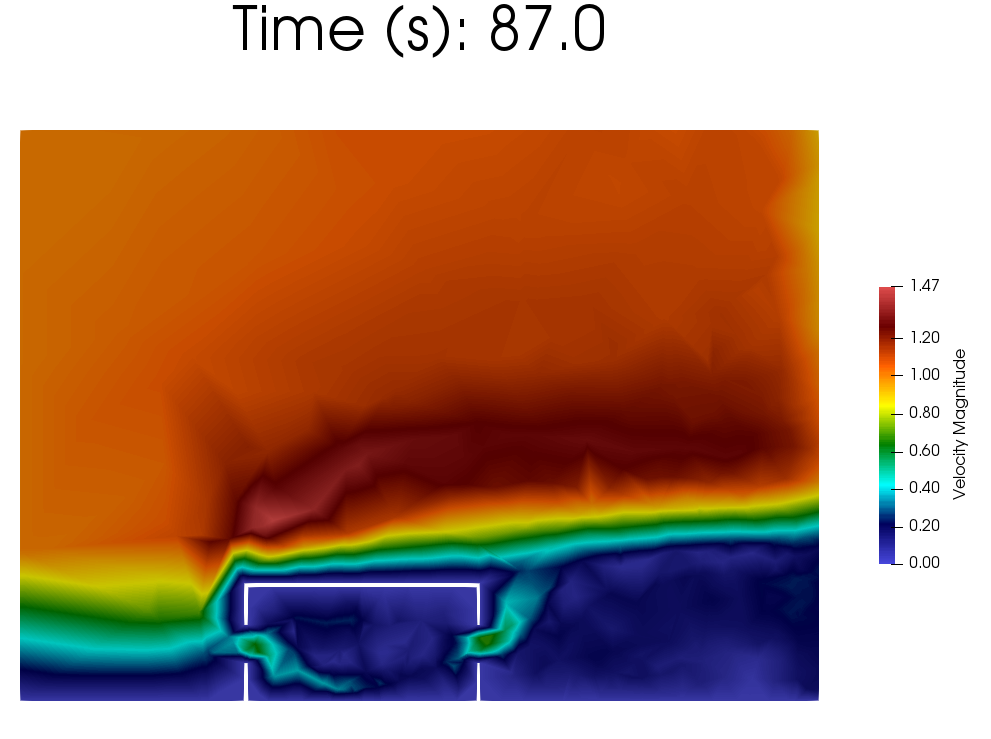}
        \caption{}
        \label{Fig:Case12c_VelocityPlane}
    \end{subfigure}
    \caption{In example \textit{3dBox\_Case12c.flml} an absorption layer is added at the outlet of the domain to avoid crash of the simulation. (a) Absorption layer, (b) Velocity magnitude in the domain and (c) Velocity magnitude on a vertical plane.}
    \label{Fig:Case12c}
\end{figure}

\noindent To avoid any unwanted recirculation and/or instabilities occurring at the domain edges, the user can add a so-called `sponge' layer which will stabilise the turbulent flow and suppress the eddies. When using this solution, it is recommended to the user to be sure that the results in the domain of interest are not affected by the `sponge' layer, i.e to be sure that the killed turbulence does not have a direct impact and repercussion on the results. The `sponge' layer is prescribed through a python script and there are two ways to define it:
\begin{itemize}
    \item \textbf{Viscosity layer:} the value of the viscosity is linearly increased at the end of the domain as shown in Figure~\ref{Fig:Case12b_Viscosity}. In the velocity field, the python script in Code~\ref{Lst:ViscosityLayer} is used to prescribe the viscosity: the viscosity increases linearly from the typical value for air of $1.511\times10^{-5}$ m\textsuperscript{2}/s to several orders of magnitude higher. The user can refer to example \textit{3dBox\_Case12b.flml}. This example runs properly and no instabilities occur anymore at the outlet as shown in Figure~\ref{Fig:Case12b}.
    \begin{Code}[language=python, caption={Python script to define an increasing viscosity at the outlet of the domain.}, label={Lst:ViscosityLayer}]
def val(X, t):
    # Function code
    # Viscosity values
    nu    = 1.5e-5    # Value of viscosity in the domain
    nuMax = 0.15      # Maximum value of the viscosity
    
    # Geometry variables
    xMax   = 21.0          # Length of the domain
    Llayer = 2.0           # Length of the layer - in meter
    xStart = xMax - Llayer # Where to start the sponge layer

    # Viscosity Ramp x-Direction ------ viscosity = a*x+b
    a = (nuMax - nu)/(xMax - xStart) # Slope
    b = nu - a * xStart

    val = nu
    # Assigning an increased value of viscosity near the outlet
    if (X[0] >= xStart):
      val = a * X[0] + b

    return val #Return value

    \end{Code}
    \item \textbf{Absorption layer:} an absorption term is prescribe at the end of the domain as shown in Figure~\ref{Fig:Case12c_Absorption}. In the velocity field, the python script in Code~\ref{Lst:AbsorptionLayer} is used to prescribe an absorption term: the absorption is equal to zero in the domain and increases linearly to $0.25$. The latest value will have to be adjusted by the user depending of the case used. The user can refer to example \textit{3dBox\_Case12c.flml}. This example runs properly and no instabilities occur anymore at the outlet as shown in Figure~\ref{Fig:Case12c}.
    \begin{Code}[language=python, caption={Python script to define an absorption term at the outlet of the domain.}, label={Lst:AbsorptionLayer}]
def val(X, t):
    # Function code
    # Absorption values
    AMin = 0.0  # Value of the absorption in the domain equal to zero.
    AMax = 0.25 # Maximum value of absorption
    
    # Geometry variables
    xMax   = 21.0          # Length of the domain
    Llayer = 2.0           # Length of the layer - in meter
    xStart = xMax - Llayer # Where to start the sponge layer

    # Absorption Ramp x-Direction ------ absorption = a*x+b
    a = (AMax - AMin)/(xMax - xStart) # Slope
    b = AMin - a * xStart

    val = [AMin, 0, 0]
    # Assigning an absorption term near the outlet
    if (X[0] >= xStart): 
      val = [a * X[0] + b, 0, 0]

    return val #Return value

    \end{Code}
\end{itemize}

\noindent The difference of what is happening at the outlet of the domain when using a viscosity or an absorption layer can be seen in Figure~\ref{Fig:Case12b} and Figure~\ref{Fig:Case12c}, and the authors leave the users to appreciate which one in the more suitable for their application.

\section{Reference temperature}
\begin{figure}
    \centering
    \begin{subfigure}{0.3\textwidth}
        \includegraphics[width=\textwidth]{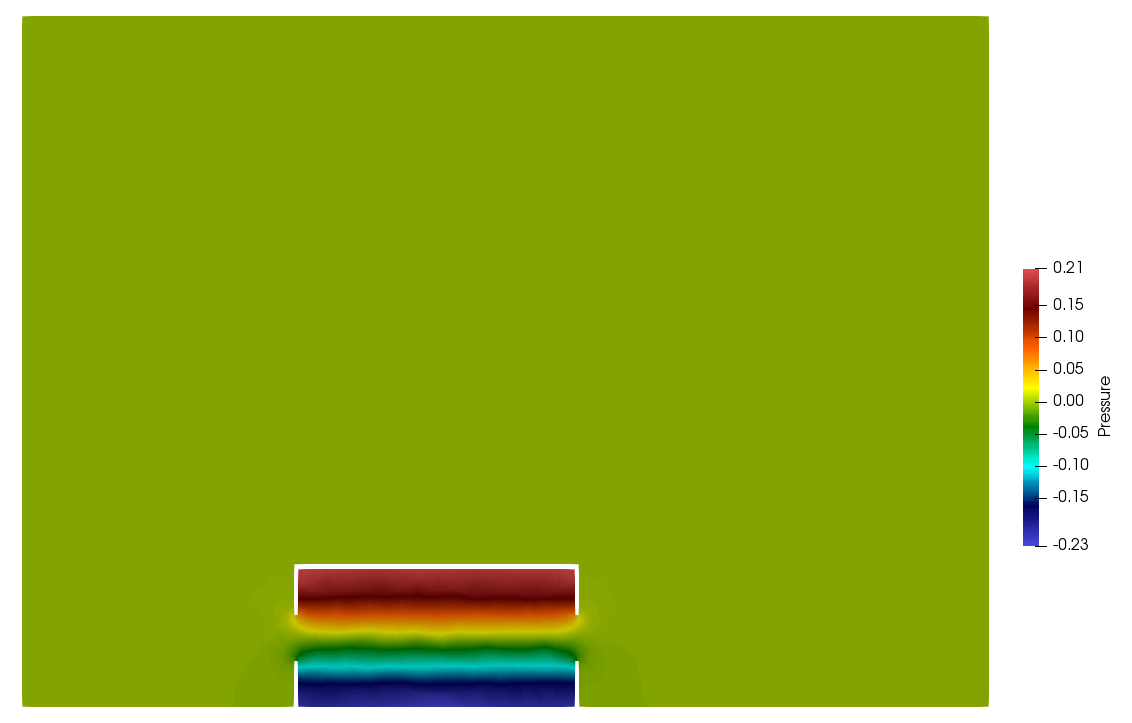}
        \caption{293 K}
        \label{Fig:Tref293K}
    \end{subfigure}
    \begin{subfigure}{0.3\textwidth}
        \includegraphics[width=\textwidth]{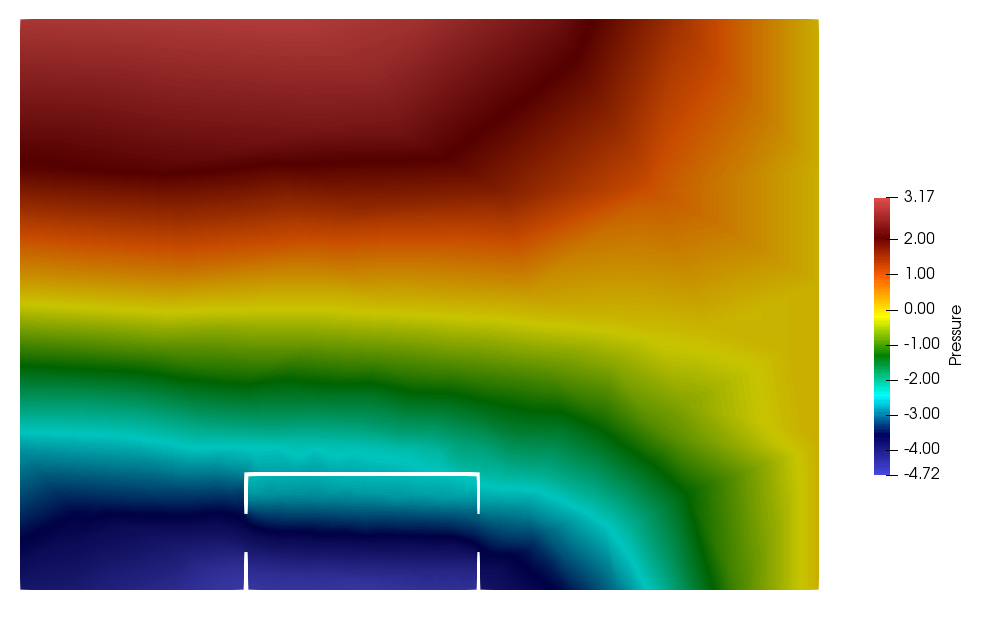}
        \caption{273 K}
        \label{Fig:Tref273K}
    \end{subfigure}
    \begin{subfigure}{0.3\textwidth}
        \includegraphics[width=\textwidth]{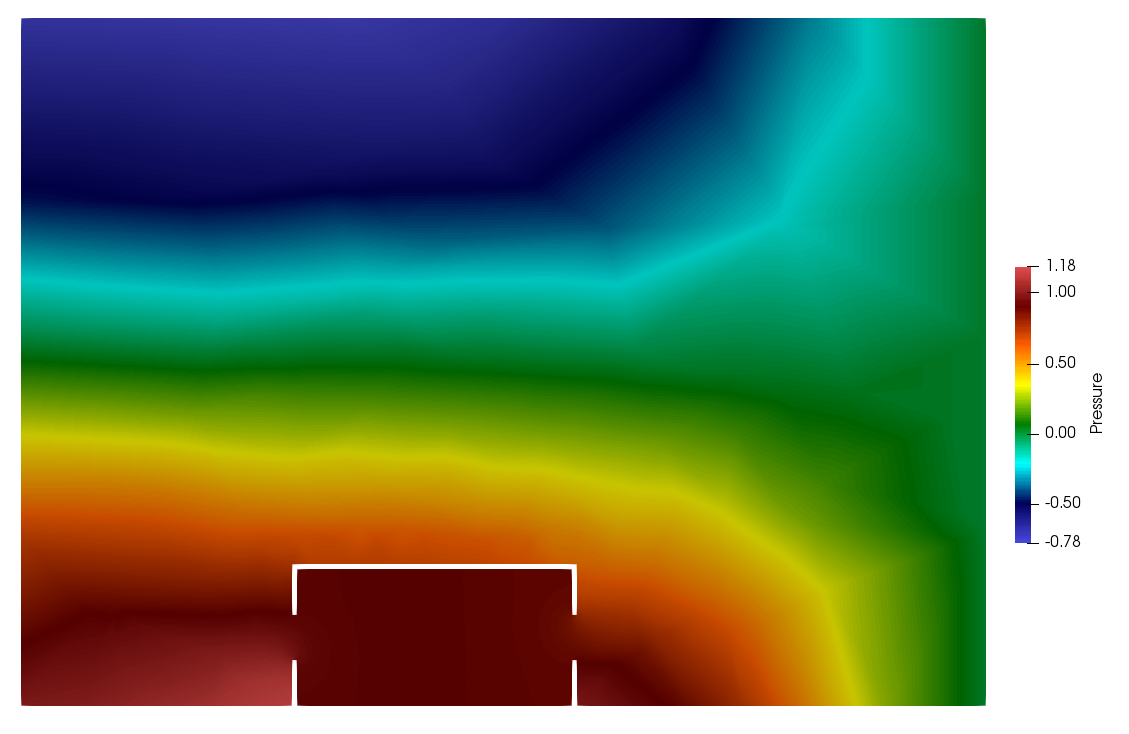}
        \caption{298 K}
        \label{Fig:Tref298K}
    \end{subfigure}
    \caption{Pressure field obtained after the first time step for different reference temperature: (a) 293 K, (b) 273 K and (c) 298 K. The ambient and initial temperature is 293 K for the three cases. Results (a) are the one expected. Results (b) and (c) will generate fake recirculation at the outlet.}
    \label{Fig:ReferenceTemperature}
\end{figure}

In \textbf{Fluidity}, the option \texttt{subtract\_out\_hydrostatic\_level} changes the buoyancy term. For the Boussinesq case, it changes to $g\rho '=g(\rho-\rho_{0})/\rho_{0}$ and this option should always be used. Note that, using Boussinesq, the reference density does not influence any of the terms in the momentum equation. It may however influence the outcome of diagnostic fields depending on density.
\noindent The \textbf{reference temperature}, used in the equation of state, needs to correspond to the \textbf{ambient temperature}. This will avoid unwanted re-circulation near the boundary where the pressure is defined, i.e. at the outlet boundary of the domain in the case presented here. Effect on the pressure field of not using the proper reference temperature is shown in Figure~\ref{Fig:ReferenceTemperature}.

\section{Walls boundary condition} \label{Sec:BCTricks}
See Section~\ref{Sec:BCWalls} for more details.

\noindent The velocity boundary conditions on solid surfaces can be applied either weakly or strongly and can represent a slip or a no-slip behaviour.
\begin{itemize}
    \item It is to be noted that when boundary conditions are applying weakly, the discrete solution will not satisfy the boundary condition exactly. Instead the solution will converge to the correct boundary condition along with the solution in the interior as the mesh is refined. An alternative way of implementing boundary conditions, so called strongly imposed boundary conditions. Although this guarantees that the Dirichlet boundary condition will be satisfied exactly, it does not at all mean that the discrete solution converges to the exact continuous solution more quickly than it would with weakly imposed boundary conditions. Strongly imposed boundary conditions may sometimes be necessary if the boundary condition needs to be imposed strictly for physical reasons. Unlike the strong form of the Dirichlet conditions, weak Dirichlet conditions do not force the solution on the boundary to be point-wise equal to the boundary condition.
\noindent If boundary conditions are applied weakly, then the following options need to be turned on in \textbf{Diamond}:
\begin{itemize}
    \item Under Pressure field: \texttt{spatial\_discretisation/continuous\_galerkin/} \newline
    \texttt{integrate\_continuity\_by\_parts}
    \item Under Velocity field: \texttt{spatial\_discretisation/continuous\_galerkin/} \newline
    \texttt{advection\_terms/integrate\_advection\_by\_parts}
\end{itemize}
    \item A no-slip boundary condition is defined by the three components of the velocity being equal to zero, while a slip boundary condition corresponds to the normal component of the velocity only being equal to zero. Both are Dirichlet type.
\end{itemize}

There are two options available to apply a Dirichlet boundary condition in \textbf{Fluidity}:
\begin{itemize}
    \item \texttt{align\_bc\_with\_cartesian}: the three components of the velocity are assigned - Always work.
    \item \texttt{align\_bc\_with\_surface}: the normal and the two tangential components of the velocity are assigned - Does not always work.
\end{itemize}
The option \texttt{align\_bc\_with\_surface} in the Dirichlet type boundary condition is not well-implemented in \textbf{Fluidity} and not always work properly.

In summary:
\begin{itemize}
    \item The option \texttt{align\_bc\_with\_cartesian} should always be preferred if possible.
    \item Only one no-slip Dirichlet \texttt{align\_bc\_with\_surface} applied strongly is allowed, while several can be used when applied weakly.
    \item For a slip boundary condition weakly applied, \texttt{align\_bc\_with\_cartesian} type for simple geometry or \texttt{no\_normal\_flow} type for any geometry should be used. The Dirichlet \texttt{align\_bc\_with\_surface} type does not work.
    \item A slip boundary condition is recommended instead of a no-slip one if the boundary layer is not going to be fully resolved with the chosen mesh, to avoid weird behaviour of the temperature field.
\end{itemize}

\section{Consistent interpolation}\label{sec:interpolation}
This section is almost a copy-paste of the \textbf{Fluidity} manual~\cite{AMCG2015}, section 7.6 and the user can refer to it for more details. Unless you are running a simulation with discontinuous Galerkin fields, the option \texttt{consistent\_interpolation} needs to be enabled in any prognostic and diagnostic field. Indeed, the application of adaptive re-meshing divides naturally into three sub-problems. The first is to decide what mesh is desired; the second is to actually generate that mesh. The third, discussed here, is how to interpolate any necessary data from the previous mesh to the adapted one. For the third problem, the \texttt{consistent\_interpolation} is almost universally used. This standard method consists of evaluating the previous solution at the locations of the nodes in the adapted mesh. The choice should be \texttt{consistent\_interpolation}, unless any of the following conditions hold:
\begin{itemize}
    \item The simulation has a discontinuous prognostic field which must be interpolated.
    \item Conservation of some field is crucial for the dynamics.
    \item The field is obtained using a time averaged algorithm in Fluidity.
\end{itemize}
\noindent In such cases, \texttt{galerkin\_projection} should be applied. In the case of time averaged fields that will avoid excessive numerical diffusion.

\section{Comparing files}
Files, like \textit{*.flml} files, can be compared using the Command~\ref{Lst:meld}:
\begin{Terminal}[caption={Use of \texttt{meld} to compare two files.}, label={Lst:meld}]
ä\colorbox{davysgrey}{
\parbox{435pt}{
\color{applegreen} \textbf{user@mypc}\color{white}\textbf{:}\color{codeblue}$\sim$
\color{white}\$ meld file1 file2
}}
\end{Terminal}
 
\noindent \texttt{meld} is not a \textbf{Fluidity} tool and needs to be installed if not already using the command:
\begin{Terminal}[caption={\texttt{meld} installation.}, label={Lst:meldInstall}]
ä\colorbox{davysgrey}{
\parbox{435pt}{
\color{applegreen} \textbf{user@mypc}\color{white}\textbf{:}\color{codeblue}$\sim$
\color{white}\$ sudo apt-get install meld
}}
\end{Terminal}

\section{Checkpointing}\label{Sec:Checkpoint}
When running a \textbf{Fluidity} simulation, checkpoint files are outputted every $n$ period, where $n$ is specified by the user under \texttt{io\_checkpointing} (see Figure~\ref{Fig:io}). Checkpoints are useful if a simulation needs to be restarted from a particular point in time. 

\noindent The files needed to checkpoint at a certain time step (\texttt{index}) are the following: 
\begin{itemize}
    \item The \textit{flml} checkpoint file:~\texttt{name\_index\_checkpoint.flml}
    \item The \texttt{CoordinateMesh} \textit{msh} file: ~\texttt{name\_CoordinateMesh\_index\_checkpoint.msh}
    \item The \texttt{PressureMesh} \textit{vtu} file: ~\texttt{name\_PressureMesh\_index\_checkpoint.vtu}
    \item The \texttt{VelocityMesh} \textit{vtu} file: ~\texttt{name\_VelocityMesh\_index\_checkpoint.vtu} 
\end{itemize}
It must be noted that if the simulation is run on parallel on $N$ processors there will be $N$ \texttt{CoordinateMesh} \textit{msh} and \textit{halo} files as well as folders with the $N$ \texttt{PressureMesh} and \texttt{VelocityMesh} \textit{vtu} files on top of the \textit{pvtu} files.

\noindent The new simulation will then be running using the command:
\begin{Terminal}[]
ä\colorbox{davysgrey}{
\parbox{435pt}{
\color{applegreen} \textbf{user@mypc}\color{white}\textbf{:}\color{codeblue}$\sim$
\color{white}\$ <<FluiditySourcePath>>/bin/fluidity -l -v3 name\_index\_checkpoint.flml \&
}}
\end{Terminal}

\noindent It should be repeated that there is currently a bug in \textbf{Fluidity} concerning the time-average field: when a simulation is run from a checkpoint the value of the previous time-average is not properly read from the checkpointed files implying that the average is then not correct. This will later be fixed but for now it is recommended to the user to specify a spin-up time higher than the checkpoint time to start a new time-average.
    \chapter{Fluidity in parallel}
\section{When should \textbf{Fluidity} be run in parallel ?}
As a rule of thumb, there should be 50,000 nodes per processor. If the number of nodes per processor is smaller than that, there will be no benefit in switching from serial to parallel, although the simulation should still run in parallel.
\section{Running on a PC}
In order for a simulation to run in parallel, the mesh first needs to be decomposed using \texttt{flredecomp} (more information on this tool is found in~\cite{AMCG2015}). A mesh is decomposed (from 1 to 4 parts for example) using Command~\ref{Lst:MeshDecomp14}. It must be noted that the number of processors used to run the ~\texttt{mpiexec} command needs to be equal to the maximum number of parts in either the original or new mesh. If a mesh was to be recomposed from 4 parts to 1, the Command~\ref{Lst:MeshDecomp41} is the one to use. Once the mesh is decomposed the simulation can be run in parallel, on 4 processors for example, using Command~\ref{Lst:FluidityParallel}.

\begin{Terminal}[caption={Decomposition of the mesh from 1 processor to 4.}, label={Lst:MeshDecomp14}]
ä\colorbox{davysgrey}{
\parbox{435pt}{
\color{applegreen} \textbf{user@mypc}\color{white}\textbf{:}\color{codeblue}$\sim$
\color{white}\$ mpiexec -n 4 <<FluiditySourcePath>>/bin/flredecomp -i 1 -o 4 foo foo\_new
}}
\end{Terminal}

\begin{Terminal}[caption={Recomposition of the mesh from 4 processors to 1.}, label={Lst:MeshDecomp41}]
ä\colorbox{davysgrey}{
\parbox{435pt}{
\color{applegreen} \textbf{user@mypc}\color{white}\textbf{:}\color{codeblue}$\sim$
\color{white}\$ mpiexec -n 4 <<FluiditySourcePath>>/bin/flredecomp -i 4 -o 1 foo foo\_new
}}
\end{Terminal}

\begin{Terminal}[caption={Running \textbf{Fluidity} in parallel.}, label={Lst:FluidityParallel}]
ä\colorbox{davysgrey}{
\parbox{435pt}{
\color{applegreen} \textbf{user@mypc}\color{white}\textbf{:}\color{codeblue}$\sim$
\color{white}\$ mpiexec -n 4 <<FluiditySourcePath>>/bin/fluidity -l -v3 foo\_new.flml \& 
}}
\end{Terminal}

\noindent Note: The tool \texttt{fldecomp} in \textbf{Fluidity} is obsolete and the tool \texttt{flredecomp} has always to be used.

\section{Running on CX1}
\textbf{Fluidity} can also be run on a computer cluster, and here the example of Imperial cluster CX1 is detailed. More information on CX1 in general is available online, particularly regarding job sizing:
\begin{itemize}
    \item \url{https://www.imperial.ac.uk/admin-services/ict/self-service/rese} \newline \url{arch-support/rcs/computing/high-throughput-computing/}
    \item \url{https://wiki.imperial.ac.uk/display/HPC/MPI+Jobs}
\end{itemize}

\noindent The following \texttt{singlenode.pbs} script shown in Code~\ref{Lst:singlenode} will run a simulation with the pre-installed \textbf{Fluidity} version on the chosen single node queue.

\begin{Code}[language=bash, caption={Bash script to run a simulation with a single node on CX1.}, label={Lst:singlenode}]
#!/bin/sh 
# Job name
#PBS -N name
# Time required in hh:mm:ss
#PBS -l walltime=72:00:00
# Ressource requirements 
#PBS -l select=1:ncpus=16:mem=30gb
# Files to contain standard error and standard output
#PBS -o stdout
#PBS -e stderr
echo Working Directory is $PBS_O_WORKDIR
cd $PBS_O_WORKDIR
rm -f stdout* stderr*
module load ese-software
module load ese-fluidity

cp -r $PBS_O_WORKDIR/* $EPHEMERAL
cd $EPHEMERAL  
pbsexec mpiexec flredecomp -i 1 -o 16 foo foo_decomp
pbsexec mpiexec fluidity foo_decomp.flml -v1 -l
\end{Code}

\noindent Apart from the normal CX1 procedure, it is worth noting the loading of the pre-installed \textbf{Fluidity} modules with \texttt{module load ese-software} and \texttt{module load ese-fluidity}. During and after the simulation, the output can be visualised directly in the \texttt{ephemeral} directory. A maximum wall time in the \textit{flml} can also be specified to match the one in the \textit{pbs} file to ensure the simulation will be terminated on a checkpoint. 

\noindent If a multinode queue is to be used, the mesh needs to be decomposed in the total number of processors used (i.e. for CX1: number of cpus $\times$ number of nodes).
    \chapter{Post-processing data obtained with Fluidity}\label{Sec:PostProcessing}

\section{\textbf{ParaView}}
\begin{figure}
    \centering
    \includegraphics[scale=0.5]{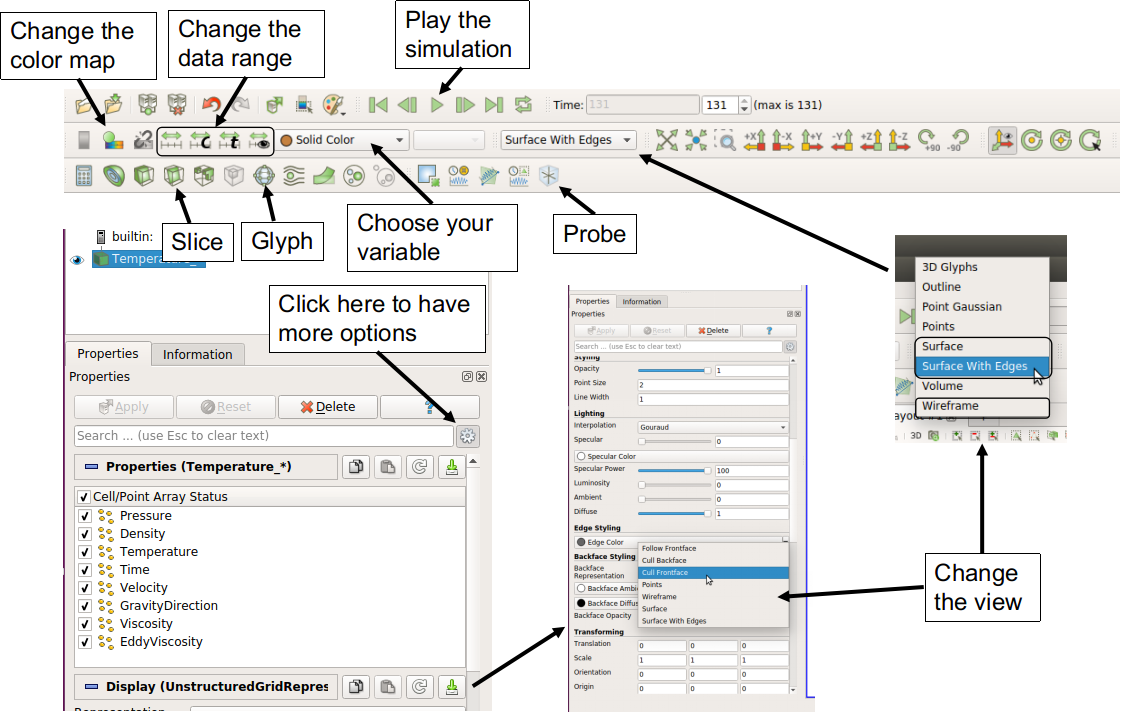}
    \caption{Options in \textbf{ParaView}.}
    \label{fig:ParaviewOptions}
\end{figure}

Simulation's results can be viewed by loading the \textit{vtu} files into \textbf{ParaView}. This software can be downloaded from \url{https://www.paraview.org/}, where a number of useful information can also be found, such as guides and tutorials. The interface is shown in Figure~\ref{fig:ParaviewOptions} and a number of useful features are described below:
\begin{itemize}
    \item \texttt{Slice}: a type of view that allows the user to view an orthogonal slice of the simulation's results.
    \item \texttt{Glyph}: a type of filter that shows the data at specific points as markers with an orientation and size defined by the input. Typically, this tool is used to display velocity vectors.
    \item \texttt{Probe}: a type of filter that gives the values of the fields at a specific location.
    \item \texttt{Surface with edge}: a type of render that shows the values of a field as well as the mesh used during the simulation. 
    \item \texttt{Follow cullback}: is found in \texttt{Backface styling} and allows the user to tune the rendering of the front and back faces (with regards to the camera), modifying the visualisation.
\end{itemize}

\noindent It is then recommended to play around with all \textbf{ParaView} features to become familiar with them.

\section{Python scripts}
The results can also be visualised and analysed using python scripts. The main python library used is the library \textbf{vtk}. In addition, the \textit{vtktools.py} and \textit{vtutools.py} python modules can also be used and imported in every script. \textit{vtktools.py} and \textit{vtutools.py} can be found in \texttt{<<FluiditySourcePath>>} under \texttt{/python/} and \texttt{/python/fluidity/} \texttt{diagnostics/}, respectively. Both are including in the materials attached to this manual. Both scripts are really useful to work, manipulate and post-process \textit{*.vtu} or \textit{*.pvtu} data. 
Lots of example of python script using the \textbf{vtk} library can also be found online:
\begin{itemize}
    \item \url{https://www.vtk.org/doc/nightly/html/pages.html}: the sections Class to Examples and Class to Tests are plenty of python and C++ script examples.
    \item \url{https://lorensen.github.io/VTKExamples/site/Python/} is also full of script using vtk functions.
\end{itemize}
Documentations for \textbf{vtk} can also be found online:
\begin{itemize}
    \item version 5.10 \url{https://www.vtk.org/doc/release/5.10/html/index.html}
    \item version 7.1 \url{https://www.vtk.org/doc/release/7.1/html/}
\end{itemize}
\noindent To take advantage of all the new \textbf{vtk} features, it is recommended to download one of the latest release available online: \url{https://www.vtk.org/} and install it following \url{https://www.vtk.org/Wiki/VTK/Configure_and_Build}. During the installation, in the \textbf{Cmake} interface, use the `advanced mode' (\texttt{t} keyboard) and make sure that \texttt{VTK WRAP PYTHON} is set to \texttt{ON}. Note that installing the \textbf{vtk} library is not trivial and a number of errors can occur. Please ask the help of an advanced user if needed. Once the latest version is installed, the user should use the executable \texttt{vtkpython} located under \texttt{<<vtkPath>>/bin/}.
\noindent Note that they are lots of major change between versions 5 and 6 of \textbf{vtk}. Please refer to the following for more details:
\begin{itemize}
    \item \url{https://www.vtk.org/Wiki/VTK/VTK_6_Migration_Guide}
    \item \url{https://www.vtk.org/Wiki/VTK/VTK_6_Migration/Overview}
\end{itemize}
\noindent \textbf{Note:} the probe tool provided in the \textit{vtktools.py} can behave weirdly when trying to probe data near/on solid walls due to inconsistent behaviour of the \textbf{vtk} library. This weird behaviour might also differ depending the version of \textbf{vtk} you are using. Have a look of the function in \textit{vtktools.py} to know why and/or ask an advanced user to explain what is happening...

\noindent The library \textbf{matplotlib} is the one recommended to plot data.
Following are two examples of python scripts using the module \textit{vtktools.py}, the library \textbf{vtk} and that are inspired by some functions in \textit{vtutools.py}. The following examples of python script will show how to:
\begin{itemize}
    \item Compute the mass flow rate at openings: \textit{MassFlowRate.py}
    \item Compute the radius of a thermal plume: \textit{PlumeRadius.py}
\end{itemize}

\section{Mass flow rate at the openings}
\subsection{Test case}
The script \textit{MassFlowRate.py} is located in \texttt{Scripts/MassFlowRate/}. The following results and discussions are based on data obtained with example \textit{3dBox\_Case11.flml}.

\subsection{Generality and method}
\subsubsection{Equation}
The mass flow rate $\dot{m}$ ($kg/s$) through an opening is defined by equation~\ref{Eq:MassFlowRate}:

\begin{equation}
    \dot{m} = \rho v S
    \label{Eq:MassFlowRate}
\end{equation}

\noindent where $\rho$ is the density of the fluid, $v$ is the magnitude of the normal component of the velocity (normal to the surface considered) and $S$ is the surface of the opening.

\subsubsection{Method} 
When calculating the mass flow rate at openings, to compensate the possible lack of mesh resolution, it is recommended to do an average over several planes through the opening's thickness or (what we think to be less accurate) to compute the mass flow rate over a plane located at the middle of the opening's thickness, i.e thickness/2. In the script \textit{MassFlowRate.py} provided the first option is chosen.
A Cartesian grid is defined through the opening, data are extracted for each points of this grid, then averaged to obtain the final mass flow rate through that particular plane. This procedure is repeated for several planes along the thickness of the opening. Finally, the final mass flow rate is computed averaging mass flow rates obtained for each planes.

\subsubsection{Input variables}
The user input variables are the followings:
\begin{itemize}
    \item \texttt{path\_simu} is the path where the simulation is located.
    \item \texttt{basename} is the name of the simulation.
    \item \texttt{vtu\_start}, \texttt{vtu\_end} and \texttt{vtu\_step} are respectively the first \textit{vtu} file, the last \textit{vtu} file and the step between those \textit{vtu} files that the user want to consider. This allows the user to have the evolution of the mass flow rate as a function of the time.
    \item \texttt{ni\_start}, \texttt{ni\_end} and \texttt{ni\_step}, where \texttt{i} stands for the \texttt{x}, \texttt{y} and \texttt{z} direction, define the size of the grids that the user want consider. This allows the user to test different grid resolutions.
\end{itemize}

\subsection{Results}
\begin{figure}
    \centering
    \includegraphics[scale=0.5]{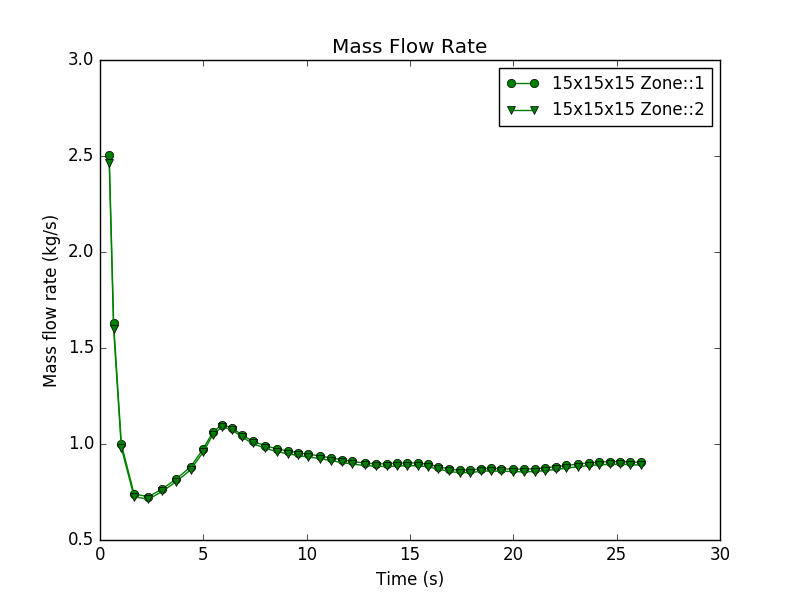}
    \caption{Mass flow rate at the inlet, i.e zone 1, and at the outlet, i.e. zone 2, of the box as a function of time and for a grid resolution $10 \times 10 \times 10$.}
    \label{Fig:MassFlowRate}
\end{figure}

The mass flow rate is evaluated at the two openings of the box. \texttt{Zone1} refers to the opening on the left, i.e. the one closer to the inlet of the domain where the velocity is prescribed, while \texttt{Zone2} refers to the opening on the right, i.e. closer to the outlet of the domain where the pressure is prescribed. In that particular case, the thickness of the openings is towards the x-direction, therefore the $x$-component of the velocity is used. The Cartesian grids where the data are extracted are aligned with $(yOz)$.
Figure~\ref{Fig:MassFlowRate} shows the computed mass flow rate at the inlet and at the outlet of the box as a function of time for a grid resolution $10 \times 10 \times 10$. Playing around with the script, the user will notice that for low grid resolution, the computed mass flow rate at the inlet is not equal to the one at the outlet. Increasing the resolution of the grid, the input and output mass flow rate tends to be more equal to each other, which is expected. Also, increasing the grid resolution tends to converge to the correct mass flow rate but at some point continuing increasing the grid resolution does not improve results. It is also to mention that the computed mass flow rate at the inlet and at the outlet are not strictly equal, due to lack of mesh resolution, and tends to be equal while refining the mesh at the openings.

\noindent The results are written in the text file \textit{MassFlowRate.txt} where the 7 columns corresponds to:
\begin{enumerate}
    \item Resolution used in the $x-$direction
    \item Resolution used in the $y-$direction
    \item Resolution used in the $z-$direction
    \item Time in seconds
    \item Mass flow rate at the inlet of the box in $kg/s$
    \item Mass flow rate at the outlet of the box in $kg/s$
    \item Error in percentage (\%) between the inlet and outlet mass flow rates
\end{enumerate}

\subsection{Numerical implementation}
The results described above are implemented in a python script and the main libraries used are the followings:
\begin{itemize}
    \item \textbf{vtk} is used to read the data obtained with \textbf{Fluidity} and extract data at specific points (Code~\ref{Lst:VTK_Probe}).
    \item \textbf{matplolib} is used to plot the data (Figure~\ref{Fig:MassFlowRate}).
\end{itemize}

\begin{Code}[language=python, firstnumber=20, caption={Function to extract data at specific coordinates using the \textbf{vtk} library.}, label={Lst:VTK_Probe}]
#-------------------------------------------------#
#-- Function to probe data at specific locations -#
#-------------------------------------------------#
def ExtractData(coordinates):

    v   = []
    t   = []
    rho = []
    r   = 0

    for fileID in range(vtu_start,vtu_end+1,vtu_step):
        filename = path_simu+basename+'_'+str(fileID)+'.vtu'
        print '       File::', basename+'_'+str(fileID)+'.vtu'
        # Read the vtu files
        data = vtktools.vtu(filename)
        v.append([])
        rho.append([])
        for j in range(len(coordinates)):
            coord_temp = vtktools.arr([coordinates[j]])
            v[r].append(data.ProbeData(coord_temp, 'Velocity')[0][0]) # X component of the velocity
            rho[r].append(data.ProbeData(coord_temp, 'Density')[0][0])

        r = r+1
        t.append(data.ProbeData(coord_temp, 'Time')[0][0])

    return v, t, rho
\end{Code}

\section{Plume radius}
\subsection{Test case}
The script \textit{PlumeRadius.py} is located in the folder \texttt{Scripts/PlumeRadius}. The script can be tested using the file \textit{Case9b\_Box\_300.vtu} (obtained from \textit{3dBox\_Case9b.flml}) located in the same folder. However, in the following, pictures and plots showing results were obtained using another set of more complicated simulations not provided with this manual. Indeed, in the case presented below, the box is in the middle of the big domain with two openings at the bottom, two openings at the top and a heat source at the bottom of the box. Moreover, if the user wants to use that script then the \texttt{path} (path of where the data are stored), the \texttt{name} (name of the simulation) and the \texttt{vtu\_start} (the number of the vtu) need to be changed in the script according to the user's requirements.
\noindent Also note that the data presented below is obtained from instantaneous data. In practice, the user should have included the time average of the velocity (and of any other interesting quantities) in the simulation and those quantities should be used to compute the time average radius of the plume. In that case, the line \texttt{Vz=up.GetField('Velocity')[:,2]} should be replaced by \texttt{Vz=up.GetField('VelocityAverage')[:,2]}, assuming that \texttt{VelocityAverage} is the name of the time average velocity field chosen by the user in \textbf{Diamond}.

\subsection{Generality}
The radius $r$ of the plume is usually defined by equation~\ref{Eq:PlumeRadius}:
\begin{equation}
    r=\alpha V_{max}
    \label{Eq:PlumeRadius}
\end{equation}
\noindent where $V_{max}$ is the time-averaged maximum velocity and $\alpha$ is a threshold usually equal to 0.02.
\noindent In the following, it is assumed that the $z$-axis is pointing up. Therefore, the maximum velocity is given by the $z$-component of the velocity, i.e $v_{z}$.

\begin{figure}
    \centering
    \includegraphics[scale=0.3]{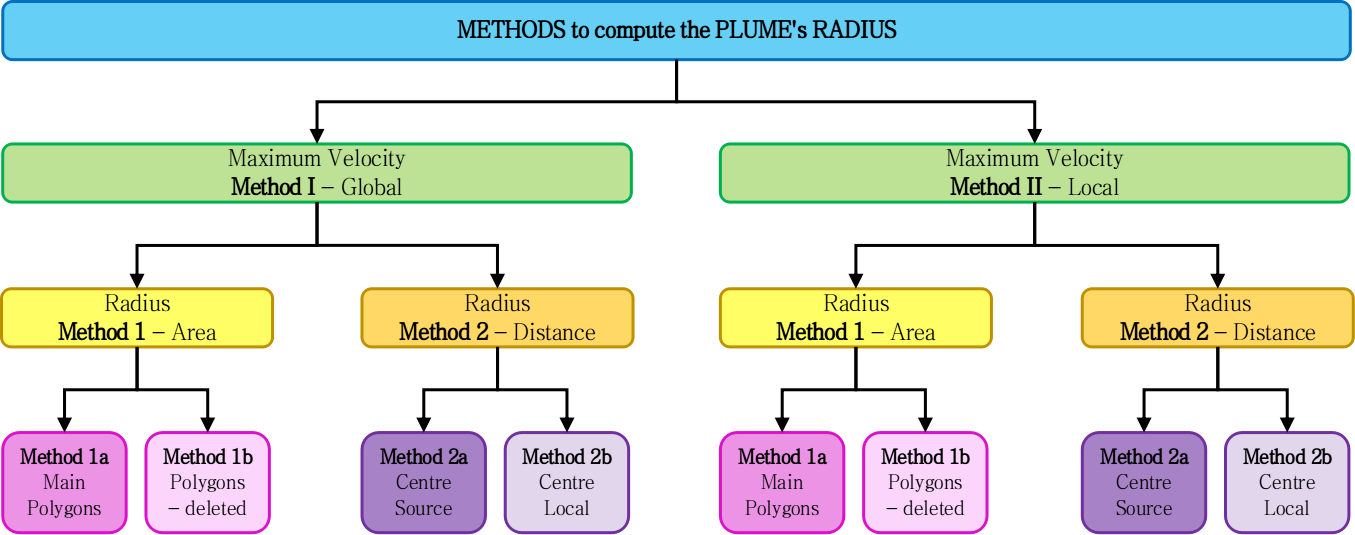}
    \caption{Summary of the different methods tested to compute the radius of the plume as a function of the height.}
    \label{Fig:PlumeRadius_Methods}
\end{figure}

\noindent The Figure~\ref{Fig:PlumeRadius_Methods} summarises the different methods tested to compute the radius of the plume as a function of the height. All these methods are described in detail in the following sections.

\subsection{Description of the methods}
\textbf{a) Algorithms} \newline
\noindent The general idea is to get the geometry of the plume at different height and deduce the radius of the plume based on the geometry. Based on equation~\ref{Eq:PlumeRadius}, two methods are tested and described in the following sections:
\begin{itemize}
    \item Method I: Based on the global maximum velocity: $V_{z,max,global}$
    \item Method II: Based on the local maximum velocity: $V_{z,max,local}(z)$
\end{itemize}

\noindent \textbf{\textit{Method I based on the global velocity}} \newline
\noindent This method consists of generating a contour based on the global maximum velocity and then creating slice cuts at every height.

\begin{algorithm}
    \begin{algorithmic}[1]
    \State Find $V_{z,max,global}$ \label{Alg:alg1_step1}
    \State Extract the 3D contour of value $\alpha V_{z,max,global}$ (Figure~\ref{Fig:Algo1_Step2}). This contour is saved in the file \textit{VelocityZ\_Contour\_Global.vtu} \label{Alg:alg1_step2}
    \For {every height $z$:}
    \State Do a slice which gives the 2D geometry of the plume at $z$ (Figure~\ref{Fig:Algo1_Step3}). \State Contours are saved in \textit{VelocityZSliceContour\_Global\_n.vtu} where $n$ is the height index. \label{Alg:alg1_step3}
    \State Get all the polygons which define the plume geometry \label{Alg:alg1_step4}
    \State Clean the geometry \label{Alg:alg1_step5}
    \State Compute the different radii \label{Alg:alg1_step6}
    \EndFor
    \end{algorithmic}
\caption{Method I}\label{Alg:Algorithm1}
\end{algorithm}

\begin{figure}
    \centering
    \begin{subfigure}{0.3\textwidth}
        \includegraphics[width=\textwidth]{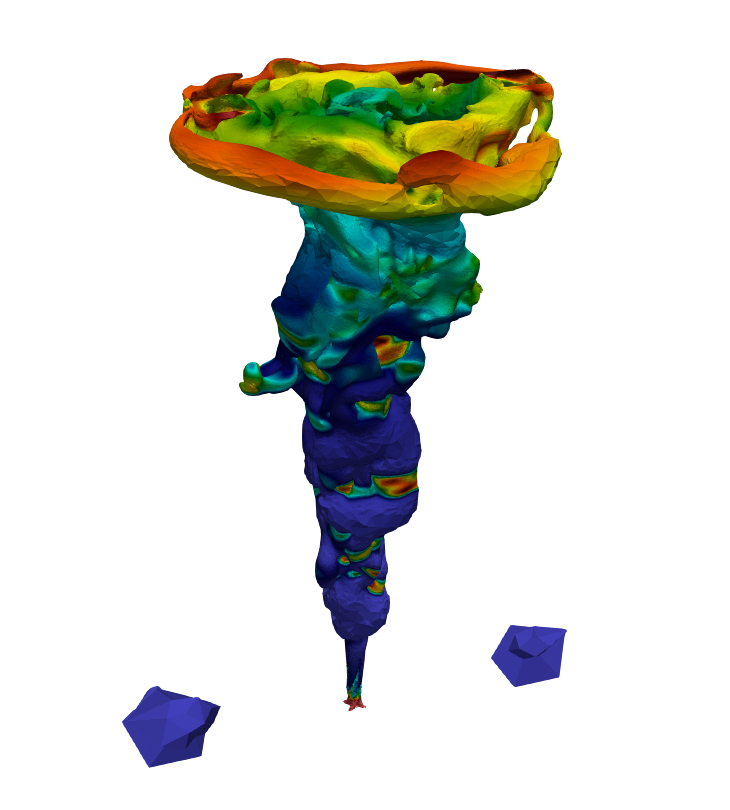}
        \caption{}
        \label{Fig:Algo1_Step2}
    \end{subfigure}
    \begin{subfigure}{0.3\textwidth}
        \includegraphics[width=\textwidth]{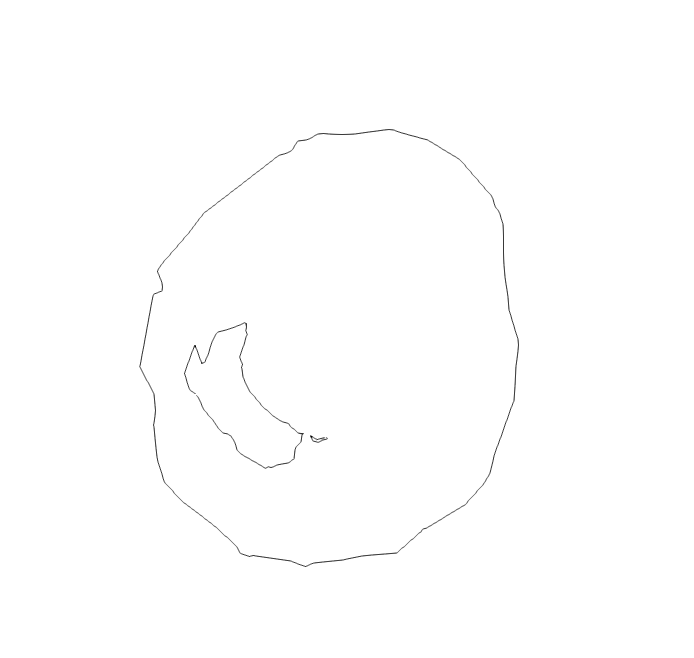}
        \caption{}
        \label{Fig:Algo1_Step3}
    \end{subfigure}
    \caption{(a) 3D contour obtained for the value $\alpha V_{z,max,global}$ (Step~\ref{Alg:alg1_step2} of Algorithm~\ref{Alg:Algorithm1}) and (b) 2D contour obtained at height $z$ (Step~\ref{Alg:alg1_step3} of Algorithm~\ref{Alg:Algorithm1}).}
    \label{Fig:Algorithm1}
\end{figure}

\noindent \textbf{\textit{Method II based on the local velocity}} \newline
\noindent This method consists of generating slices at every height, then determining the local maximum velocity and finally generating the contour of the plume.

\begin{algorithm}
    \caption{Method II}\label{Alg:Algorithm2}
    \begin{algorithmic}[1]
    \For {every height $z$:}
    \State Do a slice at the height $z$ (Figure~\ref{Fig:Algo2_Step2}). Slice is saved in \textit{VelocityZSlice\_Local\_n.vtu} where $n$ is the height index \label{Alg:alg2_step1}
    \State Find $V_{z,max,local}(z)$ \label{Alg:alg2_step2}
    \State Extract the 2D contour of value $\alpha V_{z,max,local}(z)$ (Figure~\ref{Fig:Algo2_Step3}). This contour is saved in \textit{VelocityZ\_ContourSlice\_Local\_n.vtu} where $n$ is the height index. \label{Alg:alg2_step3}
    \State Get all the polygons which define the plume geometry \label{Alg:alg2_step4}
    \State Clean the geometry \label{Alg:alg2_step5}
    \State Compute the different radii \label{Alg:alg2_step6}
    \EndFor
    \end{algorithmic}
\end{algorithm}

\begin{figure}
    \centering
    \begin{subfigure}{0.5\textwidth}
        \includegraphics[width=\textwidth]{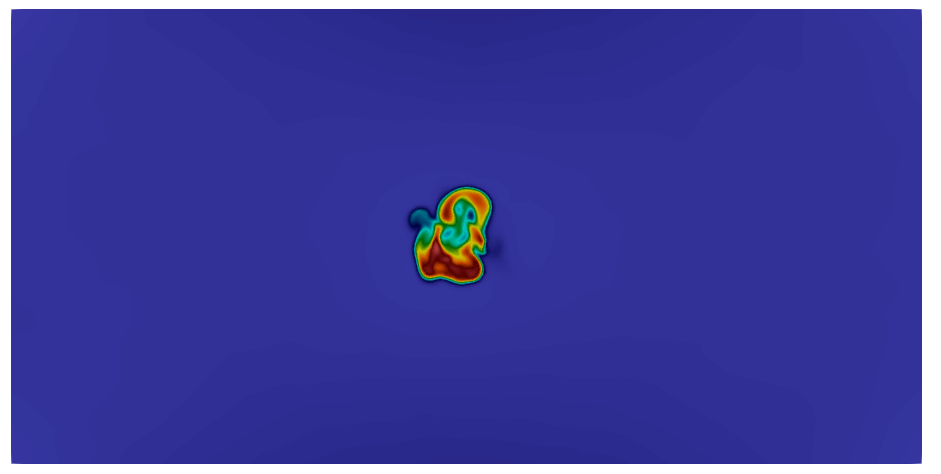}
        \caption{}
        \label{Fig:Algo2_Step2}
    \end{subfigure}
    \begin{subfigure}{0.2\textwidth}
        \includegraphics[width=\textwidth]{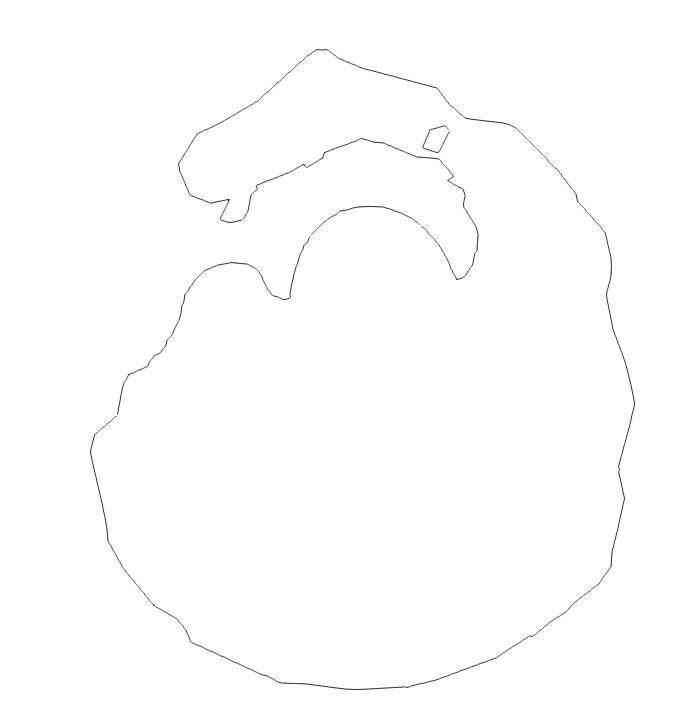}
        \caption{}
        \label{Fig:Algo2_Step3}
    \end{subfigure}
    \caption{(a) 2D slice at height $z$ (Step~\ref{Alg:alg2_step1} of Algorithm~\ref{Alg:Algorithm2}) and (b) 2D contour obtained for the value $\alpha V_{z,max,local}(z)$ (Step~\ref{Alg:alg2_step3} of Algorithm~\ref{Alg:Algorithm2}).}
    \label{Fig:Algorithm2}
\end{figure}

\begin{figure}
    \centering
    \begin{subfigure}{0.1\textwidth}
        \includegraphics[width=\textwidth]{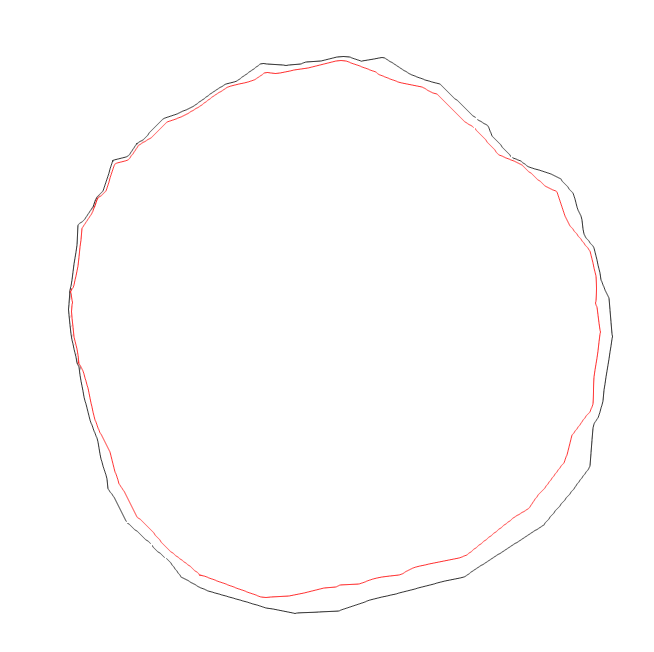}
        \caption{}
        \label{Fig:Comparison1}
    \end{subfigure}
    \begin{subfigure}{0.2\textwidth}
        \includegraphics[width=\textwidth]{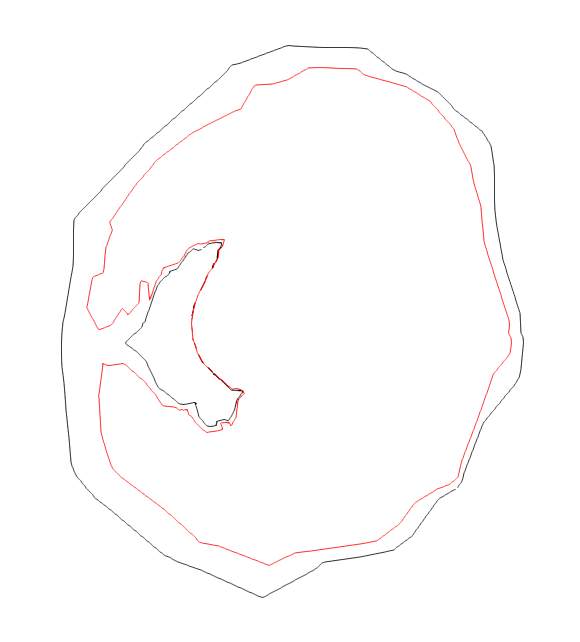}
        \caption{}
        \label{Fig:Comparison2}
    \end{subfigure}
    \begin{subfigure}{0.3\textwidth}
        \includegraphics[width=\textwidth]{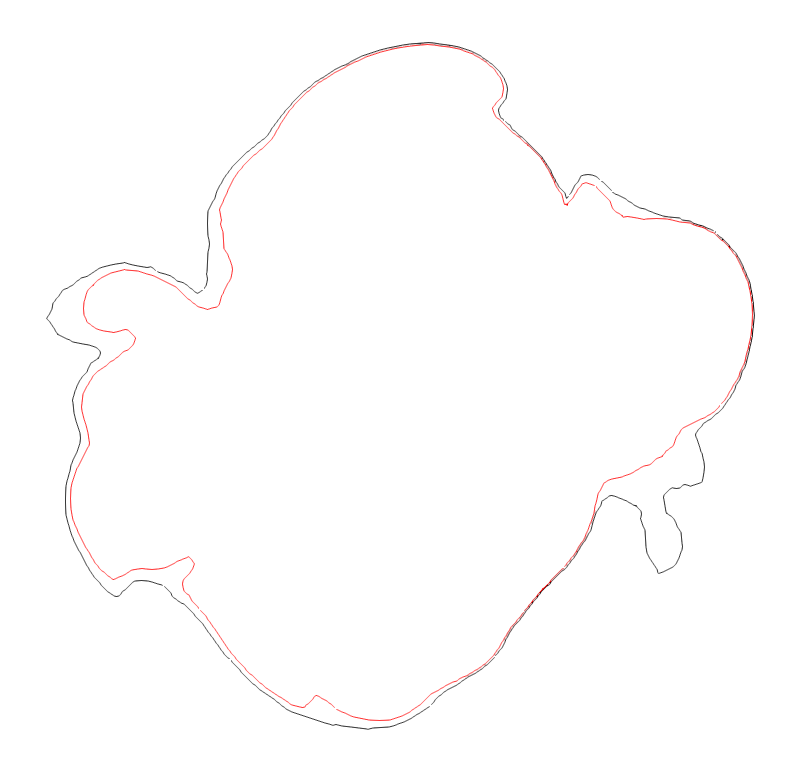}
        \caption{}
        \label{Fig:Comparison3}
    \end{subfigure}
    \caption{Contour of the plume at different heights generated by Algorithm~\ref{Alg:Algorithm1} (red lines) and Algorithm~\ref{Alg:Algorithm2} (black lines) before the geometry cleaning.}
    \label{Fig:ComparisonAlgo}
\end{figure}

\noindent \textbf{\textit{Comparison of the two methods}} \newline
\noindent Figure~\ref{Fig:ComparisonAlgo} shows the plume’s contour at different height generated by Algorithm~\ref{Alg:Algorithm1} and Algorithm~\ref{Alg:Algorithm2}. The red lines are the results obtained for Algorithm~\ref{Alg:Algorithm1}, while black lines are the results obtained using Algorithm~\ref{Alg:Algorithm2}.
It can be noted that Algorithm~\ref{Alg:Algorithm1} generally predicts a geometry of smaller area than the one found with Algorithm~\ref{Alg:Algorithm2} (as also shown in Figure~\ref{Fig:ComparisonAlgo}).

\begin{figure}
    \centering
    \begin{subfigure}{0.3\textwidth}
        \includegraphics[width=\textwidth]{PostProcess/PlumeRadiusFigures/Algo1_SliceStep3.png}
        \caption{}
        \label{Fig:MultiplePoly_Plume}
    \end{subfigure}
    \begin{subfigure}{0.4\textwidth}
        \includegraphics[width=\textwidth]{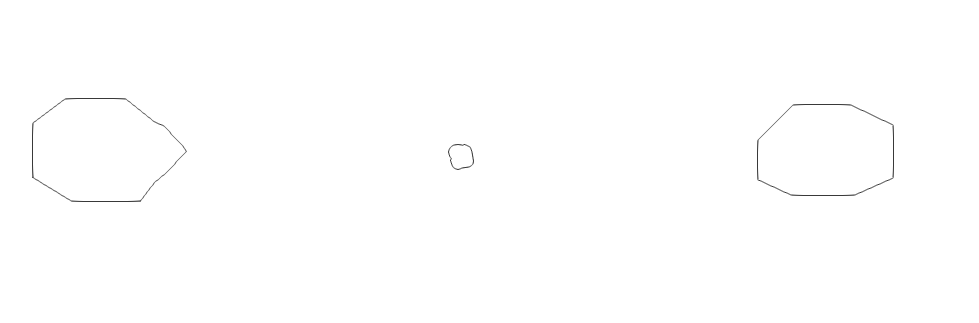}
        \caption{}
        \label{Fig:MultiplePoly_Inlet}
    \end{subfigure}
    \caption{(a) Two polygons are within the main polygon defining the plume. These two polygons need to be detected and ignored. (b) Contour of $\alpha V_{z,max}$ at a low height showing that the right and left polygons are the consequence of the inlets. These two polygons do not belong to the plume and need to be detected and ignored.}
    \label{Fig:MultiplePoly}
\end{figure}

\noindent \textbf{b) Cleaning the geometry - Step~\ref{Alg:alg2_step5} of the algorithms} \newline
\noindent The process of cleaning the geometry is explained below:
\begin{itemize}
    \item The polygons within the main plume geometry are deleted. In Figure~\ref{Fig:MultiplePoly_Plume}, for example, the polygons within the main polygon are not taken into account in the later computation of the radius. However, the area of these polygons is kept (see section 2.3.1 - method 1b to know why and have more details).
    \item When $z$ is low, the contour obtained shows the influence of the inlets. Figure~\ref{Fig:MultiplePoly_Inlet} and Figure~\ref{Fig:Algo1_Step2} are good examples showing the influence of the inlets: indeed the right and the left polygons do not define the main plume geometry. Therefore, these polygons need to be detected and deleted.
\end{itemize}

\noindent \textbf{c) Radius of the plume - Step~\ref{Alg:alg2_step6} of the algorithms} \newline
\noindent The computation of the radius is performed based on two different methods described in the following sections:
\begin{itemize}
    \item Method 1: the radius is computed based on the area of the polygons.
    \item Method 2: the radius is computed based on the distance between the center and the exterior boundary of the polygons.
\end{itemize}

\begin{figure}
    \centering
    \includegraphics[scale=0.3]{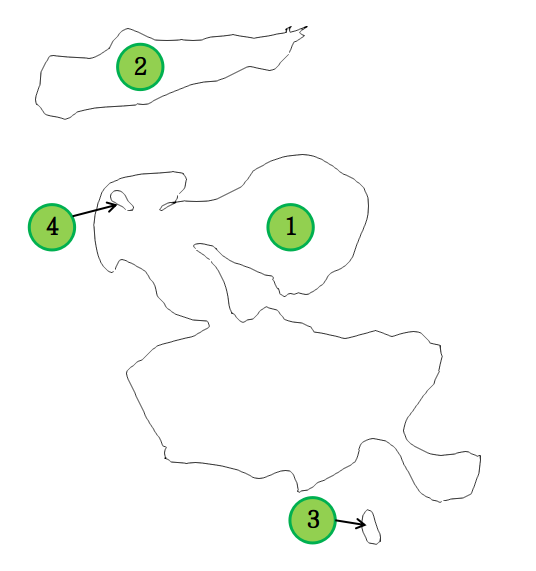}
    \caption{2D geometry of the plume at height z composed of 3 main polygons (labelled 1, 2 and 3). Polygon 1 has another polygon (labelled 4) within.}
    \label{Fig:CleanGeom}
\end{figure}

\noindent \textbf{\textit{Method 1 based on the area}} \newline
\noindent This method consists of computing the radius $r$ of the plume assuming an ideal circle, such that:
\begin{equation}
    r=\sqrt{\frac{\mathcal{A}}{\pi}}
    \label{Eq:PlumeRadiusArea}
\end{equation}

\noindent where $\mathcal{A}$ is the total area of the plume. The aim of this method is to compute $\mathcal{A}$ and then determine the radius of the plume based on equation~\ref{Eq:PlumeRadiusArea}.
This method is explained based on Figure~\ref{Fig:CleanGeom} geometry. Figure~\ref{Fig:CleanGeom} shows the 2D geometry of the plume at a certain height $z$. The geometry is composed by 3 main polygons (labelled 1, 2 and 3). The polygon 1 has also another polygon (labelled 4) within. Based on the cleaning process described in the previous paragraph, the small polygon 4 is detected. The total area $\mathcal{A}$ is therefore computed in two different ways as follows: 
\begin{itemize}
    \item \textbf{Method 1a} - polygon 4 is ignored.
    \begin{equation}
        \mathcal{A} = \mathcal{A}_{1} + \mathcal{A}_{2} + \mathcal{A}_{3}
        \label{Eq:PlumeRadiusAreaAll}
    \end{equation}
    \item \textbf{Method 1b} - polygon 4 is taken into account and subtracted to the total area.
    \begin{equation}
        \mathcal{A} = \mathcal{A}_{1} + \mathcal{A}_{2} + \mathcal{A}_{3} - \mathcal{A}_{4}
        \label{Eq:PlumeRadiusAreaIgnored}
    \end{equation}
\end{itemize}

\begin{figure}
    \centering
    \begin{subfigure}{0.3\textwidth}
        \includegraphics[width=\textwidth]{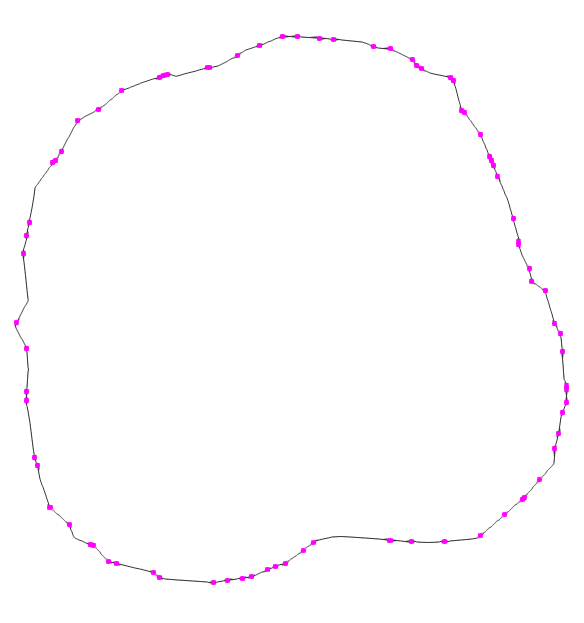}
        \caption{}
        \label{Fig:Radius_Dots_1Poly}
    \end{subfigure}
    \begin{subfigure}{0.3\textwidth}
        \includegraphics[width=\textwidth]{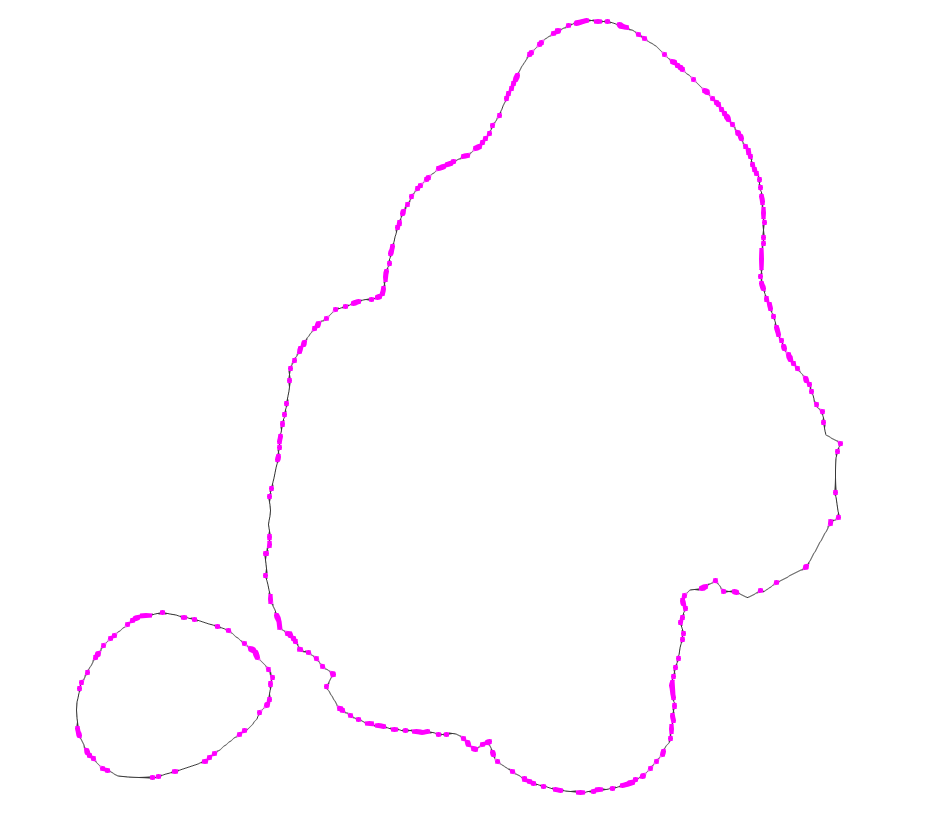}
        \caption{}
        \label{Fig:Radius_Dots_2Poly}
    \end{subfigure}
    \caption{2D geometry of the plume at two different heights. The pink dots show the locations of the nodes defining the boundary of the polygon. (a) Case of a single polygon and (b) case of multiple polygons.}
    \label{Fig:RadiusDots}
\end{figure}

\noindent \textbf{\textit{Method 2 based on the distance}} \newline
\noindent In this section, the radius is computed based on the distance between a centre point of the plume and the nodes defining the exterior boundaries of the polygons (themselves defining the plume geometry) and denoted by pink dots in Figure~\ref{Fig:RadiusDots}. A simple arithmetic mean is performed:
\begin{equation}
    r = \frac{\sum{d_{i}}}{N}
    \label{Eq:PlumeRadiusDistance}
\end{equation}
where $d_{i}$ is the distance between the node $i$ and the centre of the plume and $N$ is the total number of nodes defining the polygons.

\noindent Now the main question consists of determining the centre of the plume. For this, two approaches are tested:
\begin{itemize}
    \item \textbf{Method 2a:} the centre of the plume is assumed to be the centre of the source.
    \item \textbf{Method 2b:} the centre of the plume is taken locally at each height. For this, the centroid of the polygons defining the plume geometry is used as the local centre.
\end{itemize}

\noindent It has to be noted that this method is somehow not really accurate for the following reason:
\begin{itemize}
    \item Even if the method is acceptable in the case shown in Figure~\ref{Fig:Radius_Dots_1Poly}, it is not the case for the case shown in Figure~\ref{Fig:Radius_Dots_2Poly}. Indeed, the nodes of the small polygon will artificially weigh the mean radius towards the distance between this polygon and the centre of the plume.
    \item For both case shown in Figure~\ref{Fig:RadiusDots}, it can be observed that certain region has lot of point defining the boundary (it happens where there is sharp change in the geometry) while other regions have less nodes. Once again, the computed radius will be artificially weighted by the region where there are lots of nodes, which is not desirable.
\end{itemize}

\noindent However, comparing this method with the one using the area, the mean radii generated are quite close.

\subsection{Outputs and plots}
\subsubsection{Summary of the methods}
In summary, these methods give as output:
\begin{itemize}
    \item \textbf{Method I.1.a} $r$ based on the global velocity using the area of the polygons. 
    \item \textbf{Method I.1.b} $r$ based on the global velocity using the area of the polygons minus the area of the polygons deleted.
    \item \textbf{Method I.2.a} $r$ based on the global velocity using the distance between the centre of the source and the boundaries of the polygons.
    \item \textbf{Method I.2.b} $r$ based on the global velocity using the distance between the local centre and the boundaries of the polygons.
    \item \textbf{Method II.1.a} $r$ based on the local velocity using the area of the polygons.
    \item \textbf{Method II.1.b} $r$ based on the local velocity using the area of the polygons minus the area of the polygons deleted.
    \item \textbf{Method II.2.a} $r$ based on the local velocity using the distance between the centre of the source and the boundaries of the polygons.
    \item \textbf{Method II.2.b} $r$ based on the local velocity using the distance between the local centre and the boundaries of the polygons.
\end{itemize}

\subsubsection{Text files with data}
The data are written in text files and each files are explained below.

\begin{itemize}
\item \textbf{Radius\_Global.txt} \newline
Radii computed by method I - Global Velocity - 5 columns
\begin{enumerate}
    \item The height $z$ in meters
    \item $r$ computed by method I.2.a
    \item $r$ computed by method I.2.b
    \item $r$ computed by method I.1.a
    \item $r$ computed by method I.1.b
\end{enumerate}

\item \textbf{Radius\_Local.txt} \newline
Radii computed by method II - Local Velocity - 5 columns
\begin{enumerate}
    \item The height $z$ in meters
    \item $r$ computed by method II.2.a
    \item $r$ computed by method II.2.b
    \item $r$ computed by method II.1.a
    \item $r$ computed by method II.1.b
\end{enumerate}

\item \textbf{Centre\_Global.txt} \newline
Centre of the plume computed by method I - Global Velocity - 5 columns
\begin{enumerate}
    \item The height $z$ in meters
    \item $x$ coordinate of the local centre computed by method I.2.b
    \item $y$ coordinate of the local centre computed by method I.2.b
    \item $x$ coordinate of the source
    \item $y$ coordinate of the source
\end{enumerate}
Note that columns 4 and 5 have constant values.

\item \textbf{Centre\_Local.txt} \newline
Centre of the plume computed by method II - Local Velocity - 5 columns
\begin{enumerate}
    \item The height $z$ in meters
    \item $x$ coordinate of the local centre computed by method II.2.b
    \item $y$ coordinate of the local centre computed by method II.2.b
    \item $x$ coordinate of the source
    \item $y$ coordinate of the source
\end{enumerate}
Note that columns 4 and 5 have constant values.

\item \textbf{Location\_VelocityZ\_max.txt} \newline
Location of the velocity max in the domain - 6 columns
\begin{enumerate}
    \item $x$ coordinate of the location of the maximum velocity for method II
    \item $y$ coordinate of the location of the maximum velocity for method II
    \item $z$ coordinate of the location of the maximum velocity for method II
    \item $x$ coordinate of the location of the maximum velocity for method I
    \item $y$ coordinate of the location of the maximum velocity for method I
    \item $z$ coordinate of the location of the maximum velocity for method I
\end{enumerate}
Note that columns 4, 5 and 6 have constant values.
\end{itemize}

\subsubsection{Plots}
\begin{figure}
    \centering
    \begin{subfigure}{0.3\textwidth}
        \includegraphics[width=\textwidth]{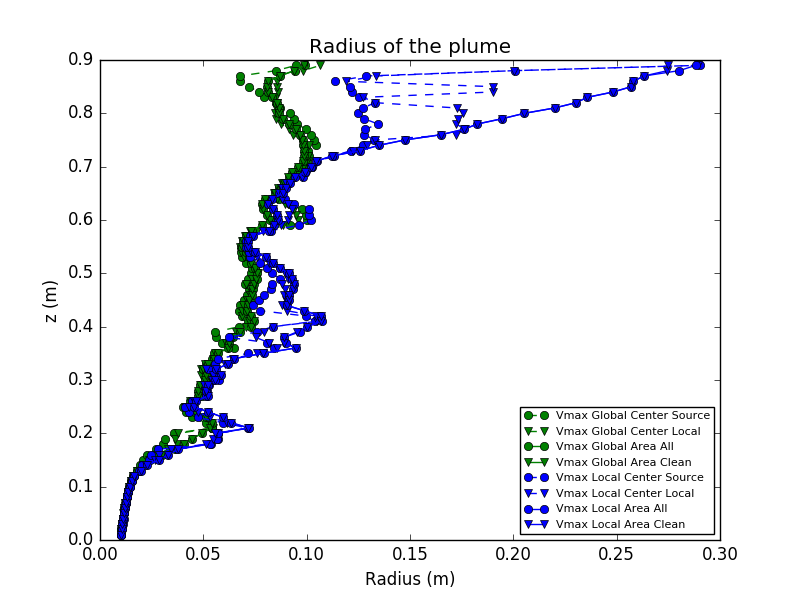}
        \caption{}
        \label{Fig:Radius_All}
    \end{subfigure}
    \begin{subfigure}{0.3\textwidth}
        \includegraphics[width=\textwidth]{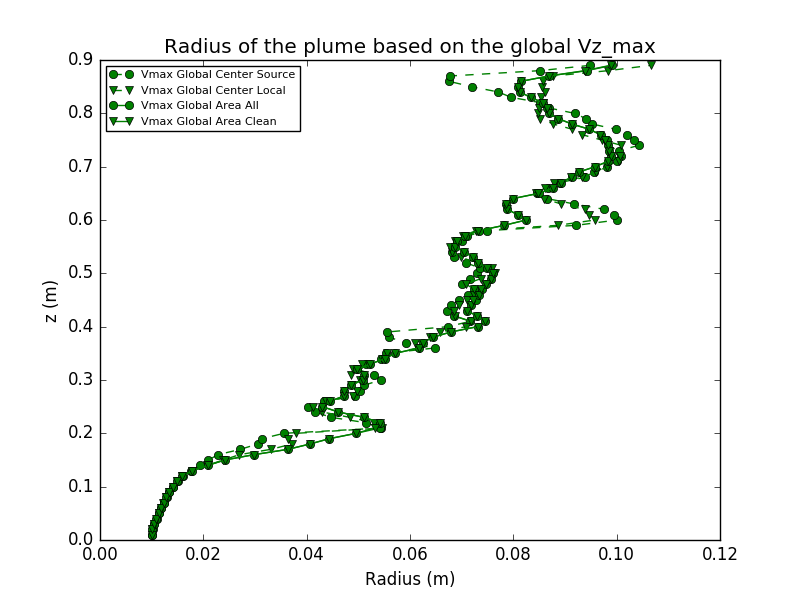}
        \caption{}
        \label{Fig:Radius_Global}
    \end{subfigure}
    \begin{subfigure}{0.3\textwidth}
        \includegraphics[width=\textwidth]{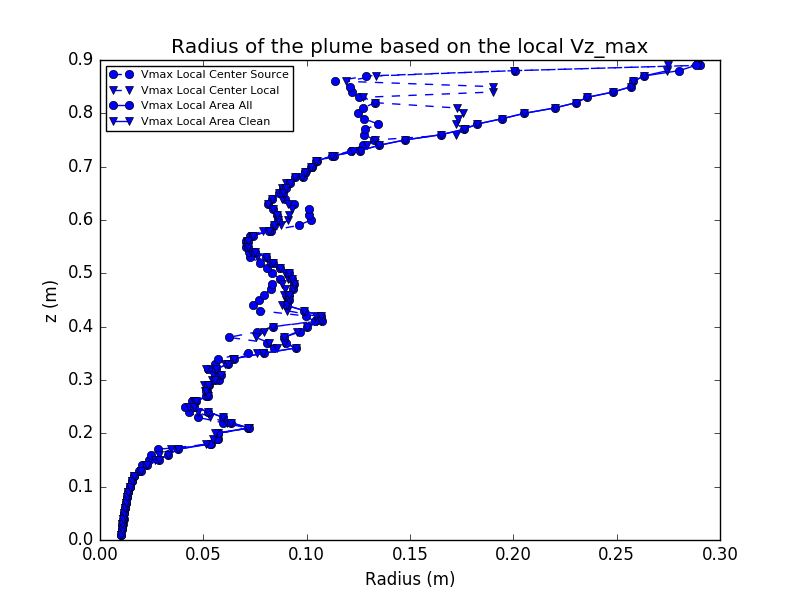}
        \caption{}
        \label{Fig:Radius_Local}
    \end{subfigure}
    \caption{Variation of the plume’s radius as a function of the height. (a) Using all the methods presented, (b) using method I and (c) using method II.}
    \label{Fig:PlotsRadius}
\end{figure}

\begin{figure}
    \centering
    \begin{subfigure}{0.4\textwidth}
        \includegraphics[width=\textwidth]{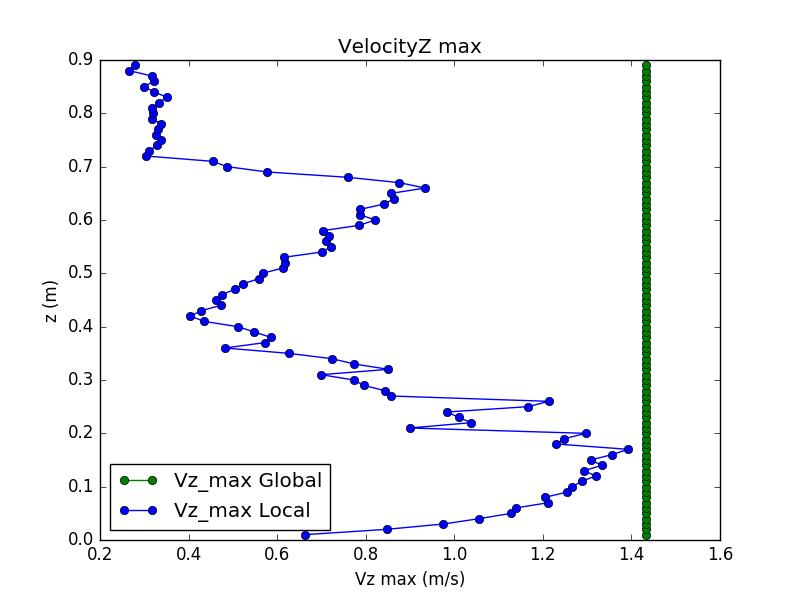}
        \caption{}
        \label{Fig:PlotVmax}
    \end{subfigure}
    \begin{subfigure}{0.4\textwidth}
        \includegraphics[width=\textwidth]{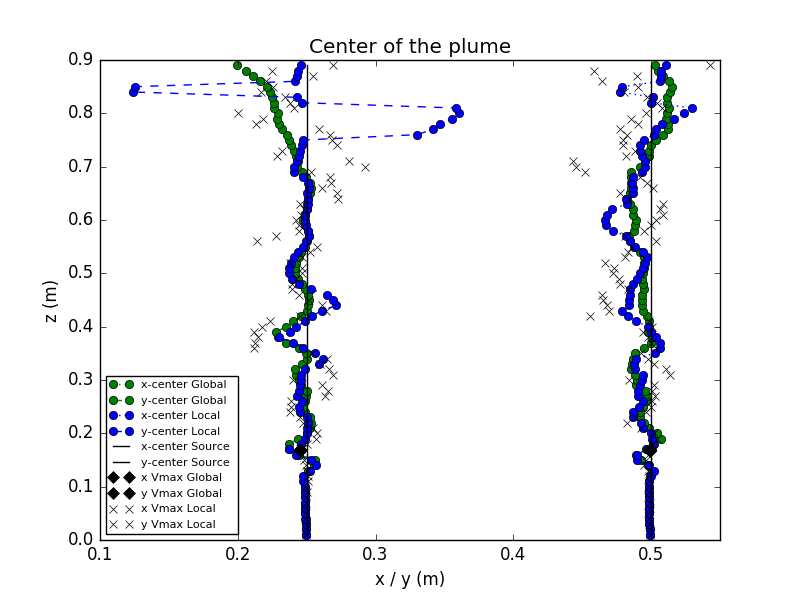}
        \caption{}
        \label{Fig:PlotCenter}
    \end{subfigure}
    \caption{(a) Variation of the maximum velocity as a function of the height for methods I and II. (b) Location of the plume’s centre as a function of the height for methods I.2 and II.2.}
    \label{Fig:PlotVmaxCenter}
\end{figure}

\begin{figure}
    \centering
    \begin{subfigure}{0.4\textwidth}
        \includegraphics[width=\textwidth]{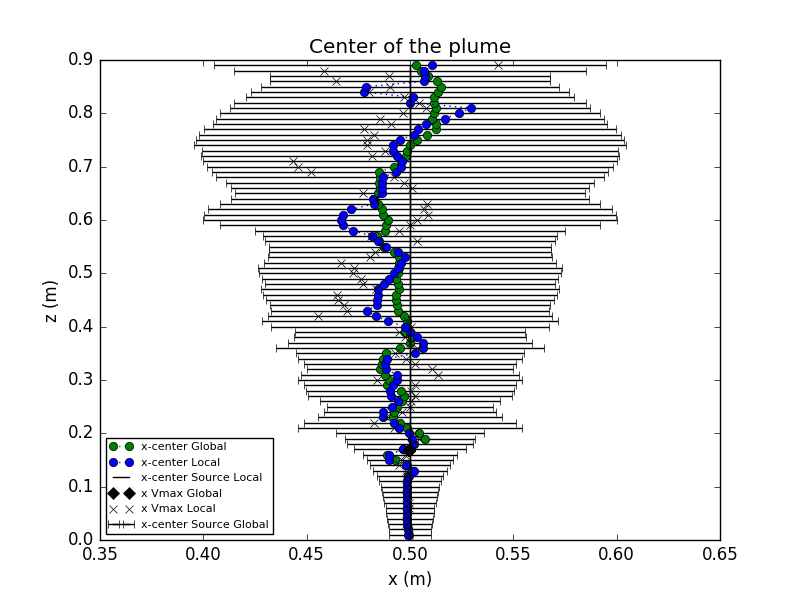}
        \caption{}
        \label{Fig:Plot_x_CenterSource_VzGlobal}
    \end{subfigure}
    \begin{subfigure}{0.4\textwidth}
        \includegraphics[width=\textwidth]{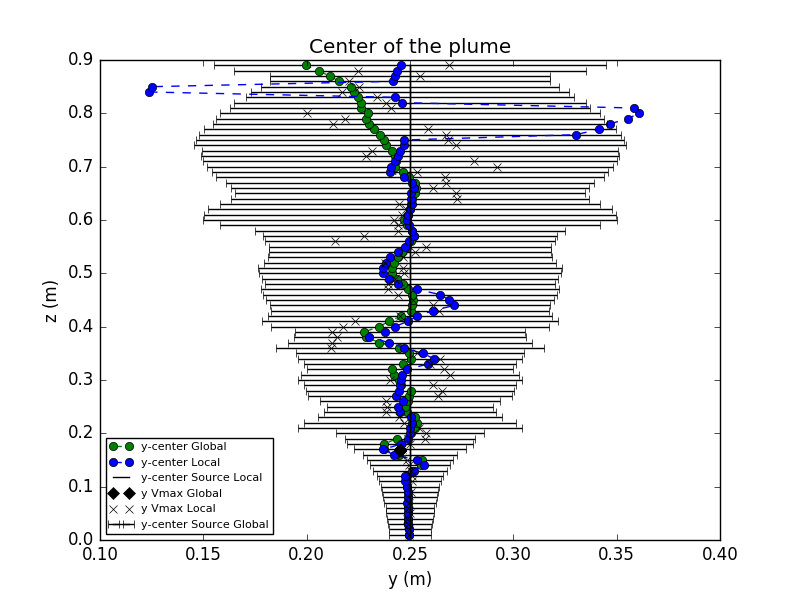}
        \caption{}
        \label{Fig:Plot_y_CenterSource_VzGlobal}
    \end{subfigure}
    \caption{(a) $x$ and (b) $y$ location of the plume’s centre. The error bars depict the computed radius at each height using the method I.2.a.}
    \label{Fig:Plot_CenterSource_VzGlobal}
\end{figure}

\begin{figure}
    \centering
    \begin{subfigure}{0.4\textwidth}
        \includegraphics[width=\textwidth]{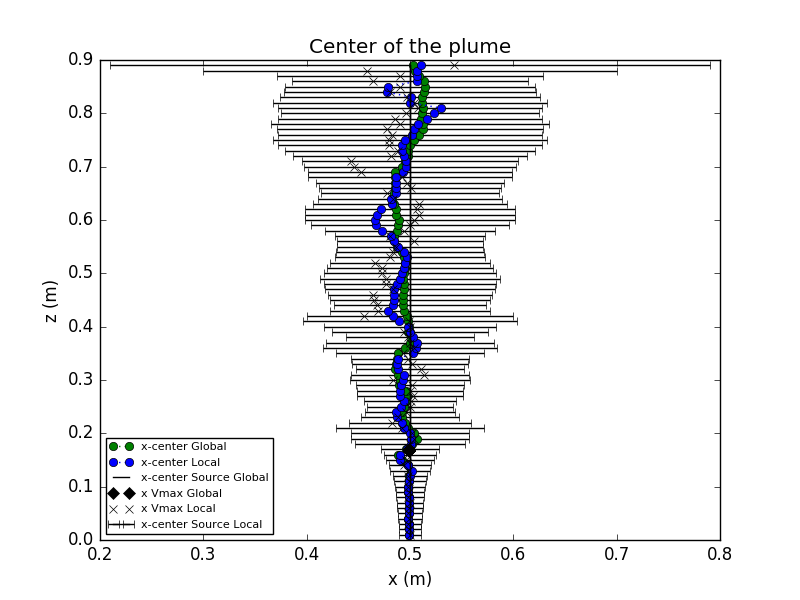}
        \caption{}
        \label{Fig:Plot_x_CenterSource_VzLocal}
    \end{subfigure}
    \begin{subfigure}{0.4\textwidth}
        \includegraphics[width=\textwidth]{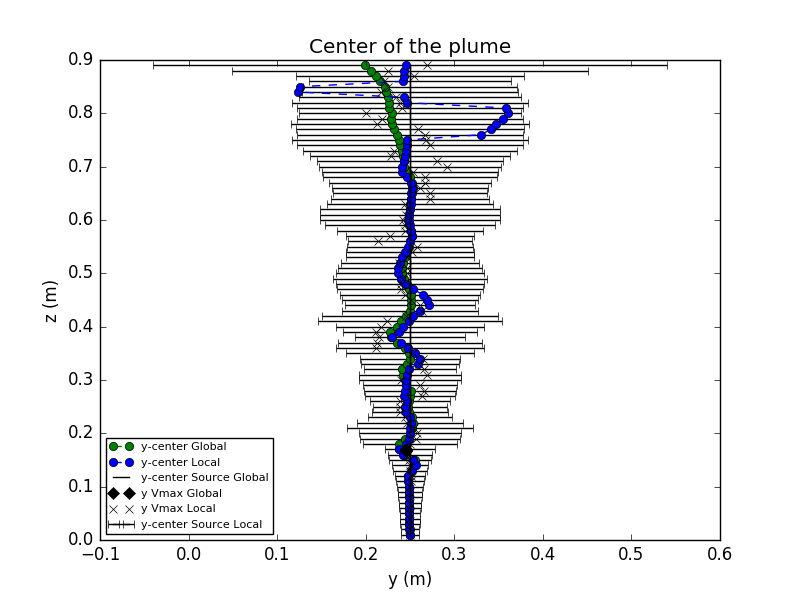}
        \caption{}
        \label{Fig:Plot_y_CenterSource_VzLocal}
    \end{subfigure}
    \caption{(a) $x$ and (b) $y$ location of the plume’s centre. The error bars depict the computed radius at each height using method II.2.a.}
    \label{Fig:Plot_CenterSource_VzLocal}
\end{figure}

\begin{figure}
    \centering
    \begin{subfigure}{0.4\textwidth}
        \includegraphics[width=\textwidth]{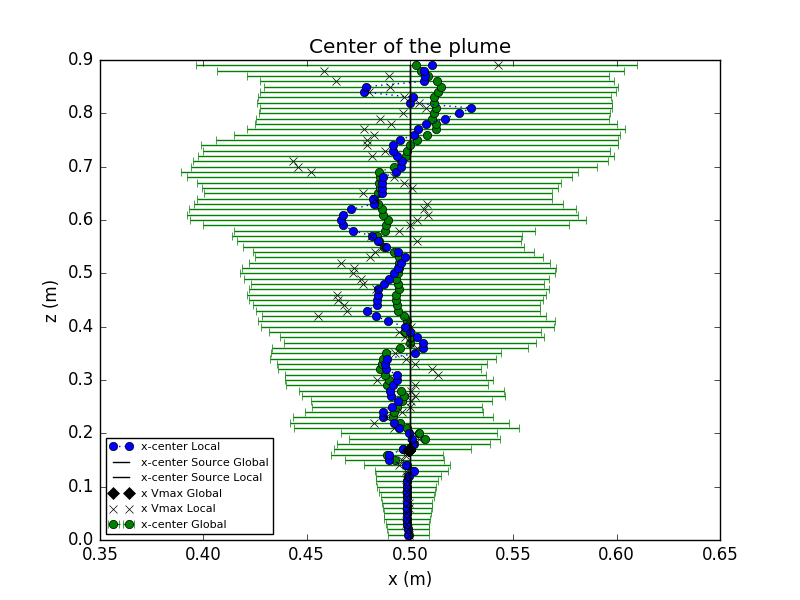}
        \caption{}
        \label{Fig:Plot_x_CenterLocal_VzGlobal}
    \end{subfigure}
    \begin{subfigure}{0.4\textwidth}
        \includegraphics[width=\textwidth]{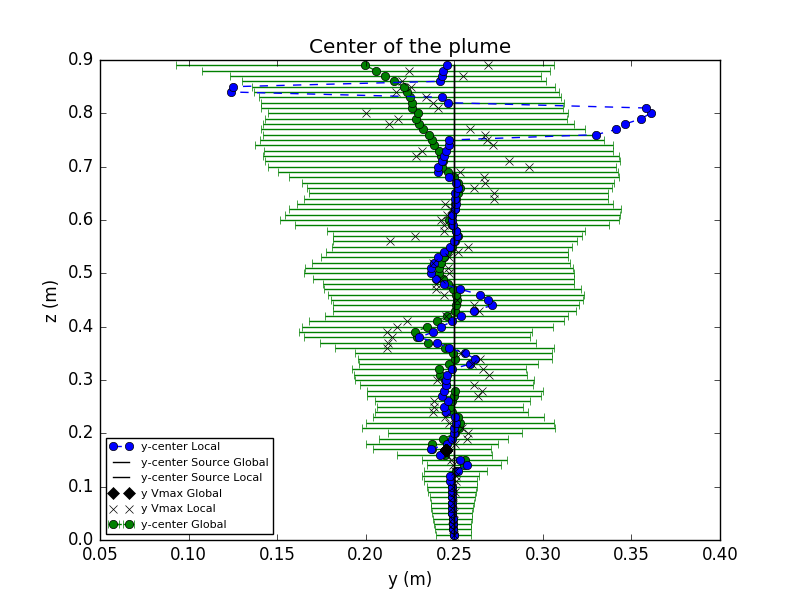}
        \caption{}
        \label{Fig:Plot_y_CenterLocal_VzGlobal}
    \end{subfigure}
    \caption{(a) $x$ and (b) $y$ location of the plume’s centre. The error bars depict the computed radius at each height using method I.2.b.}
    \label{Fig:Plot_CenterLocal_VzGlobal}
\end{figure}

\begin{figure}
    \centering
    \begin{subfigure}{0.4\textwidth}
        \includegraphics[width=\textwidth]{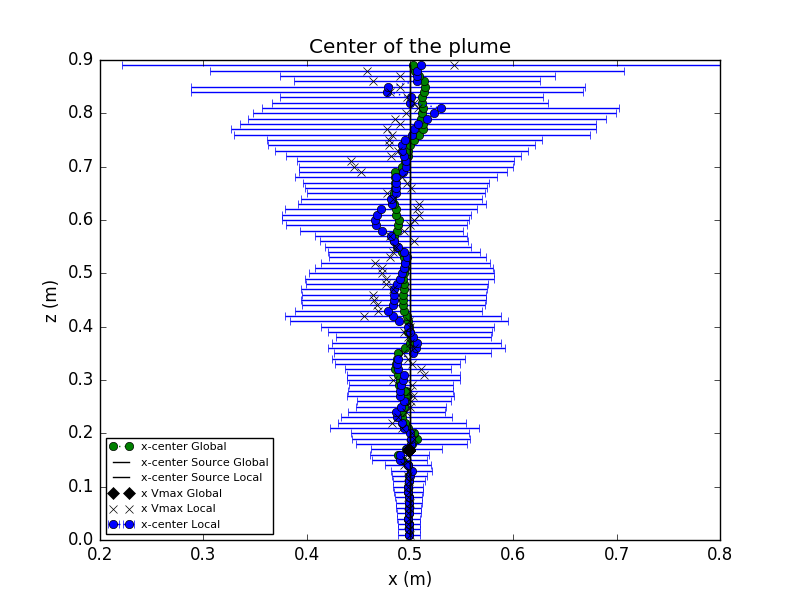}
        \caption{}
        \label{Fig:Plot_x_CenterLocal_VzLocal}
    \end{subfigure}
    \begin{subfigure}{0.4\textwidth}
        \includegraphics[width=\textwidth]{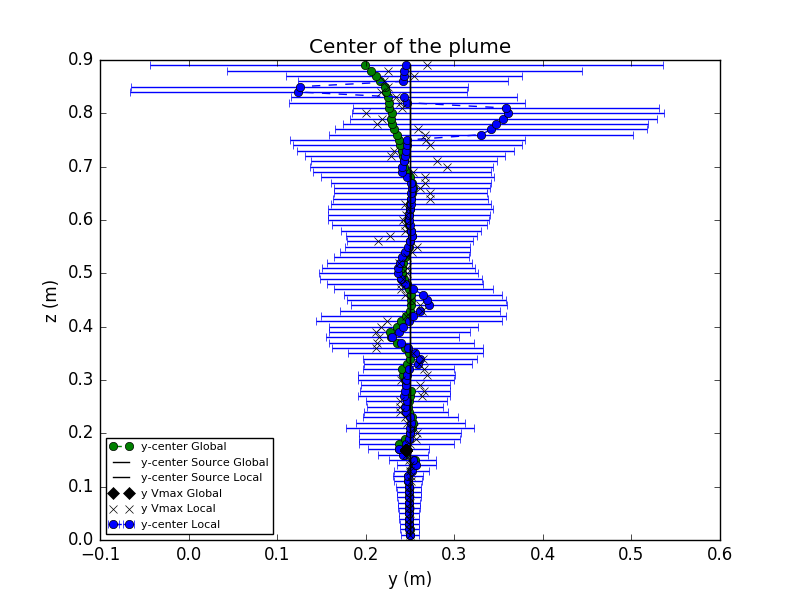}
        \caption{}
        \label{Fig:Plot_y_CenterLocal_VzLocall}
    \end{subfigure}
    \caption{(a) $x$ and (b) $y$ location of the plume’s centre. The error bars depict the computed radius at each height using method II.2.b.}
    \label{Fig:Plot_CenterLocal_VzLocal}
\end{figure}

This section shows examples of plots that can be obtained. Figure~\ref{Fig:PlotsRadius} shows the variation of the plume’s radius as a function of the height for all the methods presented above. Figure~\ref{Fig:PlotVmax} shows the variation of the maximum velocity as a function of the height for methods I and II. Figure~\ref{Fig:PlotCenter} shows the location of the plume centre as a function of the height for methods I and II. The green lines are for method I, while the blue lines are for method II. In Figure~\ref{Fig:PlotCenter}, the black crosses depicts the location of the maximum velocity while using method II. The global maximum velocity is depicted by black diamonds. Figure~\ref{Fig:Plot_CenterSource_VzGlobal} to Figure~\ref{Fig:Plot_CenterLocal_VzLocal} show the $x$ and $y$ locations of the plume centre as a function of the height for the method I.2 and II.2. The error bars depicts the radius computed by the associated method centred on the source location (Figure~\ref{Fig:Plot_CenterSource_VzGlobal} and Figure~\ref{Fig:Plot_CenterSource_VzLocal}) and the local plume’s centre (Figure~\ref{Fig:Plot_CenterLocal_VzGlobal} and Figure~\ref{Fig:Plot_CenterLocal_VzLocal}). 

\subsection{Numerical implementation}
All the methods described above are implemented in a python script and the main libraries used are:
\begin{itemize}
    \item \textbf{vtk} is used to read the data obtained with \textbf{Fluidity}, to obtain the contours (Code~\ref{Lst:VTK_Contour}), the slices (Code~\ref{Lst:VTK_Plane}) and to write data into \textit{vtu} files (Code~\ref{Lst:VTK_Write}).
    \item \textbf{shapely} is used to play around with the geometry. For example, the centroid of the polygons and the detection of polygons within others are done using this library.
    \item \textbf{matplotlib} is used to plot all the data (Figure~\ref{Fig:PlotsRadius} to Figure~\ref{Fig:Plot_CenterLocal_VzLocal}).
\end{itemize}

\noindent Codes~\ref{Lst:Radius_Method1} and~\ref{Lst:Radius_Method2} show the functions used to compute the radius.

\begin{Code}[language=python, firstnumber=27, caption={Function to generate a contour using the \textbf{vtk} library.}, label={Lst:VTK_Contour}]
#------------------------------------------------------------#
#-- Function to generate the contour based on a given value -#
#------------------------------------------------------------#
def Contour(ugrid, value):
    filter = vtk.vtkContourFilter()
    if vtk.vtkVersion.GetVTKMajorVersion() <= 5:
        filter.SetInput(ugrid)
    else:
        filter.SetInputData(ugrid)

    filter.SetNumberOfContours(1)
    filter.SetValue(0, value)
    filter.Update()
    cutContour = filter.GetOutput()

    return cutContour
\end{Code}

\begin{Code}[language=python, firstnumber=81, caption={Function to generate a plane using the \textbf{vtk} library.}, label={Lst:VTK_Plane}]
#---------------------------#
#-- Function to do a slice -#
#---------------------------#
def Plane(origin, normal, ugrid):

    plane = vtk.vtkPlane()
    plane.SetNormal(normal[0], normal[1], normal[2])
    plane.SetOrigin(origin[0], origin[1], origin[2])

    cutter = vtk.vtkCutter()
    cutter.SetCutFunction(plane)
    if vtk.vtkVersion.GetVTKMajorVersion() <= 5:
        cutter.SetInput(ugrid)
    else:
        cutter.SetInputData(ugrid)

    # Cut
    cutter.Update()
    cutPlane = cutter.GetOutput()

    # Write the data in a new grid
    uugrid = vtk.vtkUnstructuredGrid()
    # Add the points
    uugrid.SetPoints(cutPlane.GetPoints())

    return uugrid, cutPlane
\end{Code}

\begin{Code}[language=python, firstnumber=44, caption={Function to write data in a \textit{vtu} file using \textbf{vtk} library.}, label={Lst:VTK_Write}]
#---------------------------------------------#
#-- Function to write the data in a vtu file -#
#---------------------------------------------#
def WriteData(cutContour, filename):

    ugrid = vtk.vtkUnstructuredGrid()

    # Add the points
    ugrid.SetPoints(cutContour.GetPoints())

    # Add the cells
    for i in range(cutContour.GetNumberOfCells()):
        cellType = cutContour.GetCellType(i)
        cell = cutContour.GetCell(i)
        ugrid.InsertNextCell(cellType, cell.GetPointIds())

    # Add the point data
    for i in range(cutContour.GetPointData().GetNumberOfArrays()):
        ugrid.GetPointData().AddArray(cutContour.GetPointData().GetArray(i))

    # Add the cell data
    for i in range(cutContour.GetCellData().GetNumberOfArrays()):
        ugrid.GetCellData().AddArray(cutContour.GetCellData().GetArray(i))

    # Write the contour in a new vtu file
    gridwriter=vtk.vtkXMLUnstructuredGridWriter()
    gridwriter.SetFileName(filename)

    if vtk.vtkVersion.GetVTKMajorVersion() <= 5:
        gridwriter.SetInput(ugrid)
    else:
        gridwriter.SetInputData(ugrid)

    gridwriter.Write()

    return ugrid
\end{Code}

\begin{Code}[language=python, firstnumber=254, caption={Function to get the radius using Method 1.}, label={Lst:Radius_Method1}]
#-------------------------------------------------------------------#
#-- Function to find the mean radius based on the area of polygons -#
#-------------------------------------------------------------------#
#--------------------------------#
# Method 1 Area                  #
#--------------------------------#
def RadiusArea(poly,area_del):
    sum_area = 0.0
    for polyID in range(len(poly)):
        poly1 = poly[polyID]
        area = poly1.area
        sum_area = sum_area + area

    sum_area = sum_area - area_del
    mean_radius = np.sqrt((sum_area/np.pi))

    return mean_radius
\end{Code}

\begin{Code}[language=python, firstnumber=272, caption={Function to get the radius using Method 2.}, label={Lst:Radius_Method2}]
#-----------------------------------------------------------------#
#-- Function to find the mean radius after cleaning the geometry -#
#-----------------------------------------------------------------#
#--------------------------------#
# Method 2 Distance              #
#--------------------------------#
def RadiusPoly(coordinates, x0, y0):
    sum_dist = 0.0
    tot_pts = 0
    for nodeID in range(len(coordinates)-1):
        x, y = coordinates[nodeID][0], coordinates[nodeID][1]
        distance = np.sqrt(np.power(x-x0,2)+np.power(y-y0,2))
        sum_dist = sum_dist + distance
        tot_pts += 1

    mean_radius = sum_dist/tot_pts

    return mean_radius
\end{Code}

\section{\textit{*.stat} file}
\begin{figure}
    \centering
    \begin{subfigure}{0.4\textwidth}
        \includegraphics[width=\textwidth]{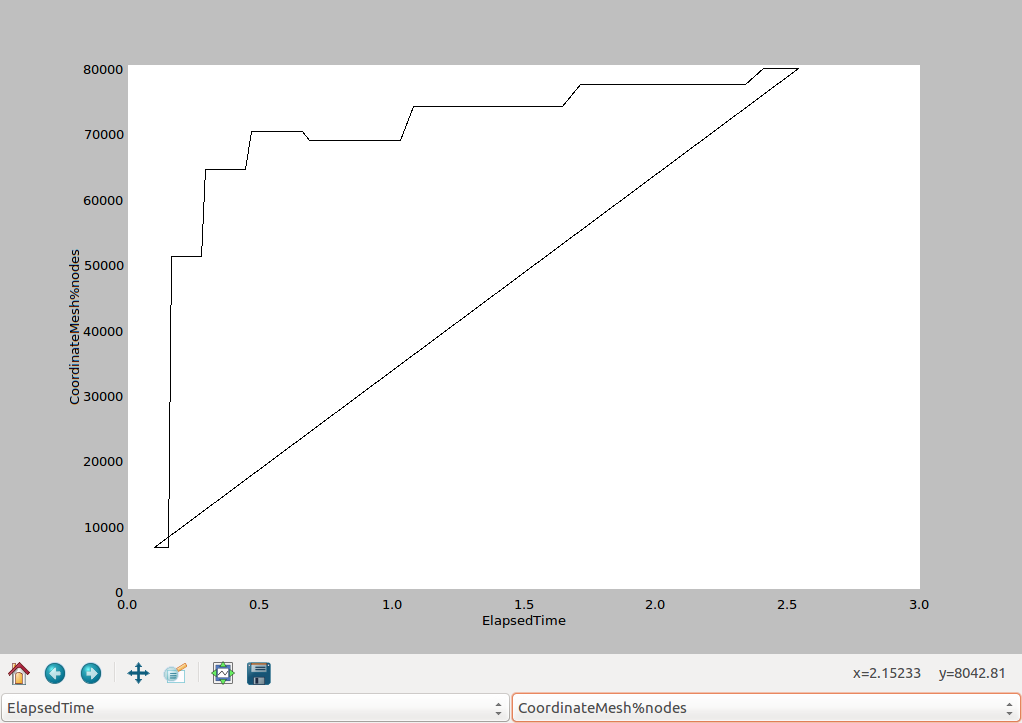}
        \caption{}
        \label{Fig:StatPlot_Nodes}
    \end{subfigure}
    \begin{subfigure}{0.4\textwidth}
        \includegraphics[width=\textwidth]{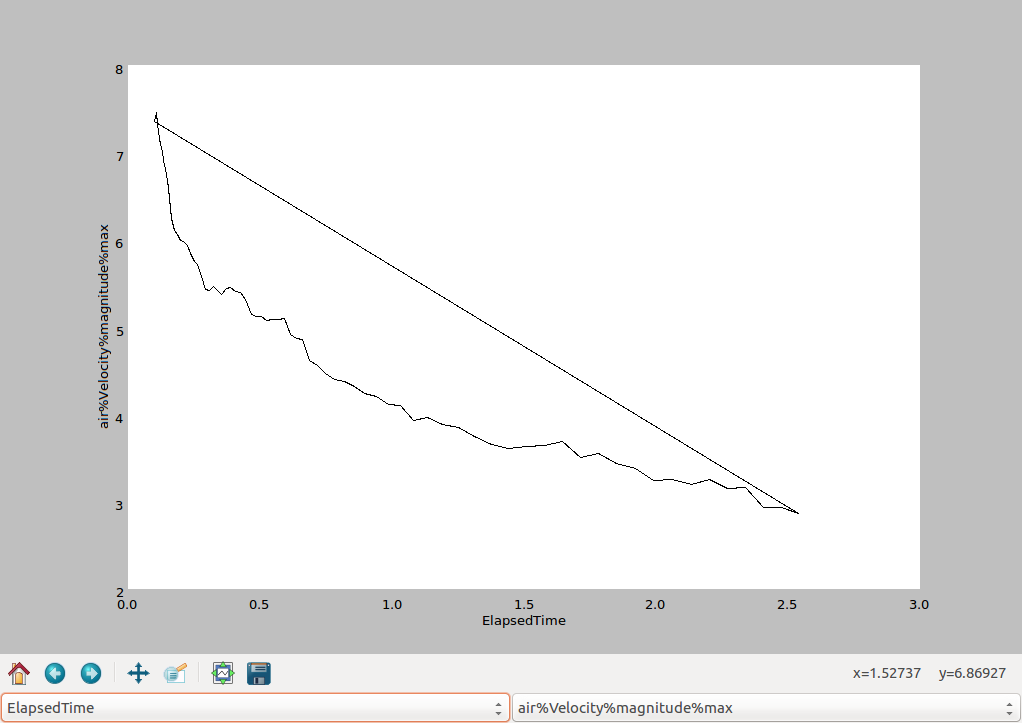}
        \caption{}
        \label{Fig:StatPlot_Velocity}
    \end{subfigure}
    \caption{(a) Number of nodes and (b) maximum velocity magnitude as a function of the time visualised with the tool \textbf{statplot}.}
    \label{Fig:Statplot}
\end{figure}

While a \textbf{Fluidity} simulation is running, a \textit{*.stat} file is generated, which is basically a text file that can be viewed with any text editor. This file records a number of interesting information such as the number of nodes, the maximum and minimum values of each fields... As shown in Figure~\ref{Fig:Statplot}, information in the \textit{*.stat} file can also be viewed directly using the graphical interface \textbf{statplot} launched by typing in a terminal:
\begin{Terminal}[caption={Visualising a \textit{stat} file using \textbf{statplot}.}, label={Lst:Statplot}]
ä\colorbox{davysgrey}{
\parbox{435pt}{
\color{applegreen} \textbf{user@mypc}\color{white}\textbf{:}\color{codeblue}$\sim$
\color{white}\$ statplot simu\_name.stat
}}
\end{Terminal}

\begin{figure}
    \centering
    \begin{subfigure}{0.4\textwidth}
        \includegraphics[width=\textwidth]{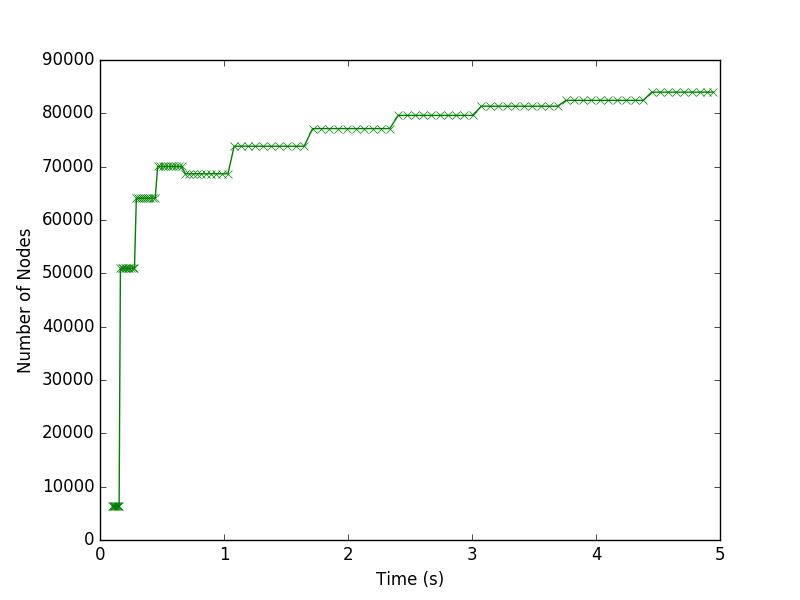}
        \caption{}
        \label{Fig:StatReader_Nodes}
    \end{subfigure}
    \begin{subfigure}{0.4\textwidth}
        \includegraphics[width=\textwidth]{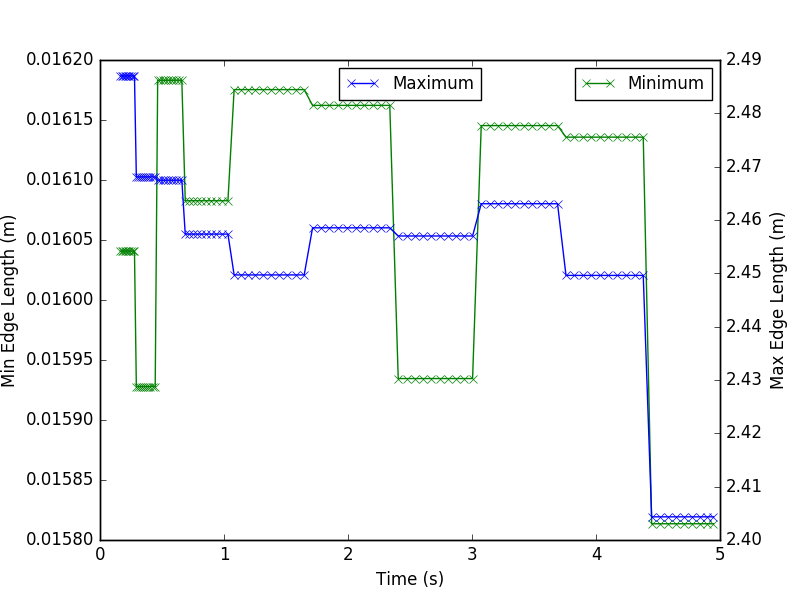}
        \caption{}
        \label{Fig:StatReader_EdgeLength}
    \end{subfigure}
    \begin{subfigure}{0.4\textwidth}
        \includegraphics[width=\textwidth]{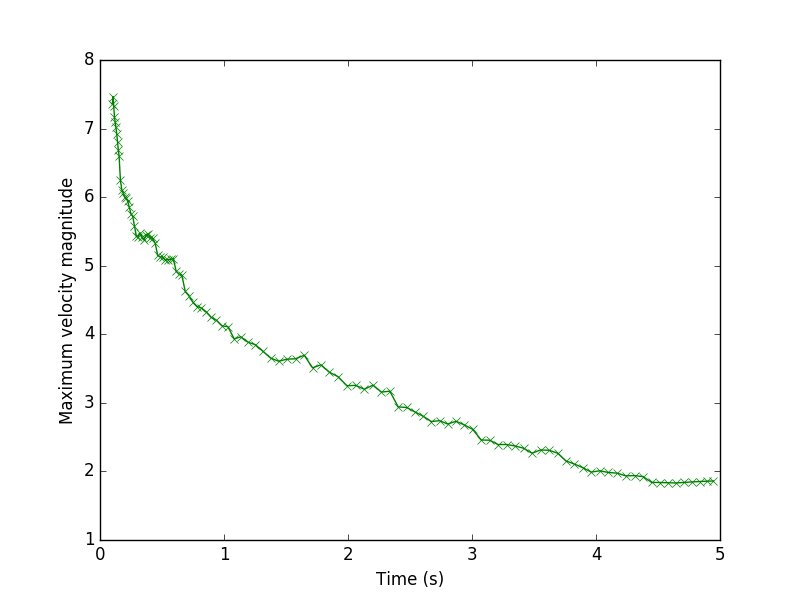}
        \caption{}
        \label{Fig:StatReader_Velocity}
    \end{subfigure}
    \caption{(a) Number of nodes, (b) minimum and maximum edge length and (c) maximum velocity magnitude as a function of the time. Data were extracted from the \textit{*.stat} file using the python script \textit{StatReader.py}.}
    \label{Fig:StatReader}
\end{figure}

\noindent The data can also be read easily using a python script: an example of python script, \textit{StatReader.py}, can be found in the folder \texttt{Scripts/StatReader/}. As shown in Code~\ref{Lst:StatReader}, Line 10, the \texttt{fluidity\_tools} library is required. The \textit{*.stat} file needs to be imported (Line 25 in Code~\ref{Lst:StatReader}), then the data can be extracted (see Lines 32 to 38 in Code~\ref{Lst:StatReader}). This allows the manipulation of the data and some plots are shown in Figure~\ref{Fig:StatReader}

\begin{Code}[language=python, firstnumber=3, caption={Extract data from the \textit{*.stat} file using a python script.}, label={Lst:StatReader}]
from numpy import *
from math  import *

import sys, os
import numpy as np
from termcolor import colored

import fluidity_tools

import matplotlib.pyplot as plt

os.system('rm -r *txt *png')

#--------------------------------#
# User Input variables           #
#--------------------------------#
path_simu = '../../Examples/Case11_T0.1_All/' # Path of the simu
basename  = 'Case11_Box'                   # Name of the simu

#--------------------------------#
# Read stat file                 #
#--------------------------------#
stat = fluidity_tools.stat_parser(path_simu+basename+'.stat')
print colored('The variables in the stat file are ::', 'red')
print stat['air'].keys()

#--------------------------------#
# Extract data from stat file    #
#--------------------------------#
Time     = stat['ElapsedTime']['value']
NbrNodes = stat['CoordinateMesh']['nodes']

EdgeLength_min  = stat['air']['EdgeLength']['min']
EdgeLength_max  = stat['air']['EdgeLength']['max']
Velocity_max    = stat['air']['Velocity%magnitude']['max']
Reynolds_uu_max = stat['air']['uuAverage']['max']
\end{Code}

\section{fltools}
A number of \textbf{Fluidity} tools exist and are detailed in the manual. Here are described the more useful ones. For information, the executable of these tools are located under the directory: \texttt{<<FluiditySourcePath>>/bin/}.

\subsection{rename\_checkpoint}
When running a simulation from a checkpoint, the new simulation will have the name of the simulation following by \texttt{\_checkpoint} and the files' numbering will re-start at 0. Therefore, it can be useful to rename these files afterwards to visualise them as continuous files in \textbf{ParaView} for example. Hence, the tool \texttt{rename\_checkpoint} allows to rename the \textbf{ParaView} files (\textit{*.vtu} or \textit{*.pvtu}) of a simulation that has been checkpointed. It is run typing in a terminal Command~\ref{Lst:RenameCheckpoint}, where \texttt{base\_filename} is the original name of the simulation before it was checkpointed and \texttt{index} is the index at which it was checkpointed. For example, if simulations were run from a file called \texttt{Case11\_Box\_10\_checkpoint.flml} the resulting checkpointed files would have to be renamed with the command \texttt{rename\_checkpoint Case11\_Box 10}.

\begin{Terminal}[caption={Use of \texttt{rename\_checkpoint} to rename checkpointed simulations.}, label={Lst:RenameCheckpoint}]
ä\colorbox{davysgrey}{
\parbox{435pt}{
\color{applegreen} \textbf{user@mypc}\color{white}\textbf{:}\color{codeblue}$\sim$
\color{white}\$ rename\_checkpoint base\_filename index
}}
\end{Terminal}

\subsection{pvtu2vtu}
When running a simulation in parallel, \textit{*.pvtu} files are generated. When visualising these files in \textbf{ParaView} and particularly when using the \texttt{cull frontface} visualisation style, the overlapping region used during parallelisation (due to the domain decomposition) are visible, making difficult to view the results. One easy way to avoid this is to convert the \textit{*.pvtu} files into \textit{*.vtu} files using the tool \texttt{pvtu2vtu} which combines the \textit{*.pvtu} obtained in parallel into \textit{*.vtu} files. This tool is run typing in a terminal Command~\ref{Lst:pvtu2vtu}, where \texttt{base\_filename} is the name of the simulation, \texttt{index} and \texttt{last index} are the first and the last index of the files to convert.

\begin{Terminal}[caption={Use of \texttt{pvtu2vtu} to convert \textit{*pvtu} files into \textit{*.vtu} files.}, label={Lst:pvtu2vtu}]
ä\colorbox{davysgrey}{
\parbox{435pt}{
\color{applegreen} \textbf{user@mypc}\color{white}\textbf{:}\color{codeblue}$\sim$
\color{white}\$ pvtu2vtu base\_filename index [last index]
}}
\end{Terminal}

\subsection{genpvd}
Finally, \texttt{genpvd} can be used to generate a \textit{*.pvd} file from the list of \textit{*.vtu} or \textit{*.pvtu} files. This will notably allow the user to have access to the simulation real time. It can be run using Command~\ref{Lst:genpvd}, where \texttt{base\_filename} is the name of the simulation. It must be noted that checkpointed files will have to be renamed before this can be done.
\begin{Terminal}[caption={Use of \texttt{genpvd} to generate a \textit{*.pvd} file.}, label={Lst:genpvd}]
ä\colorbox{davysgrey}{
\parbox{435pt}{
\color{applegreen} \textbf{user@mypc}\color{white}\textbf{:}\color{codeblue}$\sim$
\color{white}\$ genpvd base\_filename
}}
\end{Terminal}
    \chapter{Others}

\section{Past and present people using \textbf{Fluidity}}
Following are past and present people working with \textbf{Fluidity} to perform outdoor simulations and/or indoor-outdoor exchange simulations. This manual was initiated by \textcolor{red}{Dr Laetitia Mottet}, post-doc in the Earth Science and Engineering department, Imperial College London and in the Architecture department, University of Cambridge; and \textcolor{red}{Carolanne Vouriot}, PhD student in the Civil Engineering department, Imperial College London. It is also important to mention that \textbf{Fluidity} was initially developed by \textcolor{red}{Prof. Christopher Pain} \url{c.pain@imperial.ac.uk}, department of Earth Science and Engineering, Imperial College London.

\begin{table}
    \centering
    \begin{tabular}{|l l l l l|}
    \hline
    \multicolumn{5}{|c|}{Complex outdoor simulations}\\
    \hline
    \hline
    Since 2000 & E. Aristodemou & Senior Lecturer & London South Bank University & \\
    Since 2016 & \textcolor{red}{L. Mottet} & Post-doc & Imperial / Cambridge &  \\ 
    Since 2017 & H. Woordward & Post-doc & Imperial & \\
    \hline
    \end{tabular}
    \caption{People who worked or are currently working with \textbf{Fluidity} on simulations in urban environment.}
\end{table}
\begin{table}
    \centering
    \begin{tabular}{|l l l l l|}
    \hline
    \multicolumn{5}{|c|}{Box simulations of indoor - outdoor exchanges}\\
    \hline
    \hline
    Summer 2017 & Faron Hesse & MRes & ICL & \textit{H. Burridge} \\
    Summer 2017 & Jean-Etienne Debay & Intern & Cam & \textit{M. Davies Wykes} \\
    Since 2018 & \textcolor{red}{Carolanne Vouriot} & PhD Student & ICL & \textit{H. Burridge} \\
    Summer 2018 & Nouhaila Fadhi & Intern & Cam & \textit{M. Davies Wykes} \\
    Summer 2018 & Theresa Bischof & MSc & ICL & \textit{H. Burridge} \\
    Summer 2018 & Sam Charlwood & UROP & Cam & \textit{M. Davies Wykes} \\
    \hline
    \end{tabular}
    \caption{People who worked or are currently working with \textbf{Fluidity} on indoor - outdoor simulations. ICL stands for Imperial College London and Cam for the University of Cambridge, names in italic are the \textit{supervisors}.}
    \label{tab:chronology}
\end{table}

\noindent \textbf{Contacts:}  prioritise people in red if you have any questions or amendment to this manual to suggest.
\begin{itemize}
    \item Elsa Aristodemou - \url{aristode@lsbu.ac.uk}
    \item \textcolor{red}{Laetitia Mottet} - \url{l.mottet@imperial.ac.uk}
    \item Henry Burridge - \url{h.burridge@imperial.ac.uk}
    \item Megan Davies Wykes - \url{msd38@cam.ac.uk}
    \item Huw Woodward - \url{h.woodward@imperial.ac.uk}
    \item \textcolor{red}{Carolanne Vouriot} - \url{carolanne.vouriot12@imperial.ac.uk}
\end{itemize}

\section{Online open-source data set}
Following is a non-exhaustive list of open-source experimental and/or numerical data available for indoor-outdoor exchange and urban environment simulations.
\begin{itemize}
    \item \textbf{Urban environment simulations}
    \begin{itemize}
        \item \url{https://www.windforschung.de/CODASC.htm} 
        \item \url{http://mi-pub.cen.uni-hamburg.de/index.php?id=432}
        \item \url{https://mi-pub.cen.uni-hamburg.de/index.php?id=6339}
        \item \url{https://www.aij.or.jp/jpn/publish/cfdguide/index_e.htm}
        \item \url{http://www.dapple.org.uk/}
    \end{itemize}
    \item \textbf{Indoor-outdoor exchanges}
    \begin{itemize}
        \item Please contact Laetitia Mottet and/or Carolanne Vouriot if you are aware of open-source data available online, either numerical or experimental, on indoor-outdoor exchanges.
    \end{itemize}
\end{itemize}

    \bibliographystyle{unsrt}
    \bibliography{biblio.bib}

\end{document}